\documentclass[11pt,letterpaper,reqno]{amsart}
\usepackage{chngcntr}
\usepackage{algorithm}
\usepackage{algpseudocode}
\usepackage{amsmath}
\usepackage{mathtools}
\usepackage{adjustbox}
\usepackage{listings}
\usepackage{tikz}
\usepackage{xcolor}
\usepackage{extarrows}

\usepackage{pstricks}
\usepackage{pst-poly}

\usepackage{color}

\lstdefinelanguage{Mathematica}{
  morekeywords={
    Module, If, Then, Else, Which, While, For, Do, Table, Return, Block, 
    Eigenvalues, AdjacencyMatrix, Select, Total, Max, Abs, Plot, ListPlot, 
    Import, Export, Solve, Simplify, Expand, Factor, Det, MatrixForm, NullSpace
  },
  sensitive=true,
  morecomment=[l]{(*},
  morestring=[b]",
}

\usepackage{amssymb}
\usepackage{amsmath}
\usepackage{amsthm}
\usepackage{amsfonts}
\usepackage{bbm}
\usepackage{graphicx}
\usepackage[T1]{fontenc}
\usepackage{doi}
\usepackage{bookmark}
\usepackage{hyperref}
\hypersetup{pdfstartview={FitH}}
\usepackage{cleveref}

\newcommand{\exc}{\operatorname{exc}}

\newcommand{\tcr}{\textcolor{red}}

\newcommand{\tcw}{\textcolor{white}}

\newtheorem{thm}{Theorem}[section]
\newtheorem{lem}[thm]{Lemma}
\newtheorem{prop}[thm]{Proposition}
\newtheorem{cor}[thm]{Corollary}
\newtheorem{co}[thm]{Corollary}
\newtheorem{claim}{Claim}
\theoremstyle{definition}
\newtheorem{remark}[thm]{Remark}

\newtheorem{conjecture}[thm]{Conjecture}
\newtheorem{defn}[thm]{Definition}
\newtheorem{exm}[thm]{Example}
\numberwithin{equation}{section}

\makeatother

\newcommand{\sheet}[2]{{\stackrel{{#1}}{{#2}}}}

\newcommand{\col}[1]{{\raise-2pt\hbox{\tiny$\bullet$}\hskip -4.5pt \raise4pt\hbox{\tiny$\bullet$}{{#1}} \raise-2pt\hbox{\tiny$\bullet$}\hskip -4.5pt \raise4pt\hbox{\tiny$\bullet$}}}

\long\def\rem#1{}

\begin{document}

\title{Constructing Casimirs for the quantum reflection equation}

\author{Erik Brodsky}
\address{Cornell University, Ithaca, USA}
\email{eab384@cornell.edu}

\author{Leonid O. Chekhov}
\address{Steklov Mathematical Institute, Moscow, Russia, and
  Michigan State University, East Lansing, USA}
\email{chekhov@msu.edu}


\author{Quanyu Tang}
\address{School of Mathematical Sciences, University of Science and
  Technology of China, Hefei 230026, P.~R.~China}
\email{tangquanyu827@gmail.com}


\begin{abstract}
We explore the quantum reflection equation in the case of a full-dimensional $n\times n$ matrix $\mathbb A$. Using the technique of $q$-determinants, we find closed expressions for the quantum Pfaffian (in the case of even $n$) and for new quantum Casimirs that are ratios of special quantum (Sklyanin) minors of $\mathbb A$. Special attention is paid to the self-adjointness property of the newly constructed Casimirs. We put forward a conjecture that, adopting a natural ansatz for a quantum deformation of a Casimir of a classical algebra, self-adjointness implies commutativity. We also explore the relationship to networks of quantum cluster variables.
\end{abstract}

\maketitle

\tableofcontents

\section{Introduction}

Constructing Casimirs (commuting elements) of the quantum reflection equation has a long history. Started by Molev in \cite{Molev1995},  Molev, Ragoucy, and Sorba \cite{MRS2003} constructed them for $O_N$ and $Sp_{2N}$ algebras using Sklyanin determinants; for a comprehensive accounting of these earlier results, see \cite{MolevBook}. However, explicit formulas for both quantum Pfaffians $\widehat{Pf}$ and quantum Markov elements $\widehat{C}_j$ were missing even in the upper-triangular case ($O_N$ case in MRS's terminology). More recently, Jing and Zhang in 2014 \cite{Jing2014} constructed quantum Pfaffians for effectively the upper-triangular case. These results were extended recently \cite{JingZhangSklyanin} to the construction of $q$-determinants for $q$-symmetric and $q$-skew-symmetric matrices $\mathcal A$ subject to the reflection equation (the authors' nomenclature in these papers is somewhat different from that in \cite{MRS2003}: they called the symmetric matrix case $O_N$ and the skew-symmetric case $Sp_{2N}$). In the same paper \cite{JingZhangSklyanin}, Jing and Zhang derived formulas for quantum Sklyanin minors (see also Sec.2.5 in \cite{MolevBook}), which we use in this text when addressing new central elements appearing in the case of a full-size (nonrestricted) matrix $\mathbb A$ in Sec.~\ref{ss:minors}.

In this text, we explore Casimir functions of algebras governed by the quantum reflection equation (\ref{RARA}) for a general $n\times n$ matrix $\mathbb A$. Previous works on this subject were mostly concentrated either on the case of (unipotent) upper-triangular matrices $A$ (this case has deep connections with hyperbolic geometry and monodromies of Fuchsian differential equations \cite{DubrovinMazzocco2000}) with some generalization to $2\times 2$-block upper triangular matrices \cite{MRS2003} or on the case of $q$-symmetric or $q$-skew-symmetric matrices $\mathcal A$ \cite{Jing2014,JingZhang2017,JingZhangSklyanin}. For the nonrestricted (or full-size) quantum matrix $\mathbb A$, we have new Casimir functions (constructed in \cite{ChekhovMazzocco2013} in the semiclassical limit); in the present paper, we provide closed formulas for their quantum counterparts (Sec.~\ref{ss:minors}) as well as explicit formulas for quantum Pfaffians for a general $(2n)\times (2n)$ matrix $\mathbb A$ (Theorem~\ref{thm-Pf}). We also show that all constructed Casimirs are self-adjoint quantum operators and put forward Conjecture~\ref{con:comm} that under some, widely adopted and natural, quantum deformation conditions, for ``smallest'' Casimirs, i.e., for the Pfaffian $\widehat{Pf}$ and the first Markov element $\widehat{C}_1$, self-adjointness implies commutativity.

\subsection{Algebras of unipotent upper-triangular matrices and quantization}
Before discussing the quantization procedure, we first recall the semiclassical setting of unipotent upper-triangular matrices and their associated Casimir functions.

Let \( \mathcal{A}_n \) denote the space of unipotent upper-triangular \( n \times n \) matrices. That is,
\[
  \mathcal{A}_n := \left\{ A = [a_{i,j}]_{1 \le i,j \le n} \in \mathrm{Mat}_n(\mathbb{C}) \,\middle|\, A \text{ is upper triangular and } a_{i,i} = 1 \text{ for all } i \right\}.
\]
Each matrix in \( \mathcal{A}_n \) has the form:
\[
  \begin{bmatrix}
    1 & a_{1,2} & a_{1,3} & \cdots & a_{1,n} \\
    0 & 1 & a_{2,3} & \cdots & a_{2,n} \\
    \vdots & \ddots & \ddots & \ddots & \vdots \\
    0 & \cdots & 0 & 1 & a_{n-1,n} \\
    0 & \cdots & \cdots & 0 & 1
  \end{bmatrix}
\]

The Poisson brackets between $a_{i,j}$ can be presented in the standard $r$-matrix form realized in the direct product of two matrix vector spaces:
$$
\bigl\{\stackrel{1}{A}\underset{,}{\otimes}\stackrel{2}{A}\bigr\}=r_{12} \stackrel{1}{A}\otimes\stackrel{2}{A} + \stackrel{1}{A}r_{12}^{t_1}\stackrel{2}{A} -\stackrel{2}{A}r_{12}^{t_1}\stackrel{1}{A} - \stackrel{2}{A}\otimes\stackrel{1}{A}r_{12}
$$
with the classical $r$-matrix
$$
r_{12}=\sum_{i\ge j}\stackrel{1}e_{i,j}\otimes \stackrel{2}e_{j,i}\theta(i-j),\quad \theta(x)=\{0,\ x<0; 1/2,\ x=0;1,\ x>0\}.
$$

All Casimirs (i.e., algebraic combinations of $a_{i,j}$ that are central with respect to the above Poisson bracket) have been found by Bondal~\cite{Bondal2004} (see also Chekhov and Shapiro~\cite[Lemma 5.1]{Chekhov2020}).
\begin{prop}[\cite{Bondal2004}] In the semiclassical case, all commuting elements $C_i$ are generated by 
  \label{prop:ChekhovShapiro}
  \[
    \det({A} + \lambda {A}^T) = 1+\sum_{i=1}^{\lfloor n/2\rfloor} (\lambda^i+\lambda^{n-i})C_i +\lambda^n.
  \]
\end{prop}
\begin{exm}\label{egclassical3}
  Let \( n = 3 \). Consider the characteristic polynomial
  \begin{equation}\label{eq:char}
    p(\lambda) = \det({A} + \lambda {A}^T),
  \end{equation}

  where \( {A} \in \mathcal{A}_3 \) is a unipotent upper-triangular matrix:
  \[
    {A} =
    \begin{bmatrix}
      1 & a_{1,2} & a_{1,3} \\
      0 & 1 & a_{2,3} \\
      0 & 0 & 1
    \end{bmatrix}.
  \]
  Then we compute the first 2 coefficients of \(p(\lambda)\) as:
  \[
    \begin{array}{c|l}
      \lambda^k & \text{Coefficient} \\ \hline
      \lambda^3 & 1 \\
      \lambda^2 & 3 - a_{1,2}^2 - a_{1,3}^2 - a_{2,3}^2 + a_{1,2} a_{1,3} a_{2,3} 
    \end{array}
  \]This coefficient of \( \lambda^2 \), viewed as a polynomial in the entries of \( {A} \), represents the unique (up to a scalar) classical Casimir function.
\end{exm}

Next, to illustrate the notion of quantization for classical Casimir functions, we still consider the case \( n = 3 \).
\begin{exm}\label{egclassical31}
  Recall from Example~\ref{egclassical3} that the unique classical Casimir function (up to a scalar multiple) is given by
  \[
    C_1 := a_{1,2} a_{1,3} a_{2,3} -  a_{1,2}^2 -  a_{1,3}^2 -  a_{2,3}^2.
  \] 
  We seek a quantization of this expression, denoted \( \widehat{C}_1 \), in the quantum algebra considered in the skein-theoretic and quantum-Teichm\"uller literature; see, for instance, Bullock--Przytycki~\cite{BullockPrzytycki2000} and Chekhov--Fock~\cite{ChekhovFock1999Steklov}. Motivated by the Kontsevich deformation quantization construction \cite{Kont2003}, we conjecture the following form:
  \[
    \widehat{C}_1 = q^{\beta_1} a_{1,2} a_{1,3} a_{2,3} - q^{\beta_2} a_{1,2}^2 - q^{\beta_3} a_{1,3}^2 - q^{\beta_4} a_{2,3}^2,
  \]
  where \( \beta_1, \beta_2, \beta_3, \beta_4 \in \mathbb{R} \) are parameters to be determined. Instead of verifying zero commutation relations with all $a_{i,j}$, we can require the self-adjointness of $\widehat{C}_1$. Indeed, (i) this requirement fixes all $\beta_j$ in a unique way; (ii) the Casimir must be self-adjoint, otherwise $\widehat{C}_1-\widehat{C}^*_1$ is a new Casimir, which is independent and ``simpler'' than $\widehat{C}_1$, so, in the semiclassical limit, we obtain two Casimirs rather than one, which is a contradiction. 

  Since $\widehat{C}_1$ is a self-adjoint Casimir function, it should satisfy the following system of equations for the \( \beta_k \):\[
    \widehat{C}_1^* = \widehat{C}_1 \qquad\text{and}\qquad \widehat{C}_1a_{ij} = a_{ij}\widehat{C}_1 \quad \text{for } 1\leq i<j \leq 3. 
  \] Solving this system yields
  \[
    \beta_1 = -\frac{1}{2}, \quad \beta_2 = -1, \quad \beta_3 = 1, \quad \beta_4 = -1.
  \] Therefore, the self-adjoint quantum Casimir function for \( n = 3 \) is
  \begin{equation}\label{eq:quantum_casimir_for_n_3_1}
    \widehat{C}_1 = q^{-\frac{1}{2}} a_{1,2} a_{1,3} a_{2,3} - q^{-1} a_{1,2}^2 - q a_{1,3}^2 - q^{-1} a_{2,3}^2.
  \end{equation}
\end{exm}

\begin{remark}
  Note that the same $\widehat{C}_1$ can be presented in many equivalent ways; due to the cyclic symmetry of our relations (see section \ref{s:triangular}), we can, e.g., write
  $$
  \widehat{C}_1 = q^{-\frac{1}{2}} a_{1,3} a_{2,3} a_{1,2} - q^{-1} a_{1,3}^2 - q a_{2,3}^2 - q^{-1} a_{1,2}^2.
  $$
  But fixing the lexicographic ordering in the first, longest term determines the power of $q$ in all the remaining terms.
\end{remark}

\begin{remark}
  In the quantum setting, we impose the condition that the Casimirs are self-adjoint. For the lowest-order Casimirs, the Pfaffian and the first Markov element, this condition can be met simply by scaling the commuting Casimir by some power of $q$. We did however find that for the second Markov element $\widehat{C}_2$, a self-adjoint quantization can only be attained after adding a term $q\widehat{C}_1$ (see Sec.~\ref{subsecn=5}).
\end{remark}

Algebras of unipotent upper-triangular matrices, also known as symplectic groupoid algebras~\cite{Mackenzie2005}, are omnipresent in mathematical physics. Their particular Poisson leaves (and corresponding quantum counterparts) describe particular sets of geodesic functions (correspondingly, quantum geodesic functions) on Riemann surfaces of higher genera with one or two holes; very recently, the same algebras appeared in the description of Teichm\"uller spaces of closed Riemann surfaces~\cite{ChekhovShapiro2023}. At the same time, particular finite orbits of braid-group action on Markov triples were associated by Dubrovin and Mazzocco~\cite{DubrovinMazzocco2000} with algebraic solutions of the Painlev\'e VI equation, just to name few of their appearances. Quantum versions of these algebras were known to be described by the so-called \emph{reflection equation}~\cite{MolevBook}, which can be presented concisely in the $R$-matrix form\footnote{In the literature, there is another version of the quadratic $R$-matrix equation that appears, for example, in the Reshetikhin--Turaev construction (see, e.g. \cite{JordanWhite2020}) and is also called the reflection equation. Semiclassical Casimirs of this algebra are $\mathop{tr} A^k$ and are completely different from those of (\ref{RARA}).}
\begin{equation}\label{RARA}
  R_{12}\stackrel{1}{A}R_{12}^{t_1}\stackrel{2}A=\stackrel2A R_{12}^{t_1}\stackrel1A R_{12},
\end{equation}
with the so-called trigonometric quantum $R$-matrix
\begin{equation}\label{eq:R12}
  R_{12}(q):=\stackrel1I\otimes \stackrel 2I+(q-1)\sum_i\stackrel1e_{ii}\otimes\stackrel2e_{ii}+(q-q^{-1})\sum_{i>j}\stackrel1e_{ij}\otimes\stackrel2e_{ji}
\end{equation}
and its partial transpose in the first space
\begin{equation}\label{eq:Rt12}
  R^{t_1}_{12}(q):=\stackrel1I\otimes \stackrel 2I+(q-1)\sum_i\stackrel1e_{ii}\otimes\stackrel2e_{ii}+(q-q^{-1})\sum_{i>j}\stackrel1e_{ji}\otimes\stackrel2e_{ji}.
\end{equation}

Note that the inverses to the corresponding matrices are
$$
R^{-1}_{12}(q)=R_{12}(q^{-1});\quad\bigl[ R^{t_1}_{12}(q)\bigr]^{-1}=\stackrel1Q\otimes\stackrel2Q R^{t_1}_{12}(q^{-1}) \stackrel1Q{}^{-1}\otimes\stackrel2Q{}^{-1},\quad\hbox{where}\quad Q=\sum_i q^{i-n/2}e_{ii}.
$$

We identify $A$ with the coefficient matrix of a 2-form; upon the change of basis $|v\rangle\to F|v\rangle$, the corresponding change of forms is $A\to F^T A F$.

The objects in Bondal's symplectic groupoid are pairs of matrices $(B,A)$ where $A$ are unipotent upper-triangular matrices and $B\in SL_n$ are such that combinations $B^TAB$ are themselves unipotent upper-triangular. A cluster algebra (network) description of such pairs was obtained only very recently~\cite{ChekhovShapiro2023} in the semiclassical setting; the corresponding quantum version is still in the making. 

In Bondal's groupoid setting, two natural projections,
$$
p_1:(B,A)\to A\quad \hbox{and}\quad p_2: (B,A)\to B^TAB
$$
are respectively Poisson and anti-Poisson, so we may expect that every ``proper'' transformation
$$
A\to B^TAB
$$
preserving the unipotent upper-triangular structure is a Poisson morphism of the reflection equation algebra and should have its quantum counterpart. This is indeed the case for special morphisms~\cite{Bondal2004} with the matrix $B_{i,i+1}$ having block structure: it differs from the unit matrix only in the $2\times2$ diagonal block $\begin{pmatrix} a_{i,i+1}&1\\-1&0\end{pmatrix}$. The corresponding transformations
$$
A\to B^T_{i,i+1}AB_{i,i+1}
$$
are admissible morphisms preserving the upper-triangular form of $A$, satisfy the braid-group relations, are morphisms of the Poisson structure and, for geometric leaves, correspond to Dehn twists along geodesics corresponding to matrix entries $a_{i,i+1}$. 

The quantum counterpart of the above morphisms was constructed by Chekhov and Mazzocco~\cite{Ch-Maz-Advances}. In the quantum case, entries $a_{i,j}$ of $A$ become self-adjoint operators, whereas units of the diagonal should be replaced by $q^{1/2}$, where $q=e^{-i\hbar}$ is the quantization parameter. In this setting, the quantum algebra takes the form (\ref{RARA}), and the quantum braid-group transformation becomes
$$
A\to B^{\dagger}_{i,i+1}AB_{i,i+1},
$$
where $B_{i,i+1}$ is again the matrix that differs from the unit matrix only by the $2\times 2$ block $\begin{pmatrix} q^{1/2}a_{i,i+1}&1\\-q&0\end{pmatrix}$ inserted on the diagonal. $B^{\dagger}_{i,i+1}$ is the Hermitian conjugate of  $B_{i,i+1}$ with $a_{i,i+1}^*=a_{i,i+1}$, $q^*=q^{-1}$. All quantum Casimir operators are invariant under quantum braid-group transformations. 

This clearly indicates the importance of studying algebras and central elements of entries of the matrix $A$. We begin with the triangular case restricting consideration to two ``lowest'', or ``simplest'' central elements of these algebras for a general $n$: the Pfaffian $\widehat{Pf}$ (which exists for even $n$) and the first Markov element $\widehat{C}_1$; we present the closed formulas for both these elements for a general $n$ (Theorems \ref{t1} and \ref{thm-M1}). These expressions were obtained by Jing and Zhang in \cite{Jing2014} and \cite{JingZhang2017}, but we think it is still useful to show a purely combinatorial proof of commutativity for these two elements.


Note that, although we have other central elements in the case of the general algebra, $C_1$ and $Pf$ deplete the list of Casimirs that are algebraically independent in the geometric case, as all other elements are algebraic functions of these two. The case of odd $n$ corresponds to the Teichm\"uller space of Riemann surfaces with one hole, so the cycle corresponding to the hole perimeter is the only cycle invariant under the Dehn twists; after a constant shift, $C_1=e^{P/2}+e^{-P/2}:=G_P$---the geodesic function related to this cycle. In the case of even $n$, we have two holes with corresponding perimeters $P_1$ and $P_2$ and, in addition to $C_1=G_{P_1}G_{P_2}+(n-4)$, we also have the Pfaffian $Pf=2^{(n-4)/2}(G_{P_1}+G_{P_2})$, whereas all other Casimirs, which are coefficients of the characteristic polynomial, are expressed as polynomials of $C_1$ and $Pf$ (e.g., $C_2=\frac{Pf^2}{2^{n-4}}+(n-6)C_1-\frac{n^2-13n+40}{2}$ for arbitrary even $n\ge 4$).

An important observation is that $C_1$ and $Pf$ (if the latter exists) are shortest (in terms of the number of constituting monomials) among all Casimirs, all of which are polynomial expressions in matrix elements $a_{i,j}$. This observation is pivotal for the claim in Theorems~\ref{t1} and~\ref{thm-M1} that both $\widehat{Pf}$ and $\widehat{C}_1$ are self-adjoint quantum operators.

\subsection{Full-dimensional quantum algebras and Poisson reductions}
It was shown in~\cite{ChekhovMazzocco2014} that we can extend the symplectic structure of the symplectic groupoid of upper-triangular matrices to the set of pairs $(B,\mathbb A)$ with \emph{no restriction} on the form of the $n\times n$ matrix $\mathbb A$. Then the Poisson and quantum algebras of entries of the matrix $\mathbb A$ are changed dramatically, but can still be written  in $R$-matrix terms in exactly the same form (\ref{RARA}) as in the triangular case. The list of central elements in the semiclassical limit is then given by the following lemma.

\begin{lem} [\cite{ChekhovMazzocco2014}] The complete list of central elements for a full-size $n\times n$ matrix $\mathbb A$ comprises exactly $n$ linearly and algebraically independent elements: the determinant $\det \mathbb A$, $\lfloor\frac{n}{2}\rfloor$ coefficients of the $\lambda$-expansion of $\det(\mathbb A+\lambda \mathbb A^{\text{T}})$, and $\lfloor \frac{n-1}{2}\rfloor$ new Casimirs $K_i=\mathfrak M_{n-i}/\mathfrak M_{i}$, $1 \le i \le \lfloor \frac{n-1}{2} \rfloor$, where we let $\mathfrak M_k$ denote $k\times k$ minors of $\mathbb A$ located in the bottom-left corner of the matrix (all these minors vanish in the upper-triangular case). 
\end{lem}

\begin{exm}
  When $n=3$, in the semiclassical case for $\mathbb A$ we have three Casimirs: the determinant
  $$
  \det \mathbb A = a_{11}a_{22}a_{33}-a_{11}a_{23}a_{32}-a_{22}a_{13}a_{31}-a_{33}a_{12}a_{21}+a_{12}a_{23}a_{31}+a_{21}a_{32}a_{13},
  $$
  the first Markov element $C_1$ for which we have
  \begin{align*}
    \det \mathbb A +C_1=&(a_{12}+a_{21})(a_{13}+a_{31})(a_{23}+a_{32})\\ &-a_{11}(a_{23}+a_{32})^2 - a_{22}(a_{13}+a_{31})^2-a_{33}(a_{12}+a_{21})^2+4a_{11}a_{22}a_{33},
  \end{align*}
  and we have one new Casimir given by $K_1 = \mathfrak M_2/\mathfrak M_1$. It takes the form
  \begin{align*}
    K_1 &= a_{31}^{-1}(a_{31}a_{22}-a_{21}a_{32}) = a_{22} - a_{21}a_{31}^{-1}a_{32}.
  \end{align*}
\end{exm}

\begin{figure}[h]
  \centering
  \psscalebox{0.9}{
    \begin{pspicture}(-1.8,-1.8)(1.8,1.8){\psset{unit=0.7}
        \psframe[linecolor=yellow, fillstyle=solid, fillcolor=yellow](-2.5,1)(2.5,2.5)
        \psframe[linecolor=yellow, fillstyle=solid, fillcolor=yellow](-1.5,0)(2.5,1)
        \psframe[linecolor=yellow, fillstyle=solid, fillcolor=yellow](0.5,-1.5)(2.5,0)
        \psframe[linecolor=yellow, fillstyle=solid, fillcolor=yellow](2,-2.5)(2.5,-1.5)
        \psframe[linecolor=blue, fillstyle=solid, fillcolor=blue](-1.7,0.8)(-1.5,1)
        \psframe[linecolor=blue, fillstyle=solid, fillcolor=blue](0.3,-0.2)(0.5,0)
        \psframe[linecolor=blue, fillstyle=solid, fillcolor=blue](1.8,-1.7)(2,-1.5)
        \psframe[linewidth=1pt](-2.5,-2.5)(2.5,2.5)
        \pcline[linecolor=blue, linestyle=dashed, linewidth=1.5pt](-2.5,1)(-1.5,1)
        \pcline[linecolor=blue, linestyle=dashed, linewidth=1.5pt](-1.5,1)(-1.5,0)
        \pcline[linecolor=blue, linestyle=dashed, linewidth=1.5pt](-1.5,0)(.5,0)
        \pcline[linecolor=blue, linestyle=dashed, linewidth=1.5pt](.5,0)(.5,-1.5)
        \pcline[linecolor=blue, linestyle=dashed, linewidth=1.5pt](.5,-1.5)(2,-1.5)
        \pcline[linecolor=blue, linestyle=dashed, linewidth=1.5pt](2,-1.5)(2,-2.5)
        %
        \psellipse[linewidth=1.5pt](-1,-1.3)(0.3,0.5)
      }
    \end{pspicture}
  }
  \caption{\small
    The Young-diagram-type reduction of $\mathbb A$: setting $a_{i,j}=0$ for all elements below the broken line is a Poisson reduction. Small dark squares indicate pivot elements of the Young diagram.
  }
  \label{fi:Young}
\end{figure}
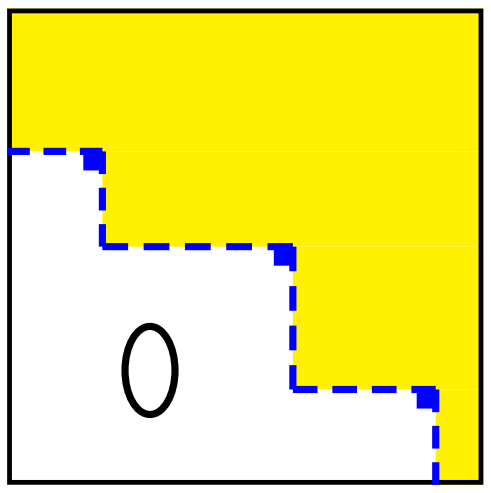

Let us mention one, presumably important, direction of research related to full-dimensional algebras (of matrix elements of unrestricted $\mathbb A$) and their central elements. These algebras admit various coisotropic reductions, some of which are discussed in Sec.~\ref{ss:reductions}, and one of the most promising is where we ``clip out'' the Young diagram based in the lower-left corner of $\mathbb A$: we set $a_{i,j}=0$ for all $(i,j)$ situated below some broken line, i.e., we require that the condition $a_{i,j}=0$ implies that $a_{k,l}=0$ for all $k\ge i$ and $l\le j$ (the corresponding Young diagram is then determined by a set of pivot elements, see Fig.~\ref{fi:Young}). For all such coisotropic reductions, Casimirs of the full algebra remain commuting elements in the new algebra (which may also acquire new Casimirs due to reduction).

We obtain an interesting class of such Poisson reductions by choosing the pivot elements $a_{i_s,i_s+1}=0$ for some subset $\{s_k\}_{k=1}^{N-1}\subseteq \{1,2,\dots,n-1\}$. If these two sets coincide ($N=n$), we obtain an upper-triangular matrix $A$; in general we obtain a block-upper-triangular matrix with $|\{s_k\}|+1=N$ square diagonal blocks $A_{I,I}$ of sizes $(s_I-s_{I-1})\times (s_I-s_{I-1})$ (where $s_0=0$, $s_N=n$) and $N-1$, generally rectangular, upper-diagonal blocks $A_{I,I+1}$. Importantly, this class of block-upper-triangular matrices admits their own braid group action $A\to B_{I,I+1}^T A B_{I,I+1}$ with matrices $B_{I,I+1}$ that are unit matrices for all diagonal blocks except the $(I,I+1)\times (I,I+1)$ matrix block of the form
$$
\begin{pmatrix} A^T_{I,I+1}A^{-T}_{I,I}&-{\mathbb I}\\A_{I,I}A^{-T}_{I,I}&\mathbb  O\end{pmatrix}
$$
The matrix $B_{I,I+1}^TAB_{I,I+1}$ then again has a block-upper-triangular form, and this transformation is a Poisson algebra morphism~\cite{ChekhovMazzocco2013}. Note that $\det A_{I,I}$ is one of the new Casimirs appearing due to reduction, and algebras of elements of inverse matrices $A^{-T}_{I,I}$ are therefore well defined.

No quantum version of the above transformations is known besides the case of blocks of size $2\times 2$, which corresponds to symplectic algebras $Sp_{2N}$ studied by Molev, Ragoucy, and Sorba~\cite{MRS2003}, so our results related to quantum Pfaffians can be relevant for quantization of these structures.


\subsection{Main results and structure of the paper}

In Sec.~\ref{s:clusters} we review the construction of $a_{i,j}$ for a full-size matrix $\mathbb A$ in terms of cluster variables of a special network constructed out of two Fock--Goncharov--Shen triangular networks \cite{FG,GS}. We show that in the semiclassical case, all central elements are monomials in (generally fractional) powers of log-canonical cluster variables constituting the network. We expect that quantum Casimirs obtained in Sec.~\ref{ss:minors} should enjoy the same property when expressed in terms of quantum cluster variables. We also show how the conjugation conditions for the matrix elements $a_{i,j}$ of the quantum matrix $\mathbb A$ naturally arise from the transport-matrix realization of $\mathbb A$.

In Sec.~\ref{s:triangular}, we give closed formulas for the first two Casimirs in the triangular case for a general $n$ (Theorems \ref{t1} and \ref{thm-M1}):
\begin{align*}
  \widehat{Pf}_\triangle &= q^{-\frac{n^2 - n}{2}} \cdot \sum_{\pi \in \Pi} (-q)^{l(\pi)} a_{i_1 j_1} a_{i_2 j_2} \cdots a_{i_n j_n} \\
  \widehat{C_1}(n) &= \sum_{k=3}^n\sum_{i_1<i_2<\cdots<i_k} (-1)^{n-k}q^{2(i_k-i_1)-n-k/2} a_{i_1i_2}a_{i_1i_k}a_{i_2i_3}\cdots a_{i_{k-1}i_k}\\
                         &\qquad +\sum_{i_1<i_2} (-1)^{n-2}q^{2(i_2-i_1)-n}a_{i_1i_2}^2
\end{align*}
We prove that these two elements are self-adjoint.

Section~\ref{sec:full_algebra} contains novel results: we present the closed expression for the quantum Pfaffian for a general (full-size) matrix $\mathbb A$ (Theorem \ref{thm-Pf}) and show that, like in the upper-triangular case \cite{Jing2014}, it can be expressed in four different ways pertaining to different (lexicographic) orderings of the same expression.
\begin{align*}
  \widehat{Pf}_\square=q^{-n(n-1)}&\sum_{\pi\in \wp} (-q)^{l(\pi)} a_{\pi_1,\pi_2}\cdots a_{\pi_{2n-1},\pi_{2n}}\\
  \widehat{Pf}_\square=q^{-n(n-1)}&\sum_{\pi\in \Pi'} (-q)^{l(\pi)} b_{\pi_1,\pi_2}\cdots b_{\pi_{2n-1},\pi_{2n}}\\
  \widehat{Pf}_\square=q^{-n(n-1)}&\sum_{\pi\in \Pi} F_n(q) (-q)^{l(\pi)} \bigl( a_{\pi_1,\pi_2}-q^{2n-1}a_{\pi_2,\pi_1}\bigr)\bigl( a_{\pi_3,\pi_4}-q^{2n-3}a_{\pi_4,\pi_3}\bigr)\cdots\times\\
                          &\times\cdots \bigl( a_{\pi_{2n-3},\pi_{2n-2}}-q^3 a_{\pi_{2n-2},\pi_{2n-3}}\bigr) \bigl( a_{\pi_{2n-1},\pi_{2n}}-q a_{\pi_{2n},\pi_{2n-1}}\bigr),\\
  \widehat{Pf}_\square=q^{-n(n-1)}&\sum_{\pi\in \widehat{\Pi}} F_n(q) (-q)^{l(\pi)} \bigl( a_{\pi_1,\pi_2}-q a_{\pi_2,\pi_1}\bigr)\bigl( a_{\pi_3,\pi_4}-q^{3}a_{\pi_4,\pi_3}\bigr)\cdots\times\\
                          &\times\cdots \bigl( a_{\pi_{2n-3},\pi_{2n-2}}-q^{2n-3} a_{\pi_{2n-2},\pi_{2n-3}}\bigr) \bigl( a_{\pi_{2n-1},\pi_{2n}}-q^{2n-1} a_{\pi_{2n},\pi_{2n-1}}\bigr),
\end{align*}
where $F_n(q)$ is a Laurent polynomial in $q$ depending only on $n$.
We then prove that this quantum Pfaffian is self-adjoint and has the form of a quantum deformation of a classical Pfaffian. We then put forward the conjecture that, at least in the case of Pfaffians $\widehat{Pf}_\square$ and first Markov elements $\widehat{C_1}$, self-adjointness implies commutativity. 

We then use the results of \cite{JingZhangSklyanin} to derive closed-form expressions for the new quantum Casimirs $\widehat{K}_i$ appearing in the case of a full-size matrix $\mathbb A$. For $0 \le m \le \lfloor \frac{n-1}{2} \rfloor$, $\widehat{K}_m$ equals (Theorem~\ref{thm-mcas})
\begin{align*}
  \widehat{M}_{n-m}^{\text{BL}}[\widehat{M}_m^{\text{BL}}]^{-1}
\end{align*}
where $\widehat{M}_m^{BL}$ is the Sklyanin minor of size $m$ in the bottom left corner, and which takes the explicit form (Lemma~\ref{lm:mbl})
\begin{align*}
  \widehat{M}_m^{BL} = \sum_{\sigma \in S_m} (-q)^{l(\sigma)} q^{\#\{\sigma(i)>n-m+i, 1 \le i \le m\}} a_{n-m+1,\sigma(1)}a_{n-m+2,\sigma(2)} \cdots a_{n,\sigma(m)}.
\end{align*}
We show that the quantum Casimirs $q^{m-n/2}\widehat{M}_{n-m}^{\text{BL}}[\widehat{M}_m^{\text{BL}}]^{-1}$ are self-adjoint (Lemma~\ref{lem-msa}).

In three appendices we collect the description of algorithms used for finding quantum deformations of Casimirs that are self-adjoint, the full list of commutation relations for $a_{i,j}$ for a full-size matrix $\mathbb A$ and verification of commutation of the quantum Pfaffian with the element $a_{2n,2}$. Appendix~\ref{app:centrality} containing the A.I.-assisted proof of centrality of the full quantum Pfaffian was added after the completion of the main part of the text. We use polarization operators $E_k,F_k,K_k$ introduced in this appendix to prove centrality of the quantum determinant (Theorem~\ref{thm-EFK} and Corollary~\ref{cor-q-det}).

\subsection{Notation}\label{subsubsec:notation}
\begin{itemize}
\item $A$---unipotent upper-triangular matrix satisfying the quantum or semiclassical reflection equation; we denote the corresponding quantum algebra by the same symbol;
\item $\mathbb A$---full-size $n\times n$ matrix satisfying quantum or semiclassical reflection equation; as in the previous case, we use the same letter to denote the corresponding quantum algebra;
\item $a_{i,j}$---matrix elements of either $A$ or $\mathbb A$;
\item $M_1$ and $M_2$ are correspondingly upper and lower $n\times n$ blocks of a $2n\times n$ quantum or semiclassical transport matrix $M$. These blocks are respectively lower and upper triangular matrices in the case of a triangular Fock--Goncharov network;
\item $C_i$ and $\widehat{C}_i$---respective semiclassical and quantum Casimir functions generated by either $\det(A+\lambda A^T)$ or $\det(\mathbb A+\lambda \mathbb A^T)$;
\item $Pf_\triangle,Pf_\square$ and $\widehat{Pf}_\triangle,\widehat{Pf}_\square$ are semiclassical and quantum Pfaffians for the respective triangular ($A$) or full ($\mathbb A$) algebras;
\item $\mathfrak M_i$---semiclassical bottom-left $i\times i$ minor of $\mathbb A$;
\item $\widehat{M}{}^{\text{BL}}_m$---quantum bottom-left $m\times m$ Sklyanin $q$-minor of $\mathbb A$;
\item $\stackrel{k}{\mathbb A}=\sum_{i,j} \stackrel 1{\mathbb I}\otimes \stackrel 2{\mathbb I}\otimes \cdots \otimes \stackrel k{e}_{i,j}\otimes \stackrel {k+1}{\mathbb I}\otimes\cdots\otimes \stackrel m{\mathbb I}a_{i,j}$, where $a_{i,j}$ is an operator in the quantum space and we have a direct product of $m$ classical $n\times n$-dimensional spaces;
\item In all formulas involving quantum space, we assume that the order of product of elements coincides with the order of product in the classical spaces, e.g., 
$$\stackrel{1}{\mathbb A}\otimes \stackrel{2}{\mathbb A}=\sum\limits_{i,j,k,l} \stackrel{1}{e}_{i,j}\otimes \stackrel{2}{e}_{k,l} a_{i,j}a_{k,l};$$
\item We use the term ``coisotropic reduction'' for any reduction of a Poisson or quantum algebra w.r.t. an ideal generated by the reduction conditions.
\end{itemize}
Fix $n\ge1$ and write $N \coloneqq [2n]=\{1,2,\dots,2n\}$.

\begin{itemize}
\item $\Pi$---the set of permutations $\pi = (\pi_1,\ldots,\pi_{2n}) \in S_{2n}$ satisfying $\pi_{2k-1}<\pi_{2k}$ for all $k$ and $\pi_{2k-1}<\pi_{2k+1}$ for $k=1,\ldots,n-1$ (in particular, $\pi_1=1$).
\item $\widehat{\Pi}$---the set of permutations $\pi=(\pi_1,\ldots,\pi_{2n}) \in S_{2n}$ satisfying $\pi_{2k-1}<\pi_{2k}$ for all $k$ and $\pi_{2k}<\pi_{2k+2}$ for $k=1,\ldots,n-1$ (in particular, $\pi_{2n}=2n$).
\item $\Pi'$---the set of permutations $\pi=(\pi_1,\ldots,\pi_{2n}) \in S_{2n}$ satisfying $\pi_{2k-1}<\pi_{2k}$ for all $k$ and with no further ordering imposed.
\item $\wp$---the entirety of $S_{2n}$.
\item $l(\pi)$---the inversion number of a permutation $\pi$, i.e., the number of pairs $(i,j)$ with $i<j$ and $\pi(i)>\pi(j)$.
\item $\exc(\pi)$---the excedance number of a permutation $\pi$, i.e., the number of $i$ with $\pi(i)>i$.


\end{itemize}

\section{Networks of cluster variables and algebras}\label{s:clusters}
\subsection{Transport matrices}\label{ss:transport}

\begin{figure}[h]
  \centering
  \psscalebox{0.9}{
    \begin{pspicture}(-6,-2.5)(4,4){\psset{unit=1}
        \newcommand{\LEFTDOWNARROW}{%
          {\psset{unit=1}
            \rput(0,0){\psline[doubleline=true,linewidth=1pt, doublesep=1pt, linecolor=black]{<-}(0,0)(.765,.45)}
          }}
        \newcommand{\DOWNARROW}{%
          {\psset{unit=1}
            \rput(0,0){\psline[doubleline=true,linewidth=1pt, doublesep=1pt, linecolor=black]{->}(0,0.1)(0,-0.566)}
            \put(0,0){\pscircle[fillstyle=solid,fillcolor=white]{.1}}
          }}
        \newcommand{\LEFTUPARROW}{%
          {\psset{unit=1}
            \rput(0,0){\psline[doubleline=true,linewidth=1pt, doublesep=1pt, linecolor=black]{->}(0,0)(-.765,.45)}
          }}
        \newcommand{\LEFTUPARROWSHORT}{%
          {\psset{unit=1}
            \rput(0,0){\psline[doubleline=true,linewidth=1pt, doublesep=1pt, linecolor=black]{->}(0,0)(-.45,.23)}
          }}
	\newcommand{\STARUP}{%
          {\psset{unit=1}
            \rput(0,0){\psline[doubleline=true,linewidth=1pt, doublesep=1pt, linecolor=black]{<-}(0.06,-0.03)(.4,-.24)}
            \rput(0,0){\psline[doubleline=true,linewidth=1pt, doublesep=1pt, linecolor=black]{<-}(0,0.1)(0,.466)}
            \rput(0,0){\psline[doubleline=true,linewidth=1pt, doublesep=1pt, linecolor=black]{->}(0,0)(-.45,-.23)}
            \put(0,0){\pscircle[fillstyle=solid,fillcolor=black]{.1}}
            \put(0,.566){\pscircle[fillstyle=solid,fillcolor=white]{.1}}
          }}
        \newcommand{\PATGEN}{%
          {\psset{unit=1}
            \rput(0,0){\psline[linecolor=blue,linewidth=2pt]{->}(0,0)(.45,.765)}
            \rput(0,0){\psline[linecolor=blue,linewidth=2pt]{->}(1,0)(0.1,0)}
            \rput(0,0){\psline[linecolor=blue,linewidth=2pt]{->}(0,0)(.45,-.765)}
            \put(0,0){\pscircle[fillstyle=solid,fillcolor=red]{.1}}
          }}
        \newcommand{\PATLEFT}{%
          {\psset{unit=1}
            \rput(0,0){\psline[linecolor=blue,linewidth=2pt,linestyle=dashed]{->}(0,0)(.45,.765)}
            \rput(0,0){\psline[linecolor=blue,linewidth=2pt]{->}(1,0)(0.1,0)}
            \rput(0,0){\psline[linecolor=blue,linewidth=2pt]{->}(0,0)(.45,-.765)}
            \put(0,0){\pscircle[fillstyle=solid,fillcolor=lightgray]{.1}}
          }}
        \newcommand{\PATRIGHT}{%
          {\psset{unit=1}
            \rput(0,0){\psline[linecolor=blue,linewidth=2pt,linestyle=dashed]{->}(0,0)(.45,-.765)}
            \put(0,0){\pscircle[fillstyle=solid,fillcolor=lightgray]{.1}}
          }}
        \newcommand{\PATBOTTOM}{%
          {\psset{unit=1}
            \rput(0,0){\psline[linecolor=blue,linewidth=2pt]{->}(0,0)(.45,.765)}
            \rput(0,0){\psline[linecolor=blue,linewidth=2pt,linestyle=dashed]{->}(1,0)(0.1,0)}
            \put(0,0){\pscircle[fillstyle=solid,fillcolor=red]{.1}}
          }}
        \newcommand{\PATTOP}{%
          {\psset{unit=1}
            \rput(0,0){\psline[linecolor=blue,linewidth=2pt]{->}(1,0)(0.1,0)}
            \rput(0,0){\psline[linecolor=blue,linewidth=2pt]{->}(0,0)(.45,-.765)}
            \put(0,0){\pscircle[fillstyle=solid,fillcolor=red]{.1}}
          }}
        \newcommand{\PATBOTRIGHT}{%
          {\psset{unit=1}
            \rput(0,0){\psline[linecolor=blue,linewidth=2pt]{->}(0,0)(.45,.765)}
            \put(0,0){\pscircle[fillstyle=solid,fillcolor=red]{.1}}
            \put(.5,0.85){\pscircle[fillstyle=solid,fillcolor=lightgray]{.1}}
          }}
        \newcommand{\PATNORTH}{%
          {\psset{unit=1}
            \rput(0,0){\psline[linecolor=blue,linewidth=2pt,linestyle=dashed]{->}(-.45,-.765)(0,0)}
            \rput(0,0){\psline[linecolor=blue,linewidth=2pt,linestyle=dashed]{<-}(.45,-.765)(0,0)}
            \put(0,0){\pscircle[fillstyle=solid,fillcolor=lightgray]{.1}}
          }}
        \newcommand{\PATSW}{%
          {\psset{unit=1}
            \rput(0,0){\psline[linecolor=blue,linewidth=2pt,linestyle=dashed]{->}(0,0)(.45,.765)}
            \rput(0,0){\psline[linecolor=blue,linewidth=2pt,linestyle=dashed]{<-}(0.1,0)(0.9,0)}
            \put(0,0){\pscircle[fillstyle=solid,fillcolor=lightgray]{.1}}
          }}
        \newcommand{\PATSE}{%
          {\psset{unit=1}
            \rput(0,0){\psline[linecolor=blue,linewidth=2pt,linestyle=dashed]{<-}(0,0)(-.45,.765)}
            \rput(0,0){\psline[linecolor=blue,linewidth=2pt,linestyle=dashed]{<-}(-0.9,0)(-0.1,0)}
            \put(0,0){\pscircle[fillstyle=solid,fillcolor=lightgray]{.1}}
          }}
        \multiput(-2.5,-0.85)(0.5,0.85){4}{\PATLEFT}
        \multiput(-2,-1.7)(1,0){4}{\PATBOTTOM}
        \put(-0.5,2.55){\PATTOP}
        \put(0,3.4){\PATNORTH}
        \multiput(-1.5,-0.85)(1,0){4}{\PATGEN}
        \multiput(-1,0)(1,0){3}{\PATGEN}
        \multiput(-.5,0.85)(1,0){2}{\PATGEN}
        \put(0,1.7){\PATGEN}
        \multiput(-1.5,-0.85)(1,0){4}{\PATGEN}
        \multiput(0.5,2.55)(0.5,-0.85){4}{\PATRIGHT}
        \put(2,-1.7){\PATBOTRIGHT}
        \put(-3,-1.7){\PATSW}
        \put(3,-1.7){\PATSE}
        \multiput(-2,-1.176)(1.0,0){5}{\STARUP}
        \multiput(-1.5,-0.335)(1.0,0){4}{\STARUP}
        \multiput(-1.0,0.5)(1.0,0){3}{\STARUP}
        \multiput(-.5,1.4)(1.0,0){2}{\STARUP}
        \put(0,2.3){\STARUP}
        \multiput(2.6,-1.4)(-0.5,.85){6}{\LEFTDOWNARROW}
        \multiput(-2.6,-1.4)(0.5,.85){6}{\LEFTUPARROWSHORT}
        \multiput(-2.5,-1.5)(1.0,0){6}{\DOWNARROW}
        \put(1.2,3.2){\makebox(0,0)[br]{\hbox{{$1$}}}}
        \put(1.7,2.4){\makebox(0,0)[br]{\hbox{{$2$}}}}
        \put(2.2,1.6){\makebox(0,0)[br]{\hbox{{$3$}}}}
        \put(2.7,0.8){\makebox(0,0)[br]{\hbox{{$4$}}}}
        \put(3.2,-0.1){\makebox(0,0)[br]{\hbox{{$5$}}}}
        \put(3.7,-1.0){\makebox(0,0)[br]{\hbox{{$6$}}}}

        \put(-1,3.2){\makebox(0,0)[tl]{\hbox{{$1'$}}}}
        \put(-1.5,2.4){\makebox(0,0)[tl]{\hbox{{$2'$}}}}
        \put(-2,1.6){\makebox(0,0)[tl]{\hbox{{$3'$}}}}
        \put(-2.5,0.8){\makebox(0,0)[tl]{\hbox{{$4'$}}}}
        \put(-3,-0.1){\makebox(0,0)[tl]{\hbox{{$5'$}}}}
        \put(-3.5,-1.0){\makebox(0,0)[tl]{\hbox{{$6'$}}}}

        \put(-2.4,-2.6){\makebox(0,0)[bc]{\hbox{{$1''$}}}}
        \put(-1.4,-2.6){\makebox(0,0)[bc]{\hbox{{$2''$}}}}
        \put(-0.4,-2.6){\makebox(0,0)[bc]{\hbox{{$3''$}}}}
        \put(0.6,-2.6){\makebox(0,0)[bc]{\hbox{{$4''$}}}}
        \put(1.6,-2.6){\makebox(0,0)[bc]{\hbox{{$5''$}}}}
        \put(2.6,-2.6){\makebox(0,0)[bc]{\hbox{{$6''$}}}}
      }
    \end{pspicture}
  }
   \caption{\small
    The Fock--Goncharov (FG) network $G$ and the corresponding quiver $G^{\text{dual}}$ for $n=6$. The network $G$ has $n=6$ sources ($1$--$6$) and $2n$ sinks ($1'$--$6'$ and $1''$--$6''$).}
  \label{fi:plab}
\end{figure}
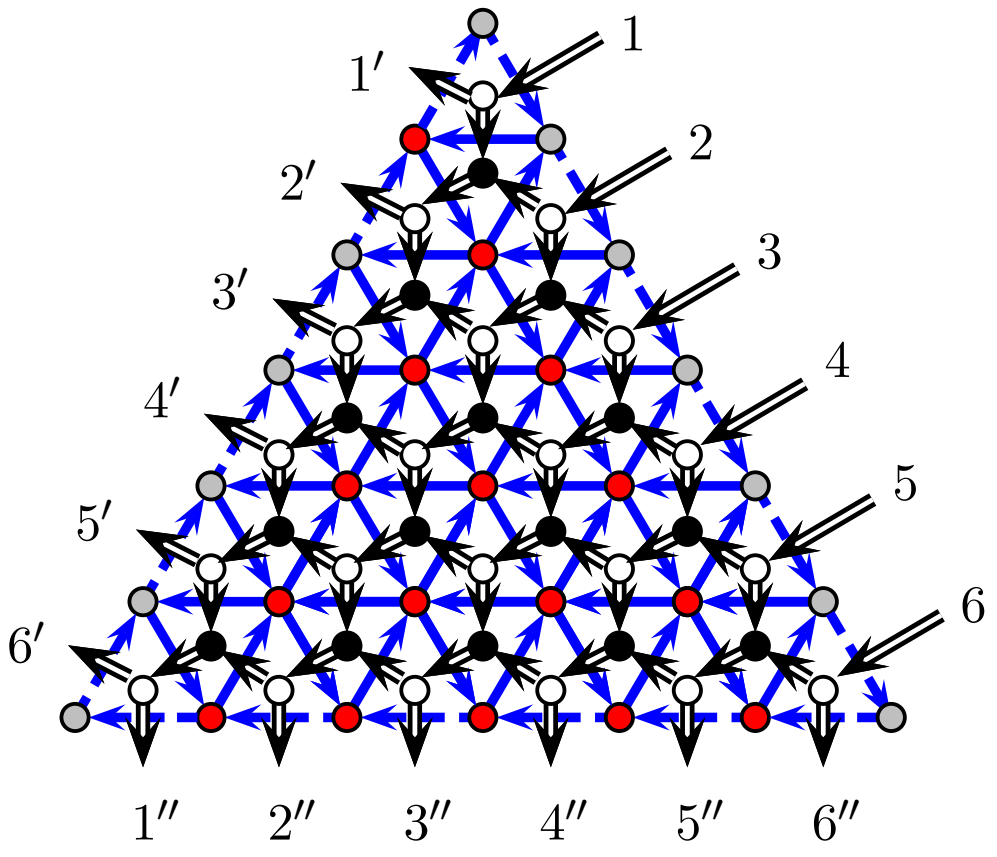

The graph $G$ that corresponds to an upper-triangular matrix $A$ is the planar oriented graph in the disk depicted in Figure~\ref{fi:plab}. We call it the Fock--Goncharov (FG) quiver \cite{FG}. Label outer vertices on the left, on the right and on the bottom sides from $1$ to $n$ as shown in the figure. Then assign variables $Z_\alpha$ to vertices of the quiver which correspond to the \emph{faces} of $G$. That is, we associate variables with vertices of the dual oriented graph $G^\text{dual}$. Three-valent vertices of the original graph $G$ are colored black and white depending on whether they have two or one incoming arrow. The variables $Z_\alpha$ labeled by the faces of $G$ (or the vertices of $G^\text{dual}$) enjoy  $q$-commutation relations encoded in the directed edges of $G^\text{dual}$: a solid arrow from $Z_1$ to $Z_2$ indicates $Z_1Z_2=qZ_2Z_1$, a dashed arrow indicates $Z_1Z_2=q^{1/2}Z_2Z_1$, a double arrow indicates $Z_1Z_2=q^{2}Z_2Z_1$, and so on.

Any maximal oriented path in the graph $G$ connects a vertex on the right side 1--n of the triangle either with a vertex of the left side $1'$--$n'$ or with a vertex on the bottom side $1''$--$n''$.  
We assign to every oriented path $P:j\leadsto i'$ from the right side to the left side or to the bottom side $P:j\leadsto i''$ the \emph{quantum weight} 
$$
\displaystyle{w(P)=\col{\mathop{\prod_{\text{ face }\alpha\text{ lies to the right}}}_{\text{ of the path } P} Z_\alpha}},
$$ 
where $\col{\bullet}$ indicates the Weyl ordering of exponentiated canonical variables: for $Z_i=e^{z_i}$, $\col{Z_1\cdots Z_s}=e^{z_1+z_2+\cdots +z_s}$. We assume all $Z_\alpha$ to be self-adjoint (Hermitian) operators. Any expression under the sign of the Weyl ordering is automatically self-adjoint.

\begin{defn}
  We define two $n\times n$ \emph{quantum transport matrices} \newline
  $\displaystyle{
    (M_1)_{i,j}=\mathop{\sum_{\text{directed path\,}{P: j\leadsto i'}}}_{\text{ from right to left}} w(P)}$\qquad and\qquad
  $\displaystyle{
    (M_2)_{i,j}=\mathop{\sum_{\text{directed path\,}{P: j\leadsto i''}}}_{\text{ from right to bottom}} w(P).}$\newline 
  Note that ${M}_1$ is a lower-triangular matrix and ${M}_2$ is an upper-triangular matrix.
\end{defn}

In fact, the above construction is valid for any plabic network: Let $\Gamma$ be a planar oriented graph in the rectangle with all sources and sinks located on the disc boundary,  $m$ univalent boundary sinks on the left labeled $1$ to $m$ top to bottom and  $n$ univalent boundary sources on the right labelled $1$ to $n$ top to bottom (note that in general $m\ne n$). All edges of  $\Gamma$ are oriented right to left or to the bottom, in particular, $\Gamma$ has no oriented cycles. The Fock--Goncharov plabic graph $G$ in Fig~\ref{fi:plab} is an example of $\Gamma$ with $n$ sources and $2n$ sinks for $n=6$.

For a general plabic graph $\Gamma$, its faces (including those adjacent to the disc boundary) are equipped with $q$-commuting weights $Z_\alpha$ whose commutation relations are governed by the dual directed graph $G^\text{dual}$. The path weights are defined as in the above example of the FG quiver, and entries $[M]_{i,j}$ of an $m\times n$ \emph{quantum transport matrix} are represented by sums of path weights over all paths $P: j\leadsto i$.

For {\bf any} planar network, the matrix $M$ satisfies the quantum Lie--Poisson $R$-matrix relation~\cite{Chekhov2020}
$$
R_{m}(q) \sheet{1}{ M}\otimes \sheet{2}{ M}=\sheet{2}{M}\otimes \sheet{1}{M} R_n(q),
$$ 
where $R_{k}(q)$ is a $k^2\times k^2$ $R$-matrix (\ref{eq:R12}).

\begin{thm}\label{th:MMsquare}\cite{Chekhov2020}
  Let the plabic directed network $\Gamma$ have $n$ sources and $m=2n$ separated sinks. The matrix $M$ then has the block form of two $n\times n$ matrices: let $M_1$ be the matrix of the top block (with paths going from right to left in the FG case) and $M_2$ be the matrix of the bottom block (with paths going from right to bottom in the FG case). The transport matrices $M_1$ and $M_2$ then satisfy the commutation relations
  $$
  \sheet{1}M_1 \otimes \sheet{2} M_2=\sheet{2}M_2\otimes \sheet{1}M_1 R_{n}(q), 
  $$
  $$
  R_{n}(q)\sheet{1}M_i \otimes \sheet{2} M_i=\sheet{2}M_i\otimes \sheet{1}M_i R_{n}(q) \quad \hbox{for $i=1,2$.}
  $$
\end{thm}

\subsubsection{Constructing an upper-triangular $A$}

Note that in the FG case,  $M_1$ is lower-triangular and $M_2$ is an upper-triangular matrix.

The main theorem of \cite{Chekhov2020} explores a special combination of quantum $M_1$ and $M_2$ satisfying relations in Theorem~\ref{th:MMsquare} for a \emph{general} transport matrix $M$: let
\begin{equation}\label{A}
  \mathbb A:=M_1^{\text{T}}M_2.
\end{equation}
Note that the transposition in the quantum case is formal: the quantum ordering is preserved, and only matrix elements are permuted. Note also that in the FG case, since $M_1^{\text{T}}$ and $M_2$ are both upper-triangular, the matrix $\mathbb A=A$ is automatically upper-triangular. We then have the theorem

\begin{thm}\label{th:A}\cite{Chekhov2020} Take any planar network $\Gamma$ with separated marked $n+n$ sinks and $n$ sources (we may have arbitrarily many unmarked sources and sinks that may even interlace the marked ones; the only requirement is that the marked sources and sinks are separated).
  Then the matrix $\mathbb A=M_1^{\text{T}}M_2$ satisfies the quantum reflection equation (\ref{RARA})
  $$
  R_{n}(q)\sheet{1}{\mathbb A} R_{n}^{\text{t}_1}(q)\sheet{2} {\mathbb A}=
  \sheet{2}{\mathbb A} R_{n}^{\text{t}_1}(q)\sheet{1} {\mathbb A}R_{n}(q)
  $$
  with the trigonometric $R$-matrix (\ref{eq:R12}), where $R_{n}^{\text{t}_1}(q)$ is a partially transposed (w.r.t. the first space) $R$-matrix (\ref{eq:Rt12}).
\end{thm}

The proof is a short direct calculation that uses only $R$-matrix relations. 

We thus conclude that we have a log-canonical coordinate representation for operators satisfying the reflection equation. Moreover all matrix elements of $\mathbb A$ are Laurent polynomials with positive coefficients of $Z_\alpha$ and $q$. 

All matrix elements of $M_1$ and $M_2$ are Weyl-ordered. Consider the FG case first. For $a_{i,j}:=[A]_{i,j}=\sum\limits_{k=i}^j [M_1]_{k,i}[M_2]_{k,j}$ we have that for $i<j$, $[M_1]_{k,i}$ commutes with  $[M_2]_{k,j}$ (all paths contributing to $[M_1]_{k,i}$ are disjoint from all paths contributing to $[M_2]_{k,j}$ for $i<j$), so the corresponding products are automatically Weyl-ordered,  
$$
a_{i,j}:=[A]_{i,j}=\sum\limits_{k=i}^j \col{[M_1]_{k,i}[M_2]_{k,j}},
$$ 
and $a_{i,j}$ is therefore self-adjoint. For $i=j$, the corresponding two paths share the common starting edge, and then 
$$
a_{i,i}=[A]_{i,i}=q^{1/2} \col{[M_1]_{i,i}[M_2]_{i,i}},\quad\hbox{therefore,}\quad a_{i,i}^*=q^{-1}a_{i,i}.
$$ 
This explains the appearance of $q^{1/2}$ factors on the diagonal of the quantum matrix $A$. 

A full-dimensional (not upper-triangular) matrix $\mathbb A$ can be obtained by the adjoint action by any transport matrix:
\begin{thm}\label{th:A-gen}\cite{Chekhov2020}
  Any matrix $\mathbb A:=M_\gamma^{\text{T}} A M_\gamma$, where $M_\gamma$ is a (transport) matrix satisfying commutation relations of Theorem \ref{th:MMsquare} and such that its elements commute with those of $A=M_1^{\text{T}} M_2$, satisfies the quantum reflection equation of Theorem \ref{th:A}. Moreover, this statement remains valid for the normalized transport matrices $M_i$ with $\det M_i=1$.
\end{thm}

The proof is again a direct computation; note also that if we represent $A=M_1^{\text{T}} M_2$, then $M_1'=M_1M_\gamma$ and $M_2'=M_2 M_\gamma$ satisfy commutation relations from Theorem~\ref{th:MMsquare}, and $\mathbb A:={M_1'}^{\text{T}} M'_2$ then satisfies the quantum reflection equation.

\subsection{Conjugation law for elements of \texorpdfstring{$\mathbb A$}{A} via the network}

\begin{lem}\label{lm:conj}
  The conjugation law
  \begin{align*}
    a_{i,j}^*=
    \begin{cases}
      a_{i,j} & \text{for $i>j$;} \\
      q^{-1}a_{i,i} & \text{for $i=j$;} \\
      a_{i,j}-(q-q^{-1})a_{j,i} & \text{for $i<j$}
    \end{cases}
  \end{align*}
  of $\mathbb{A}$ is consistent with the network construction of $\mathbb{A}$.
\end{lem}
\begin{proof}
  Let $\mathcal N$ be any planar network with separated $n$ sources and $2n$ sinks whose $2n\times n$ transport matrix $M$ is composed out of the upper $n\times n$ block $M_1$ and the lower $n\times n$ block $M_2$. For quantum cluster variables, all weights of paths are assumed to be Weyl-ordered, so all matrix elements $M_{i,j}$ are self-adjoint quantum operators.

  In~\cite{Chekhov2020} it was proven that matrix elements of any such network (which may even contain cycles) enjoy a closed quantum algebra: Set $k<l$. Then
  \begin{align*}
    &[ M_{k,i},M_{l,j}]=(q-q^{-1})  M_{k,j} M_{l,i}\quad\hbox{for}\ i<j\\
    &[M_{k,j}, M_{l,i}]=0\quad\hbox{for}\ i<j\\
    & M_{k,i} M_{l,i}=qM_{l,i} M_{k,i}
  \end{align*}
  Since $[M_1]_{k,i}=[M]_{k,i}$ and $[M_2]_{l,j}=[ M]_{n+l,j}$, the algebra of matrix elements of $M_1$ and $M_2$ reads
  \begin{align*}
    &\bigl[ [M_1]_{k,i},[M_2]_{l,j}\bigr]=(q-q^{-1})  [M_1]_{k,j} [M_2]_{l,i}\quad\hbox{for}\ i<j\ \hbox{and}\ \forall k,l \\
    &\bigl[[M_1]_{k,j}, [M_2]_{l,i}\bigr]=0\quad\hbox{for}\ i<j  \ \hbox{and}\ \forall k,l\\
    & [M_1]_{k,i} [M_2]_{l,i}=q[M_2]_{l,i} [M_1]_{k,i}\ \forall k,l
  \end{align*}
  For $a_{i,j}=\sum_k [M_1]_{k,i}[M_2]_{k,j}$ we have
  \begin{equation}
    a^*_{i,j}=\sum_k [M_2]_{k,j}[M_1]_{k,i}=
    \begin{cases} \sum_k [M_1]_{k,i} [M_2]_{k,j}=a_{i,j}&\hbox{for}\ i>j;\\ q^{-1} \sum_k [M_1]_{k,i} [M_2]_{k,j}=q^{-1} a_{i,i}&\hbox{for}\ i=j\\ \sum_k [M_1]_{k,i} [M_2]_{k,j}-(q-q^{-1})\sum_k [M_1]_{k,j} [M_2]_{k,i}\\
      \qquad\qquad \qquad =a_{i,j}-(q-q^{-1})a_{j,i}&\hbox{for}\ i<j
    \end{cases}
  \end{equation}
  The above conjugation laws are therefore consistent with the quantum algebra of $\mathbb A$.
\end{proof}

\subsection{Casimirs of the FG quiver} \label{ss:FGCasimirs}

The complete set of Casimir operators for the FG quiver are $\lfloor\frac{n}{2}\rfloor$ products of cluster variables. One may find details in \cite{Chekhov2020}, here we only mention cases that play a prominent role in our construction.

We need only Casimirs for a \emph{reduced} quiver in which we eliminate one of the three sets of frozen variables. This case is relevant when we do not consider, say, sinks at the bottom of the FG triangular quiver. The remaining $n(n+1)/2-1$ variables are those parameterizing, say, the transport matrix $M_1$. We then have $\lfloor \frac{n-1}{2} \rfloor$ Casimirs, which now contain cluster variables in both positive and negative powers. (``reflection'' from the bottom boundary ``inverts'' the powers of cluster variables after the reflection; the next reflection returns positive powers, so, for such Casimirs to be defined, we need two reflections, which explains why we lose one Casimir in the case of even $n$). In the case $n=6$, we have two such Casimirs depicted in Fig.~\ref{fi:Cas-incomplete}.

\begin{figure}[tb]
  \centering
  \begin{pspicture}(-2.7,-1.5)(2.7,2.5){\psset{unit=0.8}
      \newcommand{\PATGEN}{%
        {\psset{unit=1}
          \rput(0,0){\psline[linecolor=blue,linewidth=2pt]{->}(0,0)(.45,.765)}
          \rput(0,0){\psline[linecolor=blue,linewidth=2pt]{->}(1,0)(0.1,0)}
          \rput(0,0){\psline[linecolor=blue,linewidth=2pt]{->}(0,0)(.45,-.765)}
          \put(0,0){\pscircle[fillstyle=solid,fillcolor=lightgray]{.1}}
        }}
      \newcommand{\PATLEFT}{%
        {\psset{unit=1}
          \rput(0,0){\psline[linecolor=blue,linewidth=2pt,linestyle=dashed]{->}(0,0)(.45,.765)}
          \rput(0,0){\psline[linecolor=blue,linewidth=2pt]{->}(1,0)(0.1,0)}
          \rput(0,0){\psline[linecolor=blue,linewidth=2pt]{->}(0,0)(.45,-.765)}
          \put(0,0){\pscircle[fillstyle=solid,fillcolor=lightgray]{.1}}
        }}
      \newcommand{\PATRIGHT}{%
        {\psset{unit=1}
          \rput(0,0){\psline[linecolor=blue,linewidth=2pt,linestyle=dashed]{->}(0,0)(.45,-.765)}
          \put(0,0){\pscircle[fillstyle=solid,fillcolor=lightgray]{.1}}
        }}
      \newcommand{\PATBOTTOM}{%
        {\psset{unit=1}
          \rput(0,0){\psline[linecolor=blue,linewidth=2pt]{->}(0,0)(.45,.765)}
          \rput(0,0){\psline[linecolor=blue,linewidth=2pt,linestyle=dashed]{->}(1,0)(0.1,0)}
          \put(0,0){\pscircle[fillstyle=solid,fillcolor=lightgray]{.1}}
        }}
      \newcommand{\PATTOP}{%
        {\psset{unit=1}
          \rput(0,0){\psline[linecolor=blue,linewidth=2pt]{->}(1,0)(0.1,0)}
          \rput(0,0){\psline[linecolor=blue,linewidth=2pt]{->}(0,0)(.45,-.765)}
          \put(0,0){\pscircle[fillstyle=solid,fillcolor=lightgray]{.1}}
        }}
      \newcommand{\PATBOTRIGHT}{%
        {\psset{unit=1}
          \rput(0,0){\psline[linecolor=blue,linewidth=2pt]{->}(0,0)(.45,.765)}
          \put(0,0){\pscircle[fillstyle=solid,fillcolor=lightgray]{.1}}
          \put(.5,0.85){\pscircle[fillstyle=solid,fillcolor=lightgray]{.1}}
        }}
      \newcommand{\ODIN}{%
        {\psset{unit=1}
          \put(0,0){\pscircle[fillstyle=solid,fillcolor=white,linecolor=white]{.15}}
          \put(0,0){\makebox(0,0)[cc]{\hbox{\tcw{\large$\mathbf 1$}}}}
          \put(0,0){\makebox(0,0)[cc]{\hbox{\tcr{$\mathbf 1$}}}}
        }}
      \newcommand{\MINUSODIN}{%
        {\psset{unit=1}
          \put(-0.05,-0.03){\pscircle[fillstyle=solid,fillcolor=white,linecolor=white]{.2}}
          \put(0,0){\makebox(0,0)[cc]{\hbox{\tcw{\large-$\mathbf 1$}}}}
          \put(0,0){\makebox(0,0)[cc]{\hbox{\textcolor{black}{-$\mathbf 1$}}}}
        }}
      \newcommand{\DVA}{%
        {\psset{unit=1}
          \put(0,0){\pscircle[fillstyle=solid,fillcolor=white,linecolor=white]{.15}}
          \put(0,0){\makebox(0,0)[cc]{\hbox{\tcw{\large$\mathbf 2$}}}}
          \put(0,0){\makebox(0,0)[cc]{\hbox{\tcr{$\mathbf 2$}}}}
        }}
      \newcommand{\MINUSDVA}{%
        {\psset{unit=1}
          \put(-0.05,-0.03){\pscircle[fillstyle=solid,fillcolor=white,linecolor=white]{.2}}
          \put(0,0){\makebox(0,0)[cc]{\hbox{\tcw{\large -$\mathbf 2$}}}}
          \put(0,0){\makebox(0,0)[cc]{\hbox{\textcolor{black}{-$\mathbf 2$}}}}
        }}
      \newcommand{\TRI}{%
        {\psset{unit=1}
          \put(0,0){\pscircle[fillstyle=solid,fillcolor=white,linecolor=white]{.15}}
          \put(0,0){\makebox(0,0)[cc]{\hbox{\tcw{\large$\mathbf 3$}}}}
          \put(0,0){\makebox(0,0)[cc]{\hbox{\tcr{$\mathbf 3$}}}}
        }}
      \newcommand{\ZERO}{%
        {\psset{unit=1}
          \put(0,0){\pscircle[fillstyle=solid,fillcolor=white,linecolor=white]{.15}}
          \put(0,0){\makebox(0,0)[cc]{\hbox{\tcw{\large$\mathbf 0$}}}}
          \put(0,0){\makebox(0,0)[cc]{\hbox{{$\mathbf 0$}}}}
        }}
      \multiput(-2.5,-0.85)(0.5,0.85){4}{\PATLEFT}
      \multiput(-2,-1.7)(1,0){4}{\PATBOTTOM}
      \put(-0.5,2.55){\PATTOP}
      \multiput(-1.5,-0.85)(1,0){4}{\PATGEN}
      \multiput(-1,0)(1,0){3}{\PATGEN}
      \multiput(-.5,0.85)(1,0){2}{\PATGEN}
      \put(0,1.7){\PATGEN}
      \multiput(-1.5,-0.85)(1,0){4}{\PATGEN}
      \multiput(0.5,2.55)(0.5,-0.85){4}{\PATRIGHT}
      \put(2,-1.7){\PATBOTRIGHT}
      \psframe[linecolor=white,fillstyle=solid, fillcolor=white](-3,-2.2)(3,-.9)
      \multiput(-1.5,-0.85)(0.5,0.85){4}{\psline[linecolor=red,linewidth=2pt]{->}(0,0)(.45,.765)}
      \multiput(2,-1.7)(-4,0){2}{\psline[linecolor=black,linewidth=3pt]{->}(0,0)(.45,.765)}
      \multiput(1.95,-1.785)(-4,0){2}{\psline[linecolor=white,linewidth=2pt]{->}(0,0)(.45,.765)}
      \multiput(-0.5,2.55)(0.5,-0.85){4}{\psline[linecolor=red,linewidth=2pt]{->}(0,0)(.45,-.765)}
      \multiput(-2.45,-0.935)(4,0){2}{\psline[linecolor=black,linewidth=3pt]{->}(0,0)(.45,-.765)}
      \multiput(-2.5,-0.85)(4,0){2}{\psline[linecolor=white,linewidth=2pt]{->}(0,0)(.45,-.765)}
      \multiput(-2.5,-0.85)(1,0){5}{\psline[linecolor=magenta,linewidth=2pt]{->}(1,0)(.1,0)}
      \multiput(-0.5,2.55)(1,0){1}{\psline[linecolor=red,linewidth=2pt]{->}(1,0)(.1,0)}
      \multiput(-.5,0.85)(1,0){2}{\ODIN}
      \multiput(-1,0)(2,0){2}{\ODIN}
      \multiput(-.5,-0.85)(1,0){2}{\MINUSODIN}
      \multiput(-2.5,-0.85)(5,0){2}{\MINUSDVA}
      \multiput(-1.5,-0.85)(3,0){2}{\ZERO}
      \multiput(-0.5,2.55)(1,0){2}{\DVA}
      \put(0,1.7){\DVA}
      %
    }
  \end{pspicture}
  \begin{pspicture}(-2.7,-1.5)(2.7,2.5){\psset{unit=0.8}
      \newcommand{\PATGEN}{%
        {\psset{unit=1}
          \rput(0,0){\psline[linecolor=blue,linewidth=2pt]{->}(0,0)(.45,.765)}
          \rput(0,0){\psline[linecolor=blue,linewidth=2pt]{->}(1,0)(0.1,0)}
          \rput(0,0){\psline[linecolor=blue,linewidth=2pt]{->}(0,0)(.45,-.765)}
          \put(0,0){\pscircle[fillstyle=solid,fillcolor=lightgray]{.1}}
        }}
      \newcommand{\PATLEFT}{%
        {\psset{unit=1}
          \rput(0,0){\psline[linecolor=blue,linewidth=2pt,linestyle=dashed]{->}(0,0)(.45,.765)}
          \rput(0,0){\psline[linecolor=blue,linewidth=2pt]{->}(1,0)(0.1,0)}
          \rput(0,0){\psline[linecolor=blue,linewidth=2pt]{->}(0,0)(.45,-.765)}
          \put(0,0){\pscircle[fillstyle=solid,fillcolor=lightgray]{.1}}
        }}
      \newcommand{\PATRIGHT}{%
        {\psset{unit=1}
          \rput(0,0){\psline[linecolor=blue,linewidth=2pt,linestyle=dashed]{->}(0,0)(.45,-.765)}
          \put(0,0){\pscircle[fillstyle=solid,fillcolor=lightgray]{.1}}
        }}
      \newcommand{\PATBOTTOM}{%
        {\psset{unit=1}
          \rput(0,0){\psline[linecolor=blue,linewidth=2pt]{->}(0,0)(.45,.765)}
          \rput(0,0){\psline[linecolor=blue,linewidth=2pt,linestyle=dashed]{->}(1,0)(0.1,0)}
          \put(0,0){\pscircle[fillstyle=solid,fillcolor=lightgray]{.1}}
        }}
      \newcommand{\PATTOP}{%
        {\psset{unit=1}
          \rput(0,0){\psline[linecolor=blue,linewidth=2pt]{->}(1,0)(0.1,0)}
          \rput(0,0){\psline[linecolor=blue,linewidth=2pt]{->}(0,0)(.45,-.765)}
          \put(0,0){\pscircle[fillstyle=solid,fillcolor=lightgray]{.1}}
        }}
      \newcommand{\PATBOTRIGHT}{%
        {\psset{unit=1}
          \rput(0,0){\psline[linecolor=blue,linewidth=2pt]{->}(0,0)(.45,.765)}
          \put(0,0){\pscircle[fillstyle=solid,fillcolor=lightgray]{.1}}
          \put(.5,0.85){\pscircle[fillstyle=solid,fillcolor=lightgray]{.1}}
        }}
      \newcommand{\ODIN}{%
        {\psset{unit=1}
          \put(0,0){\pscircle[fillstyle=solid,fillcolor=white,linecolor=white]{.15}}
          \put(0,0){\makebox(0,0)[cc]{\hbox{\tcw{\large$\mathbf 1$}}}}
          \put(0,0){\makebox(0,0)[cc]{\hbox{\tcr{$\mathbf 1$}}}}
        }}
      \newcommand{\MINUSODIN}{%
        {\psset{unit=1}
          \put(-0.05,-0.03){\pscircle[fillstyle=solid,fillcolor=white,linecolor=white]{.2}}
          \put(0,0){\makebox(0,0)[cc]{\hbox{\tcw{\large-$\mathbf 1$}}}}
          \put(0,0){\makebox(0,0)[cc]{\hbox{\textcolor{black}{-$\mathbf 1$}}}}
        }}
      \newcommand{\DVA}{%
        {\psset{unit=1}
          \put(0,0){\pscircle[fillstyle=solid,fillcolor=white,linecolor=white]{.15}}
          \put(0,0){\makebox(0,0)[cc]{\hbox{\tcw{\large$\mathbf 2$}}}}
          \put(0,0){\makebox(0,0)[cc]{\hbox{\tcr{$\mathbf 2$}}}}
        }}
      \newcommand{\MINUSDVA}{%
        {\psset{unit=1}
          \put(-0.05,-0.03){\pscircle[fillstyle=solid,fillcolor=white,linecolor=white]{.2}}
          \put(0,0){\makebox(0,0)[cc]{\hbox{\tcw{\large -$\mathbf 2$}}}}
          \put(0,0){\makebox(0,0)[cc]{\hbox{\textcolor{black}{-$\mathbf 2$}}}}
        }}
      \newcommand{\TRI}{%
        {\psset{unit=1}
          \put(0,0){\pscircle[fillstyle=solid,fillcolor=white,linecolor=white]{.15}}
          \put(0,0){\makebox(0,0)[cc]{\hbox{\tcw{\large$\mathbf 3$}}}}
          \put(0,0){\makebox(0,0)[cc]{\hbox{\tcr{$\mathbf 3$}}}}
        }}
      \newcommand{\ZERO}{%
        {\psset{unit=1}
          \put(0,0){\pscircle[fillstyle=solid,fillcolor=white,linecolor=white]{.15}}
          \put(0,0){\makebox(0,0)[cc]{\hbox{\tcw{\large$\mathbf 0$}}}}
          \put(0,0){\makebox(0,0)[cc]{\hbox{{$\mathbf 0$}}}}
        }}
      \multiput(-2.5,-0.85)(0.5,0.85){4}{\PATLEFT}
      \multiput(-2,-1.7)(1,0){4}{\PATBOTTOM}
      \put(-0.5,2.55){\PATTOP}
      \multiput(-1.5,-0.85)(1,0){4}{\PATGEN}
      \multiput(-1,0)(1,0){3}{\PATGEN}
      \multiput(-.5,0.85)(1,0){2}{\PATGEN}
      \put(0,1.7){\PATGEN}
      \multiput(-1.5,-0.85)(1,0){4}{\PATGEN}
      \multiput(0.5,2.55)(0.5,-0.85){4}{\PATRIGHT}
      \put(2,-1.7){\PATBOTRIGHT}
      \psframe[linecolor=white,fillstyle=solid, fillcolor=white](-3,-2.2)(3,-.95)
      \multiput(-0.5,-.85)(0.5,0.85){3}{\psline[linecolor=red,linewidth=2pt]{->}(0,0)(.45,.765)}
      \multiput(1.5,-.85)(0.5,0.85){1}{\psline[linecolor=magenta,linewidth=2pt]{->}(0,0)(.45,.765)}
      \multiput(-1,1.7)(0.5,-0.85){3}{\psline[linecolor=red,linewidth=2pt]{->}(0,0)(.45,-.765)}
      \multiput(-2,0)(0.5,-0.85){1}{\psline[linecolor=magenta,linewidth=2pt]{->}(0,0)(.45,-.765)}
      \multiput(-2,0)(1,0){4}{\psline[linecolor=magenta,linewidth=2pt]{->}(1,0)(.1,0)}
      \multiput(-1,1.7)(1,0){2}{\psline[linecolor=red,linewidth=2pt]{->}(1,0)(.1,0)}
      \multiput(1,-1.7)(-2,0){2}{\psline[linecolor=black,linewidth=3pt]{->}(0,0)(.45,.765)}
      \multiput(0.95,-1.785)(-2,0){2}{\psline[linecolor=white,linewidth=2pt]{->}(0,0)(.45,.765)}
      \multiput(-1.45,-0.935)(2,0){2}{\psline[linecolor=black,linewidth=3pt]{->}(0,0)(.45,-.765)}
      \multiput(-1.5,-0.85)(2,0){2}{\psline[linecolor=white,linewidth=2pt]{->}(0,0)(.45,-.765)}
      %
      \multiput(0,1.7)(0,-1.7){2}{\ODIN}
      \multiput(-0.5,0.85)(0,-1.7){2}{\ODIN}
      \multiput(0.5,0.85)(0,-1.7){2}{\ODIN}
      \multiput(-1.5,-0.85)(3,0){2}{\MINUSODIN}
      \multiput(-1,0)(2,0){2}{\MINUSODIN}
      \multiput(-2,0)(4,0){2}{\MINUSDVA}
      \multiput(-1,1.7)(2,0){2}{\DVA}
      %
    }
  \end{pspicture}
  \caption{\small
    Two central elements of the reduced quiver for $n=6$. }
  \label{fi:Cas-incomplete}
\end{figure}
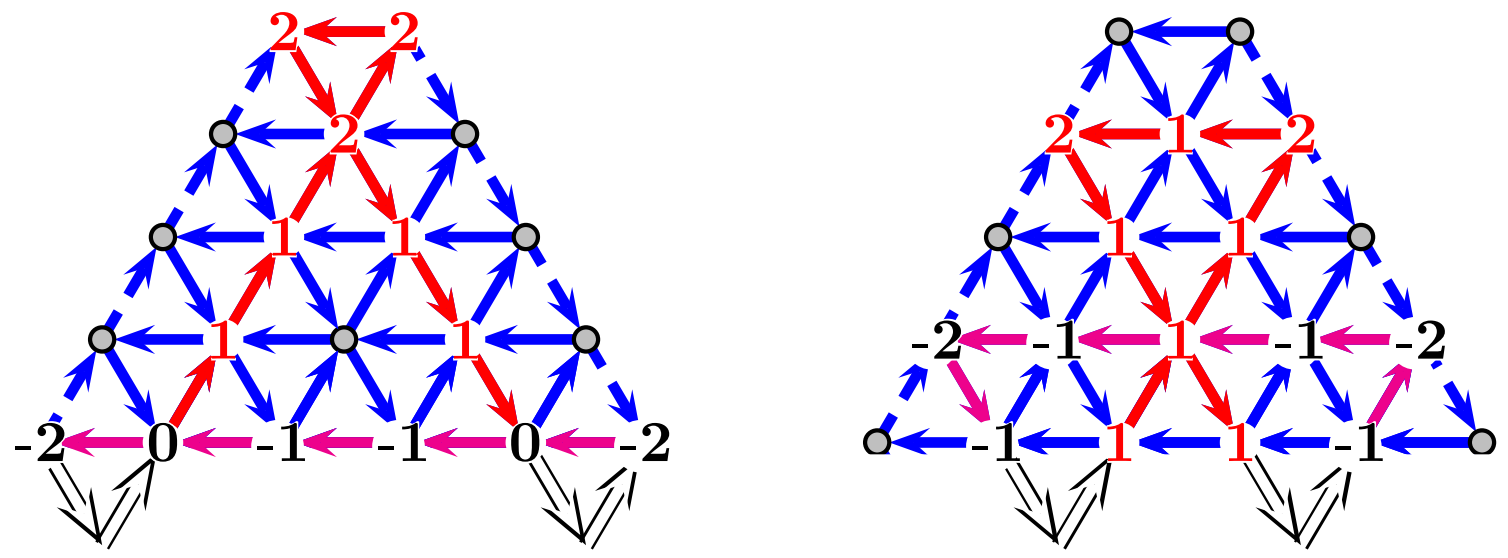

\begin{figure}[tb]
  \centering
  \begin{pspicture}(-3,-2)(3,2){\psset{unit=0.8}
      \newcommand{\PATGEN}{%
        {\psset{unit=1}
          \rput(0,0){\psline[linecolor=blue,linewidth=2pt]{->}(0,0)(.45,.765)}
          \rput(0,0){\psline[linecolor=blue,linewidth=2pt]{->}(1,0)(0.1,0)}
          \rput(0,0){\psline[linecolor=blue,linewidth=2pt]{->}(0,0)(.45,-.765)}
          \put(0,0){\pscircle[fillstyle=solid,fillcolor=lightgray]{.1}}
        }}
      \newcommand{\PATLEFT}{%
        {\psset{unit=1}
          \rput(0,0){\psline[linecolor=blue,linewidth=2pt,linestyle=dashed]{->}(0,0)(.45,.765)}
          \rput(0,0){\psline[linecolor=blue,linewidth=2pt]{->}(1,0)(0.1,0)}
          \rput(0,0){\psline[linecolor=blue,linewidth=2pt]{->}(0,0)(.45,-.765)}
          \put(0,0){\pscircle[fillstyle=solid,fillcolor=lightgray]{.1}}
        }}
      \newcommand{\PATRIGHT}{%
        {\psset{unit=1}
          \rput(0,0){\psline[linecolor=blue,linewidth=2pt,linestyle=dashed]{->}(0,0)(.45,-.765)}
          \put(0,0){\pscircle[fillstyle=solid,fillcolor=lightgray]{.1}}
        }}
      \newcommand{\PATBOTTOM}{%
        {\psset{unit=1}
          \rput(0,0){\psline[linecolor=blue,linewidth=2pt]{->}(0,0)(.45,.765)}
          \rput(0,0){\psline[linecolor=blue,linewidth=2pt,linestyle=dashed]{->}(1,0)(0.1,0)}
          \put(0,0){\pscircle[fillstyle=solid,fillcolor=lightgray]{.1}}
        }}
      \newcommand{\PATTOP}{%
        {\psset{unit=1}
          \rput(0,0){\psline[linecolor=blue,linewidth=2pt]{->}(1,0)(0.1,0)}
          \rput(0,0){\psline[linecolor=blue,linewidth=2pt]{->}(0,0)(.45,-.765)}
          \put(0,0){\pscircle[fillstyle=solid,fillcolor=lightgray]{.1}}
        }}
      \newcommand{\PATBOTRIGHT}{%
        {\psset{unit=1}
          \rput(0,0){\psline[linecolor=blue,linewidth=2pt]{->}(0,0)(.45,.765)}
          \put(0,0){\pscircle[fillstyle=solid,fillcolor=lightgray]{.1}}
          \put(.5,0.85){\pscircle[fillstyle=solid,fillcolor=lightgray]{.1}}
        }}
      \multiput(-2.5,-0.85)(0.5,0.85){4}{\PATLEFT}
      \multiput(-2,-1.7)(1,0){4}{\PATBOTTOM}
      \put(-0.5,2.55){\PATTOP}
      \multiput(-1.5,-0.85)(1,0){4}{\PATGEN}
      \multiput(-1,0)(1,0){3}{\PATGEN}
      \multiput(-.5,0.85)(1,0){2}{\PATGEN}
      \put(0,1.7){\PATGEN}
      \multiput(-1.5,-0.85)(1,0){4}{\PATGEN}
      \multiput(0.5,2.55)(0.5,-0.85){4}{\PATRIGHT}
      \put(2,-1.7){\PATBOTRIGHT}
      \put(-0.8,2.8){\makebox(0,0)[tr]{\hbox{{ $2$}}}}
      \put(-1.3, 1.95){\makebox(0,0)[tr]{\hbox{{ $3$}}}}
      \put(-1.8,1.1){\makebox(0,0)[tr]{\hbox{{ $4$}}}}
      \put(-2.3,0.25){\makebox(0,0)[tr]{\hbox{{ $5$}}}}
      \put(-2.8,-0.6){\makebox(0,0)[tr]{\hbox{{ $6$}}}}
      \put(-2,-2){\makebox(0,0)[tc]{\hbox{{ $2$}}}}
      \put(-1,-2){\makebox(0,0)[tc]{\hbox{{ $3$}}}}
      \put(0,-2){\makebox(0,0)[tc]{\hbox{{ $4$}}}}
      \put(1,-2){\makebox(0,0)[tc]{\hbox{{ $5$}}}}
      \put(2,-2){\makebox(0,0)[tc]{\hbox{{ $6$}}}}
      %
      %
    }
  \end{pspicture}
  \caption{\small
    The amalgamation of the quiver corresponding to the triangle $\Sigma_{0,1,3}$. Pairwise amalgamated vertices are indicated by the same index. (The example in the figure corresponds to $n=6$). Frozen variables at corners are amalgamated into a special Casimir and are removed.
  }
  \label{fi:amalgamation}
\end{figure}
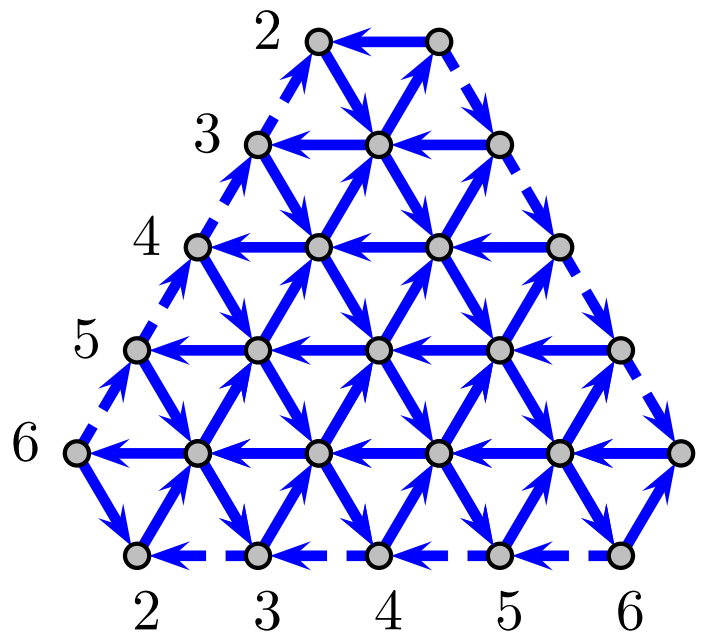

\subsection{Casimirs for the upper-triangular matrices}\label{ss:Casimirs-triangle}

Entries of the matrix $A:= M_1^{\text{T}}M_2$ depend on all variables of the FG-quiver, but due to the transposition, two sets of the frozen variables become amalgamated, that is, only their products appear in the entries of the matrix  $A$. We explicitly show this amalgamation in Fig.~\ref{fi:amalgamation}. Moreover, for the {\bf normalized} matrices $M_i$, the dependence on the three corner variables also disappears; we may then amalgamate all these variables into one frozen variable and remove it.

\begin{figure}[H]
  \centering
  \begin{pspicture}(-2.4,-2)(2.4,2.5){\psset{unit=0.7}
      \newcommand{\PATGEN}{%
        {\psset{unit=1}
          \rput(0,0){\psline[linecolor=blue,linewidth=2pt]{->}(0,0)(.45,.765)}
          \rput(0,0){\psline[linecolor=blue,linewidth=2pt]{->}(1,0)(0.1,0)}
          \rput(0,0){\psline[linecolor=blue,linewidth=2pt]{->}(0,0)(.45,-.765)}
          \put(0,0){\pscircle[fillstyle=solid,fillcolor=lightgray]{.1}}
        }}
      \newcommand{\PATLEFT}{%
        {\psset{unit=1}
          \rput(0,0){\psline[linecolor=blue,linewidth=2pt,linestyle=dashed]{->}(0,0)(.45,.765)}
          \rput(0,0){\psline[linecolor=blue,linewidth=2pt]{->}(1,0)(0.1,0)}
          \rput(0,0){\psline[linecolor=blue,linewidth=2pt]{->}(0,0)(.45,-.765)}
          \put(0,0){\pscircle[fillstyle=solid,fillcolor=lightgray]{.1}}
        }}
      \newcommand{\PATRIGHT}{%
        {\psset{unit=1}
          \rput(0,0){\psline[linecolor=blue,linewidth=2pt,linestyle=dashed]{->}(0,0)(.45,-.765)}
          \put(0,0){\pscircle[fillstyle=solid,fillcolor=lightgray]{.1}}
        }}
      \newcommand{\PATBOTTOM}{%
        {\psset{unit=1}
          \rput(0,0){\psline[linecolor=blue,linewidth=2pt]{->}(0,0)(.45,.765)}
          \rput(0,0){\psline[linecolor=blue,linewidth=2pt,linestyle=dashed]{->}(1,0)(0.1,0)}
          \put(0,0){\pscircle[fillstyle=solid,fillcolor=lightgray]{.1}}
        }}
      \newcommand{\PATTOP}{%
        {\psset{unit=1}
          \rput(0,0){\psline[linecolor=blue,linewidth=2pt]{->}(1,0)(0.1,0)}
          \rput(0,0){\psline[linecolor=blue,linewidth=2pt]{->}(0,0)(.45,-.765)}
          \put(0,0){\pscircle[fillstyle=solid,fillcolor=lightgray]{.1}}
        }}
      \newcommand{\PATBOTRIGHT}{%
        {\psset{unit=1}
          \rput(0,0){\psline[linecolor=blue,linewidth=2pt]{->}(0,0)(.45,.765)}
          \put(0,0){\pscircle[fillstyle=solid,fillcolor=lightgray]{.1}}
          \put(.5,0.85){\pscircle[fillstyle=solid,fillcolor=lightgray]{.1}}
        }}
      \newcommand{\ODIN}{%
        {\psset{unit=1}
          \put(0,0){\pscircle[fillstyle=solid,fillcolor=white,linecolor=white]{.15}}
          \put(0,0){\makebox(0,0)[cc]{\hbox{\tcw{\large$\mathbf 1$}}}}
          \put(0,0){\makebox(0,0)[cc]{\hbox{\tcr{$\mathbf 1$}}}}
        }}
      \multiput(-2.5,-0.85)(0.5,0.85){4}{\PATLEFT}
      \multiput(-2,-1.7)(1,0){4}{\PATBOTTOM}
      \put(-0.5,2.55){\PATTOP}
      \multiput(-1.5,-0.85)(1,0){4}{\PATGEN}
      \multiput(-1,0)(1,0){3}{\PATGEN}
      \multiput(-.5,0.85)(1,0){2}{\PATGEN}
      \put(0,1.7){\PATGEN}
      \multiput(-1.5,-0.85)(1,0){4}{\PATGEN}
      \multiput(0.5,2.55)(0.5,-0.85){4}{\PATRIGHT}
      \put(2,-1.7){\PATBOTRIGHT}
      %
      \put(-0.8,2.8){\makebox(0,0)[tr]{\hbox{{ $2$}}}}
      \put(-1.3, 1.95){\makebox(0,0)[tr]{\hbox{{ $3$}}}}
      \put(-1.8,1.1){\makebox(0,0)[tr]{\hbox{{ $4$}}}}
      \put(-2.3,0.25){\makebox(0,0)[tr]{\hbox{{ $5$}}}}
      \put(-2.8,-0.6){\makebox(0,0)[tr]{\hbox{{ $6$}}}}
      \put(-2,-2){\makebox(0,0)[tc]{\hbox{{ $2$}}}}
      \put(-1,-2){\makebox(0,0)[tc]{\hbox{{ $3$}}}}
      \put(0,-2){\makebox(0,0)[tc]{\hbox{{ $4$}}}}
      \put(1,-2){\makebox(0,0)[tc]{\hbox{{ $5$}}}}
      \put(2,-2){\makebox(0,0)[tc]{\hbox{{ $6$}}}}
      %
      %
      \multiput(-0.5,2.55)(0.5,-0.85){5}{\psline[linecolor=red,linewidth=2pt]{->}(0,0)(.45,-.765)}
      \multiput(-2.5,-0.85)(0.5,-0.85){1}{\psline[linecolor=red,linewidth=2pt]{->}(0,0)(.45,-.765)}
      \multiput(-.5,2.55)(0.5,-0.85){6}{\ODIN}
      \multiput(-2.5,-0.85)(0.5,-0.85){2}{\ODIN}
      %
    }
  \end{pspicture}
  \begin{pspicture}(-2.4,-2)(2.4,2.5){\psset{unit=0.7}
      \newcommand{\PATGEN}{%
        {\psset{unit=1}
          \rput(0,0){\psline[linecolor=blue,linewidth=2pt]{->}(0,0)(.45,.765)}
          \rput(0,0){\psline[linecolor=blue,linewidth=2pt]{->}(1,0)(0.1,0)}
          \rput(0,0){\psline[linecolor=blue,linewidth=2pt]{->}(0,0)(.45,-.765)}
          \put(0,0){\pscircle[fillstyle=solid,fillcolor=lightgray]{.1}}
        }}
      \newcommand{\PATLEFT}{%
        {\psset{unit=1}
          \rput(0,0){\psline[linecolor=blue,linewidth=2pt,linestyle=dashed]{->}(0,0)(.45,.765)}
          \rput(0,0){\psline[linecolor=blue,linewidth=2pt]{->}(1,0)(0.1,0)}
          \rput(0,0){\psline[linecolor=blue,linewidth=2pt]{->}(0,0)(.45,-.765)}
          \put(0,0){\pscircle[fillstyle=solid,fillcolor=lightgray]{.1}}
        }}
      \newcommand{\PATRIGHT}{%
        {\psset{unit=1}
          \rput(0,0){\psline[linecolor=blue,linewidth=2pt,linestyle=dashed]{->}(0,0)(.45,-.765)}
          \put(0,0){\pscircle[fillstyle=solid,fillcolor=lightgray]{.1}}
        }}
      \newcommand{\PATBOTTOM}{%
        {\psset{unit=1}
          \rput(0,0){\psline[linecolor=blue,linewidth=2pt]{->}(0,0)(.45,.765)}
          \rput(0,0){\psline[linecolor=blue,linewidth=2pt,linestyle=dashed]{->}(1,0)(0.1,0)}
          \put(0,0){\pscircle[fillstyle=solid,fillcolor=lightgray]{.1}}
        }}
      \newcommand{\PATTOP}{%
        {\psset{unit=1}
          \rput(0,0){\psline[linecolor=blue,linewidth=2pt]{->}(1,0)(0.1,0)}
          \rput(0,0){\psline[linecolor=blue,linewidth=2pt]{->}(0,0)(.45,-.765)}
          \put(0,0){\pscircle[fillstyle=solid,fillcolor=lightgray]{.1}}
        }}
      \newcommand{\PATBOTRIGHT}{%
        {\psset{unit=1}
          \rput(0,0){\psline[linecolor=blue,linewidth=2pt]{->}(0,0)(.45,.765)}
          \put(0,0){\pscircle[fillstyle=solid,fillcolor=lightgray]{.1}}
          \put(.5,0.85){\pscircle[fillstyle=solid,fillcolor=lightgray]{.1}}
        }}
      \newcommand{\ODIN}{%
        {\psset{unit=1}
          \put(0,0){\pscircle[fillstyle=solid,fillcolor=white,linecolor=white]{.15}}
          \put(0,0){\makebox(0,0)[cc]{\hbox{\tcw{\large$\mathbf 1$}}}}
          \put(0,0){\makebox(0,0)[cc]{\hbox{\tcr{$\mathbf 1$}}}}
        }}
      \multiput(-2.5,-0.85)(0.5,0.85){4}{\PATLEFT}
      \multiput(-2,-1.7)(1,0){4}{\PATBOTTOM}
      \put(-0.5,2.55){\PATTOP}
      \multiput(-1.5,-0.85)(1,0){4}{\PATGEN}
      \multiput(-1,0)(1,0){3}{\PATGEN}
      \multiput(-.5,0.85)(1,0){2}{\PATGEN}
      \put(0,1.7){\PATGEN}
      \multiput(-1.5,-0.85)(1,0){4}{\PATGEN}
      \multiput(0.5,2.55)(0.5,-0.85){4}{\PATRIGHT}
      \put(2,-1.7){\PATBOTRIGHT}
      %
      \put(-0.8,2.8){\makebox(0,0)[tr]{\hbox{{ $2$}}}}
      \put(-1.3, 1.95){\makebox(0,0)[tr]{\hbox{{ $3$}}}}
      \put(-1.8,1.1){\makebox(0,0)[tr]{\hbox{{ $4$}}}}
      \put(-2.3,0.25){\makebox(0,0)[tr]{\hbox{{ $5$}}}}
      \put(-2.8,-0.6){\makebox(0,0)[tr]{\hbox{{ $6$}}}}
      \put(-2,-2){\makebox(0,0)[tc]{\hbox{{ $2$}}}}
      \put(-1,-2){\makebox(0,0)[tc]{\hbox{{ $3$}}}}
      \put(0,-2){\makebox(0,0)[tc]{\hbox{{ $4$}}}}
      \put(1,-2){\makebox(0,0)[tc]{\hbox{{ $5$}}}}
      \put(2,-2){\makebox(0,0)[tc]{\hbox{{ $6$}}}}
      %
      %
      %
      \multiput(-1,1.7)(0.5,-0.85){4}{\psline[linecolor=red,linewidth=2pt]{->}(0,0)(.45,-.765)}
      \multiput(-2,0)(0.5,-0.85){2}{\psline[linecolor=red,linewidth=2pt]{->}(0,0)(.45,-.765)}
      \multiput(-1,1.7)(0.5,-0.85){5}{\ODIN}
      \multiput(-2,0)(0.5,-0.85){3}{\ODIN}
      %
    }
  \end{pspicture}
  \begin{pspicture}(-2.4,-2)(2.4,2.5){\psset{unit=0.7}
      \newcommand{\PATGEN}{%
        {\psset{unit=1}
          \rput(0,0){\psline[linecolor=blue,linewidth=2pt]{->}(0,0)(.45,.765)}
          \rput(0,0){\psline[linecolor=blue,linewidth=2pt]{->}(1,0)(0.1,0)}
          \rput(0,0){\psline[linecolor=blue,linewidth=2pt]{->}(0,0)(.45,-.765)}
          \put(0,0){\pscircle[fillstyle=solid,fillcolor=lightgray]{.1}}
        }}
      \newcommand{\PATLEFT}{%
        {\psset{unit=1}
          \rput(0,0){\psline[linecolor=blue,linewidth=2pt,linestyle=dashed]{->}(0,0)(.45,.765)}
          \rput(0,0){\psline[linecolor=blue,linewidth=2pt]{->}(1,0)(0.1,0)}
          \rput(0,0){\psline[linecolor=blue,linewidth=2pt]{->}(0,0)(.45,-.765)}
          \put(0,0){\pscircle[fillstyle=solid,fillcolor=lightgray]{.1}}
        }}
      \newcommand{\PATRIGHT}{%
        {\psset{unit=1}
          \rput(0,0){\psline[linecolor=blue,linewidth=2pt,linestyle=dashed]{->}(0,0)(.45,-.765)}
          \put(0,0){\pscircle[fillstyle=solid,fillcolor=lightgray]{.1}}
        }}
      \newcommand{\PATBOTTOM}{%
        {\psset{unit=1}
          \rput(0,0){\psline[linecolor=blue,linewidth=2pt]{->}(0,0)(.45,.765)}
          \rput(0,0){\psline[linecolor=blue,linewidth=2pt,linestyle=dashed]{->}(1,0)(0.1,0)}
          \put(0,0){\pscircle[fillstyle=solid,fillcolor=lightgray]{.1}}
        }}
      \newcommand{\PATTOP}{%
        {\psset{unit=1}
          \rput(0,0){\psline[linecolor=blue,linewidth=2pt]{->}(1,0)(0.1,0)}
          \rput(0,0){\psline[linecolor=blue,linewidth=2pt]{->}(0,0)(.45,-.765)}
          \put(0,0){\pscircle[fillstyle=solid,fillcolor=lightgray]{.1}}
        }}
      \newcommand{\PATBOTRIGHT}{%
        {\psset{unit=1}
          \rput(0,0){\psline[linecolor=blue,linewidth=2pt]{->}(0,0)(.45,.765)}
          \put(0,0){\pscircle[fillstyle=solid,fillcolor=lightgray]{.1}}
          \put(.5,0.85){\pscircle[fillstyle=solid,fillcolor=lightgray]{.1}}
        }}
      \newcommand{\ODIN}{%
        {\psset{unit=1}
          \put(0,0){\pscircle[fillstyle=solid,fillcolor=white,linecolor=white]{.15}}
          \put(0,0){\makebox(0,0)[cc]{\hbox{\tcw{\large$\mathbf 1$}}}}
          \put(0,0){\makebox(0,0)[cc]{\hbox{\tcr{$\mathbf 1$}}}}
        }}
      \multiput(-2.5,-0.85)(0.5,0.85){4}{\PATLEFT}
      \multiput(-2,-1.7)(1,0){4}{\PATBOTTOM}
      \put(-0.5,2.55){\PATTOP}
      \multiput(-1.5,-0.85)(1,0){4}{\PATGEN}
      \multiput(-1,0)(1,0){3}{\PATGEN}
      \multiput(-.5,0.85)(1,0){2}{\PATGEN}
      \put(0,1.7){\PATGEN}
      \multiput(-1.5,-0.85)(1,0){4}{\PATGEN}
      \multiput(0.5,2.55)(0.5,-0.85){4}{\PATRIGHT}
      \put(2,-1.7){\PATBOTRIGHT}
      \put(-0.8,2.8){\makebox(0,0)[tr]{\hbox{{ $2$}}}}
      \put(-1.3, 1.95){\makebox(0,0)[tr]{\hbox{{ $3$}}}}
      \put(-1.8,1.1){\makebox(0,0)[tr]{\hbox{{ $4$}}}}
      \put(-2.3,0.25){\makebox(0,0)[tr]{\hbox{{ $5$}}}}
      \put(-2.8,-0.6){\makebox(0,0)[tr]{\hbox{{ $6$}}}}
      \put(-2,-2){\makebox(0,0)[tc]{\hbox{{ $2$}}}}
      \put(-1,-2){\makebox(0,0)[tc]{\hbox{{ $3$}}}}
      \put(0,-2){\makebox(0,0)[tc]{\hbox{{ $4$}}}}
      \put(1,-2){\makebox(0,0)[tc]{\hbox{{ $5$}}}}
      \put(2,-2){\makebox(0,0)[tc]{\hbox{{ $6$}}}}
      %
      %
      %
      %
      \multiput(-1.5,0.85)(0.5,-0.85){3}{\psline[linecolor=red,linewidth=2pt]{->}(0,0)(.45,-.765)}
      \multiput(-1.5,0.85)(0.5,-0.85){4}{\ODIN}
      %
    }
  \end{pspicture}
  \caption{\small
    Three central elements of the full-rank $\mathcal A_6$ quiver obtained upon amalgamation. 
    Amalgamated variables on the left and lower boundaries are indicated by numbers.
  }
  \label{fi:Casimirs-A}
\end{figure}
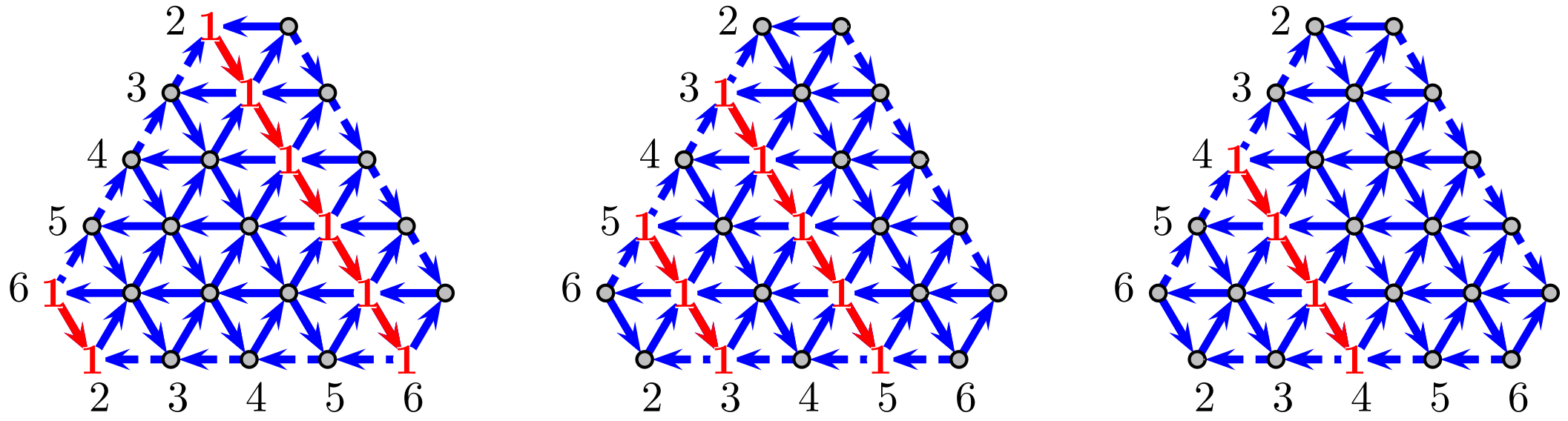

\begin{figure}[tb]
  \centering
  \begin{pspicture}(-2.4,-2)(2.4,2.5){\psset{unit=0.7}
      \newcommand{\PATGEN}{%
        {\psset{unit=1}
          \rput(0,0){\psline[linecolor=blue,linewidth=2pt]{->}(0,0)(.45,.765)}
          \rput(0,0){\psline[linecolor=blue,linewidth=2pt]{->}(1,0)(0.1,0)}
          \rput(0,0){\psline[linecolor=blue,linewidth=2pt]{->}(0,0)(.45,-.765)}
          \put(0,0){\pscircle[fillstyle=solid,fillcolor=lightgray]{.1}}
        }}
      \newcommand{\PATLEFT}{%
        {\psset{unit=1}
          \rput(0,0){\psline[linecolor=blue,linewidth=2pt,linestyle=dashed]{->}(0,0)(.45,.765)}
          \rput(0,0){\psline[linecolor=blue,linewidth=2pt]{->}(1,0)(0.1,0)}
          \rput(0,0){\psline[linecolor=blue,linewidth=2pt]{->}(0,0)(.45,-.765)}
          \put(0,0){\pscircle[fillstyle=solid,fillcolor=lightgray]{.1}}
        }}
      \newcommand{\PATRIGHT}{%
        {\psset{unit=1}
          \rput(0,0){\psline[linecolor=blue,linewidth=2pt,linestyle=dashed]{->}(0,0)(.45,-.765)}
          \put(0,0){\pscircle[fillstyle=solid,fillcolor=lightgray]{.1}}
        }}
      \newcommand{\PATBOTTOM}{%
        {\psset{unit=1}
          \rput(0,0){\psline[linecolor=blue,linewidth=2pt]{->}(0,0)(.45,.765)}
          \rput(0,0){\psline[linecolor=blue,linewidth=2pt,linestyle=dashed]{->}(1,0)(0.1,0)}
          \put(0,0){\pscircle[fillstyle=solid,fillcolor=lightgray]{.1}}
        }}
      \newcommand{\PATTOP}{%
        {\psset{unit=1}
          \rput(0,0){\psline[linecolor=blue,linewidth=2pt]{->}(1,0)(0.1,0)}
          \rput(0,0){\psline[linecolor=blue,linewidth=2pt]{->}(0,0)(.45,-.765)}
          \put(0,0){\pscircle[fillstyle=solid,fillcolor=lightgray]{.1}}
        }}
      \newcommand{\PATBOTRIGHT}{%
        {\psset{unit=1}
          \rput(0,0){\psline[linecolor=blue,linewidth=2pt]{->}(0,0)(.45,.765)}
          \put(0,0){\pscircle[fillstyle=solid,fillcolor=lightgray]{.1}}
          \put(.5,0.85){\pscircle[fillstyle=solid,fillcolor=lightgray]{.1}}
        }}
      \newcommand{\ODIN}{%
        {\psset{unit=1}
          \put(0,0){\pscircle[fillstyle=solid,fillcolor=white,linecolor=white]{.15}}
          \put(0,0){\makebox(0,0)[cc]{\hbox{\tcw{\large$\mathbf 1$}}}}
          \put(0,0){\makebox(0,0)[cc]{\hbox{\tcr{$\mathbf 1$}}}}
        }}
      \newcommand{\DVA}{%
        {\psset{unit=1}
          \put(0,0){\pscircle[fillstyle=solid,fillcolor=white,linecolor=white]{.15}}
          \put(0,0){\makebox(0,0)[cc]{\hbox{\tcw{\large$\mathbf 2$}}}}
          \put(0,0){\makebox(0,0)[cc]{\hbox{\tcr{$\mathbf 2$}}}}
        }}
      \newcommand{\TRI}{%
        {\psset{unit=1}
          \put(0,0){\pscircle[fillstyle=solid,fillcolor=white,linecolor=white]{.15}}
          \put(0,0){\makebox(0,0)[cc]{\hbox{\tcw{\large$\mathbf 3$}}}}
          \put(0,0){\makebox(0,0)[cc]{\hbox{\tcr{$\mathbf 3$}}}}
        }}
      \multiput(-2.5,-0.85)(0.5,0.85){4}{\PATLEFT}
      \multiput(-2,-1.7)(1,0){4}{\PATBOTTOM}
      \put(-0.5,2.55){\PATTOP}
      \multiput(-1.5,-0.85)(1,0){4}{\PATGEN}
      \multiput(-1,0)(1,0){3}{\PATGEN}
      \multiput(-.5,0.85)(1,0){2}{\PATGEN}
      \put(0,1.7){\PATGEN}
      \multiput(-1.5,-0.85)(1,0){4}{\PATGEN}
      \multiput(0.5,2.55)(0.5,-0.85){4}{\PATRIGHT}
      \put(2,-1.7){\PATBOTRIGHT}
      %
      \put(-0.8,2.8){\makebox(0,0)[tr]{\hbox{{ $2$}}}}
      \put(-1.3, 1.95){\makebox(0,0)[tr]{\hbox{{ $3$}}}}
      \put(-1.8,1.1){\makebox(0,0)[tr]{\hbox{{ $4$}}}}
      \put(-2.3,0.25){\makebox(0,0)[tr]{\hbox{{ $5$}}}}
      \put(-2.8,-0.6){\makebox(0,0)[tr]{\hbox{{ $6$}}}}
      \put(-2,-2){\makebox(0,0)[tc]{\hbox{{ $2$}}}}
      \put(-1,-2){\makebox(0,0)[tc]{\hbox{{ $3$}}}}
      \put(0,-2){\makebox(0,0)[tc]{\hbox{{ $4$}}}}
      \put(1,-2){\makebox(0,0)[tc]{\hbox{{ $5$}}}}
      \put(2,-2){\makebox(0,0)[tc]{\hbox{{ $6$}}}}
      %
      \multiput(-1,-1.7)(0.5,0.85){4}{\psline[linecolor=red,linewidth=2pt]{->}(0,0)(.45,.765)}
      \multiput(-1,1.7)(1,0){2}{\psline[linecolor=red,linewidth=2pt]{->}(1,0)(.1,0)}
      \multiput(-1,-1.7)(0.5,0.85){4}{\ODIN}
      \put(1,1.7){\DVA}
      \multiput(-1,1.7)(1,0){2}{\ODIN}
      %
    }
  \end{pspicture}
  \begin{pspicture}(-2.4,-2)(2.4,2.5){\psset{unit=0.7}
      %
      \newcommand{\PATGEN}{%
        {\psset{unit=1}
          \rput(0,0){\psline[linecolor=blue,linewidth=2pt]{->}(0,0)(.45,.765)}
          \rput(0,0){\psline[linecolor=blue,linewidth=2pt]{->}(1,0)(0.1,0)}
          \rput(0,0){\psline[linecolor=blue,linewidth=2pt]{->}(0,0)(.45,-.765)}
          \put(0,0){\pscircle[fillstyle=solid,fillcolor=lightgray]{.1}}
        }}
      \newcommand{\PATLEFT}{%
        {\psset{unit=1}
          \rput(0,0){\psline[linecolor=blue,linewidth=2pt,linestyle=dashed]{->}(0,0)(.45,.765)}
          \rput(0,0){\psline[linecolor=blue,linewidth=2pt]{->}(1,0)(0.1,0)}
          \rput(0,0){\psline[linecolor=blue,linewidth=2pt]{->}(0,0)(.45,-.765)}
          \put(0,0){\pscircle[fillstyle=solid,fillcolor=lightgray]{.1}}
        }}
      \newcommand{\PATRIGHT}{%
        {\psset{unit=1}
          \rput(0,0){\psline[linecolor=blue,linewidth=2pt,linestyle=dashed]{->}(0,0)(.45,-.765)}
          \put(0,0){\pscircle[fillstyle=solid,fillcolor=lightgray]{.1}}
        }}
      \newcommand{\PATBOTTOM}{%
        {\psset{unit=1}
          \rput(0,0){\psline[linecolor=blue,linewidth=2pt]{->}(0,0)(.45,.765)}
          \rput(0,0){\psline[linecolor=blue,linewidth=2pt,linestyle=dashed]{->}(1,0)(0.1,0)}
          \put(0,0){\pscircle[fillstyle=solid,fillcolor=lightgray]{.1}}
        }}
      \newcommand{\PATTOP}{%
        {\psset{unit=1}
          \rput(0,0){\psline[linecolor=blue,linewidth=2pt]{->}(1,0)(0.1,0)}
          \rput(0,0){\psline[linecolor=blue,linewidth=2pt]{->}(0,0)(.45,-.765)}
          \put(0,0){\pscircle[fillstyle=solid,fillcolor=lightgray]{.1}}
        }}
      \newcommand{\PATBOTRIGHT}{%
        {\psset{unit=1}
          \rput(0,0){\psline[linecolor=blue,linewidth=2pt]{->}(0,0)(.45,.765)}
          \put(0,0){\pscircle[fillstyle=solid,fillcolor=lightgray]{.1}}
          \put(.5,0.85){\pscircle[fillstyle=solid,fillcolor=lightgray]{.1}}
        }}
      \newcommand{\ODIN}{%
        {\psset{unit=1}
          \put(0,0){\pscircle[fillstyle=solid,fillcolor=white,linecolor=white]{.15}}
          \put(0,0){\makebox(0,0)[cc]{\hbox{\tcw{\large$\mathbf 1$}}}}
          \put(0,0){\makebox(0,0)[cc]{\hbox{\tcr{$\mathbf 1$}}}}
        }}
      \newcommand{\DVA}{%
        {\psset{unit=1}
          \put(0,0){\pscircle[fillstyle=solid,fillcolor=white,linecolor=white]{.15}}
          \put(0,0){\makebox(0,0)[cc]{\hbox{\tcw{\large$\mathbf 2$}}}}
          \put(0,0){\makebox(0,0)[cc]{\hbox{\tcr{$\mathbf 2$}}}}
        }}
      \newcommand{\TRI}{%
        {\psset{unit=1}
          \put(0,0){\pscircle[fillstyle=solid,fillcolor=white,linecolor=white]{.15}}
          \put(0,0){\makebox(0,0)[cc]{\hbox{\tcw{\large$\mathbf 3$}}}}
          \put(0,0){\makebox(0,0)[cc]{\hbox{\tcr{$\mathbf 3$}}}}
        }}
      \multiput(-2.5,-0.85)(0.5,0.85){4}{\PATLEFT}
      \multiput(-2,-1.7)(1,0){4}{\PATBOTTOM}
      \put(-0.5,2.55){\PATTOP}
      \multiput(-1.5,-0.85)(1,0){4}{\PATGEN}
      \multiput(-1,0)(1,0){3}{\PATGEN}
      \multiput(-.5,0.85)(1,0){2}{\PATGEN}
      \put(0,1.7){\PATGEN}
      \multiput(-1.5,-0.85)(1,0){4}{\PATGEN}
      \multiput(0.5,2.55)(0.5,-0.85){4}{\PATRIGHT}
      \put(2,-1.7){\PATBOTRIGHT}
      %
      \put(-0.8,2.8){\makebox(0,0)[tr]{\hbox{{ $2$}}}}
      \put(-1.3, 1.95){\makebox(0,0)[tr]{\hbox{{ $3$}}}}
      \put(-1.8,1.1){\makebox(0,0)[tr]{\hbox{{ $4$}}}}
      \put(-2.3,0.25){\makebox(0,0)[tr]{\hbox{{ $5$}}}}
      \put(-2.8,-0.6){\makebox(0,0)[tr]{\hbox{{ $6$}}}}
      \put(-2,-2){\makebox(0,0)[tc]{\hbox{{ $2$}}}}
      \put(-1,-2){\makebox(0,0)[tc]{\hbox{{ $3$}}}}
      \put(0,-2){\makebox(0,0)[tc]{\hbox{{ $4$}}}}
      \put(1,-2){\makebox(0,0)[tc]{\hbox{{ $5$}}}}
      \put(2,-2){\makebox(0,0)[tc]{\hbox{{ $6$}}}}
      %
      \multiput(1,-1.7)(0.5,0.85){2}{\psline[linecolor=red,linewidth=2pt]{->}(0,0)(.45,.765)}
      \multiput(-2,0)(1,0){4}{\psline[linecolor=red,linewidth=2pt]{->}(1,0)(.1,0)}
      \multiput(1,-1.7)(0.5,0.85){2}{\ODIN}
      \put(2,0){\DVA}
      \multiput(-2,0)(1,0){4}{\ODIN}
      %
    }
  \end{pspicture}
  \caption{\small
    Two (out of five) new central elements appearing in the case $n=6$ due to amalgamation (equal numbers on the left and bottom boundaries indicate amalgamated variables) and containing remaining frozen variables on the left boundary. The two Casimirs shown in the figure appear in the expression for the Casimir $\mathfrak C_2$ in Fig.~\ref{fi:fullCasimirs}.)
  }
  \label{fi:diagonal-Casimirs}
\end{figure}
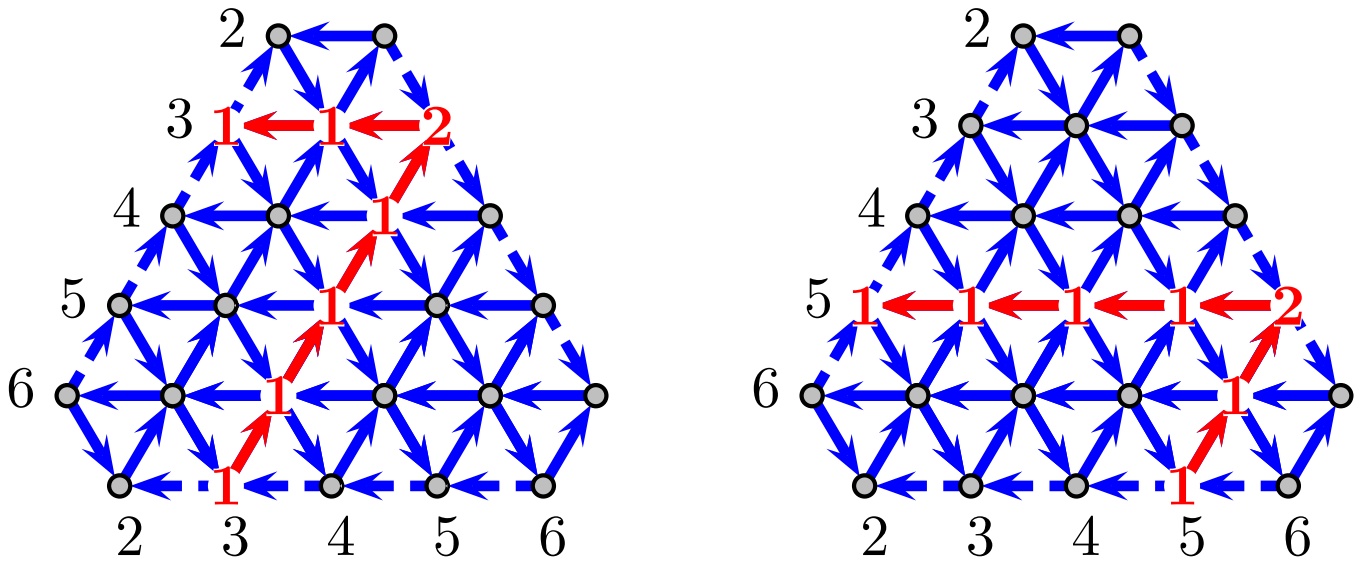

We begin with $\lfloor \frac{n}{2} \rfloor$ Casimirs that do not contain remaining (frozen) variables on the right boundary. We depict these three Casimirs for $n=6$ in Fig.~\ref{fi:Casimirs-A}. Besides Casimirs shown in the figure,  amalgamation results in the appearance of $n-1$ additional Casimirs; in Fig.~\ref{fi:diagonal-Casimirs}, we present two (out of five) such Casimirs, namely those that enter the expression for the Casimir $\mathfrak C_2$ in the case $n=6$ (for more details, see Sec.~6 in \cite{Chekhov2020}); these additional $n-1$ Casimirs can be used to set all diagonal entries of $A$ (which are central elements themselves) to be equal to $1$ (in the classical case) or $q^{1/2}$ (in the quantum case). We can therefore fix values of all remaining frozen variables on the right boundary in the FG case.

\subsection{Casimirs for the full size matrix \texorpdfstring{$\mathbb A$}{A}}\label{ss:Casimirs-quadrangle}

Now consider the network $\Gamma$ and the corresponding quiver obtained by gluing two FG networks
$$
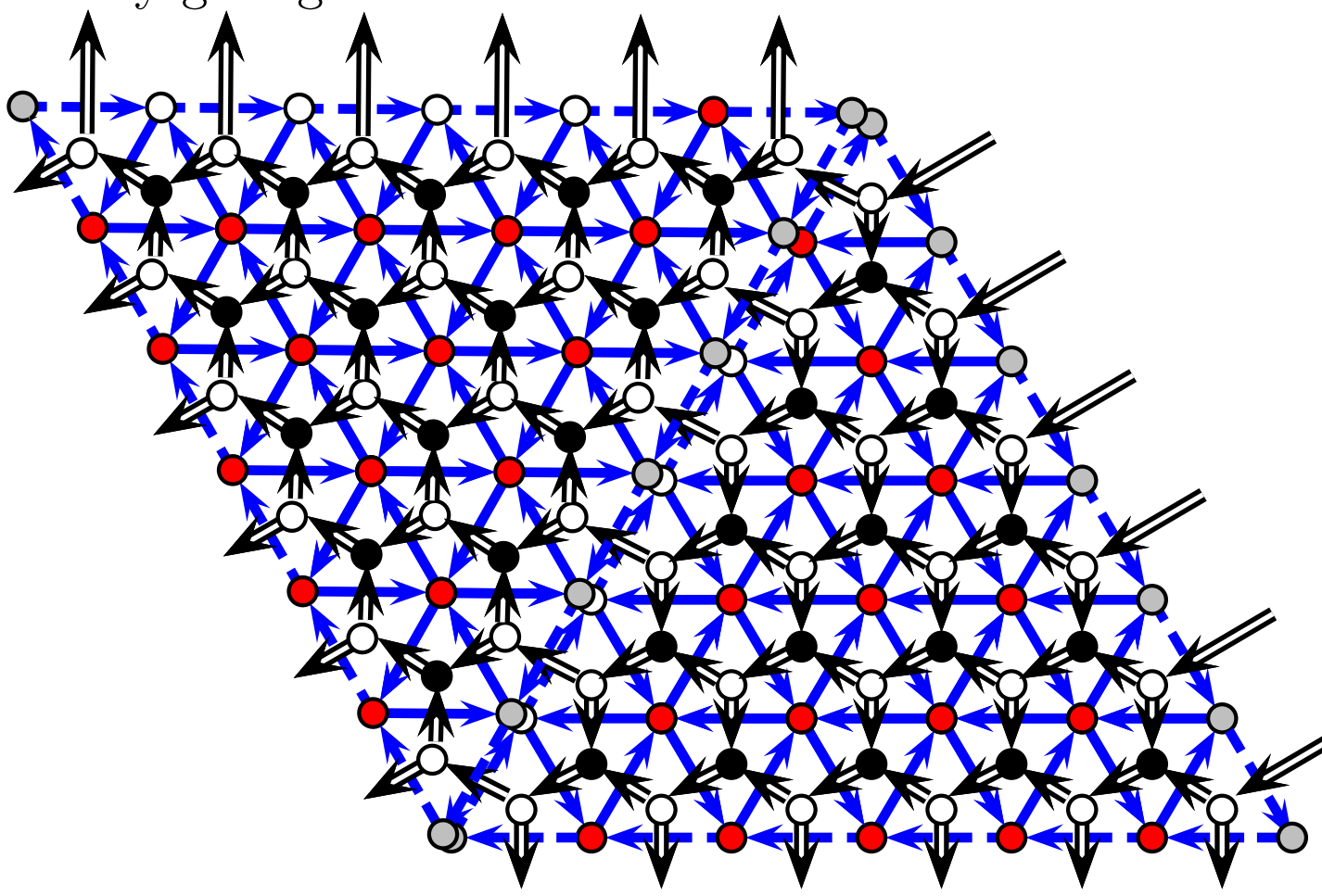
\begin{pspicture}(-6,-2.5)(4,4){\psset{unit=1}
    \newcommand{\LEFTDOWNARROW}{%
      {\psset{unit=1}
        \rput(0,0){\psline[doubleline=true,linewidth=1pt, doublesep=1pt, linecolor=black]{<-}(0,0)(.765,.45)}
      }}
    \newcommand{\DOWNARROW}{%
      {\psset{unit=1}
        \rput(0,0){\psline[doubleline=true,linewidth=1pt, doublesep=1pt, linecolor=black]{->}(0,0.1)(0,-0.566)}
        \put(0,0){\pscircle[fillstyle=solid,fillcolor=white]{.1}}
      }}
    \newcommand{\LEFTUPARROW}{%
      {\psset{unit=1}
        \rput(0,0){\psline[doubleline=true,linewidth=1pt, doublesep=1pt, linecolor=black]{->}(0,0)(-.765,.45)}
      }}
    \newcommand{\LEFTUPARROWSHORT}{%
      {\psset{unit=1}
        \rput(0,0){\psline[doubleline=true,linewidth=1pt, doublesep=1pt, linecolor=black]{->}(0,0)(-.45,.23)}
      }}
    \newcommand{\STARUP}{%
      {\psset{unit=1}
	\rput(0,0){\psline[doubleline=true,linewidth=1pt, doublesep=1pt, linecolor=black]{<-}(0.06,-0.03)(.4,-.24)}
	\rput(0,0){\psline[doubleline=true,linewidth=1pt, doublesep=1pt, linecolor=black]{<-}(0,0.1)(0,.466)}
	\rput(0,0){\psline[doubleline=true,linewidth=1pt, doublesep=1pt, linecolor=black]{->}(0,0)(-.45,-.23)}
	\put(0,0){\pscircle[fillstyle=solid,fillcolor=black]{.1}}
	\put(0,.566){\pscircle[fillstyle=solid,fillcolor=white]{.1}}
      }}
    \newcommand{\PATGEN}{%
      {\psset{unit=1}
        \rput(0,0){\psline[linecolor=blue,linewidth=2pt]{->}(0,0)(.45,.765)}
        \rput(0,0){\psline[linecolor=blue,linewidth=2pt]{->}(1,0)(0.1,0)}
        \rput(0,0){\psline[linecolor=blue,linewidth=2pt]{->}(0,0)(.45,-.765)}
        \put(0,0){\pscircle[fillstyle=solid,fillcolor=red]{.1}}
      }}
    \newcommand{\PATLEFT}{%
      {\psset{unit=1}
        \rput(0,0){\psline[linecolor=blue,linewidth=2pt,linestyle=dashed]{->}(0,0)(.45,.765)}
        \rput(0,0){\psline[linecolor=blue,linewidth=2pt]{->}(1,0)(0.1,0)}
        \rput(0,0){\psline[linecolor=blue,linewidth=2pt]{->}(0,0)(.45,-.765)}
        \put(0,0){\pscircle[fillstyle=solid,fillcolor=white]{.1}}
      }}
    \newcommand{\PATRIGHT}{%
      {\psset{unit=1}
        \rput(0,0){\psline[linecolor=blue,linewidth=2pt,linestyle=dashed]{->}(0,0)(.45,-.765)}
        \put(0,0){\pscircle[fillstyle=solid,fillcolor=lightgray]{.1}}
      }}
    \newcommand{\PATBOTTOM}{%
      {\psset{unit=1}
        \rput(0,0){\psline[linecolor=blue,linewidth=2pt]{->}(0,0)(.45,.765)}
        \rput(0,0){\psline[linecolor=blue,linewidth=2pt,linestyle=dashed]{->}(1,0)(0.1,0)}
        \put(0,0){\pscircle[fillstyle=solid,fillcolor=red]{.1}}
      }}
    \newcommand{\PATTOP}{%
      {\psset{unit=1}
        \rput(0,0){\psline[linecolor=blue,linewidth=2pt]{->}(1,0)(0.1,0)}
        \rput(0,0){\psline[linecolor=blue,linewidth=2pt]{->}(0,0)(.45,-.765)}
        \put(0,0){\pscircle[fillstyle=solid,fillcolor=red]{.1}}
      }}
    \newcommand{\PATBOTRIGHT}{%
      {\psset{unit=1}
        \rput(0,0){\psline[linecolor=blue,linewidth=2pt]{->}(0,0)(.45,.765)}
        \put(0,0){\pscircle[fillstyle=solid,fillcolor=red]{.1}}
        \put(.5,0.85){\pscircle[fillstyle=solid,fillcolor=lightgray]{.1}}
      }}
    \newcommand{\PATNORTH}{%
      {\psset{unit=1}
        \rput(0,0){\psline[linecolor=blue,linewidth=2pt,linestyle=dashed]{->}(-.45,-.765)(0,0)}
        \rput(0,0){\psline[linecolor=blue,linewidth=2pt,linestyle=dashed]{<-}(.45,-.765)(0,0)}
        \put(0,0){\pscircle[fillstyle=solid,fillcolor=lightgray]{.1}}
      }}
    \newcommand{\PATSW}{%
      {\psset{unit=1}
        \rput(0,0){\psline[linecolor=blue,linewidth=2pt,linestyle=dashed]{->}(0,0)(.45,.765)}
        \rput(0,0){\psline[linecolor=blue,linewidth=2pt,linestyle=dashed]{<-}(0.1,0)(0.9,0)}
        \put(0,0){\pscircle[fillstyle=solid,fillcolor=lightgray]{.1}}
      }}
    \newcommand{\PATSE}{%
      {\psset{unit=1}
        \rput(0,0){\psline[linecolor=blue,linewidth=2pt,linestyle=dashed]{<-}(0,0)(-.45,.765)}
        \rput(0,0){\psline[linecolor=blue,linewidth=2pt,linestyle=dashed]{<-}(-0.9,0)(-0.1,0)}
        \put(0,0){\pscircle[fillstyle=solid,fillcolor=lightgray]{.1}}
      }}
    \multiput(-2.5,-0.85)(0.5,0.85){4}{\PATLEFT}
    \multiput(-2,-1.7)(1,0){4}{\PATBOTTOM}
    \put(-0.5,2.55){\PATTOP}
    \put(0,3.4){\PATNORTH}
    \multiput(-1.5,-0.85)(1,0){4}{\PATGEN}
    \multiput(-1,0)(1,0){3}{\PATGEN}
    \multiput(-.5,0.85)(1,0){2}{\PATGEN}
    \put(0,1.7){\PATGEN}
    \multiput(-1.5,-0.85)(1,0){4}{\PATGEN}
    \multiput(0.5,2.55)(0.5,-0.85){4}{\PATRIGHT}
    \put(2,-1.7){\PATBOTRIGHT}
    \put(-3,-1.7){\PATSW}
    \put(3,-1.7){\PATSE}
    \multiput(-2,-1.176)(1.0,0){5}{\STARUP}
    \multiput(-1.5,-0.335)(1.0,0){4}{\STARUP}
    \multiput(-1.0,0.5)(1.0,0){3}{\STARUP}
    \multiput(-.5,1.4)(1.0,0){2}{\STARUP}
    \put(0,2.3){\STARUP}
    \multiput(2.6,-1.4)(-0.5,.85){6}{\LEFTDOWNARROW}
    \multiput(-2.6,-1.4)(0.5,.85){6}{\LEFTUPARROWSHORT}
    \multiput(-2.5,-1.5)(1.0,0){6}{\DOWNARROW}
    \rput{-60}(-2.35,2.2){{\multiput(-2.5,-0.85)(0.5,0.85){4}{\PATLEFT}}
      \multiput(-2,-1.7)(1,0){4}{\PATBOTTOM}
      \put(-0.5,2.55){\PATTOP}
      \put(0,3.4){\PATNORTH}
      \multiput(-1.5,-0.85)(1,0){4}{\PATGEN}
      \multiput(-1,0)(1,0){3}{\PATGEN}
      \multiput(-.5,0.85)(1,0){2}{\PATGEN}
      \put(0,1.7){\PATGEN}
      \multiput(-1.5,-0.85)(1,0){4}{\PATGEN}
      \multiput(0.5,2.55)(0.5,-0.85){4}{\PATRIGHT}
      \put(2,-1.7){\PATBOTRIGHT}
      \put(-3,-1.7){\PATSW}
      \put(3,-1.7){\PATSE}
      \multiput(-2,-1.176)(1.0,0){5}{\STARUP}
      \multiput(-1.5,-0.335)(1.0,0){4}{\STARUP}
      \multiput(-1.0,0.5)(1.0,0){3}{\STARUP}
      \multiput(-.5,1.4)(1.0,0){2}{\STARUP}
      \put(0,2.3){\STARUP}
      \multiput(-2.6,-1.4)(0.5,.85){6}{\LEFTUPARROW}
      \multiput(-2.5,-1.5)(1.0,0){6}{\DOWNARROW}
      %
    }}
  %
\end{pspicture}
$$
The matrix $M_1$ is the transport matrix from the right side to the top (it is still lower triangular); the matrix $M_2$ is the transport matrix from right to left; it is now full-dimensional. Sinks on the bottom side are irrelevant and we amalgamate all the corresponding variables. To simplify the picture, we erase lines of the network; amalgamated vertices are assigned the same numbers (from $1$ to $6$). We indicate by the same color vertices contributing to three Casimirs of the same form as in Fig.~\ref{fi:Casimirs-A}, these are Casimirs that contribute to $\det(\mathbb A+\lambda \mathbb A^\text{T})$, i.e., to the quantum Pfaffian and Markov elements, and they amazingly depend only on cluster variables of the left FG triangle. Observe also that the number of variables $Z_\alpha$ coincides with the number of variables $a_{ij}$: there are $(n+1)(n-1)$ variables in the middle (i.e., neither top nor bottom) rows of the network, plus $1$ from the variable labeled $1$ above, for a total of $n^2$. Since the dimension of the Poisson algebra and number of Casimirs is the same in each setting, there is a local isomorphism between the two coordinate sets $\{a_{i,j}\}$ and $\{Z_\alpha\}$. 
\(
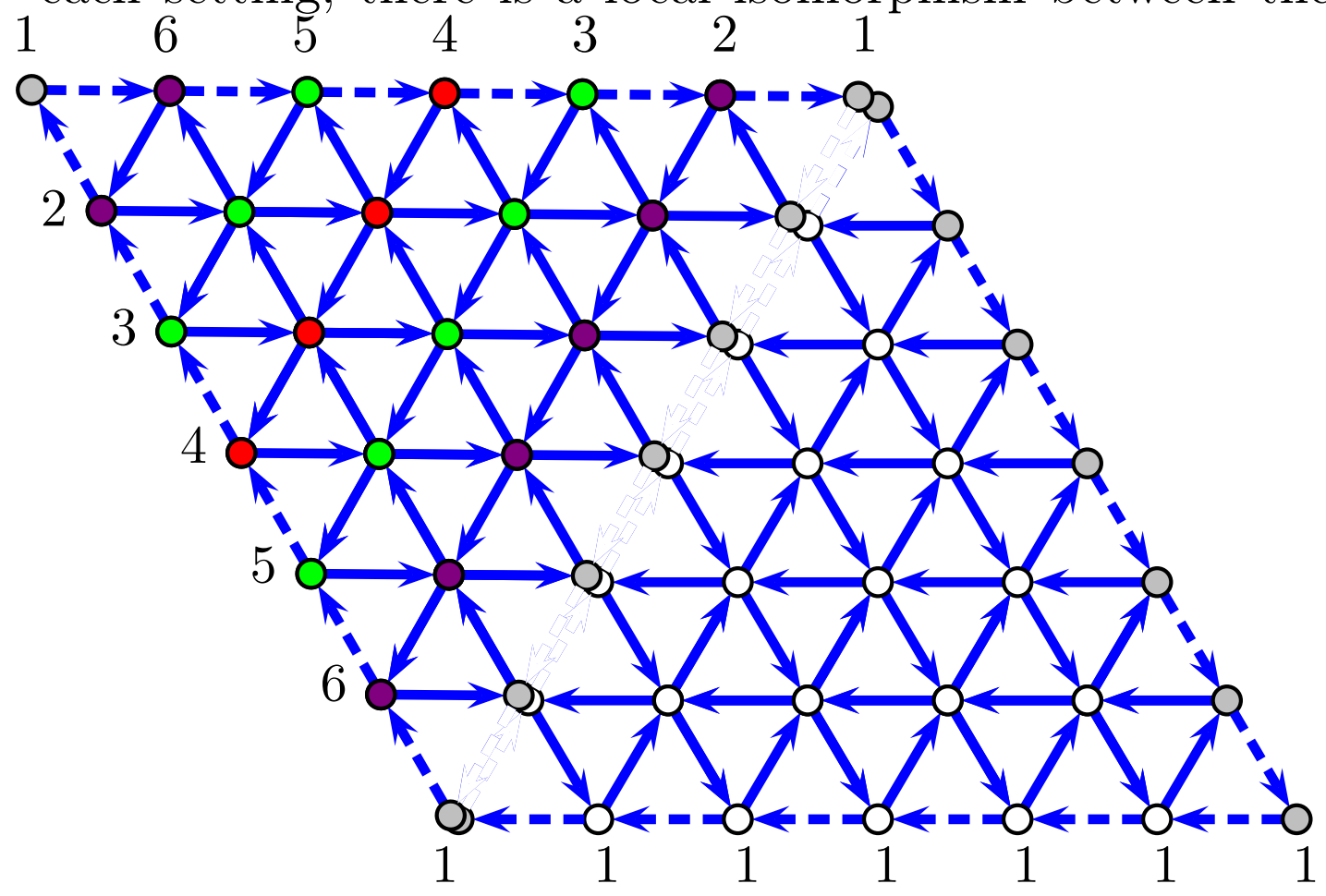
\begin{pspicture}(-6,-3)(4,4){\psset{unit=1}
    \newcommand{\LEFTDOWNARROW}{%
      {\psset{unit=1}
        \rput(0,0){\psline[doubleline=true,linewidth=1pt, doublesep=1pt, linecolor=black]{<-}(0,0)(.765,.45)}
      }}
    \newcommand{\DOWNARROW}{%
      {\psset{unit=1}
        \rput(0,0){\psline[doubleline=true,linewidth=1pt, doublesep=1pt, linecolor=black]{->}(0,0.1)(0,-0.566)}
        \put(0,0){\pscircle[fillstyle=solid,fillcolor=white]{.1}}
      }}
    \newcommand{\LEFTUPARROW}{%
      {\psset{unit=1}
        \rput(0,0){\psline[doubleline=true,linewidth=1pt, doublesep=1pt, linecolor=black]{->}(0,0)(-.765,.45)}
      }}
    \newcommand{\LEFTUPARROWSHORT}{%
      {\psset{unit=1}
        \rput(0,0){\psline[doubleline=true,linewidth=1pt, doublesep=1pt, linecolor=black]{->}(0,0)(-.45,.23)}
      }}
    \newcommand{\STARUP}{%
      {\psset{unit=1}
	\rput(0,0){\psline[doubleline=true,linewidth=1pt, doublesep=1pt, linecolor=black]{<-}(0.06,-0.03)(.4,-.24)}
	\rput(0,0){\psline[doubleline=true,linewidth=1pt, doublesep=1pt, linecolor=black]{<-}(0,0.1)(0,.466)}
	\rput(0,0){\psline[doubleline=true,linewidth=1pt, doublesep=1pt, linecolor=black]{->}(0,0)(-.45,-.23)}
	\put(0,0){\pscircle[fillstyle=solid,fillcolor=white]{.1}}
	\put(0,.566){\pscircle[fillstyle=solid,fillcolor=white]{.1}}
      }}
    \newcommand{\PATGEN}{%
      {\psset{unit=1}
        \rput(0,0){\psline[linecolor=blue,linewidth=2pt]{->}(0,0)(.45,.765)}
        \rput(0,0){\psline[linecolor=blue,linewidth=2pt]{->}(1,0)(0.1,0)}
        \rput(0,0){\psline[linecolor=blue,linewidth=2pt]{->}(0,0)(.45,-.765)}
        \put(0,0){\pscircle[fillstyle=solid,fillcolor=white]{.1}}
      }}
    \newcommand{\PATLEFT}{%
      {\psset{unit=1}
        \rput(0,0){\psline[linecolor=blue,linewidth=2pt,linestyle=dashed]{->}(0.05,0.085)(.45,.765)}
        \rput(0,0){\psline[linecolor=blue,linewidth=2pt]{->}(1,0)(0.1,0)}
        \rput(0,0){\psline[linecolor=blue,linewidth=2pt]{->}(0,0)(.45,-.765)}
        \put(0,0){\pscircle[fillstyle=solid,fillcolor=white]{.1}}
      }}
    \newcommand{\PATRIGHT}{%
      {\psset{unit=1}
        \rput(0,0){\psline[linecolor=blue,linewidth=2pt,linestyle=dashed]{->}(0.05,-0.085)(.45,-.765)}
        \put(0,0){\pscircle[fillstyle=solid,fillcolor=lightgray]{.1}}
      }}
    \newcommand{\PATBOTTOM}{%
      {\psset{unit=1}
        \rput(0,0){\psline[linecolor=blue,linewidth=2pt]{->}(0,0)(.45,.765)}
        \rput(0,0){\psline[linecolor=blue,linewidth=2pt,linestyle=dashed]{->}(0.9,0)(0.1,0)}
        \put(0,0){\pscircle[fillstyle=solid,fillcolor=white]{.1}}
      }}
    \newcommand{\PATTOP}{%
      {\psset{unit=1}
        \rput(0,0){\psline[linecolor=blue,linewidth=2pt]{->}(1,0)(0.1,0)}
        \rput(0,0){\psline[linecolor=blue,linewidth=2pt]{->}(0,0)(.45,-.765)}
        \put(0,0){\pscircle[fillstyle=solid,fillcolor=white]{.1}}
      }}
    \newcommand{\PATBOTRIGHT}{%
      {\psset{unit=1}
        \rput(0,0){\psline[linecolor=blue,linewidth=2pt]{->}(0,0)(.45,.765)}
        \put(0,0){\pscircle[fillstyle=solid,fillcolor=white]{.1}}
        \put(.5,0.85){\pscircle[fillstyle=solid,fillcolor=lightgray]{.1}}
      }}
    \newcommand{\PATNORTH}{%
      {\psset{unit=1}
        \rput(0,0){\psline[linecolor=blue,linewidth=2pt,linestyle=dashed]{->}(-.45,-.765)(-0.05,-0.085)}
        \rput(0,0){\psline[linecolor=blue,linewidth=2pt,linestyle=dashed]{<-}(.45,-.765)(0.05,-0.085)}
        \put(0,0){\pscircle[fillstyle=solid,fillcolor=lightgray]{.1}}
      }}
    \newcommand{\PATSW}{%
      {\psset{unit=1}
        \rput(0,0){\psline[linecolor=blue,linewidth=2pt,linestyle=dashed]{->}(0.05,0.085)(.45,.765)}
        \rput(0,0){\psline[linecolor=blue,linewidth=2pt,linestyle=dashed]{<-}(0.1,0)(0.9,0)}
        \put(0,0){\pscircle[fillstyle=solid,fillcolor=lightgray]{.1}}
      }}
    \newcommand{\PATSE}{%
      {\psset{unit=1}
        \rput(0,0){\psline[linecolor=blue,linewidth=2pt,linestyle=dashed]{<-}(-0.05,0.085)(-.45,.765)}
        \rput(0,0){\psline[linecolor=blue,linewidth=2pt,linestyle=dashed]{<-}(-0.9,0)(-0.1,0)}
        \put(0,0){\pscircle[fillstyle=solid,fillcolor=lightgray]{.1}}
      }}
    \multiput(-2.5,-0.85)(0.5,0.85){4}{\PATLEFT}
    \multiput(-2,-1.7)(1,0){4}{\PATBOTTOM}
    \put(-0.5,2.55){\PATTOP}
    \put(0,3.4){\PATNORTH}
    \multiput(-1.5,-0.85)(1,0){4}{\PATGEN}
    \multiput(-1,0)(1,0){3}{\PATGEN}
    \multiput(-.5,0.85)(1,0){2}{\PATGEN}
    \put(0,1.7){\PATGEN}
    \multiput(-1.5,-0.85)(1,0){4}{\PATGEN}
    \multiput(0.5,2.55)(0.5,-0.85){4}{\PATRIGHT}
    \put(2,-1.7){\PATBOTRIGHT}
    \put(-3,-1.7){\PATSW}
    \put(3,-1.7){\PATSE}
    \multiput(-3,-1.7)(0.5,0.85){6}{\psline[linecolor=white,linewidth=2pt,linestyle=dashed]{->}(0.05,0.085)(.45,.765)}
    %
    \rput{-60}(-2.35,2.2){{\multiput(-2.5,-0.85)(0.5,0.85){4}{\PATLEFT}}
      \multiput(-2,-1.7)(1,0){4}{\PATBOTTOM}
      \put(-0.5,2.55){\PATTOP}
      \put(0,3.4){\PATNORTH}
      \multiput(-1.5,-0.85)(1,0){4}{\PATGEN}
      \multiput(-1,0)(1,0){3}{\PATGEN}
      \multiput(-.5,0.85)(1,0){2}{\PATGEN}
      \put(0,1.7){\PATGEN}
      \multiput(-1.5,-0.85)(1,0){4}{\PATGEN}
      \multiput(0.5,2.55)(0.5,-0.85){4}{\PATRIGHT}
      \put(2,-1.7){\PATBOTRIGHT}
      \put(-3,-1.7){\PATSW}
      \put(3,-1.7){\PATSE}
      \multiput(-0.5,2.55)(0.5,-0.85){6}{\pscircle[fillstyle=solid,fillcolor=violet]{.1}}
      \multiput(-1,1.7)(0.5,-0.85){5}{\pscircle[fillstyle=solid,fillcolor=green]{.1}}
      \multiput(-1.5,0.85)(0.5,-0.85){4}{\pscircle[fillstyle=solid,fillcolor=red]{.1}}
      \multiput(-2,0)(0.5,-0.85){3}{\pscircle[fillstyle=solid,fillcolor=green]{.1}}
      \multiput(-2.5,-0.85)(0.5,-0.85){2}{\pscircle[fillstyle=solid,fillcolor=violet]{.1}}
      \multiput(0,3.4)(0.5,-0.85){6}{\psline[linecolor=white,linewidth=2pt,linestyle=dashed]{->}(0.05,-0.085)(.45,-.765)}
    }
    \put(0,3.8){\makebox(0,0)[br]{\hbox{{ $1$}}}}
    \put(-1,3.8){\makebox(0,0)[br]{\hbox{{ $2$}}}}
    \put(-2,3.8){\makebox(0,0)[br]{\hbox{{ $3$}}}}
    \put(-3,3.8){\makebox(0,0)[br]{\hbox{{ $4$}}}}
    \put(-4,3.8){\makebox(0,0)[br]{\hbox{{ $5$}}}}
    \put(-5,3.8){\makebox(0,0)[br]{\hbox{{ $6$}}}}
    \put(-6,3.8){\makebox(0,0)[br]{\hbox{{ $1$}}}}
    \put(-5.8,2.8){\makebox(0,0)[tr]{\hbox{{ $2$}}}}
    \put(-5.3, 1.95){\makebox(0,0)[tr]{\hbox{{ $3$}}}}
    \put(-4.8,1.1){\makebox(0,0)[tr]{\hbox{{ $4$}}}}
    \put(-4.3,0.25){\makebox(0,0)[tr]{\hbox{{ $5$}}}}
    \put(-3.8,-0.6){\makebox(0,0)[tr]{\hbox{{ $6$}}}}
    \put(-3,-1.9){\makebox(0,0)[tr]{\hbox{{ $1$}}}}
    \put(-2,-1.9){\makebox(0,0)[tc]{\hbox{{ $1$}}}}
    \put(-1,-1.9){\makebox(0,0)[tc]{\hbox{{ $1$}}}}
    \put(0,-1.9){\makebox(0,0)[tc]{\hbox{{ $1$}}}}
    \put(1,-1.9){\makebox(0,0)[tc]{\hbox{{ $1$}}}}
    \put(2,-1.9){\makebox(0,0)[tc]{\hbox{{ $1$}}}}
    \put(3,-1.9){\makebox(0,0)[tc]{\hbox{{ $1$}}}}
    
    %
    
  }
\end{pspicture}
\)

In the case $n=6$, we have three more Casimirs in addition to the three Casimirs in the left FG triangle: one is just the product of \emph{all} variables; we do not depict it. Two more are products of monomials of cluster variables in the left FG triangle that were Casimirs used for making $A$ unipotent in the FG case (see Fig.~\ref{fi:diagonal-Casimirs} for examples of these Casimirs) and Casimirs in Fig.~\ref{fi:Cas-incomplete} in the right FG triangle. Powers of variables on the boundary separating two FG triangles must match for making these variables amalgamated pairwise into new (nonfrozen) variables.

We use the color code: white/grey vertex for variables not contributing to the Casimir; green for variables with power $1$, double green circles for variables with power $2$, red for variables with power $-1$, and double red circles for variables with power $-2$. Then the two new Casimir operators in Fig.~\ref{fi:diagonal-Casimirs} are depicted in Fig.~\ref{fi:fullCasimirs} below.

\begin{figure}[H]
  \centering
  \begin{pspicture}(-5,-2.5)(2,4){\psset{unit=1}
      \newcommand{\PATGEN}{%
        {\psset{unit=1}
          \rput(0,0){\psline[linecolor=blue,linewidth=2pt]{->}(0,0)(.45,.765)}
          \rput(0,0){\psline[linecolor=blue,linewidth=2pt]{->}(1,0)(0.1,0)}
          \rput(0,0){\psline[linecolor=blue,linewidth=2pt]{->}(0,0)(.45,-.765)}
          \put(0,0){\pscircle[fillstyle=solid,fillcolor=white]{.1}}
        }}
      \newcommand{\PATLEFT}{%
        {\psset{unit=1}
          \rput(0,0){\psline[linecolor=blue,linewidth=2pt,linestyle=dashed]{->}(0.05,0.085)(.45,.765)}
          \rput(0,0){\psline[linecolor=blue,linewidth=2pt]{->}(1,0)(0.1,0)}
          \rput(0,0){\psline[linecolor=blue,linewidth=2pt]{->}(0,0)(.45,-.765)}
          \put(0,0){\pscircle[fillstyle=solid,fillcolor=white]{.1}}
        }}
      \newcommand{\PATRIGHT}{%
        {\psset{unit=1}
          \rput(0,0){\psline[linecolor=blue,linewidth=2pt,linestyle=dashed]{->}(0.05,-0.085)(.45,-.765)}
          \put(0,0){\pscircle[fillstyle=solid,fillcolor=lightgray]{.1}}
        }}
      \newcommand{\PATBOTTOM}{%
        {\psset{unit=1}
          \rput(0,0){\psline[linecolor=blue,linewidth=2pt]{->}(0,0)(.45,.765)}
          \rput(0,0){\psline[linecolor=blue,linewidth=2pt,linestyle=dashed]{->}(0.9,0)(0.1,0)}
          \put(0,0){\pscircle[fillstyle=solid,fillcolor=white]{.1}}
        }}
      \newcommand{\PATTOP}{%
        {\psset{unit=1}
          \rput(0,0){\psline[linecolor=blue,linewidth=2pt]{->}(1,0)(0.1,0)}
          \rput(0,0){\psline[linecolor=blue,linewidth=2pt]{->}(0,0)(.45,-.765)}
          \put(0,0){\pscircle[fillstyle=solid,fillcolor=white]{.1}}
        }}
      \newcommand{\PATBOTRIGHT}{%
        {\psset{unit=1}
          \rput(0,0){\psline[linecolor=blue,linewidth=2pt]{->}(0,0)(.45,.765)}
          \put(0,0){\pscircle[fillstyle=solid,fillcolor=white]{.1}}
          \put(.5,0.85){\pscircle[fillstyle=solid,fillcolor=lightgray]{.1}}
        }}
      \newcommand{\PATNORTH}{%
        {\psset{unit=1}
          \rput(0,0){\psline[linecolor=blue,linewidth=2pt,linestyle=dashed]{->}(-.45,-.765)(-0.05,-0.085)}
          \rput(0,0){\psline[linecolor=blue,linewidth=2pt,linestyle=dashed]{<-}(.45,-.765)(0.05,-0.085)}
          \put(0,0){\pscircle[fillstyle=solid,fillcolor=lightgray]{.1}}
        }}
      \newcommand{\PATSW}{%
        {\psset{unit=1}
          \rput(0,0){\psline[linecolor=blue,linewidth=2pt,linestyle=dashed]{->}(0.05,0.085)(.45,.765)}
          \rput(0,0){\psline[linecolor=blue,linewidth=2pt,linestyle=dashed]{<-}(0.1,0)(0.9,0)}
          \put(0,0){\pscircle[fillstyle=solid,fillcolor=lightgray]{.1}}
        }}
      \newcommand{\PATSE}{%
        {\psset{unit=1}
          \rput(0,0){\psline[linecolor=blue,linewidth=2pt,linestyle=dashed]{<-}(-0.05,0.085)(-.45,.765)}
          \rput(0,0){\psline[linecolor=blue,linewidth=2pt,linestyle=dashed]{<-}(-0.9,0)(-0.1,0)}
          \put(0,0){\pscircle[fillstyle=solid,fillcolor=lightgray]{.1}}
        }}
      \multiput(-0.5,2.55)(1,0){2}{\pscircle[fillstyle=solid,fillcolor=green]{.2}}
      \multiput(-0.5,2.55)(0.5,-0.85){2}{\pscircle[fillstyle=solid,fillcolor=green]{.2}}
      \multiput(-2.5,-0.85)(5,0){2}{\pscircle[fillstyle=solid,fillcolor=red]{.2}}
      \multiput(-2.5,-0.85)(0.5,0.85){4}{\PATLEFT}
      \multiput(-2,-1.7)(1,0){4}{\PATBOTTOM}
      \put(-0.5,2.55){\PATTOP}
      \put(0,3.4){\PATNORTH}
      \multiput(-1.5,-0.85)(1,0){4}{\PATGEN}
      \multiput(-1,0)(1,0){3}{\PATGEN}
      \multiput(-.5,0.85)(1,0){2}{\PATGEN}
      \put(0,1.7){\PATGEN}
      \multiput(-1.5,-0.85)(1,0){4}{\PATGEN}
      \multiput(0.5,2.55)(0.5,-0.85){4}{\PATRIGHT}
      \put(2,-1.7){\PATBOTRIGHT}
      \put(-3,-1.7){\PATSW}
      \put(3,-1.7){\PATSE}
      \multiput(-1.5,-0.85)(0.5,0.85){5}{\pscircle[fillstyle=solid,fillcolor=green]{.1}}
      \multiput(1,0)(-0.5,0.85){2}{\pscircle[fillstyle=solid,fillcolor=green]{.1}}
      \multiput(-0.5,2.55)(0.5,-0.85){1}{\pscircle[fillstyle=solid,fillcolor=green]{.1}}
      \multiput(-2.5,-0.85)(1,0){6}{\pscircle[fillstyle=solid,fillcolor=red]{.1}}
      \multiput(-1.5,-0.85)(0.5,-0.85){1}{\pscircle[fillstyle=solid,fillcolor=white]{.1}}
      \multiput(1.5,-0.85)(0.5,-0.85){1}{\pscircle[fillstyle=solid,fillcolor=white]{.1}}
      \multiput(-3,-1.7)(0.5,0.85){6}{\psline[linecolor=white,linewidth=2pt,linestyle=dashed]{->}(0.05,0.085)(.45,.765)}
      \rput{-60}(-2.35,2.2){{\multiput(-2.5,-0.85)(0.5,0.85){4}{\PATLEFT}}
        \multiput(-2,-1.7)(1,0){4}{\PATBOTTOM}
        \put(-0.5,2.55){\PATTOP}
        \put(0,3.4){\PATNORTH}
        \multiput(-1.5,-0.85)(1,0){4}{\PATGEN}
        \multiput(-1,0)(1,0){3}{\PATGEN}
        \multiput(-.5,0.85)(1,0){2}{\PATGEN}
        \put(0,1.7){\PATGEN}
        \multiput(-1.5,-0.85)(1,0){4}{\PATGEN}
        \multiput(0.5,2.55)(0.5,-0.85){4}{\PATRIGHT}
        \put(2,-1.7){\PATBOTRIGHT}
        \put(-3,-1.7){\PATSW}
        \put(3,-1.7){\PATSE}
        \multiput(-2,-1.7)(0.5,0.85){6}{\pscircle[fillstyle=solid,fillcolor=green]{.1}}
        \multiput(-0.5,2.55)(0.5,-0.85){1}{\pscircle[fillstyle=solid,fillcolor=green]{.1}}
        \multiput(-2.5,-0.85)(1,0){6}{\pscircle[fillstyle=solid,fillcolor=red]{.1}}
        \multiput(2,-1.7)(0.5,-0.85){1}{\pscircle[fillstyle=solid,fillcolor=red]{.1}}
        \multiput(-1.5,-0.85)(0.5,-0.85){1}{\pscircle[fillstyle=solid,fillcolor=white]{.1}}
        \multiput(0,3.4)(0.5,-0.85){6}{\psline[linecolor=white,linewidth=2pt,linestyle=dashed]{->}(0.05,-0.085)(.45,-.765)}
      }
      \multiput(-0.5,2.55)(0.5,-0.85){1}{\pscircle[linewidth=2pt,linecolor=green]{.2}}
      \multiput(-2.5,-0.85)(5,0){1}{\pscircle[linewidth=2pt,linecolor=red]{.2}}
      \put(0,3.8){\makebox(0,0)[br]{\hbox{{ $1$}}}}
      \put(-1,3.8){\makebox(0,0)[br]{\hbox{{ $2$}}}}
      \put(-2,3.8){\makebox(0,0)[br]{\hbox{{ $3$}}}}
      \put(-3,3.8){\makebox(0,0)[br]{\hbox{{ $4$}}}}
      \put(-4,3.8){\makebox(0,0)[br]{\hbox{{ $5$}}}}
      \put(-5,3.8){\makebox(0,0)[br]{\hbox{{ $6$}}}}
      \put(-6,3.8){\makebox(0,0)[br]{\hbox{{ $1$}}}}
      \put(-5.8,2.8){\makebox(0,0)[tr]{\hbox{{ $2$}}}}
      \put(-5.3, 1.95){\makebox(0,0)[tr]{\hbox{{ $3$}}}}
      \put(-4.8,1.1){\makebox(0,0)[tr]{\hbox{{ $4$}}}}
      \put(-4.3,0.25){\makebox(0,0)[tr]{\hbox{{ $5$}}}}
      \put(-3.8,-0.6){\makebox(0,0)[tr]{\hbox{{ $6$}}}}
      \put(-3,-1.9){\makebox(0,0)[tr]{\hbox{{ $1$}}}}
      \put(-2,-1.9){\makebox(0,0)[tc]{\hbox{{ $1$}}}}
      \put(-1,-1.9){\makebox(0,0)[tc]{\hbox{{ $1$}}}}
      \put(0,-1.9){\makebox(0,0)[tc]{\hbox{{ $1$}}}}
      \put(1,-1.9){\makebox(0,0)[tc]{\hbox{{ $1$}}}}
      \put(2,-1.9){\makebox(0,0)[tc]{\hbox{{ $1$}}}}
      \put(3,-1.9){\makebox(0,0)[tc]{\hbox{{ $1$}}}}
      
      %
      
    }
  \end{pspicture}
  \begin{pspicture}(-5,-2.5)(2,4){\psset{unit=1}
      \newcommand{\PATGEN}{%
        {\psset{unit=1}
          \rput(0,0){\psline[linecolor=blue,linewidth=2pt]{->}(0,0)(.45,.765)}
          \rput(0,0){\psline[linecolor=blue,linewidth=2pt]{->}(1,0)(0.1,0)}
          \rput(0,0){\psline[linecolor=blue,linewidth=2pt]{->}(0,0)(.45,-.765)}
          \put(0,0){\pscircle[fillstyle=solid,fillcolor=white]{.1}}
        }}
      \newcommand{\PATLEFT}{%
        {\psset{unit=1}
          \rput(0,0){\psline[linecolor=blue,linewidth=2pt,linestyle=dashed]{->}(0.05,0.085)(.45,.765)}
          \rput(0,0){\psline[linecolor=blue,linewidth=2pt]{->}(1,0)(0.1,0)}
          \rput(0,0){\psline[linecolor=blue,linewidth=2pt]{->}(0,0)(.45,-.765)}
          \put(0,0){\pscircle[fillstyle=solid,fillcolor=white]{.1}}
        }}
      \newcommand{\PATRIGHT}{%
        {\psset{unit=1}
          \rput(0,0){\psline[linecolor=blue,linewidth=2pt,linestyle=dashed]{->}(0.05,-0.085)(.45,-.765)}
          \put(0,0){\pscircle[fillstyle=solid,fillcolor=lightgray]{.1}}
        }}
      \newcommand{\PATBOTTOM}{%
        {\psset{unit=1}
          \rput(0,0){\psline[linecolor=blue,linewidth=2pt]{->}(0,0)(.45,.765)}
          \rput(0,0){\psline[linecolor=blue,linewidth=2pt,linestyle=dashed]{->}(0.9,0)(0.1,0)}
          \put(0,0){\pscircle[fillstyle=solid,fillcolor=white]{.1}}
        }}
      \newcommand{\PATTOP}{%
        {\psset{unit=1}
          \rput(0,0){\psline[linecolor=blue,linewidth=2pt]{->}(1,0)(0.1,0)}
          \rput(0,0){\psline[linecolor=blue,linewidth=2pt]{->}(0,0)(.45,-.765)}
          \put(0,0){\pscircle[fillstyle=solid,fillcolor=white]{.1}}
        }}
      \newcommand{\PATBOTRIGHT}{%
        {\psset{unit=1}
          \rput(0,0){\psline[linecolor=blue,linewidth=2pt]{->}(0,0)(.45,.765)}
          \put(0,0){\pscircle[fillstyle=solid,fillcolor=white]{.1}}
          \put(.5,0.85){\pscircle[fillstyle=solid,fillcolor=lightgray]{.1}}
        }}
      \newcommand{\PATNORTH}{%
        {\psset{unit=1}
          \rput(0,0){\psline[linecolor=blue,linewidth=2pt,linestyle=dashed]{->}(-.45,-.765)(-0.05,-0.085)}
          \rput(0,0){\psline[linecolor=blue,linewidth=2pt,linestyle=dashed]{<-}(.45,-.765)(0.05,-0.085)}
          \put(0,0){\pscircle[fillstyle=solid,fillcolor=lightgray]{.1}}
        }}
      \newcommand{\PATSW}{%
        {\psset{unit=1}
          \rput(0,0){\psline[linecolor=blue,linewidth=2pt,linestyle=dashed]{->}(0.05,0.085)(.45,.765)}
          \rput(0,0){\psline[linecolor=blue,linewidth=2pt,linestyle=dashed]{<-}(0.1,0)(0.9,0)}
          \put(0,0){\pscircle[fillstyle=solid,fillcolor=lightgray]{.1}}
        }}
      \newcommand{\PATSE}{%
        {\psset{unit=1}
          \rput(0,0){\psline[linecolor=blue,linewidth=2pt,linestyle=dashed]{<-}(-0.05,0.085)(-.45,.765)}
          \rput(0,0){\psline[linecolor=blue,linewidth=2pt,linestyle=dashed]{<-}(-0.9,0)(-0.1,0)}
          \put(0,0){\pscircle[fillstyle=solid,fillcolor=lightgray]{.1}}
        }}
      \multiput(-1,1.7)(2,0){2}{\pscircle[fillstyle=solid,fillcolor=green]{.2}}
      \multiput(-2,0)(4,0){2}{\pscircle[fillstyle=solid,fillcolor=red]{.2}}
      \multiput(-2.5,-0.85)(0.5,0.85){4}{\PATLEFT}
      \multiput(-2,-1.7)(1,0){4}{\PATBOTTOM}
      \put(-0.5,2.55){\PATTOP}
      \put(0,3.4){\PATNORTH}
      \multiput(-1.5,-0.85)(1,0){4}{\PATGEN}
      \multiput(-1,0)(1,0){3}{\PATGEN}
      \multiput(-.5,0.85)(1,0){2}{\PATGEN}
      \put(0,1.7){\PATGEN}
      \multiput(-1.5,-0.85)(1,0){4}{\PATGEN}
      \multiput(0.5,2.55)(0.5,-0.85){4}{\PATRIGHT}
      \put(2,-1.7){\PATBOTRIGHT}
      \put(-3,-1.7){\PATSW}
      \put(3,-1.7){\PATSE}
      \multiput(-2,0)(1,0){5}{\pscircle[fillstyle=solid,fillcolor=red]{.1}}
      \multiput(-1,1.7)(0.5,-0.85){4}{\pscircle[fillstyle=solid,fillcolor=green]{.1}}
      \multiput(1,1.7)(-0.5,-0.85){4}{\pscircle[fillstyle=solid,fillcolor=green]{.1}}
      \multiput(0,1.7)(-0.5,0.85){1}{\pscircle[fillstyle=solid,fillcolor=green]{.1}}
      \multiput(-1.5,-0.85)(3,0){2}{\pscircle[fillstyle=solid,fillcolor=red]{.1}}
      \multiput(-3,-1.7)(0.5,0.85){6}{\psline[linecolor=white,linewidth=2pt,linestyle=dashed]{->}(0.05,0.085)(.45,.765)}
      %
      \rput{-60}(-2.35,2.2){{\multiput(-2.5,-0.85)(0.5,0.85){4}{\PATLEFT}}
        \multiput(-2,-1.7)(1,0){4}{\PATBOTTOM}
        \put(-0.5,2.55){\PATTOP}
        \put(0,3.4){\PATNORTH}
        \multiput(-1.5,-0.85)(1,0){4}{\PATGEN}
        \multiput(-1,0)(1,0){3}{\PATGEN}
        \multiput(-.5,0.85)(1,0){2}{\PATGEN}
        \put(0,1.7){\PATGEN}
        \multiput(-1.5,-0.85)(1,0){4}{\PATGEN}
        \multiput(0.5,2.55)(0.5,-0.85){4}{\PATRIGHT}
        \put(2,-1.7){\PATBOTRIGHT}
        \put(-3,-1.7){\PATSW}
        \put(3,-1.7){\PATSE}
        \multiput(-1,-1.7)(0.5,0.85){5}{\pscircle[fillstyle=solid,fillcolor=green]{.1}}
        \multiput(0,1.7)(-1,0){2}{\pscircle[fillstyle=solid,fillcolor=green]{.1}}
        \multiput(-2,0)(1,0){5}{\pscircle[fillstyle=solid,fillcolor=red]{.1}}
        \multiput(1,-1.7)(0.5,0.85){2}{\pscircle[fillstyle=solid,fillcolor=red]{.1}}
        \multiput(0,0)(0.5,-0.85){1}{\pscircle[fillstyle=solid,fillcolor=white]{.1}}
        \multiput(0,3.4)(0.5,-0.85){6}{\psline[linecolor=white,linewidth=2pt,linestyle=dashed]{->}(0.05,-0.085)(.45,-.765)}
      }
      \multiput(-1,1.7)(0.5,-0.85){1}{\pscircle[linewidth=2pt,linecolor=green]{.2}}
      \multiput(-2,0)(5,0){1}{\pscircle[linewidth=2pt,linecolor=red]{.2}}
      \put(0,3.8){\makebox(0,0)[br]{\hbox{{ $1$}}}}
      \put(-1,3.8){\makebox(0,0)[br]{\hbox{{ $2$}}}}
      \put(-2,3.8){\makebox(0,0)[br]{\hbox{{ $3$}}}}
      \put(-3,3.8){\makebox(0,0)[br]{\hbox{{ $4$}}}}
      \put(-4,3.8){\makebox(0,0)[br]{\hbox{{ $5$}}}}
      \put(-5,3.8){\makebox(0,0)[br]{\hbox{{ $6$}}}}
      \put(-6,3.8){\makebox(0,0)[br]{\hbox{{ $1$}}}}
      \put(-5.8,2.8){\makebox(0,0)[tr]{\hbox{{ $2$}}}}
      \put(-5.3, 1.95){\makebox(0,0)[tr]{\hbox{{ $3$}}}}
      \put(-4.8,1.1){\makebox(0,0)[tr]{\hbox{{ $4$}}}}
      \put(-4.3,0.25){\makebox(0,0)[tr]{\hbox{{ $5$}}}}
      \put(-3.8,-0.6){\makebox(0,0)[tr]{\hbox{{ $6$}}}}
      \put(-3,-1.9){\makebox(0,0)[tr]{\hbox{{ $1$}}}}
      \put(-2,-1.9){\makebox(0,0)[tc]{\hbox{{ $1$}}}}
      \put(-1,-1.9){\makebox(0,0)[tc]{\hbox{{ $1$}}}}
      \put(0,-1.9){\makebox(0,0)[tc]{\hbox{{ $1$}}}}
      \put(1,-1.9){\makebox(0,0)[tc]{\hbox{{ $1$}}}}
      \put(2,-1.9){\makebox(0,0)[tc]{\hbox{{ $1$}}}}
      \put(3,-1.9){\makebox(0,0)[tc]{\hbox{{ $1$}}}}
      
      %
      
    }
  \end{pspicture}
  \caption{\small
    Two new central elements, $\mathfrak C_1$ and $\mathfrak C_2$, of the full-size quiver for $n=6$. }
  \label{fi:fullCasimirs}
\end{figure}
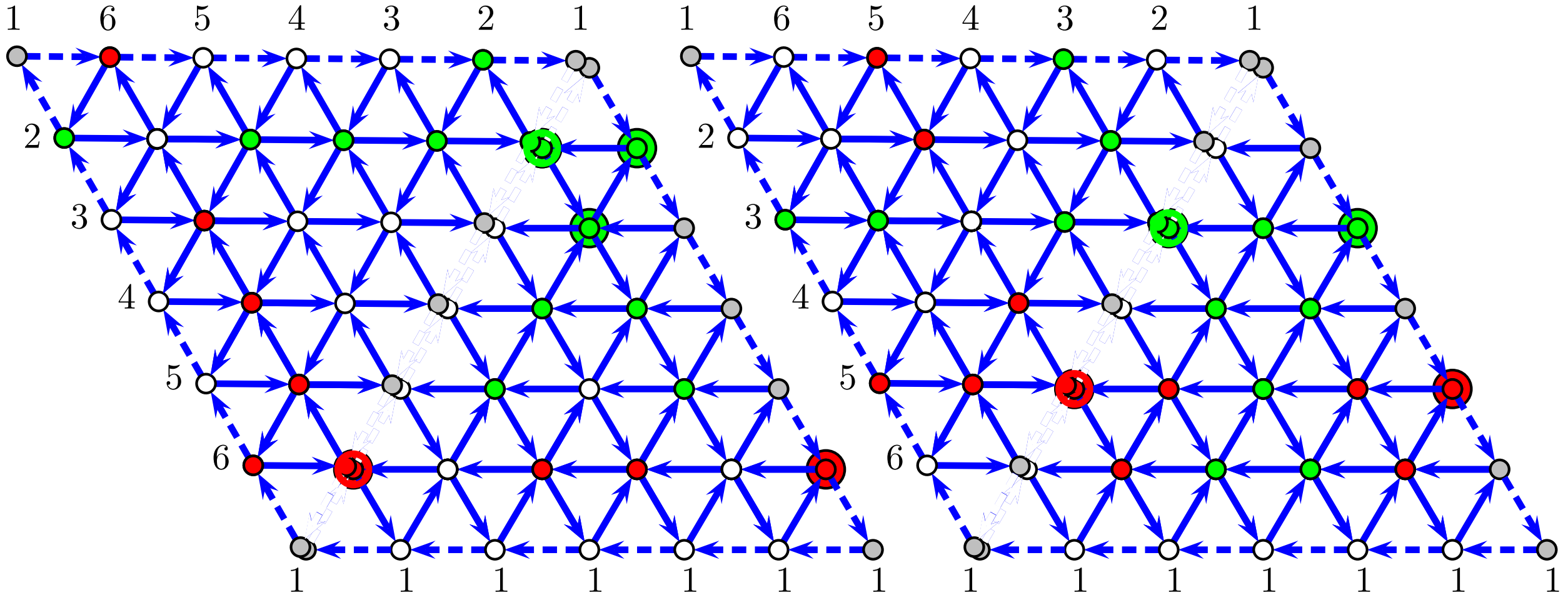

In the semiclassical case, we have the following statement.
\begin{thm}
  For normalized transport matrices $M_i$, all Casimirs ${\mathfrak M}_{n-k}/{\mathfrak M}_k$, $k=1,\dots,\lfloor \frac{n-1}{2}\rfloor$, that are ratios of $k\times k$ minors ${\mathfrak M}_k$ of the matrix $\mathbb A$ located at the lower-left corner are expressed as products of (generally fractional) powers of Casimirs $\mathfrak C_i$, $i=1,\dots,\lfloor \frac{n-1}{2}\rfloor$, of the type depicted in Fig.~\ref{fi:fullCasimirs}.
\end{thm}

The {\it proof} is constructive: we have three transport matrices: $M_1$ and $M_2$ in the left FG triangle, and $\widetilde M_1$ in the right FG triangle, and the full-size matrix $\mathbb A$ can be written as
$$
\mathbb A=\bigl(M_1\widetilde M_1\bigr)^{\text{T}}M_2\widetilde M_1=\bigl(\widetilde M_1{}^{\text{T}}M_1^{\text{T}}M_2\bigr)\widetilde M_1.
$$
The three left matrices (matrices in the parentheses) in the above expression are upper-triangular, so when evaluating minors $\mathfrak M_k$ of $\mathbb A$, only their diagonal elements are relevant. So, we first concentrate on the minors $\widetilde{\mathfrak M}_k$ of the only lower-triangular matrix $\widetilde M_1$.

We evaluate non-normalized $\widetilde{\mathfrak M}_k$ using the Lindstr\"om theorem~\cite{Lindstrom1973}, which expresses each minor $\widetilde{\mathfrak M}_k$ in the directed acyclic planar network as a sum over products of weights of all collections of $k$ vertex-disjoint paths going from $k$ sources to $k$ sinks; it is easy to see that in our case such set of paths is unique for each $k$, so the minor $\widetilde{\mathfrak M}_k$ of $\widetilde M_1$ is a monomial in cluster variables of the right FG triangle. In order to extract the complete dependence on cluster variables of the right FG triangle, we then have to consider the ratio of two monomials, $\widetilde{\mathfrak M}_{n-k}/\widetilde{\mathfrak M}_k$, multiply it by the product $\prod_{i=k+1}^{n-k}[\widetilde M_1]_{i,i}$ of diagonal elements of $\widetilde M_1$, and finally normalize the whole expression by dividing it by $[\det \widetilde M_1]^{2(n-2k)/n}$. Note that the resulting power of the variable at the uppermost vertex of the FG triangle is then zero for any $k$, and the rest of powers matches a linear combination of powers of variables in $\mathfrak C_i$.

\begin{exm}\label{ex:full-quiver}
  Let us consider $\widetilde{\mathfrak M}_4/\widetilde{\mathfrak M}_2$ for the case $n=6$. For a non-normalized $\widetilde M_1$, the corresponding minors, the product $[\widetilde M_1]_{3,3}[\widetilde M_1]_{4,4}$, and the determinant $\det \widetilde M_1$ are all monomial expressions; numbers indicate the powers of the respective variables in each of these expressions. We indicate by color in each diagram cluster variables that do not commute with the corresponding monomial.
  $$
  
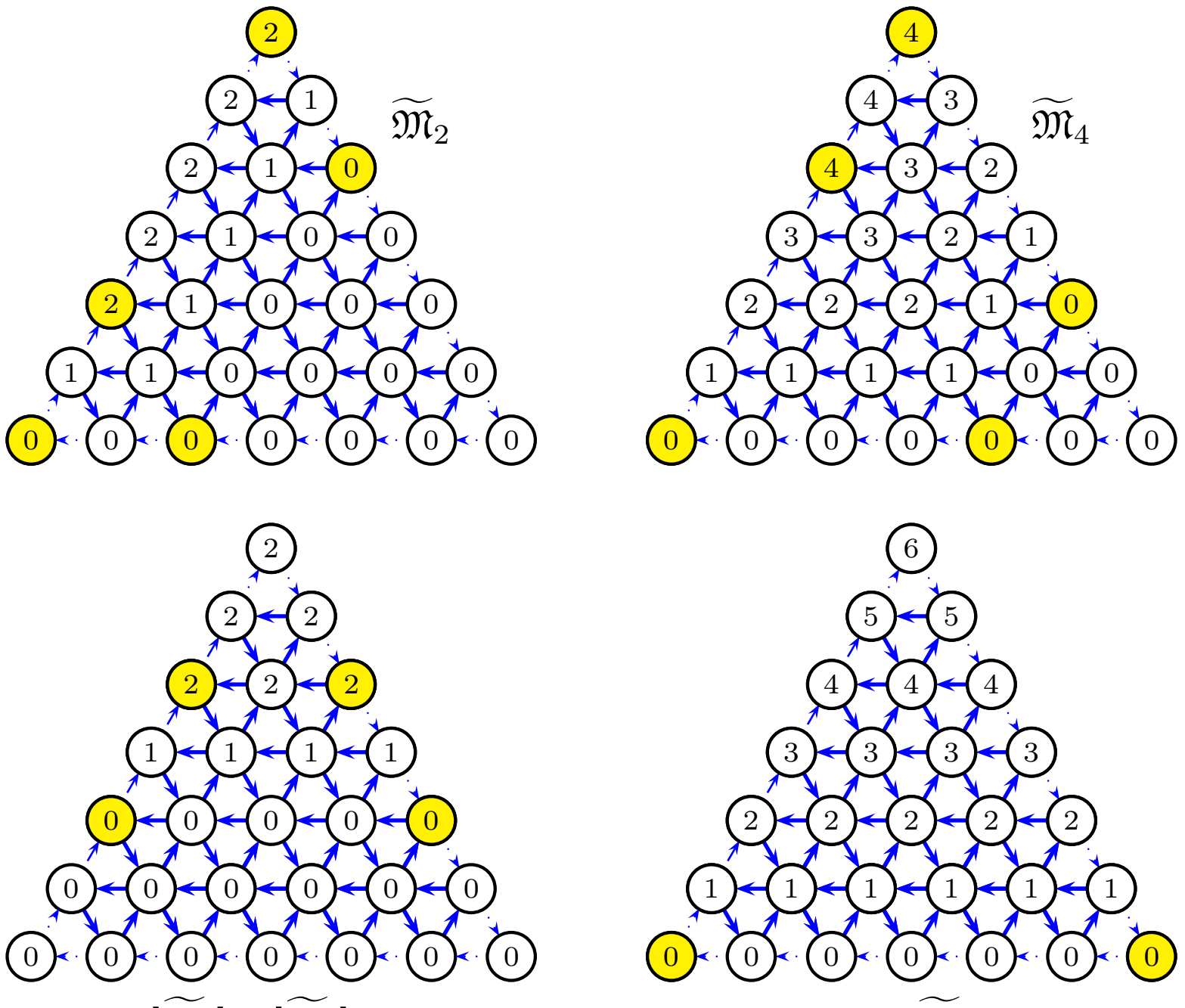
\begin{pspicture}(-6,-4.5)(6,4.5){\psset{unit=0.7}
      \newcommand{\PATGEN}{%
        {\psset{unit=1}
          \rput(0,0){\psline[linecolor=blue,linewidth=1pt]{->}(0,0)(.35,.6)}
          \rput(0,0){\psline[linecolor=blue,linewidth=1pt]{->}(0.7,0)(0.3,0)}
          \rput(0,0){\psline[linecolor=blue,linewidth=1pt]{->}(0,0)(.35,-.6)}
          \put(0,0){\pscircle[fillstyle=solid,fillcolor=white]{.3}}
          \put(0,0){\makebox(0,0){\hbox{{\tiny $0$}}}}
        }}
      \newcommand{\PATLEFT}{%
        {\psset{unit=1}
          \rput(0,0){\psline[linecolor=blue,linewidth=0.5pt,linestyle=dashed]{->}(0.05,0.085)(.35,.6)}
          \rput(0,0){\psline[linecolor=blue,linewidth=1pt]{->}(0.7,0)(0.3,0)}
          \rput(0,0){\psline[linecolor=blue,linewidth=1pt]{->}(0,0)(.35,-.6)}
          \put(0,0){\pscircle[fillstyle=solid,fillcolor=white]{.3}}
          \put(0,0){\makebox(0,0){\hbox{{\tiny $0$}}}}
        }}
      \newcommand{\PATRIGHT}{%
        {\psset{unit=1}
          \rput(0,0){\psline[linecolor=blue,linewidth=0.5pt,linestyle=dotted]{->}(0.05,-0.085)(.35,-.6)}
          \put(0,0){\pscircle[fillstyle=solid,fillcolor=white]{.3}}
          \put(0,0){\makebox(0,0){\hbox{{\tiny $0$}}}}
        }}
      \newcommand{\PATBOTTOM}{%
        {\psset{unit=1}
          \rput(0,0){\psline[linecolor=blue,linewidth=1pt]{->}(0,0)(.35,.6)}
          \rput(0,0){\psline[linecolor=blue,linewidth=0.5pt,linestyle=dotted]{->}(0.7,0)(0.3,0)}
          \put(0,0){\pscircle[fillstyle=solid,fillcolor=white]{.3}}
          \put(0,0){\makebox(0,0){\hbox{{\tiny $0$}}}}
        }}
      \newcommand{\PATTOP}{%
        {\psset{unit=1}
          \rput(0,0){\psline[linecolor=blue,linewidth=1pt]{->}(0.7,0)(0.3,0)}
          \rput(0,0){\psline[linecolor=blue,linewidth=1pt]{->}(0,0)(.35,-.6)}
          \put(0,0){\pscircle[fillstyle=solid,fillcolor=white]{.3}}
          \put(0,0){\makebox(0,0){\hbox{{\tiny $0$}}}}
        }}
      \newcommand{\PATBOTRIGHT}{%
        {\psset{unit=1}
          \rput(0,0){\psline[linecolor=blue,linewidth=1pt]{->}(0,0)(.35,.6)}
          \put(0,0){\pscircle[fillstyle=solid,fillcolor=white]{.3}}
          \put(0,0){\makebox(0,0){\hbox{{\tiny $0$}}}}
          \put(.5,0.85)
          {\pscircle[fillstyle=solid,fillcolor=white]{.3}}
          \put(0.5,0.85){\makebox(0,0){\hbox{{\tiny $0$}}}}
        }}
      \newcommand{\PATNORTH}{%
        {\psset{unit=1}
          \rput(0,0){\psline[linecolor=blue,linewidth=0.5pt,linestyle=dotted]{->}(-.35,-.6)(-0.15,-0.255)}
          \rput(0,0){\psline[linecolor=blue,linewidth=0.5pt,linestyle=dotted]{<-}(.35,-.6)(0.15,-0.255)}
          \put(0,0){\pscircle[fillstyle=solid,fillcolor=white]{.3}}
          \put(0,0){\makebox(0,0){\hbox{{\tiny $0$}}}}
        }}
      \newcommand{\PATSW}{%
        {\psset{unit=1}
          \rput(0,0){\psline[linecolor=blue,linewidth=0.5pt,linestyle=dotted]{->}(0.15,0.255)(.35,.6)}
          \rput(0,0){\psline[linecolor=blue,linewidth=0.5pt,linestyle=dotted]{<-}(0.3,0)(0.7,0)}
          \put(0,0){\pscircle[fillstyle=solid,fillcolor=white]{.3}}
          \put(0,0){\makebox(0,0){\hbox{{\tiny $0$}}}}
        }}
      \newcommand{\PATSE}{%
        {\psset{unit=1}
          \rput(0,0){\psline[linecolor=blue,linewidth=0.5pt,linestyle=dotted]{<-}(-0.15,0.255)(-.35,.6)}
          \rput(0,0){\psline[linecolor=blue,linewidth=0.5pt,linestyle=dotted]{<-}(-0.7,0)(-0.3,0)}
          \put(0,0){\pscircle[fillstyle=solid,fillcolor=white]{.3}}
          \put(0,0){\makebox(0,0){\hbox{{\tiny $0$}}}}
        }}
      \rput(-4,3.5){
        \multiput(-2.5,-0.85)(0.5,0.85){4}{\PATLEFT}
        \multiput(-2,-1.7)(1,0){4}{\PATBOTTOM}
        \put(-0.5,2.55){\PATTOP}
        \put(0,3.4){\PATNORTH}
        \multiput(-1.5,-0.85)(1,0){4}{\PATGEN}
        \multiput(-1,0)(1,0){3}{\PATGEN}
        \multiput(-.5,0.85)(1,0){2}{\PATGEN}
        \put(0,1.7){\PATGEN}
        \multiput(-1.5,-0.85)(1,0){4}{\PATGEN}
        \multiput(0.5,2.55)(0.5,-0.85){4}{\PATRIGHT}
        \put(2,-1.7){\PATBOTRIGHT}
        \put(-3,-1.7){\PATSW}
        \put(3,-1.7){\PATSE}
        \multiput(-2,0)(0.5,0.85){5}{\pscircle[fillstyle=solid,fillcolor=white]{.3}
          \put(0,0){\makebox(0,0){\hbox{{\tiny $2$}}}}} \multiput(-2.5,-0.85)(1,0){2}{\pscircle[fillstyle=solid,fillcolor=white]{.3}
          \put(0,0){\makebox(0,0){\hbox{{\tiny $1$}}}}}
        \multiput(-1,0)(0.5,0.85){4}{\pscircle[fillstyle=solid,fillcolor=white]{.3}
          \put(0,0){\makebox(0,0){\hbox{{\tiny $1$}}}}}
        \put(0,3.4){\pscircle[fillstyle=solid,fillcolor=yellow]{.3}}
        \put(0,3.4){\makebox(0,0){\hbox{{\tiny $2$}}}}
        \put(-2,0){\pscircle[fillstyle=solid,fillcolor=yellow]{.3}}
        \put(-2,0){\makebox(0,0){\hbox{{\tiny $2$}}}}
        \put(1,1.7){\pscircle[fillstyle=solid,fillcolor=yellow]{.3}}
        \put(1,1.7){\makebox(0,0){\hbox{{\tiny $0$}}}}
        \put(-3,-1.7){\pscircle[fillstyle=solid,fillcolor=yellow]{.3}}
        \put(-3,-1.7){\makebox(0,0){\hbox{{\tiny $0$}}}}
        \put(-1,-1.7){\pscircle[fillstyle=solid,fillcolor=yellow]{.3}}
        \put(-1,-1.7){\makebox(0,0){\hbox{{\tiny $0$}}}}
        \put(1.5,2){\makebox(0,0)[bl]{\hbox{{\small $\widetilde{\mathfrak M}_2$}}}}
      }
      \rput(4,3.5){
        \multiput(-2.5,-0.85)(0.5,0.85){4}{\PATLEFT}
        \multiput(-2,-1.7)(1,0){4}{\PATBOTTOM}
        \put(-0.5,2.55){\PATTOP}
        \put(0,3.4){\PATNORTH}
        \multiput(-1.5,-0.85)(1,0){4}{\PATGEN}
        \multiput(-1,0)(1,0){3}{\PATGEN}
        \multiput(-.5,0.85)(1,0){2}{\PATGEN}
        \put(0,1.7){\PATGEN}
        \multiput(-1.5,-0.85)(1,0){4}{\PATGEN}
        \multiput(0.5,2.55)(0.5,-0.85){4}{\PATRIGHT}
        \put(2,-1.7){\PATBOTRIGHT}
        \put(-3,-1.7){\PATSW}
        \put(3,-1.7){\PATSE}
        \multiput(-1,1.7)(0.5,0.85){3}{\pscircle[fillstyle=solid,fillcolor=white]{.3}
          \put(0,0){\makebox(0,0){\hbox{{\tiny $4$}}}}} 
        \multiput(-1.5,0.85)(1,0){2}{\pscircle[fillstyle=solid,fillcolor=white]{.3}
          \put(0,0){\makebox(0,0){\hbox{{\tiny $3$}}}}}
        \multiput(0,1.7)(0.5,0.85){2}{\pscircle[fillstyle=solid,fillcolor=white]{.3}
          \put(0,0){\makebox(0,0){\hbox{{\tiny $3$}}}}}
        \multiput(-2,0)(1,0){3}{\pscircle[fillstyle=solid,fillcolor=white]{.3}
          \put(0,0){\makebox(0,0){\hbox{{\tiny $2$}}}}}
        \multiput(0.5,0.85)(0.5,0.85){2}{\pscircle[fillstyle=solid,fillcolor=white]{.3}
          \put(0,0){\makebox(0,0){\hbox{{\tiny $2$}}}}}
        \multiput(-2.5,-0.85)(1,0){4}{\pscircle[fillstyle=solid,fillcolor=white]{.3}
          \put(0,0){\makebox(0,0){\hbox{{\tiny $1$}}}}}
        \multiput(1,0)(0.5,0.85){2}{\pscircle[fillstyle=solid,fillcolor=white]{.3}
          \put(0,0){\makebox(0,0){\hbox{{\tiny $1$}}}}}
        \put(0,3.4){\pscircle[fillstyle=solid,fillcolor=yellow]{.3}}
        \put(0,3.4){\makebox(0,0){\hbox{{\tiny $4$}}}}
        \put(-1,1.7){\pscircle[fillstyle=solid,fillcolor=yellow]{.3}}
        \put(-1,1.7){\makebox(0,0){\hbox{{\tiny $4$}}}}
        \put(2,0){\pscircle[fillstyle=solid,fillcolor=yellow]{.3}}
        \put(2,0){\makebox(0,0){\hbox{{\tiny $0$}}}}
        \put(-3,-1.7){\pscircle[fillstyle=solid,fillcolor=yellow]{.3}}
        \put(-3,-1.7){\makebox(0,0){\hbox{{\tiny $0$}}}}
        \put(1,-1.7){\pscircle[fillstyle=solid,fillcolor=yellow]{.3}}
        \put(1,-1.7){\makebox(0,0){\hbox{{\tiny $0$}}}}
        \put(1.5,2){\makebox(0,0)[bl]{\hbox{{\small $\widetilde{\mathfrak M}_4$}}}}
      }
      \rput(-4,-3.2){
        \multiput(-2.5,-0.85)(0.5,0.85){4}{\PATLEFT}
        \multiput(-2,-1.7)(1,0){4}{\PATBOTTOM}
        \put(-0.5,2.55){\PATTOP}
        \put(0,3.4){\PATNORTH}
        \multiput(-1.5,-0.85)(1,0){4}{\PATGEN}
        \multiput(-1,0)(1,0){3}{\PATGEN}
        \multiput(-.5,0.85)(1,0){2}{\PATGEN}
        \put(0,1.7){\PATGEN}
        \multiput(-1.5,-0.85)(1,0){4}{\PATGEN}
        \multiput(0.5,2.55)(0.5,-0.85){4}{\PATRIGHT}
        \put(2,-1.7){\PATBOTRIGHT}
        \put(-3,-1.7){\PATSW}
        \put(3,-1.7){\PATSE}
        \multiput(-1.5,0.85)(1,0){4}{\pscircle[fillstyle=solid,fillcolor=white]{.3}
          \put(0,0){\makebox(0,0){\hbox{{\tiny $1$}}}}}
        \multiput(-1,1.7)(1,0){3}{\pscircle[fillstyle=solid,fillcolor=white]{.3}
          \put(0,0){\makebox(0,0){\hbox{{\tiny $2$}}}}}
        \multiput(-0.5,2.55)(1,0){2}{\pscircle[fillstyle=solid,fillcolor=white]{.3}
          \put(0,0){\makebox(0,0){\hbox{{\tiny $2$}}}}}
        \multiput(0,3.4)(0.5,0.85){1}{\pscircle[fillstyle=solid,fillcolor=white]{.3}
          \put(0,0){\makebox(0,0){\hbox{{\tiny $2$}}}}}
        \put(-1,1.7){\pscircle[fillstyle=solid,fillcolor=yellow]{.3}}
        \put(-1,1.7){\makebox(0,0){\hbox{{\tiny $2$}}}}
        \put(1,1.7){\pscircle[fillstyle=solid,fillcolor=yellow]{.3}}
        \put(1,1.7){\makebox(0,0){\hbox{{\tiny $2$}}}}
        \put(-2,0){\pscircle[fillstyle=solid,fillcolor=yellow]{.3}}
        \put(-2,0){\makebox(0,0){\hbox{{\tiny $0$}}}}
        \put(2,0){\pscircle[fillstyle=solid,fillcolor=yellow]{.3}}
        \put(2,0){\makebox(0,0){\hbox{{\tiny $0$}}}}
        \put(0,-2.2){\makebox(0,0)[ct]{\hbox{{\small $[\widetilde M_1]_{3,3} [\widetilde M_1]_{4,4}$}}}}
      }
      \rput(4,-3.2){
        \multiput(-2.5,-0.85)(0.5,0.85){4}{\PATLEFT}
        \multiput(-2,-1.7)(1,0){4}{\PATBOTTOM}
        \put(-0.5,2.55){\PATTOP}
        \put(0,3.4){\PATNORTH}
        \multiput(-1.5,-0.85)(1,0){4}{\PATGEN}
        \multiput(-1,0)(1,0){3}{\PATGEN}
        \multiput(-.5,0.85)(1,0){2}{\PATGEN}
        \put(0,1.7){\PATGEN}
        \multiput(-1.5,-0.85)(1,0){4}{\PATGEN}
        \multiput(0.5,2.55)(0.5,-0.85){4}{\PATRIGHT}
        \put(2,-1.7){\PATBOTRIGHT}
        \put(-3,-1.7){\PATSW}
        \put(3,-1.7){\PATSE}
        \multiput(-2.5,-0.85)(1,0){6}{\pscircle[fillstyle=solid,fillcolor=white]{.3}
          \put(0,0){\makebox(0,0){\hbox{{\tiny $1$}}}}}
        \multiput(-2,0)(1,0){5}{\pscircle[fillstyle=solid,fillcolor=white]{.3}
          \put(0,0){\makebox(0,0){\hbox{{\tiny $2$}}}}}
        \multiput(-1.5,0.85)(1,0){4}{\pscircle[fillstyle=solid,fillcolor=white]{.3}
          \put(0,0){\makebox(0,0){\hbox{{\tiny $3$}}}}}
        \multiput(-1,1.7)(1,0){3}{\pscircle[fillstyle=solid,fillcolor=white]{.3}
          \put(0,0){\makebox(0,0){\hbox{{\tiny $4$}}}}}
        \multiput(-0.5,2.55)(1,0){2}{\pscircle[fillstyle=solid,fillcolor=white]{.3}
          \put(0,0){\makebox(0,0){\hbox{{\tiny $5$}}}}}
        \multiput(0,3.4)(0.5,0.85){1}{\pscircle[fillstyle=solid,fillcolor=white]{.3}
          \put(0,0){\makebox(0,0){\hbox{{\tiny $6$}}}}}
        \put(-3,-1.7){\pscircle[fillstyle=solid,fillcolor=yellow]{.3}}
        \put(-3,-1.7){\makebox(0,0){\hbox{{\tiny $0$}}}}
        \put(3,-1.7){\pscircle[fillstyle=solid,fillcolor=yellow]{.3}}
        \put(3,-1.7){\makebox(0,0){\hbox{{\tiny $0$}}}}
        \put(0,-2.2){\makebox(0,0)[ct]{\hbox{{\small $\det\widetilde M_1$}}}}
      }
    }
  \end{pspicture}
  $$
  The contribution of the left FG triangle into the normalized ratio $\mathfrak M_4/\mathfrak M_2$ is then 
  $$
  \widetilde{\mathfrak M}_4/\widetilde{\mathfrak M}_2 [\widetilde M_1]_{3,3}[\widetilde M_1]_{4,4}\cdot [\det \widetilde M_1]^{-2/3}.
  $$
  To obtain only integer powers of all cluster variables let us take the third power of this expression. Then, combining all terms, we obtain
  $$
  
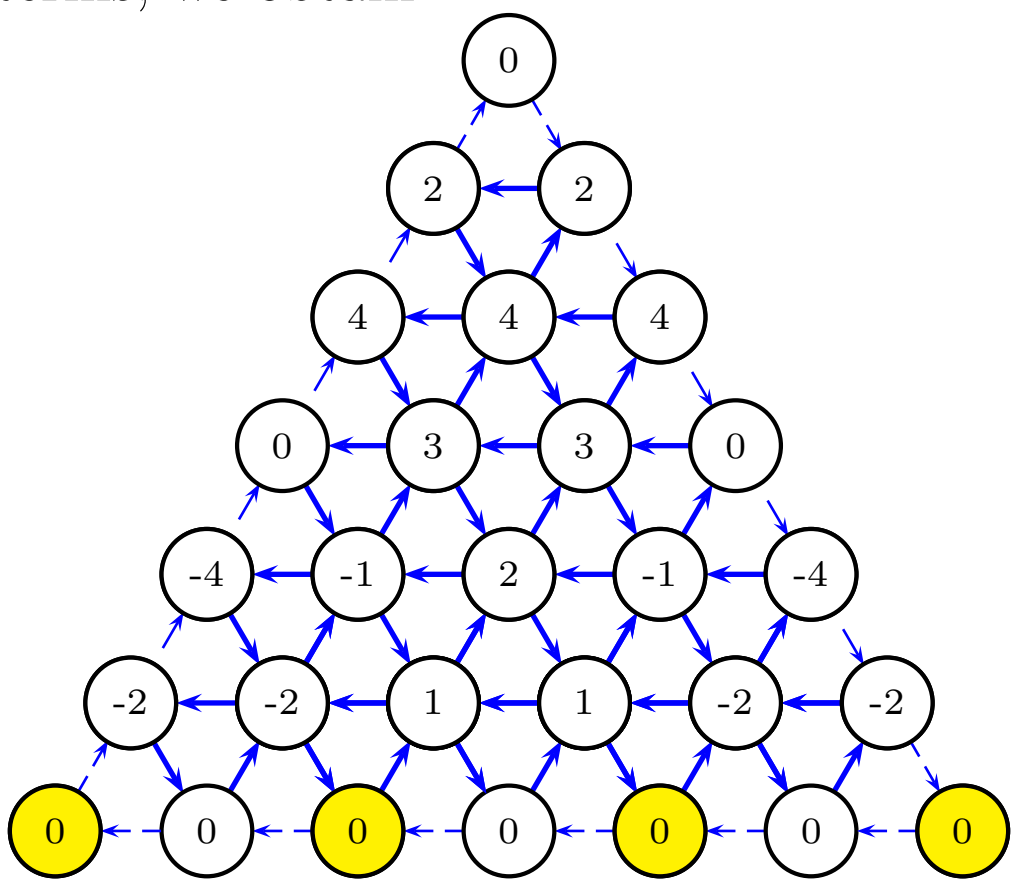
\begin{pspicture}(-3,-2.7)(3,3.5){\psset{unit=1}
      \newcommand{\PATGEN}{%
        {\psset{unit=1}
          \rput(0,0){\psline[linecolor=blue,linewidth=1pt]{->}(0,0)(.35,.6)}
          \rput(0,0){\psline[linecolor=blue,linewidth=1pt]{->}(0.7,0)(0.3,0)}
          \rput(0,0){\psline[linecolor=blue,linewidth=1pt]{->}(0,0)(.35,-.6)}
          \put(0,0){\pscircle[fillstyle=solid,fillcolor=white]{.3}}
          \put(0,0){\makebox(0,0){\hbox{{\tiny $0$}}}}
        }}
      \newcommand{\PATLEFT}{%
        {\psset{unit=1}
          \rput(0,0){\psline[linecolor=blue,linewidth=0.5pt,linestyle=dashed]{->}(0.05,0.085)(.35,.6)}
          \rput(0,0){\psline[linecolor=blue,linewidth=1pt]{->}(0.7,0)(0.3,0)}
          \rput(0,0){\psline[linecolor=blue,linewidth=1pt]{->}(0,0)(.35,-.6)}
          \put(0,0){\pscircle[fillstyle=solid,fillcolor=white]{.3}}
          \put(0,0){\makebox(0,0){\hbox{{\tiny $0$}}}}
        }}
      \newcommand{\PATRIGHT}{%
        {\psset{unit=1}
          \rput(0,0){\psline[linecolor=blue,linewidth=0.5pt,linestyle=dashed]{->}(0.05,-0.085)(.35,-.6)}
          \put(0,0){\pscircle[fillstyle=solid,fillcolor=white]{.3}}
          \put(0,0){\makebox(0,0){\hbox{{\tiny $0$}}}}
        }}
      \newcommand{\PATBOTTOM}{%
        {\psset{unit=1}
          \rput(0,0){\psline[linecolor=blue,linewidth=1pt]{->}(0,0)(.35,.6)}
          \rput(0,0){\psline[linecolor=blue,linewidth=0.5pt,linestyle=dashed]{->}(0.7,0)(0.3,0)}
          \put(0,0){\pscircle[fillstyle=solid,fillcolor=white]{.3}}
          \put(0,0){\makebox(0,0){\hbox{{\tiny $0$}}}}
        }}
      \newcommand{\PATTOP}{%
        {\psset{unit=1}
          \rput(0,0){\psline[linecolor=blue,linewidth=1pt]{->}(0.7,0)(0.3,0)}
          \rput(0,0){\psline[linecolor=blue,linewidth=1pt]{->}(0,0)(.35,-.6)}
          \put(0,0){\pscircle[fillstyle=solid,fillcolor=white]{.3}}
          \put(0,0){\makebox(0,0){\hbox{{\tiny $0$}}}}
        }}
      \newcommand{\PATBOTRIGHT}{%
        {\psset{unit=1}
          \rput(0,0){\psline[linecolor=blue,linewidth=1pt]{->}(0,0)(.35,.6)}
          \put(0,0){\pscircle[fillstyle=solid,fillcolor=white]{.3}}
          \put(0,0){\makebox(0,0){\hbox{{\tiny $0$}}}}
          \put(.5,0.85)
          {\pscircle[fillstyle=solid,fillcolor=white]{.3}}
          \put(0.5,0.85){\makebox(0,0){\hbox{{\tiny $0$}}}}
        }}
      \newcommand{\PATNORTH}{%
        {\psset{unit=1}
          \rput(0,0){\psline[linecolor=blue,linewidth=0.5pt,linestyle=dashed]{->}(-.35,-.6)(-0.15,-0.255)}
          \rput(0,0){\psline[linecolor=blue,linewidth=0.5pt,linestyle=dashed]{<-}(.35,-.6)(0.15,-0.255)}
          \put(0,0){\pscircle[fillstyle=solid,fillcolor=white]{.3}}
          \put(0,0){\makebox(0,0){\hbox{{\tiny $0$}}}}
        }}
      \newcommand{\PATSW}{%
        {\psset{unit=1}
          \rput(0,0){\psline[linecolor=blue,linewidth=0.5pt,linestyle=dashed]{->}(0.15,0.255)(.35,.6)}
          \rput(0,0){\psline[linecolor=blue,linewidth=0.5pt,linestyle=dashed]{<-}(0.3,0)(0.7,0)}
          \put(0,0){\pscircle[fillstyle=solid,fillcolor=white]{.3}}
          \put(0,0){\makebox(0,0){\hbox{{\tiny $0$}}}}
        }}
      \newcommand{\PATSE}{%
        {\psset{unit=1}
          \rput(0,0){\psline[linecolor=blue,linewidth=0.5pt,linestyle=dashed]{<-}(-0.15,0.255)(-.35,.6)}
          \rput(0,0){\psline[linecolor=blue,linewidth=0.5pt,linestyle=dashed]{<-}(-0.7,0)(-0.3,0)}
          \put(0,0){\pscircle[fillstyle=solid,fillcolor=white]{.3}}
          \put(0,0){\makebox(0,0){\hbox{{\tiny $0$}}}}
        }}
      \multiput(-2.5,-0.85)(0.5,0.85){4}{\PATLEFT}
      \multiput(-2,-1.7)(1,0){4}{\PATBOTTOM}
      \put(-0.5,2.55){\PATTOP}
      \put(0,3.4){\PATNORTH}
      \multiput(-1.5,-0.85)(1,0){4}{\PATGEN}
      \multiput(-1,0)(1,0){3}{\PATGEN}
      \multiput(-.5,0.85)(1,0){2}{\PATGEN}
      \put(0,1.7){\PATGEN}
      \multiput(-1.5,-0.85)(1,0){4}{\PATGEN}
      \multiput(0.5,2.55)(0.5,-0.85){4}{\PATRIGHT}
      \put(2,-1.7){\PATBOTRIGHT}
      \put(-3,-1.7){\PATSW}
      \put(3,-1.7){\PATSE}
      \multiput(-2.5,-0.85)(1,0){6}{\pscircle[fillstyle=solid,fillcolor=white]{.3}
        \put(0,0){\makebox(0,0){\hbox{{\tiny -$2$}}}}}
      \multiput(-0.5,-0.85)(1,0){2}{\pscircle[fillstyle=solid,fillcolor=white]{.3}
        \put(0,0){\makebox(0,0){\hbox{{\tiny $1$}}}}}
      \multiput(-2,0)(4,0){2}{\pscircle[fillstyle=solid,fillcolor=white]{.3}
        \put(0,0){\makebox(0,0){\hbox{{\tiny -$4$}}}}}
      \multiput(-1,0)(2,0){2}{\pscircle[fillstyle=solid,fillcolor=white]{.3}
        \put(0,0){\makebox(0,0){\hbox{{\tiny -$1$}}}}}
      \multiput(0,0)(4,0){1}{\pscircle[fillstyle=solid,fillcolor=white]{.3}
        \put(0,0){\makebox(0,0){\hbox{{\tiny $2$}}}}}
      \multiput(-0.5,0.85)(1,0){2}{\pscircle[fillstyle=solid,fillcolor=white]{.3}
        \put(0,0){\makebox(0,0){\hbox{{\tiny $3$}}}}}
      \multiput(-1,1.7)(1,0){3}{\pscircle[fillstyle=solid,fillcolor=white]{.3}
        \put(0,0){\makebox(0,0){\hbox{{\tiny $4$}}}}}
      \multiput(-0.5,2.55)(1,0){2}{\pscircle[fillstyle=solid,fillcolor=white]{.3}
        \put(0,0){\makebox(0,0){\hbox{{\tiny $2$}}}}}
      \multiput(-3,-1.7)(2,0){4}{\pscircle[fillstyle=solid,fillcolor=yellow]{.3}
        \put(0,0){\makebox(0,0){\hbox{{\tiny $0$}}}}}
      \put(0,-2.2){\makebox(0,0)[tc]{\hbox{{\small $\Bigl[(\mathfrak M_4/\mathfrak M_2) [\widetilde M_1]_{3,3}[\widetilde M_1]_{4,4}\Bigr]^3\cdot \det \widetilde M_1^{-2}$}}}}
    }
  \end{pspicture}
  $$
  Note that the variable obtained by amalgamating all frozen variables on the lower boundary commutes with this element, 
  and we come to a conclusion that the ratio of minors $\mathcal M_4/\mathcal M_2$ of the normalized matrix $\mathbb A$ reads
  $$
  K_2:=\bigl(\mathfrak M_4/\mathfrak M_2\bigr)^3=\mathfrak C_1\mathfrak C_2^2
  $$
  We leave it as an exercise to the reader to check that
  $$
  K_1:=\bigl(\mathfrak M_5/\mathfrak M_1\bigr)^3=\mathfrak C_1^2\mathfrak C_2.
  $$
\end{exm}

\section{Triangular case}\label{s:triangular}
In this section, we assume $a_{i,j}=0$ for all $i>j$, and $a_{i,i}=q^{1/2}$ for all $i$. We denote this algebra by $A_n$. 

This case of upper-triangular matrices $A$ is perhaps the most famous one due to its applications in geometry and physics. We have the following relations~\cite{ChekhovFock1999Steklov, NelsonRegge1991CMP, NelsonRegge1991PLB}. For $1 \le i<j<k \le n$,
\begin{align}\label{eq:triangle_3_relations}
  a_{i,j}a_{j,k} &= q^{-1}a_{j,k}a_{i,j}+q^{-1/2}(q-q^{-1})a_{i,k} \\
  a_{i,k}a_{i,j} &= q^{-1}a_{i,j}a_{i,k}+q^{-1/2}(q-q^{-1})a_{j,k} \\
  a_{j,k}a_{i,k} &= q^{-1}a_{i,k}a_{j,k}+q^{-1/2}(q-q^{-1})a_{i,j}
\end{align}
and for $1 \le i<j<k<l \le n$,
\begin{align}\label{eq:triangle_4_relations}
  [a_{i,j},a_{k,l}] &= 0 \\
  [a_{i,l},a_{j,k}] &= 0 \\
  a_{j,l} a_{i,k} &= a_{i,k} a_{j,l} - (q - q^{-1})a_{i,l} a_{j,k} + (q-q^{-1})a_{i,j} a_{k,l}
\end{align}

We can represent these relations diagrammatically as follows. Consider a circle with $n$ marked points labeled $1,2,\ldots,n$ in counterclockwise order. We represent $a_{ij}$ with an arc connecting points $i$ and $j$, and represent a monomial $a_{i_1,j_1}a_{i_2,j_2} \cdots a_{i_k,j_k}$ by drawing the respective arcs $i_1$ to $j_1$, $i_2$ to $j_2$, etc., and labeling these arcs with numbers $1,\ldots,k$ in order. For instance, the relation $a_{i,j}a_{j,k} = q^{-1}a_{j,k}a_{i,j}+q^{-1/2}(q-q^{-1})a_{i,k}$ has diagrammatic representation

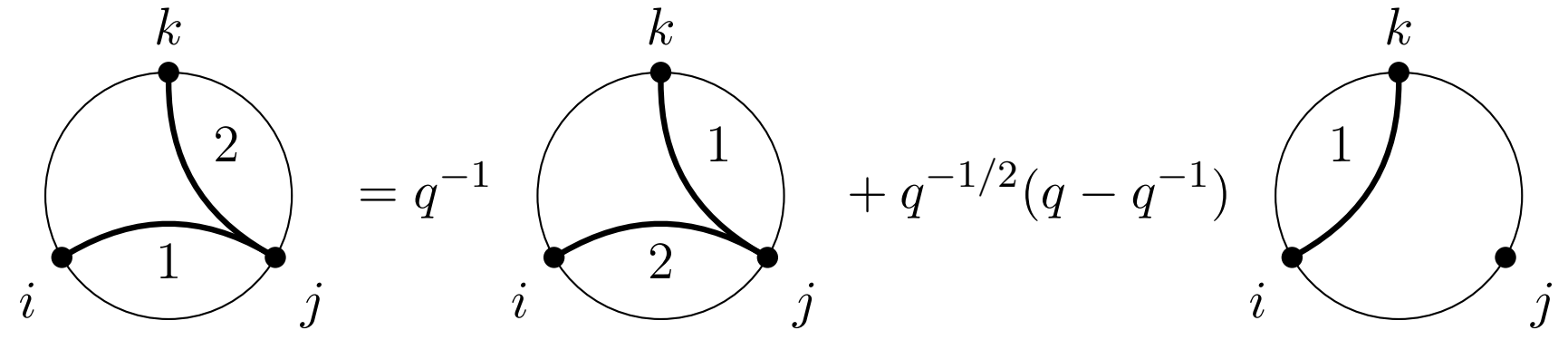
\begin{align*}
  \begin{adjustbox}{trim= 0 1cm 0 0}
    \begin{tikzpicture}[scale=0.6,every circle node/.style={minimum size=4pt, inner sep=0pt, outer sep=0pt,fill,draw}]
      \draw (0,0) circle (1.5cm);
      \draw (90:1.5) node[circle,label=above:$k$] (k) {};
      \draw (210:1.5) node[circle,label=below left:$i$] (i) {};
      \draw (330:1.5) node[circle,label=below right:$j$] (j) {};
      \draw[very thick,smooth] (i) edge[bend left] node[below] {1} (j);
      \draw[very thick,smooth] (j) edge[bend left] node[above right] {2} (k);
    \end{tikzpicture}
  \end{adjustbox}
  &=
    q^{-1}\begin{adjustbox}{trim= 0 1cm 0 0}
      \begin{tikzpicture}[scale=0.6,every circle node/.style={minimum size=4pt, inner sep=0pt, outer sep=0pt,fill,draw}]
        \draw (0,0) circle (1.5cm);
        \draw (90:1.5) node[circle,label=above:$k$] (k) {};
        \draw (210:1.5) node[circle,label=below left:$i$] (i) {};
        \draw (330:1.5) node[circle,label=below right:$j$] (j) {};
        \draw[very thick,smooth] (i) edge[bend left] node[below] {2} (j);
        \draw[very thick,smooth] (j) edge[bend left] node[above right] {1} (k);
      \end{tikzpicture}
    \end{adjustbox}
    +q^{-1/2}(q-q^{-1})\begin{adjustbox}{trim= 0 1cm 0 0}
      \begin{tikzpicture}[scale=0.6,every circle node/.style={minimum size=4pt, inner sep=0pt, outer sep=0pt,fill,draw}]
        \draw (0,0) circle (1.5cm);
        \draw (90:1.5) node[circle,label=above:$k$] (k) {};
        \draw (210:1.5) node[circle,label=below left:$i$] (i) {};
        \draw (330:1.5) node[circle,label=below right:$j$] (j) {};
        \draw[very thick,smooth] (i) edge[bend right] node[above left] {1} (k);
      \end{tikzpicture}
    \end{adjustbox}
\end{align*}
\vspace{1cm}

Observe that the first three relations exhibit a cyclic symmetry, i.e., if we remove the labels from the points on the circle, the resulting diagrams are rotations of one another. Thus, the first three relations can all be viewed as a single relation which relies solely on the cyclic order present in the diagram. The same phenomenon occurs for the 4-term relations, so we can express the relations of the upper-triangular algebra as

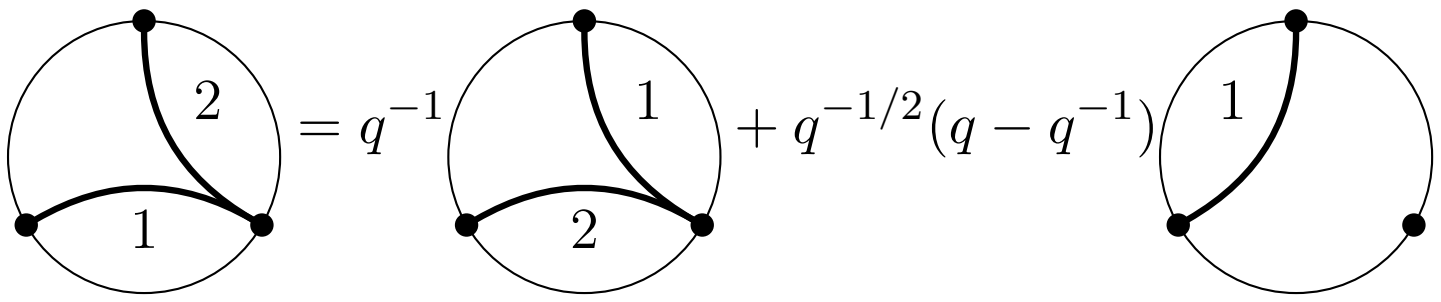
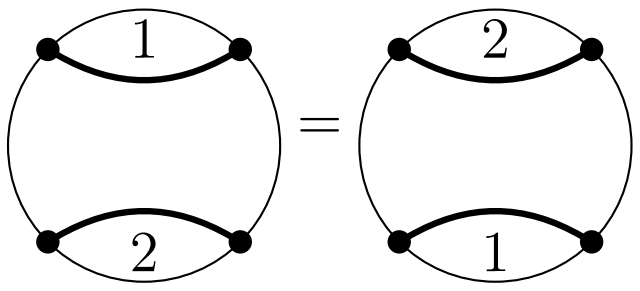
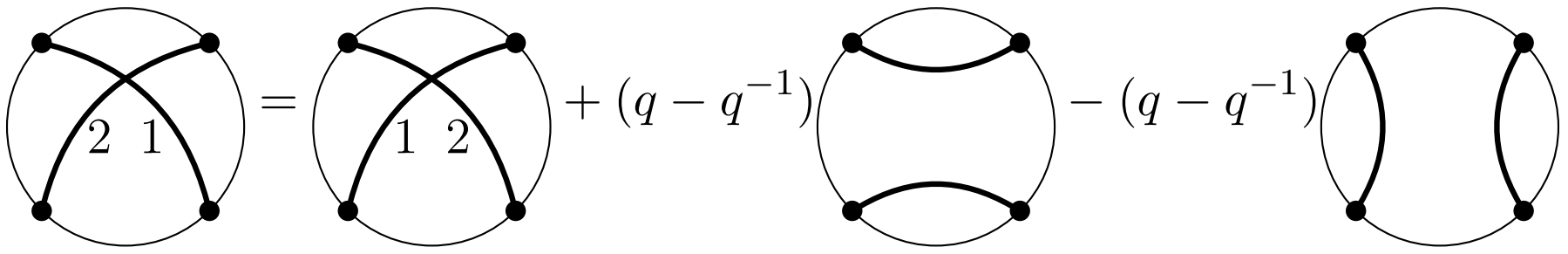
\begin{align}
  \begin{adjustbox}{trim= 0 1cm 0 0}
    \begin{tikzpicture}[scale=0.6,every circle node/.style={minimum size=4pt, inner sep=0pt, outer sep=0pt,fill,draw}]
      \draw (0,0) circle (1.5cm);
      \draw (90:1.5) node[circle] (k) {};
      \draw (210:1.5) node[circle] (i) {};
      \draw (330:1.5) node[circle] (j) {};
      \draw[very thick,smooth] (i) edge[bend left] node[below] {1} (j);
      \draw[very thick,smooth] (j) edge[bend left] node[above right] {2} (k);
    \end{tikzpicture}
  \end{adjustbox}
  &=
    q^{-1}\begin{adjustbox}{trim= 0 1cm 0 0}
      \begin{tikzpicture}[scale=0.6,every circle node/.style={minimum size=4pt, inner sep=0pt, outer sep=0pt,fill,draw}]
        \draw (0,0) circle (1.5cm);
        \draw (90:1.5) node[circle] (k) {};
        \draw (210:1.5) node[circle] (i) {};
        \draw (330:1.5) node[circle] (j) {};
        \draw[very thick,smooth] (i) edge[bend left] node[below] {2} (j);
        \draw[very thick,smooth] (j) edge[bend left] node[above right] {1} (k);
      \end{tikzpicture}
    \end{adjustbox}
    +q^{-1/2}(q-q^{-1})\begin{adjustbox}{trim= 0 1cm 0 0}
      \begin{tikzpicture}[scale=0.6,every circle node/.style={minimum size=4pt, inner sep=0pt, outer sep=0pt,fill,draw}]
        \draw (0,0) circle (1.5cm);
        \draw (90:1.5) node[circle] (k) {};
        \draw (210:1.5) node[circle] (i) {};
        \draw (330:1.5) node[circle] (j) {};
        \draw[very thick,smooth] (i) edge[bend right] node[above left] {1} (k);
      \end{tikzpicture}
    \end{adjustbox}
  \\[1cm]
  \begin{adjustbox}{trim= 0 1cm 0 0}
    \begin{tikzpicture}[scale=0.6,every circle node/.style={minimum size=4pt, inner sep=0pt, outer sep=0pt,fill,draw}]
      \draw (0,0) circle (1.5cm);
      \draw (135:1.5) node[circle] (i) {};
      \draw (225:1.5) node[circle] (j) {};
      \draw (315:1.5) node[circle] (k) {};
      \draw (45:1.5) node[circle] (l) {};
      \draw[very thick,smooth] (i) edge[bend right] node[above] {1} (l);
      \draw[very thick,smooth] (j) edge[bend left] node[below] {2} (k);
    \end{tikzpicture}
  \end{adjustbox}
  &=
    \begin{adjustbox}{trim= 0 1cm 0 0}
      \begin{tikzpicture}[scale=0.6,every circle node/.style={minimum size=4pt, inner sep=0pt, outer sep=0pt,fill,draw}]
        \draw (0,0) circle (1.5cm);
        \draw (135:1.5) node[circle] (i) {};
        \draw (225:1.5) node[circle] (j) {};
        \draw (315:1.5) node[circle] (k) {};
        \draw (45:1.5) node[circle] (l) {};
        \draw[very thick,smooth] (i) edge[bend right] node[above] {2} (l);
        \draw[very thick,smooth] (j) edge[bend left] node[below] {1} (k);
      \end{tikzpicture}
    \end{adjustbox}
  \\[1cm]
  \begin{adjustbox}{trim= 0 1cm 0 0}
    \begin{tikzpicture}[scale=0.6,every circle node/.style={minimum size=4pt, inner sep=0pt, outer sep=0pt,fill,draw}]
      \draw (0,0) circle (1.5cm);
      \draw (135:1.5) node[circle] (i) {};
      \draw (225:1.5) node[circle] (j) {};
      \draw (315:1.5) node[circle] (k) {};
      \draw (45:1.5) node[circle] (l) {};
      \draw[very thick,smooth] (i) edge[bend left] node[below] {1} (k);
      \draw[very thick,smooth] (j) edge[bend left] node[below] {2} (l);
    \end{tikzpicture}
  \end{adjustbox}
  &=
    \begin{adjustbox}{trim= 0 1cm 0 0}
      \begin{tikzpicture}[scale=0.6,every circle node/.style={minimum size=4pt, inner sep=0pt, outer sep=0pt,fill,draw}]
        \draw (0,0) circle (1.5cm);
        \draw (135:1.5) node[circle] (i) {};
        \draw (225:1.5) node[circle] (j) {};
        \draw (315:1.5) node[circle] (k) {};
        \draw (45:1.5) node[circle] (l) {};
        \draw[very thick,smooth] (i) edge[bend left] node[below] {2} (k);
        \draw[very thick,smooth] (j) edge[bend left] node[below] {1} (l);
      \end{tikzpicture}
    \end{adjustbox}
    +(q-q^{-1})\begin{adjustbox}{trim= 0 1cm 0 0}
      \begin{tikzpicture}[scale=0.6,every circle node/.style={minimum size=4pt, inner sep=0pt, outer sep=0pt,fill,draw}]
        \draw (0,0) circle (1.5cm);
        \draw (135:1.5) node[circle] (i) {};
        \draw (225:1.5) node[circle] (j) {};
        \draw (315:1.5) node[circle] (k) {};
        \draw (45:1.5) node[circle] (l) {};
        \draw[very thick,smooth] (i) edge[bend right] (l);
        \draw[very thick,smooth] (j) edge[bend left] (k);
      \end{tikzpicture}
    \end{adjustbox}
    -(q-q^{-1})\begin{adjustbox}{trim= 0 1cm 0 0}
      \begin{tikzpicture}[scale=0.6,every circle node/.style={minimum size=4pt, inner sep=0pt, outer sep=0pt,fill,draw}]
        \draw (0,0) circle (1.5cm);
        \draw (135:1.5) node[circle] (i) {};
        \draw (225:1.5) node[circle] (j) {};
        \draw (315:1.5) node[circle] (k) {};
        \draw (45:1.5) node[circle] (l) {};
        \draw[very thick,smooth] (i) edge[bend left] (j);
        \draw[very thick,smooth] (k) edge[bend left] (l);
      \end{tikzpicture}
    \end{adjustbox}
\end{align}
\vspace{1cm}

In the triangular case, $a_{i,j}^*=a_{i,j}$ for all $1\le i<j\le n$ and $q^*=q^{-1}$.




\begin{prop}\label{prop:triangle_generators}
  $A_n$ is generated by the set
  \begin{align*}
    \left\{a_{i,i+1} : 1 \le i \le n-1 \right\}.
  \end{align*}
\end{prop}
\begin{proof}
  For any $1 \le i \le n-2$, by equation \ref{eq:triangle_3_relations} with $i=i$, $j=i+1$, $k=i+2$, we have
  \begin{align*}
    a_{i,i+1}a_{i+1,i+2} &= q^{-1}a_{i+1,i+2}a_{i,i+1} + q^{-1/2}(q-q^{-1})a_{i,i+2}
  \end{align*}
  so, provided $q\ne 1,-1,0$, $a_{i,i+2}$ is expressible in terms of the generating set for any $i$. Similarly, we can write any element of the form $a_{i,i+3}$ in terms of $a_{i,i+2}$ and $a_{i+1,i+3}$, so elements of the form $a_{i,i+3}$ are expressible in terms of the generating set. We can repeat this argument for elements of the form $a_{i,i+4}$, etc., until every $a_{i,j}$ is expressed in terms of the generating set.
\end{proof}

\subsection{Self-adjoint quantum Casimir functions}

\begin{prop}
  If $C$ is a Casimir, then $C^*$ is also a Casimir.
\end{prop}

\begin{proof}
  Observe that for any algebra element $a$,
  \begin{align*}
    C^*a = (a^*C)^* = (Ca^*)^* = aC^*,
  \end{align*}
  and since $\{a_{i,j}\}$ and $\{a^*_{i,j}\}$ span the same quantum spaces for $A$ or $\mathbb A$, $C^*$ is always a Casimir of the corresponding algebra. 
\end{proof}

In~\eqref{eq:quantum_casimir_for_n_3_1}, we derived the self-adjoint quantum Casimir function for $n=3$. These expressions were obtained in the $n=4$ case in \cite{BullockPrzytycki2000} and \cite{ChekhovFock1999Steklov}. For \( n \geq 4 \), the structure of quantum Casimir functions becomes increasingly complex, making manual computation impractical. Therefore, it is necessary to design algorithmic procedures to assist in their construction and verification. The algorithmic procedures used in this section are summarized  in Appendix~\ref{appendix:algorithms}. Below, we illustrate the method in the case $n = 5$.

\subsubsection{Lexicographically Ordered Monomials}

In this section, all noncommutative monomials will be written in \emph{lexicographically ordered} form unless explicitly stated otherwise.

\begin{defn}
  A monomial of the form
  \[
    a_{i_1, j_1}  a_{i_2, j_2}  \cdots  a_{i_r, j_r}
  \]
  is said to be \emph{lexicographically ordered} if the sequence of index pairs satisfies
  \[
    (i_1, j_1) \le (i_2, j_2) \le \cdots \le (i_r, j_r)
  \]
  under the lexicographic order on pairs: $(i, j) \leq (k, l)$ if either $i < k$, or $i = k$ and $j \leq l$.
\end{defn}

\begin{exm}
  The monomial $a_{1,2}  a_{1,3}  a_{2,4}  a_{4,5}$ is lexicographically ordered, since \((1,2) \leq (1,3) \leq (2,4) \leq (4,5)\). On the other hand, the monomial $a_{2,4}  a_{1,3}  a_{1,2}$ is \emph{not} lexicographically ordered, because $(2,4) > (1,3)$.
\end{exm}

\begin{defn}\label{def:deformatiom}
Let $f(a_{i,j})\in \mathbb Z[a_{i,j}]$ be a semiclassical polynomial function with integer coefficients. We call a \emph{quantum deformation} of $f$ a quantum expression in which every monomial of $f$ with unit coefficient is replaced by an ordered product of quantum operators weighted by $q^\alpha$ with $\alpha\in \mathbb Z/2$. Note that even for terms that were identical in the classical case, both the order of operators and the power $\alpha$ can be different in the quantum case. 
\end{defn}

\begin{remark}
Based on studies in the triangular case, originally, the above definition included that the quantum ordering is  some prescribed (``lexicographic'') order whereas the integer coefficients were replaced by sign-definite Laurent polynomials in $q^{1/2}$. However, our studies in the case of the full algebra $\mathbb A_n$ showed (see Example~\ref{exm-Markov}) that in order to satisfy quantum deformation conditions, we should allow different orderings for monomials that are identical in the classical limit.
\end{remark}

\subsubsection{Computing self-adjoint quantum Casimir functions for \texorpdfstring{$n=5$}{n=5}}
\label{subsecn=5}

Analogously to Examples~\ref{egclassical3} and~\ref{egclassical31}, we first compute the characteristic polynomial $p(\lambda) = \det({A} + \lambda {A}^T)$, where \( {A} \in \mathcal{A}_5 \) is a unipotent upper-triangular matrix. By extracting the coefficients of \( \lambda^4 \) and \( \lambda^3 \) in this determinant, we obtain two classical Casimir functions, denoted
\[
  C_1 := [\lambda^4](p(\lambda)), \quad C_2 := [\lambda^3](p(\lambda)).
\]We then quantize these functions by replacing each monomial coefficient with a power of \( q \), introducing parameters \( \beta_1, \dots, \beta_m \). This yields two quantum polynomials: \(\widehat{C}_1\) and \(\widehat{C}_2\).
Substituting \( \widehat{C}_1 \) into Algorithm~\ref{algorithm2}, we solve the resulting system of equations and obtain a solution for the parameters \( \beta_k \), ensuring that \( \widehat{C}_1 \) is a self-adjoint quantum Casimir function:
\begin{align*}
\widehat{C}_1 &= q^{1/2} a_{1,2} a_{1,5} a_{2,3} a_{3,4} a_{4,5} 
    - q^{-1} a_{1,2} a_{1,4} a_{2,3} a_{3,4} 
    - q a_{1,2} a_{1,5} a_{2,3} a_{3,5} \\
    &\quad - q a_{1,2} a_{1,5} a_{2,4} a_{4,5} 
    - q a_{1,3} a_{1,5} a_{3,4} a_{4,5} 
    - q^{-1} a_{2,3} a_{2,5} a_{3,4} a_{4,5} \\
    &\quad + q^{-5/2} a_{1,2} a_{1,3} a_{2,3} 
    + q^{-1/2} a_{1,2} a_{1,4} a_{2,4} 
    + q^{3/2} a_{1,2} a_{1,5} a_{2,5} 
    + q^{-1/2} a_{1,3} a_{1,4} a_{3,4} \\
    &\quad 
    + q^{3/2} a_{1,3} a_{1,5} a_{3,5}+ q^{3/2} a_{1,4} a_{1,5} a_{4,5} + q^{-5/2} a_{2,3} a_{2,4} a_{3,4}  
    + q^{-1/2} a_{2,3} a_{2,5} a_{3,5}
     \\
    &\quad + q^{-1/2} a_{2,4} a_{2,5} a_{4,5} 
    + q^{-5/2} a_{3,4} a_{3,5} a_{4,5} 
    - q^{-3} a_{1,2}^2 
    - q^{-1} a_{1,3}^2 
    - q a_{1,4}^2 \\
    &\quad - q^{3} a_{1,5}^2 
    - q^{-3} a_{2,3}^2 
    - q^{-1} a_{2,4}^2 
    - q a_{2,5}^2 
    - q^{-3} a_{3,4}^2 
    - q^{-1} a_{3,5}^2 
    - q^{-3} a_{4,5}^2.
\end{align*}
Each term in this expression corresponds graphically to a closed cycle, or polygon, of chords inscribed into a circle without self-intersections.

Due to the length of the expression for \( C_2 \), we omit its explicit form in this paper. Note that it contains about a hundred terms and is much longer than $C_1$.

When applying the same procedure to \( \widehat{C}_2 \), Algorithm~\ref{algorithm2} returns no solution. This indicates that no self-adjoint deformation quantization of \( C_2 \) alone exists. Interestingly, if we remove the self-adjointness condition \( P_c = 0 \) from Algorithm~\ref{algorithm2}, then a solution for the parameters \( \beta_k \) does exist, yielding a non-self-adjoint quantum Casimir function. We denote this function also by \( \widehat{C}_2 \), with the understanding that it is defined only up to a scalar multiple of \( q^\ell \) for some \( \ell \in \mathbb{R} \).

\medskip

If we however consider the linear combination
\[
\Xi_2 := q \widehat{C}_1 + q^k \widehat{C}_2,
\]
we find that, for $k=0$, $\Xi_2$
 is self-adjoint. We summarize the result in the following proposition.

\begin{prop}\label{thmn=52}
For \( n = 5 \), there exist exactly two linearly independent self-adjoint quantum Casimir functions. They are given by:
\[
\Xi_1=\widehat{C}_1 \quad \text{and} \quad \Xi_2 = q \widehat{C}_1 + \widehat{C}_2,
\]
where \( \widehat{C}_2 \) is a non-self-adjoint quantum Casimir function obtained by removing the conjugation condition in Algorithm~\ref{algorithm2}.
\end{prop}
\begin{proof}
It is straightforward to check that $\widehat{C}_1$ and $\widehat{C}_2$  are deformation quantizations of $C_1$ and $C_2$, respectively.   The verification that $\Xi_1$ and $\Xi_2$ are self-adjoint  Casimir elements is computer-assisted and was carried out in  \texttt{Mathematica}; the corresponding notebook is available at \cite{repo}.
\end{proof}
\begin{remark}
This observation suggests that while individual classical Casimir functions may not always admit self-adjoint deformation quantization, suitable algebraic combinations of their quantum analogues may restore self-adjointness. This phenomenon appears naturally in the case $n = 5$, and we assume that similar structures arise for larger $n$ as well.
\end{remark}

\begin{remark}[Hierarchy of Casimirs]
  For fixed $n$, define an ordering on the (algebraically independent) semiclassical Casimirs by $C \le D$ if there is some lexicographically ordered monomial $m$ such that $m$ has nonzero coefficient in $D$, and either $m$ has nonzero coefficient in $C$, or a lexicographically ordered monomial in $m^*$ (this is a conjugation in the quantum algebra) has nonzero coefficient in $C$.

  Now assume that $\widehat{C}$ is a quantum deformation of a semiclassical Casimir $C$ which is minimal under this ordering. Observe that $\widehat{C}-\widehat{C}^*$ is itself a Casimir. Assume that $\widehat{C}-\widehat{C}^*$ is nonzero. Then the coefficient of every monomial in its lexicographically ordered expression must be divisible by $(q-1)$, since $\lim_{q \to 1} \widehat{C}-\widehat{C}^*=C-C=0$. Thus, $\widehat{C}-\widehat{C}^*=(q-1)^k\widehat{D}$, where $D \coloneqq \lim_{q \to 1} \widehat{D} \ne 0$. Since $\widehat{D}$ is a Casimir, $D$ must be a semiclassical Casimir. Since $C$ is minimal, $D$ must necessarily be a nonzero scalar multiple of $C$.

  Thus, if $m$ is a monomial in $C$ such that the coefficient of $m$ in $\widehat{C}-\widehat{C}^*$ is zero, $\widehat{C}-\widehat{C}^*$ must be zero.

  One can compute in the triangular case that the Pfaffian and the first Markov element are minimal under this ordering (we will often refer to them as ``simplest'' or ``lowest'' Casimirs). Thus, the above strategy suffices to prove that our quantizations are self-adjoint.

  In the full-dimensional case, the Pfaffian is still a lowest Casimir, but the first Markov element shares monomials with the quantum determinant (defined in Definition \ref{def:qdet}) and is therefore not a lowest Casimir. Indeed, we find that for $n=3$, a lexicographically ordered self-adjoint Casimir deformation of the first Markov element must involve a nontrivial multiple of the quantum determinant (see Example~\ref{exm-Markov}.)
\end{remark}


\subsection{The quantum Pfaffian \texorpdfstring{$\widehat{Pf}_\triangle$}{Pf_triangle}}\label{ss:Pf-trian}

\subsubsection{The quantum Pfaffian as a self-adjoint Casimir function}

Jing and Zhang considered in \cite{Jing2014} the algebra $B$ generated by \({\mathfrak b}_{ij}\) for \(1 \leq i < j \leq n\) modulo the ideal generated by the relations \begin{equation}\label{eq:JZ-relations}
    {\mathfrak b}_{ij} {\mathfrak b}_{kl} + (-q) {\mathfrak b}_{ik} {\mathfrak b}_{jl} + (-q)^2 {\mathfrak b}_{il} {\mathfrak b}_{jk} = {\mathfrak b}_{kl} {\mathfrak b}_{ij} + (-q)^{-1} {\mathfrak b}_{jl} {\mathfrak b}_{ik} + (-q)^{-2} {\mathfrak b}_{jk} {\mathfrak b}_{il},
\end{equation} where \(i < j < k < l\). Then~\cite[Theorem~2.8]{Jing2014} gives the quantum Pfaffian in \(B\) as\begin{equation}\label{12}
    \operatorname{Pf}_q= \sum_{\pi \in \Pi} (-q)^{l(\pi)} [i_1, j_1] [i_2, j_2] \cdots [i_n, j_n],
\end{equation}
where \([i,j]={\mathfrak b}_{ij}\) for $i<j$.

Clearly, the algebra $B$ is very close to the algebra we consider in the triangular case. In fact, the only difference is in the overall $q$-factors: for $a_{i,j}$, it is just another way of writing the commutation relations between $a_{i,k}$ and $a_{j,l}$:
$$
  -q^{-1}a_{i,j} a_{kl} + a_{i,k} a_{j,l} -q a_{i,l} a_{j,k} = -q a_{k,l} a_{i,j} + a_{j,l} a_{i,k}  -q^{-1} a_{j,k} a_{i,l}.
$$
We therefore have the following theorem, which states that in our algebra $A$, as long as the dimension is even, one can find an analogue of the quantum Pfaffian, which is a self-adjoint Casimir function.

\begin{defn}\label{def:qpfaffian}
With $\Pi$ and $l(\pi)$ as in Section \ref{subsubsec:notation}, define the \emph{quantum Pfaffian} $\widehat{Pf}_\triangle$ of $A_{2n}$ by
\[
    \widehat{Pf}_\triangle
    :=
    q^{-\frac{n^2-n}{2}}
    \sum_{\pi\in\Pi} (-q)^{l(\pi)} a_{i_1 j_1} a_{i_2 j_2}\cdots a_{i_n j_n}.
\]
\end{defn}

\begin{thm}\label{t1}
The quantum Pfaffian $\widehat{Pf}_\triangle$ from Definition~\ref{def:qpfaffian}
is a self-adjoint Casimir element of $A_{2n}$.
\end{thm}

\begin{proof}
\noindent\textbf{Step 1.} \emph{\(\widehat{Pf}_\triangle\) is a Casimir element\footnote{Throughout this paper, we use the terms Casimir function, Casimir element, and Casimir interchangeably.}.}

By Proposition~\ref{prop:triangle_generators}, $A_{2n}$ is generated by 
\[
    \{a_{i,i+1} : 1 \le i \le 2n-1\}.
\]
Hence, it suffices to show that $\widehat{Pf}_\triangle$ commutes with $a_{i,i+1}$ for all $1 \le i \le 2n-1$. Fix $i$, and observe that every monomial in $\widehat{Pf}_\triangle$ either commutes with $a_{i,i+1}$ or appears in one of the following expressions, where $A$, $B$, and $C$ are monomial expressions consisting of $a_{j,k}$ for $j,k \not\in \{i,i+1\}$:
\begin{itemize}
    \item $Aa_{j,i}Ba_{k,i+1}C - qAa_{j,i+1}Ba_{k,i}C$, where $1 \le j \le k < i$.
    \item $Aa_{j,i}Ba_{i+1,k}C - qAa_{j,i+1}Ba_{i,k}C$, where $1 \le j < i < i+1 < k \le n$.
    \item $Aa_{i,j}a_{i+1,k}C - qAa_{i,k}a_{i+1,j}C$, where $i+1 < j < k \le n$.
\end{itemize}
We claim that each expression of the form above commutes with $a_{i,i+1}$. Let \(A,B,C \in A_{2n}\) be arbitrary elements that commute with \(a_{i,i+1}\).  
Note that \(a_{i,i+1}\) commutes with \(a_{j,k}\) whenever both \(j,k\) are different from \(i\) and \(i+1\).  
Thus every monomial in \(\widehat{Pf}_\triangle\) either commutes with \(a_{i,i+1}\) automatically (in the case where \(\pi\) pairs \(i\) with \(i+1\)), or can be presented in one of the above three forms. In all three cases, we also draw the relevant parts of the corresponding chord diagrams (numbers assigned to chords indicate their order in the corresponding quantum monomial).

\begin{claim}\label{claim1}
For all integers \(1 \leq k < \ell < i\), the element
\begin{align*}
    Aa_{k,i}Ba_{\ell,i+1}C - qAa_{k,i+1}Ba_{\ell,i}C
    \qquad
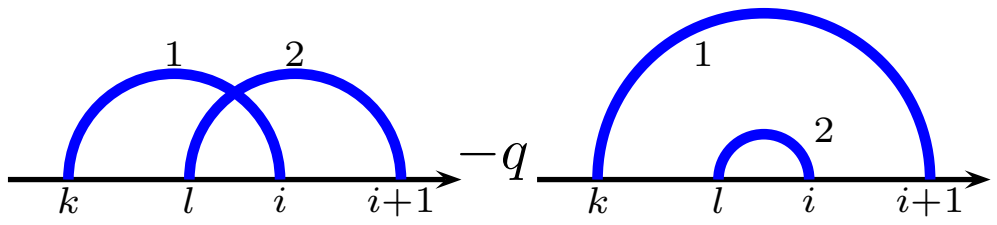
       \begin{pspicture}(-3,1)(3,0)
\newcommand{\TWOCROSSARCS}[2]{
\psline[linewidth=1pt]{->}(-1.5,0)(1.5,0)
\psarc[linecolor=blue, linewidth=2pt](-0.4,0){0.7}{0}{180}
\psarc[linecolor=blue, linewidth=2pt](0.4,0){0.7}{0}{180}
\rput(-0.4,0.75){\makebox(0,0)[cb]{\tiny $#1$}}
\rput(0.4,0.75){\makebox(0,0)[cb]{\tiny $#2$}}
\rput(-1.1,-0.05){\makebox(0,0)[ct]{\tiny $k$}}
\rput(-0.3,-0.05){\makebox(0,0)[ct]{\tiny $l$}}
\rput(0.3,-0.05){\makebox(0,0)[ct]{\tiny $i$}}
\rput(1.1,-0.05){\makebox(0,0)[ct]{\tiny $i{+}1$}}
}
\newcommand{\TWODOLLARCS}[2]{
\psline[linewidth=1pt]{->}(-1.5,0)(1.5,0)
\psarc[linecolor=blue, linewidth=2pt](0,0){1.1}{0}{180}
\psarc[linecolor=blue, linewidth=2pt](0,0){0.3}{0}{180}
\rput(-0.4,0.75){\makebox(0,0)[cb]{\tiny $#1$}}
\rput(0.4,0.25){\makebox(0,0)[cb]{\tiny $#2$}}
\rput(-1.1,-0.05){\makebox(0,0)[ct]{\tiny $k$}}
\rput(-0.3,-0.05){\makebox(0,0)[ct]{\tiny $l$}}
\rput(0.3,-0.05){\makebox(0,0)[ct]{\tiny $i$}}
\rput(1.1,-0.05){\makebox(0,0)[ct]{\tiny $i{+}1$}}
}
\newcommand{\TWOARCS}[2]{
\psline[linewidth=1pt]{->}(-1.5,0)(1.5,0)
\psarc[linecolor=blue, linewidth=2pt](-0.7,0){0.4}{0}{180}
\psarc[linecolor=blue, linewidth=2pt](0.7,0){0.4}{0}{180}
\rput(-0.7,0.45){\makebox(0,0)[cb]{\tiny $#1$}}
\rput(0.7,0.45){\makebox(0,0)[cb]{\tiny $#2$}}
\rput(-1.1,-0.05){\makebox(0,0)[ct]{\tiny $m$}}
\rput(-0.3,-0.05){\makebox(0,0)[ct]{\tiny $i$}}
\rput(0.3,-0.05){\makebox(0,0)[ct]{\tiny $i{+}1$}}
\rput(1.1,-0.05){\makebox(0,0)[ct]{\tiny $l$}}
}
\rput(-1.5,0){
\TWOCROSSARCS{1}{2}}
\rput(2,0){\rput(-1.8,0){\makebox(0,0)[cb]{$-q$}}\TWODOLLARCS{1}{2}
}
\end{pspicture}
\end{align*}
commutes with \(a_{i,i+1}\).
\end{claim}

\begin{claim}\label{claim2}
For all integers \(1 \leq k < i\) and \(i+1 < \ell \leq n\), the element
\begin{align*}
    Aa_{k,i}Ba_{i+1,\ell}C - qAa_{k,i+1}Ba_{i,\ell}C
    \qquad
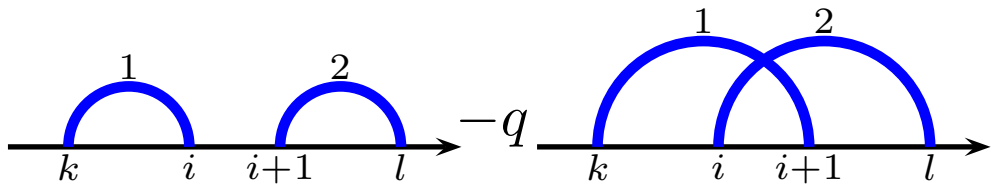
       \begin{pspicture}(-3,1)(3,0)
\newcommand{\TWOCROSSARCS}[2]{
\psline[linewidth=1pt]{->}(-1.5,0)(1.5,0)
\psarc[linecolor=blue, linewidth=2pt](-0.4,0){0.7}{0}{180}
\psarc[linecolor=blue, linewidth=2pt](0.4,0){0.7}{0}{180}
\rput(-0.4,0.75){\makebox(0,0)[cb]{\tiny $#1$}}
\rput(0.4,0.75){\makebox(0,0)[cb]{\tiny $#2$}}
\rput(-1.1,-0.05){\makebox(0,0)[ct]{\tiny $k$}}
\rput(1.1,-0.05){\makebox(0,0)[ct]{\tiny $l$}}
\rput(-0.3,-0.05){\makebox(0,0)[ct]{\tiny $i$}}
\rput(0.3,-0.05){\makebox(0,0)[ct]{\tiny $i{+}1$}}
}
\newcommand{\TWODOLLARCS}[2]{
\psline[linewidth=1pt]{->}(-1.5,0)(1.5,0)
\psarc[linecolor=blue, linewidth=2pt](0,0){1.1}{0}{180}
\psarc[linecolor=blue, linewidth=2pt](0,0){0.3}{0}{180}
\rput(-0.4,0.75){\makebox(0,0)[cb]{\tiny $#1$}}
\rput(0.4,0.25){\makebox(0,0)[cb]{\tiny $#2$}}
\rput(-1.1,-0.05){\makebox(0,0)[ct]{\tiny $k$}}
\rput(1.1,-0.05){\makebox(0,0)[ct]{\tiny $l$}}
\rput(-0.3,-0.05){\makebox(0,0)[ct]{\tiny $i$}}
\rput(0.3,-0.05){\makebox(0,0)[ct]{\tiny $i{+}1$}}
}
\newcommand{\TWOARCS}[2]{
\psline[linewidth=1pt]{->}(-1.5,0)(1.5,0)
\psarc[linecolor=blue, linewidth=2pt](-0.7,0){0.4}{0}{180}
\psarc[linecolor=blue, linewidth=2pt](0.7,0){0.4}{0}{180}
\rput(-0.7,0.45){\makebox(0,0)[cb]{\tiny $#1$}}
\rput(0.7,0.45){\makebox(0,0)[cb]{\tiny $#2$}}
\rput(-1.1,-0.05){\makebox(0,0)[ct]{\tiny $k$}}
\rput(-0.3,-0.05){\makebox(0,0)[ct]{\tiny $i$}}
\rput(0.3,-0.05){\makebox(0,0)[ct]{\tiny $i{+}1$}}
\rput(1.1,-0.05){\makebox(0,0)[ct]{\tiny $l$}}
}
\rput(-1.5,0){
\TWOARCS{1}{2}}
\rput(2,0){\rput(-1.8,0){\makebox(0,0)[cb]{$-q$}}\TWOCROSSARCS{1}{2}
}
\end{pspicture}
\end{align*}
commutes with \(a_{i,i+1}\).
\end{claim}

\begin{claim}\label{claim3}
For all integers \(i+1 < k < \ell \leq n\), the element
\begin{align*}
    Aa_{i,k}a_{i+1,\ell}C - qAa_{i,\ell}a_{i+1,k}C
    \qquad
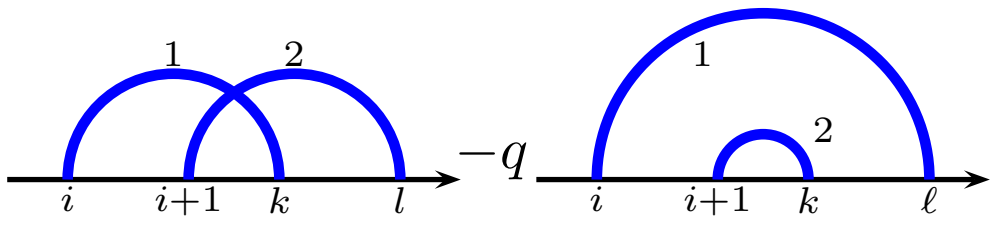
       \begin{pspicture}(-3,1)(3,0)
\newcommand{\TWOCROSSARCS}[2]{
\psline[linewidth=1pt]{->}(-1.5,0)(1.5,0)
\psarc[linecolor=blue, linewidth=2pt](-0.4,0){0.7}{0}{180}
\psarc[linecolor=blue, linewidth=2pt](0.4,0){0.7}{0}{180}
\rput(-0.4,0.75){\makebox(0,0)[cb]{\tiny $#1$}}
\rput(0.4,0.75){\makebox(0,0)[cb]{\tiny $#2$}}
\rput(0.3,-0.05){\makebox(0,0)[ct]{\tiny $k$}}
\rput(1.1,-0.05){\makebox(0,0)[ct]{\tiny $l$}}
\rput(-1.1,-0.05){\makebox(0,0)[ct]{\tiny $i$}}
\rput(-0.3,-0.05){\makebox(0,0)[ct]{\tiny $i{+}1$}}
}
\newcommand{\TWODOLLARCS}[2]{
\psline[linewidth=1pt]{->}(-1.5,0)(1.5,0)
\psarc[linecolor=blue, linewidth=2pt](0,0){1.1}{0}{180}
\psarc[linecolor=blue, linewidth=2pt](0,0){0.3}{0}{180}
\rput(-0.4,0.75){\makebox(0,0)[cb]{\tiny $#1$}}
\rput(0.4,0.25){\makebox(0,0)[cb]{\tiny $#2$}}
\rput(0.3,-0.05){\makebox(0,0)[ct]{\tiny $k$}}
\rput(1.1,-0.05){\makebox(0,0)[ct]{\tiny $\ell$}}
\rput(-1.1,-0.05){\makebox(0,0)[ct]{\tiny $i$}}
\rput(-0.3,-0.05){\makebox(0,0)[ct]{\tiny $i{+}1$}}
}
\newcommand{\TWOARCS}[2]{
\psline[linewidth=1pt]{->}(-1.5,0)(1.5,0)
\psarc[linecolor=blue, linewidth=2pt](-0.7,0){0.4}{0}{180}
\psarc[linecolor=blue, linewidth=2pt](0.7,0){0.4}{0}{180}
\rput(-0.7,0.45){\makebox(0,0)[cb]{\tiny $#1$}}
\rput(0.7,0.45){\makebox(0,0)[cb]{\tiny $#2$}}
\rput(0.3,-0.05){\makebox(0,0)[ct]{\tiny $k$}}
\rput(1.1,-0.05){\makebox(0,0)[ct]{\tiny $i$}}
\rput(-0.3,-0.05){\makebox(0,0)[ct]{\tiny $i{+}1$}}
\rput(-1.1,-0.05){\makebox(0,0)[ct]{\tiny $\ell$}}
}
\rput(-1.5,0){
\TWOCROSSARCS{1}{2}}
\rput(2,0){\rput(-1.8,0){\makebox(0,0)[cb]{$-q$}}\TWODOLLARCS{1}{2}
}
\end{pspicture}
\end{align*}
commutes with \(a_{i,i+1}\). Note that in the corresponding terms of \(\widehat{Pf}_\triangle\), \(a_{i,k}\) and \(a_{i+1,\ell}\) must be neighbors, as well as \(a_{i,\ell}\) and \(a_{i+1,k}\).
\end{claim}




Proof of Claim \ref{claim1}:
\begin{align*}
  &a_{i,i+1}\left( Aa_{k,i}Ba_{\ell,i+1}C - qAa_{k,i+1}Ba_{\ell,i}C \right) \\
  &= Aa_{i,i+1}a_{k,i}Ba_{\ell,i+1}C - qAa_{i,i+1}a_{k,i+1}Ba_{\ell,i}C \\
  &= A(qa_{k,i}a_{i,i+1}-q^{1/2}(q-q^{-1})a_{k,i+1})Ba_{\ell,i+1}C - qA(q^{-1}a_{k,i+1}a_{i,i+1}+q^{-1/2}(q-q^{-1})a_{k,i})Ba_{\ell,i}C \\
  &= (q-q^{-1})(-q^{1/2}Aa_{k,i+1}Ba_{\ell,i+1}C-q^{1/2}Aa_{k,i}Ba_{\ell,i}C) + qAa_{k,i}Ba_{i,i+1}a_{\ell,i+1}C -Aa_{k,i+1}Ba_{i,i+1}a_{\ell,i}C \\
  &= (q-q^{-1})(-q^{1/2}Aa_{k,i+1}Ba_{\ell,i+1}C-q^{1/2}Aa_{k,i}Ba_{\ell,i}C) + qAa_{k,i}B(q^{-1}a_{\ell,i+1}a_{i,i+1}+q^{-1/2}(q-q^{-1})a_{\ell,i})C \\
  &\hspace{3em} -Aa_{k,i+1}B(qa_{\ell,i}a_{i,i+1}-q^{1/2}(q-q^{-1})a_{\ell,i+1})C \\
  &= (q-q^{-1})(-q^{1/2}Aa_{k,i+1}Ba_{\ell,i+1}C-q^{1/2}Aa_{k,i}Ba_{\ell,i}C+q^{1/2}Aa_{k,i+1}Ba_{\ell,i+1}C+q^{1/2}Aa_{k,i}Ba_{\ell,i}C) \\
  &\hspace{3em} +Aa_{k,i}Ba_{\ell,i+1}a_{i,i+1}C - qAa_{k,i+1}Ba_{\ell,i}a_{i,i+1}C \\
  &= \left( Aa_{k,i}Ba_{\ell,i+1}C - qAa_{k,i+1}Ba_{\ell,i}C \right) a_{i,i+1}.
\end{align*}

Proof of Claim \ref{claim2}:
\begin{align*}
  &a_{i,i+1}\left( Aa_{k,i}Ba_{i+1,\ell}C - qAa_{k,i+1}Ba_{i,\ell}C \right)\\
  &= Aa_{i,i+1}a_{k,i}Ba_{i+1,\ell}C - qAa_{i,i+1}a_{k,i+1}Ba_{i,\ell}C \\
  &= A(qa_{k,i}a_{i,i+1}-q^{1/2}(q-q^{-1})a_{k,i+1})Ba_{i+1,\ell}C - qA(q^{-1}a_{k,i+1}a_{i,i+1}+q^{-1/2}(q-q^{-1})a_{k,i})Ba_{i,\ell}C \\
  &= (q-q^{-1})(-q^{1/2}Aa_{k,i+1}Ba_{i+1,\ell}C-q^{1/2}Aa_{k,i}Ba_{i,\ell}C) + qAa_{k,i}Ba_{i,i+1}a_{i+1,\ell}C - Aa_{k,i+1}Ba_{i,i+1}a_{i,\ell}C \\
  &= (q-q^{-1})(-q^{1/2}Aa_{k,i+1}Ba_{i+1,\ell}C-q^{1/2}Aa_{k,i}Ba_{i,\ell}C) + qAa_{k,i}B(q^{-1}a_{i+1,\ell}a_{i,i+1}+q^{-1/2}(q-q^{-1})a_{i,\ell})C \\
  &\hspace{3em} - Aa_{k,i+1}B(qa_{i,\ell}a_{i,i+1}-q^{1/2}(q-q^{-1})a_{i+1,\ell})C \\
  &= (q-q^{-1})(-q^{1/2}Aa_{k,i+1}Ba_{i+1,\ell}C-q^{1/2}Aa_{k,i}Ba_{i,\ell}C + q^{1/2}Aa_{k,i+1}Ba_{i+1,\ell}C+q^{1/2}Aa_{k,i}Ba_{i,\ell}C) \\
  &\hspace{3em} + Aa_{k,i}Ba_{i+1,\ell}a_{i,i+1}C - qAa_{k,i+1}Ba_{i,\ell}a_{i,i+1}C \\
  &= \left( Aa_{k,i}Ba_{i+1,\ell}C - qAa_{k,i+1}Ba_{i,\ell}C \right) a_{i,i+1},
\end{align*}

Proof of Claim \ref{claim3}:
\begin{align*}
  &a_{i,i+1} \left( Aa_{i,k}a_{i+1,\ell}C - qAa_{i,\ell}a_{i+1,k}C \right) \\
  &= Aa_{i,i+1}a_{i,k}a_{i+1,\ell}C - qAa_{i,i+1}a_{i,\ell}a_{i+1,k}C \\
  &= A(qa_{i,k}a_{i,i+1}-q^{1/2}(q-q^{-1})a_{i+1,k})a_{i+1,\ell}C - qA(qa_{i,\ell}a_{i,i+1}-q^{1/2}(q-q^{-1})a_{i+1,\ell})a_{i+1,k}C \\
  &= (q-q^{-1})(-q^{1/2}Aa_{i+1,k}a_{i+1,\ell}C+q^{3/2}Aa_{i+1,\ell}a_{i+1,k}C) + qAa_{i,k}a_{i,i+1}a_{i+1,\ell}C -q^2Aa_{i,\ell}a_{i,i+1}a_{i+1,k}C \\
  &= (q-q^{-1})(-q^{1/2}Aa_{i+1,k}a_{i+1,\ell}C+q^{3/2}Aa_{i+1,\ell}a_{i+1,k}C) + qAa_{i,k}(q^{-1}a_{i+1,\ell}a_{i,i+1}+q^{-1/2}(q-q^{-1})a_{i,\ell})C \\
  &\hspace{3em} -q^2Aa_{i,\ell}(q^{-1}a_{i+1,k}a_{i,i+1}+q^{-1/2}(q-q^{-1})a_{i,k})C \\
  &= (q-q^{-1})(-q^{1/2}Aa_{i+1,k}a_{i+1,\ell}C+q^{3/2}Aa_{i+1,\ell}a_{i+1,k}C + q^{1/2}Aa_{i,k}a_{i,\ell}C-q^{3/2}Aa_{i,\ell}a_{i,k}C) \\
  &\hspace{3em} +Aa_{i,k}a_{i+1,\ell}a_{i,i+1}C-qAa_{i,\ell}a_{i+1,k}a_{i,i+1}C \\
  &= (Aa_{i,k}a_{i+1,\ell}C - qAa_{i,\ell}a_{i+1,k}C) a_{i,i+1}.
\end{align*}
To see that the last step holds, observe that 
\begin{align*}
  -q^{1/2}Aa_{i+1,k}a_{i+1,\ell}C+q^{3/2}Aa_{i+1,\ell}a_{i+1,k}C &= -q^{1/2}A(qa_{i+1,\ell}a_{i+1,k}-q^{1/2}(q-q^{-1})a_{k,\ell})C + q^{3/2}Aa_{i+1,\ell}a_{i+1,k}C \\
 &= q(q-q^{-1})Aa_{k,\ell}C,
\end{align*}
and
\begin{align*}
  q^{1/2}Aa_{i,k}a_{i,\ell}C-q^{3/2}Aa_{i,\ell}a_{i,k}C &= q^{1/2}A(qa_{i,\ell}a_{i,k}-q^{1/2}(q-q^{-1})a_{k,\ell})C - q^{3/2}Aa_{i,\ell}a_{i,k}C \\
&= -q(q-q^{-1})Aa_{k,\ell}C.
\end{align*}
Hence,
\begin{align*}
  (q-q^{-1})(-q^{1/2}Aa_{i+1,k}a_{i+1,\ell}C+q^{3/2}Aa_{i+1,\ell}a_{i+1,k}C+q^{1/2}Aa_{i,k}a_{i,\ell}C-q^{3/2}Aa_{i,\ell}a_{i,k}C) = 0.
\end{align*}

These three claims imply that \(\widehat{Pf}_\triangle\) commutes with each generator \(a_{i,i+1}\), and hence with all of \(A_{2n}\). Thus \(\widehat{Pf}_\triangle\) is a Casimir function.

\medskip
\noindent\textbf{Step 2.} \emph{\(\widehat{Pf}_\triangle\) is self-adjoint.}

We now prove that \(\widehat{Pf}_\triangle\) is self-adjoint, that is,
\begin{equation}\label{eq:selfadjoint}
    \widehat{Pf}_\triangle - \widehat{Pf}_\triangle^* = 0.
\end{equation}
Consider the monomial
\[
    a_{1, n+1} \, a_{2, n+2} \cdots a_{n, 2n},
\]
which appears in \(\widehat{Pf}_\triangle\) and has the maximal possible number of chord crossings among all monomials in \(\widehat{Pf}_\triangle\).
Since simplification via the quantum relations only decreases the number of crossings, the only possible sources of this monomial in \(\widehat{Pf}_\triangle - \widehat{Pf}_\triangle^*\) are:
\begin{itemize}
    \item the original monomial \(a_{1, n+1} \, a_{2, n+2} \cdots a_{n, 2n}\) coming from \(\widehat{Pf}_\triangle\), and
    \item the conjugate monomial \(a_{n, 2n} \, a_{n-1, 2n-1} \cdots a_{1, n+1} \in \widehat{Pf}_\triangle^*\), which, after applying the quantum commutation relations, produces the same ordered product \(a_{1, n+1} \, a_{2, n+2} \cdots a_{n, 2n}\) modulo monomials with a lesser number of chord crossings.
\end{itemize}
No other monomials in \(\widehat{Pf}_\triangle\) or \(\widehat{Pf}_\triangle^*\) can give rise to this specific monomial in the simplification process, and the two contributions cancel each other. Hence the coefficient of\(a_{1, n+1} \, a_{2, n+2} \cdots a_{n, 2n}\) in \(\widehat{Pf}_\triangle - \widehat{Pf}_\triangle^*\) is zero. 

Now observe that every (lexicographically ordered) monomial in \(\widehat{Pf}_\triangle^*\) appears in $\widehat{Pf}_\triangle$. Since the coefficient of $a_{1, n+1} \, a_{2, n+2} \cdots a_{n, 2n}$ in \(\widehat{Pf}_\triangle - \widehat{Pf}_\triangle^*\) is zero, \(\widehat{Pf}_\triangle - \widehat{Pf}_\triangle^*\) must be strictly simpler than $\widehat{Pf}_\triangle$. But $\widehat{Pf}_\triangle$ is the simplest nonzero Casimir, so \(\widehat{Pf}_\triangle - \widehat{Pf}_\triangle^*\) must be identically zero.
\end{proof}

\begin{remark}
We also have a direct proof of the self-adjointness of the quantum Pfaffian. One of the theorems of Jing and Zhang~\cite[Theorem 2.10]{Jing2014} claims that the other way of writing the Pfaffian is 
$$
\widehat{Pf}_\triangle=\sum_{\pi\in\Pi'}(-q)^{\ell(\pi)}a_{\pi_1,\pi_2}\cdots a_{\pi_{2n-1},\pi_{2n}},
$$
and these two forms of writing the Pfaffian differ only by the overall factor. This theorem of Jing and Zhang is a corollary of a more general Theorem~\ref{thm-Pf} establishing this relation for the full algebra $\mathbb A_{2n}$ (see Section~\ref{sec:full_algebra} below); if we restrict to the upper-triangular case, we obtain exactly the expression in Theorem~\ref{t1} (up to a reciprocal polynomial in the $q$ factor, which is automatically self-adjoint).   
\end{remark}

\subsubsection{Examples of the quantum Pfaffian}
\begin{exm}\label{eg22}
When $n=2$, Theorem \ref{t1} gives:\[
\begin{aligned}
    \widehat{Pf}_\triangle = q \cdot a_{14} a_{23} - a_{13} a_{24} + q^{-1} \cdot a_{12} a_{34}.
\end{aligned}
\]
\end{exm}

\begin{exm}\label{example2n=6}
When $n=3$, Theorem \ref{t1} gives:
\[
\begin{aligned}
\widehat{Pf}_\triangle &= q^{-3} \cdot a_{12} a_{34} a_{56} - q^{-2}  \cdot a_{12} a_{35} a_{46} +q^{-1}  \cdot a_{12} a_{36} a_{45} - q^{-2}  \cdot a_{13} a_{24} a_{56} \\
&\quad +q^{-1}  \cdot a_{13} a_{25} a_{46}  - a_{13} a_{26} a_{45}  +q^{-1} \cdot  a_{14} a_{23} a_{56} - a_{14} a_{25} a_{36} \\
&\quad + q  \cdot a_{14} a_{26} a_{35}  - a_{15} a_{23} a_{46} + q \cdot  a_{15} a_{24} a_{36} - q^2 \cdot  a_{15} a_{26} a_{34} \\
&\quad + q  \cdot a_{16} a_{23} a_{45}- q^2  \cdot a_{16} a_{24} a_{35} + q^3 \cdot  a_{16} a_{25} a_{34}.
\end{aligned}
\]
Graphically, the same expression reads
$$
\begin{pspicture}(-7,3)(7,-2.5){\psset{unit=0.7}
\newcommand{\ARC}[1]{
\psarc[linecolor=white, linewidth=3pt](0,0){#1}{0}{180}
\psarc[linecolor=blue, linewidth=2pt](0,0){#1}{0}{180}
}
\rput(-8,2.5){\psline[linewidth=1pt](-1.5,0)(1.5,0)
\rput(-1.9,0){\makebox(0,0)[cb]{\small $-1$}}\rput(-0.5,0){\ARC{0.75}}\rput(0,0){\ARC{0.75}}\rput(0.5,0){\ARC{0.75}}
}
\rput(-4,2.5){\psline[linewidth=1pt](-1.5,0)(1.5,0)
\rput(-1.9,0){\makebox(0,0)[cb]{\small $+q^{-1}$}}\rput(-0.75,0){\ARC{0.5}}\rput(0,0){\ARC{0.75}}\rput(0.75,0){\ARC{0.5}}
}
\rput(0,2.5){\psline[linewidth=1pt](-1.5,0)(1.5,0)
\rput(-1.9,0){\makebox(0,0)[cb]{\small $+q$}}\rput(-0.25,0){\ARC{1}}\rput(-0.25,0){\ARC{0.5}}\rput(0.5,0){\ARC{0.75}}
}
\rput(4,2.5){\psline[linewidth=1pt](-1.5,0)(1.5,0)
\rput(-1.9,0){\makebox(0,0)[cb]{\small $+q$}}\rput(-0.5,0){\ARC{0.75}}\rput(0.25,0){\ARC{1}}\rput(0.25,0){\ARC{0.5}}
}
\rput(8,2.5){\psline[linewidth=1pt](-1.5,0)(1.5,0)
\rput(-1.9,0){\makebox(0,0)[cb]{\small $-q^2$}}\rput(0,0){\ARC{1.25}}\rput(-0.25,0){\ARC{0.5}}\rput(0.25,0){\ARC{0.5}}
}
\rput(-8,0){\psline[linewidth=1pt](-1.5,0)(1.5,0)
\rput(-1.9,0){\makebox(0,0)[cb]{\small $-q^{-2}$}}\rput(-0.75,0){\ARC{0.5}}\rput(-0.25,0){\ARC{0.5}}\rput(1,0){\ARC{0.25}}
}
\rput(-4,0){\psline[linewidth=1pt](-1.5,0)(1.5,0)
\rput(-1.9,0){\makebox(0,0)[cb]{\small $-q^{-2}$}}\rput(-1,0){\ARC{0.25}}\rput(0.25,0){\ARC{0.5}}\rput(0.75,0){\ARC{0.5}}
}
\rput(0,0){\psline[linewidth=1pt](-1.5,0)(1.5,0)
\rput(-1.9,0){\makebox(0,0)[cb]{\small $-1$}}\rput(-0.25,0){\ARC{1}}\rput(-0.5,0){\ARC{0.25}}\rput(0.75,0){\ARC{0.5}}
}
\rput(4,0){\psline[linewidth=1pt](-1.5,0)(1.5,0)
\rput(-1.9,0){\makebox(0,0)[cb]{\small $-q^{2}$}}\rput(-0.25,0){\ARC{1}}\rput(0.25,0){\ARC{1}}\rput(0,0){\ARC{0.25}}
}
\rput(8,0){\psline[linewidth=1pt](-1.5,0)(1.5,0)
\rput(-1.9,0){\makebox(0,0)[cb]{\small $-1$}}\rput(-0.75,0){\ARC{0.5}}\rput(0.25,0){\ARC{1}}\rput(0.25,0){\ARC{0.25}}
}
\rput(-8,-2.5){\psline[linewidth=1pt](-1.5,0)(1.5,0)
\rput(-1.9,0){\makebox(0,0)[cb]{\small $+q^{3}$}}\rput(0,0){\ARC{1.25}}\rput(0,0){\ARC{0.75}}\rput(0,0){\ARC{0.25}}
}
\rput(-4,-2.5){\psline[linewidth=1pt](-1.5,0)(1.5,0)
\rput(-1.9,0){\makebox(0,0)[cb]{\small $+q^{-1}$}}\rput(-0.5,0){\ARC{0.75}}\rput(-0.5,0){\ARC{0.25}}\rput(1,0){\ARC{0.25}}
}
\rput(0,-2.5){\psline[linewidth=1pt](-1.5,0)(1.5,0)
\rput(-1.9,0){\makebox(0,0)[cb]{\small $+q^{-1}$}}\rput(-1,0){\ARC{0.25}}\rput(0.5,0){\ARC{0.75}}\rput(0.5,0){\ARC{0.25}}
}
\rput(4,-2.5){\psline[linewidth=1pt](-1.5,0)(1.5,0)
\rput(-1.9,0){\makebox(0,0)[cb]{\small $+q$}}\rput(0,0){\ARC{1.25}}\rput(-0.5,0){\ARC{0.25}}\rput(0.5,0){\ARC{0.25}}
}
\rput(8,-2.5){\psline[linewidth=1pt](-1.5,0)(1.5,0)
\rput(-1.9,0){\makebox(0,0)[cb]{\small $+q^{-3}$}}\rput(-1,0){\ARC{0.25}}\rput(0,0){\ARC{0.25}}\rput(1,0){\ARC{0.25}}
}
}
\end{pspicture}
$$

\end{exm}

\subsection{The first Markov element}

\begin{defn}\label{def:qM1}
    Define the \emph{first Markov element} as the following sum over all closed nonintersecting cycles.
    \begin{align*}
        \widehat{C_1} \coloneqq &\sum_{k=3}^n\sum_{i_1<i_2<\cdots<i_k} (-1)^{n-k}q^{2(i_k-i_1)-n-k/2} a_{i_1i_2}a_{i_1i_k}a_{i_2i_3}\cdots a_{i_{k-1}i_k}\\
        &+\sum_{i_1<i_2} (-1)^{n-2}q^{2(i_2-i_1)-n}a_{i_1i_2}^2
    \end{align*}
    For ease of notation, we write $[[i_1,i_2,\dots,i_k]]:=a_{i_1i_2}a_{i_1i_k}a_{i_2i_3}\cdots a_{i_{k-1}i_k}$ for $1 \le i_1<i_2<\cdots<i_k \le n$.
\end{defn}

\begin{thm}\label{thm-M1}
    The first Markov element is a self-adjoint Casimir function of $A_n$.
\end{thm}

\begin{proof}
We first present the chord diagram calculation for a general case where $i_1<i<i+1<i_k$. Let $i_j=m$ be the largest index in $[[i_1\dots i_k]]$ that is strictly smaller than $i$ and $i_s=l$ be the smallest index strictly greater than $i+1$. Consider the combination of three cycles
    $$
    q^\gamma [[i_1,\dots, m,i,i+1,l,\dots,i_k]]-q^{\alpha} [[i_1,\dots, m,i,l,\dots,i_k]]-q^{\beta}[[i_1,\dots, m,i+1,l,\dots,i_k]],
    $$
    with all other indices coinciding for all three expressions. Graphically, we present this sum as
    $$
    
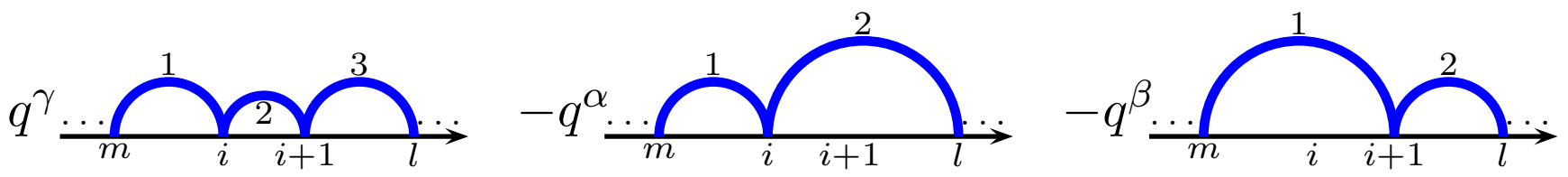
\begin{pspicture}(-5,1)(5,-0.5)
\newcommand{\ARCSLEFT}[3]{
\psline[linewidth=1pt]{->}(-1.5,0)(1.5,0)
\psarc[linecolor=blue, linewidth=2pt](-0.7,0){0.4}{0}{180}
\psarc[linecolor=blue, linewidth=2pt](0,0){0.3}{0}{180}
\psarc[linecolor=blue, linewidth=2pt](0.4,0){0.7}{0}{180}
\rput(-0.7,0.45){\makebox(0,0)[cb]{\tiny $#1$}}
\rput(0.0,0.25){\makebox(0,0)[ct]{\tiny $#2$}}
\rput(0.4,0.75){\makebox(0,0)[cb]{\tiny $#3$}}
\rput(-1.1,-0.05){\makebox(0,0)[ct]{\tiny $m$}}
\rput(-0.3,-0.05){\makebox(0,0)[ct]{\tiny $i$}}
\rput(0.3,-0.05){\makebox(0,0)[ct]{\tiny $i{+}1$}}
\rput(1.1,-0.05){\makebox(0,0)[ct]{\tiny $l$}}
\rput(-1.5,0.1){\makebox(0,0)[lc]{\tiny$\dots$}}
}
\newcommand{\ARCS}[3]{
\psline[linewidth=1pt]{->}(-1.5,0)(1.5,0)
\psarc[linecolor=blue, linewidth=2pt](-0.7,0){0.4}{0}{180}
\psarc[linecolor=blue, linewidth=2pt](0,0){0.3}{0}{180}
\psarc[linecolor=blue, linewidth=2pt](0.7,0){0.4}{0}{180}
\rput(-0.7,0.45){\makebox(0,0)[cb]{\tiny $#1$}}
\rput(0.0,0.25){\makebox(0,0)[ct]{\tiny $#2$}}
\rput(0.7,0.45){\makebox(0,0)[cb]{\tiny $#3$}}
\rput(-1.1,-0.05){\makebox(0,0)[ct]{\tiny $m$}}
\rput(-0.3,-0.05){\makebox(0,0)[ct]{\tiny $i$}}
\rput(0.3,-0.05){\makebox(0,0)[ct]{\tiny $i{+}1$}}
\rput(1.1,-0.05){\makebox(0,0)[ct]{\tiny $l$}}
\rput(-1.5,0.1){\makebox(0,0)[lc]{\tiny $\dots$}}
\rput(1.5,0.1){\makebox(0,0)[rc]{\tiny $\dots$}}
}
\newcommand{\TWOARCS}[2]{
\psline[linewidth=1pt]{->}(-1.5,0)(1.5,0)
\psarc[linecolor=blue, linewidth=2pt](-0.7,0){0.4}{0}{180}
\psarc[linecolor=blue, linewidth=2pt](0.4,0){0.7}{0}{180}
\rput(-0.7,0.45){\makebox(0,0)[cb]{\tiny $#1$}}
\rput(0.4,0.75){\makebox(0,0)[cb]{\tiny $#2$}}
\rput(-1.1,-0.05){\makebox(0,0)[ct]{\tiny $m$}}
\rput(-0.3,-0.05){\makebox(0,0)[ct]{\tiny $i$}}
\rput(0.3,-0.05){\makebox(0,0)[ct]{\tiny $i{+}1$}}
\rput(1.1,-0.05){\makebox(0,0)[ct]{\tiny $l$}}
\rput(-1.5,0.1){\makebox(0,0)[lc]{\tiny $\dots$}}
\rput(1.5,0.1){\makebox(0,0)[rc]{\tiny $\dots$}}
}
\newcommand{\TWOARCSRIGHT}[2]{
\psline[linewidth=1pt]{->}(-1.5,0)(1.5,0)
\psarc[linecolor=blue, linewidth=2pt](-0.4,0){0.7}{0}{180}
\psarc[linecolor=blue, linewidth=2pt](0.7,0){0.4}{0}{180}
\rput(-0.4,0.75){\makebox(0,0)[cb]{\tiny $#1$}}
\rput(0.7,0.45){\makebox(0,0)[cb]{\tiny $#2$}}
\rput(-1.1,-0.05){\makebox(0,0)[ct]{\tiny $m$}}
\rput(-0.3,-0.05){\makebox(0,0)[ct]{\tiny $i$}}
\rput(0.3,-0.05){\makebox(0,0)[ct]{\tiny $i{+}1$}}
\rput(1.1,-0.05){\makebox(0,0)[ct]{\tiny $l$}}
\rput(-1.5,0.1){\makebox(0,0)[lc]{\tiny $\dots$}}
\rput(1.5,0.1){\makebox(0,0)[rc]{\tiny $\dots$}}
}

\rput(-4,0){\rput(-1.7,0){\makebox(0,0)[cb]{$q^{\gamma}$}}\ARCS{1}{2}{3}}
\rput(0,0){\rput(-1.8,0){\makebox(0,0)[cb]{$-q^{\alpha}$}}\TWOARCS{1}{2}}
\rput(4,0){\rput(-1.8,0){\makebox(0,0)[cb]{$-q^{\beta}$}}\TWOARCSRIGHT{1}{2}}
%
\end{pspicture}
$$
indicating only relevant terms and keeping in mind that the rest of the terms commute with $a_{i,i+1}$. Commuting $a_{i,i+1}$ with the first term, we obtain
 $$
    
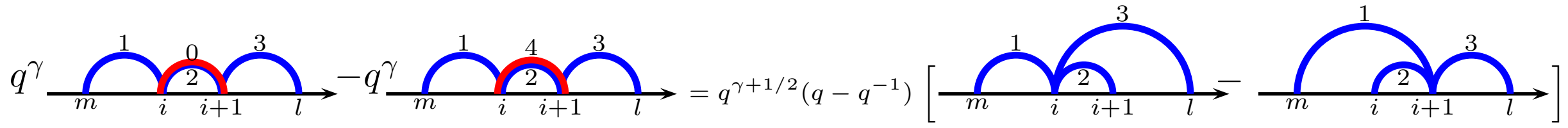
\begin{pspicture}(-7,1)(7,-0.5)
\newcommand{\ARCSLEFT}[3]{
\psline[linewidth=1pt]{->}(-1.5,0)(1.5,0)
\psarc[linecolor=blue, linewidth=2pt](-0.7,0){0.4}{0}{180}
\psarc[linecolor=blue, linewidth=2pt](0,0){0.3}{0}{180}
\psarc[linecolor=blue, linewidth=2pt](0.4,0){0.7}{0}{180}
\rput(-0.7,0.45){\makebox(0,0)[cb]{\tiny $#1$}}
\rput(0.0,0.25){\makebox(0,0)[ct]{\tiny $#2$}}
\rput(0.4,0.75){\makebox(0,0)[cb]{\tiny $#3$}}
\rput(-1.1,-0.05){\makebox(0,0)[ct]{\tiny $m$}}
\rput(-0.3,-0.05){\makebox(0,0)[ct]{\tiny $i$}}
\rput(0.3,-0.05){\makebox(0,0)[ct]{\tiny $i{+}1$}}
\rput(1.1,-0.05){\makebox(0,0)[ct]{\tiny $l$}}
}
\newcommand{\ARCSRIGHT}[3]{
\psline[linewidth=1pt]{->}(-1.5,0)(1.5,0)
\psarc[linecolor=blue, linewidth=2pt](-0.4,0){0.7}{0}{180}
\psarc[linecolor=blue, linewidth=2pt](0,0){0.3}{0}{180}
\psarc[linecolor=blue, linewidth=2pt](0.7,0){0.4}{0}{180}
\rput(-0.4,0.75){\makebox(0,0)[cb]{\tiny $#1$}}
\rput(0.0,0.25){\makebox(0,0)[ct]{\tiny $#2$}}
\rput(0.7,0.45){\makebox(0,0)[cb]{\tiny $#3$}}
\rput(-1.1,-0.05){\makebox(0,0)[ct]{\tiny $m$}}
\rput(-0.3,-0.05){\makebox(0,0)[ct]{\tiny $i$}}
\rput(0.3,-0.05){\makebox(0,0)[ct]{\tiny $i{+}1$}}
\rput(1.1,-0.05){\makebox(0,0)[ct]{\tiny $l$}}
}
\newcommand{\ARCS}[3]{
\psline[linewidth=1pt]{->}(-1.5,0)(1.5,0)
\psarc[linecolor=blue, linewidth=2pt](-0.7,0){0.4}{0}{180}
\psarc[linecolor=blue, linewidth=2pt](0,0){0.3}{0}{180}
\psarc[linecolor=blue, linewidth=2pt](0.7,0){0.4}{0}{180}
\rput(-0.7,0.45){\makebox(0,0)[cb]{\tiny $#1$}}
\rput(0.0,0.25){\makebox(0,0)[ct]{\tiny $#2$}}
\rput(0.7,0.45){\makebox(0,0)[cb]{\tiny $#3$}}
\rput(-1.1,-0.05){\makebox(0,0)[ct]{\tiny $m$}}
\rput(-0.3,-0.05){\makebox(0,0)[ct]{\tiny $i$}}
\rput(0.3,-0.05){\makebox(0,0)[ct]{\tiny $i{+}1$}}
\rput(1.1,-0.05){\makebox(0,0)[ct]{\tiny $l$}}
}
\newcommand{\TWOARCS}[2]{
\psline[linewidth=1pt]{->}(-1.5,0)(1.5,0)
\psarc[linecolor=blue, linewidth=2pt](-0.7,0){0.4}{0}{180}
\psarc[linecolor=blue, linewidth=2pt](0.4,0){0.7}{0}{180}
\rput(-0.7,0.45){\makebox(0,0)[cb]{\tiny $#1$}}
\rput(0.4,0.75){\makebox(0,0)[cb]{\tiny $#2$}}
\rput(-1.1,-0.05){\makebox(0,0)[ct]{\tiny $m$}}
\rput(-0.3,-0.05){\makebox(0,0)[ct]{\tiny $i$}}
\rput(0.3,-0.05){\makebox(0,0)[ct]{\tiny $i{+}1$}}
\rput(1.1,-0.05){\makebox(0,0)[ct]{\tiny $l$}}
}
\newcommand{\TWOARCSRIGHT}[2]{
\psline[linewidth=1pt]{->}(-1.5,0)(1.5,0)
\psarc[linecolor=blue, linewidth=2pt](-0.4,0){0.7}{0}{180}
\psarc[linecolor=blue, linewidth=2pt](0.7,0){0.4}{0}{180}
\rput(-0.4,0.75){\makebox(0,0)[cb]{\tiny $#1$}}
\rput(0.7,0.45){\makebox(0,0)[cb]{\tiny $#2$}}
\rput(-1.1,-0.05){\makebox(0,0)[ct]{\tiny $m$}}
\rput(-0.3,-0.05){\makebox(0,0)[ct]{\tiny $i$}}
\rput(0.3,-0.05){\makebox(0,0)[ct]{\tiny $i{+}1$}}
\rput(1.1,-0.05){\makebox(0,0)[ct]{\tiny $l$}}
}

\rput(-6,0){\rput(-1.7,0){\makebox(0,0)[cb]{$q^{\gamma}$}}\ARCS{1}{2}{3}\psarc[linecolor=red, linewidth=2pt](0,0){0.33}{0}{180}\rput(0.0,0.35){\makebox(0,0)[cb]{\tiny $0$}}
}
\rput(-2.5,0){\rput(-1.7,0){\makebox(0,0)[cb]{$-q^{\gamma}$}}\ARCS{1}{2}{3}\psarc[linecolor=red, linewidth=2pt](0,0){0.35}{0}{180}\rput(0.0,0.4){\makebox(0,0)[cb]{\tiny $4$}}
}
\rput(3.2,0){\rput(-2.8,0){\makebox(0,0)[cc]{\tiny $=q^{\gamma+1/2}(q-q^{-1})\biggm[$}}\ARCSLEFT{1}{2}{3}}
\rput(6.5,0){\rput(-1.8,0){\makebox(0,0)[cb]{$-$}}\ARCSRIGHT{1}{2}{3}\rput(1.6,0){\makebox(0,0)[cc]{\tiny $\biggm]$}}}
\end{pspicture}
$$
Combining these terms with four terms appearing in the commutators of $a_{i,i+1}$ with the remaining two terms, we obtain
 $$
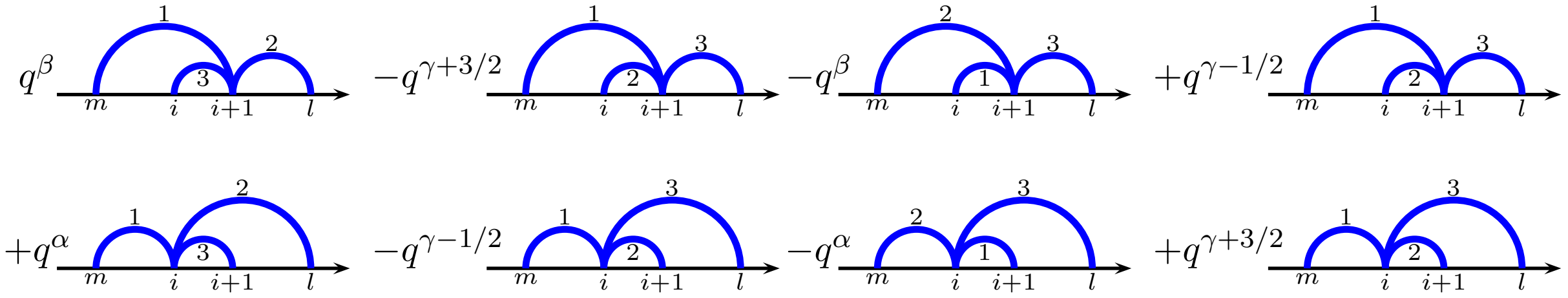
\begin{pspicture}(-4,1)(4,-0.5)
\newcommand{\ARCSLEFT}[3]{
\psline[linewidth=1pt]{->}(-1.5,0)(1.5,0)
\psarc[linecolor=blue, linewidth=2pt](-0.7,0){0.4}{0}{180}
\psarc[linecolor=blue, linewidth=2pt](0,0){0.3}{0}{180}
\psarc[linecolor=blue, linewidth=2pt](0.4,0){0.7}{0}{180}
\rput(-0.7,0.45){\makebox(0,0)[cb]{\tiny $#1$}}
\rput(0.0,0.25){\makebox(0,0)[ct]{\tiny $#2$}}
\rput(0.4,0.75){\makebox(0,0)[cb]{\tiny $#3$}}
\rput(-1.1,-0.05){\makebox(0,0)[ct]{\tiny $m$}}
\rput(-0.3,-0.05){\makebox(0,0)[ct]{\tiny $i$}}
\rput(0.3,-0.05){\makebox(0,0)[ct]{\tiny $i{+}1$}}
\rput(1.1,-0.05){\makebox(0,0)[ct]{\tiny $l$}}
}
\newcommand{\ARCSRIGHT}[3]{
\psline[linewidth=1pt]{->}(-1.5,0)(1.5,0)
\psarc[linecolor=blue, linewidth=2pt](-0.4,0){0.7}{0}{180}
\psarc[linecolor=blue, linewidth=2pt](0,0){0.3}{0}{180}
\psarc[linecolor=blue, linewidth=2pt](0.7,0){0.4}{0}{180}
\rput(-0.4,0.75){\makebox(0,0)[cb]{\tiny $#1$}}
\rput(0.0,0.25){\makebox(0,0)[ct]{\tiny $#2$}}
\rput(0.7,0.45){\makebox(0,0)[cb]{\tiny $#3$}}
\rput(-1.1,-0.05){\makebox(0,0)[ct]{\tiny $m$}}
\rput(-0.3,-0.05){\makebox(0,0)[ct]{\tiny $i$}}
\rput(0.3,-0.05){\makebox(0,0)[ct]{\tiny $i{+}1$}}
\rput(1.1,-0.05){\makebox(0,0)[ct]{\tiny $l$}}
}
\rput(-2.2,0){\rput(-1.7,0){\makebox(0,0)[cb]{$q^{\beta}$}}\ARCSRIGHT{1}{3}{2}
}
\rput(2.2,0){\rput(-2,0){\makebox(0,0)[cb]{$-q^{\gamma+3/2}$}}\ARCSRIGHT{1}{2}{3}}
\end{pspicture}
\begin{pspicture}(-4,1)(4,-0.5)
\newcommand{\ARCSLEFT}[3]{
\psline[linewidth=1pt]{->}(-1.5,0)(1.5,0)
\psarc[linecolor=blue, linewidth=2pt](-0.7,0){0.4}{0}{180}
\psarc[linecolor=blue, linewidth=2pt](0,0){0.3}{0}{180}
\psarc[linecolor=blue, linewidth=2pt](0.4,0){0.7}{0}{180}
\rput(-0.7,0.45){\makebox(0,0)[cb]{\tiny $#1$}}
\rput(0.0,0.25){\makebox(0,0)[ct]{\tiny $#2$}}
\rput(0.4,0.75){\makebox(0,0)[cb]{\tiny $#3$}}
\rput(-1.1,-0.05){\makebox(0,0)[ct]{\tiny $m$}}
\rput(-0.3,-0.05){\makebox(0,0)[ct]{\tiny $i$}}
\rput(0.3,-0.05){\makebox(0,0)[ct]{\tiny $i{+}1$}}
\rput(1.1,-0.05){\makebox(0,0)[ct]{\tiny $l$}}
}
\newcommand{\ARCSRIGHT}[3]{
\psline[linewidth=1pt]{->}(-1.5,0)(1.5,0)
\psarc[linecolor=blue, linewidth=2pt](-0.4,0){0.7}{0}{180}
\psarc[linecolor=blue, linewidth=2pt](0,0){0.3}{0}{180}
\psarc[linecolor=blue, linewidth=2pt](0.7,0){0.4}{0}{180}
\rput(-0.4,0.75){\makebox(0,0)[cb]{\tiny $#1$}}
\rput(0.0,0.25){\makebox(0,0)[ct]{\tiny $#2$}}
\rput(0.7,0.45){\makebox(0,0)[cb]{\tiny $#3$}}
\rput(-1.1,-0.05){\makebox(0,0)[ct]{\tiny $m$}}
\rput(-0.3,-0.05){\makebox(0,0)[ct]{\tiny $i$}}
\rput(0.3,-0.05){\makebox(0,0)[ct]{\tiny $i{+}1$}}
\rput(1.1,-0.05){\makebox(0,0)[ct]{\tiny $l$}}
}
\rput(-2.2,0){\rput(-1.7,0){\makebox(0,0)[cb]{$-q^{\beta}$}}\ARCSRIGHT{2}{1}{3}
}
\rput(2.2,0){\rput(-2,0){\makebox(0,0)[cb]{$+q^{\gamma-1/2}$}}\ARCSRIGHT{1}{2}{3}}
\end{pspicture}
$$
 $$
\begin{pspicture}(-4,1)(4,-0.5)
\newcommand{\ARCSLEFT}[3]{
\psline[linewidth=1pt]{->}(-1.5,0)(1.5,0)
\psarc[linecolor=blue, linewidth=2pt](-0.7,0){0.4}{0}{180}
\psarc[linecolor=blue, linewidth=2pt](0,0){0.3}{0}{180}
\psarc[linecolor=blue, linewidth=2pt](0.4,0){0.7}{0}{180}
\rput(-0.7,0.45){\makebox(0,0)[cb]{\tiny $#1$}}
\rput(0.0,0.25){\makebox(0,0)[ct]{\tiny $#2$}}
\rput(0.4,0.75){\makebox(0,0)[cb]{\tiny $#3$}}
\rput(-1.1,-0.05){\makebox(0,0)[ct]{\tiny $m$}}
\rput(-0.3,-0.05){\makebox(0,0)[ct]{\tiny $i$}}
\rput(0.3,-0.05){\makebox(0,0)[ct]{\tiny $i{+}1$}}
\rput(1.1,-0.05){\makebox(0,0)[ct]{\tiny $l$}}
}
\newcommand{\ARCSRIGHT}[3]{
\psline[linewidth=1pt]{->}(-1.5,0)(1.5,0)
\psarc[linecolor=blue, linewidth=2pt](-0.4,0){0.7}{0}{180}
\psarc[linecolor=blue, linewidth=2pt](0,0){0.3}{0}{180}
\psarc[linecolor=blue, linewidth=2pt](0.7,0){0.4}{0}{180}
\rput(-0.4,0.75){\makebox(0,0)[cb]{\tiny $#1$}}
\rput(0.0,0.25){\makebox(0,0)[ct]{\tiny $#2$}}
\rput(0.7,0.45){\makebox(0,0)[cb]{\tiny $#3$}}
\rput(-1.1,-0.05){\makebox(0,0)[ct]{\tiny $m$}}
\rput(-0.3,-0.05){\makebox(0,0)[ct]{\tiny $i$}}
\rput(0.3,-0.05){\makebox(0,0)[ct]{\tiny $i{+}1$}}
\rput(1.1,-0.05){\makebox(0,0)[ct]{\tiny $l$}}
}
\rput(-2.2,0){\rput(-1.7,0){\makebox(0,0)[cb]{$+q^{\alpha}$}}\ARCSLEFT{1}{3}{2}
}
\rput(2.2,0){\rput(-2,0){\makebox(0,0)[cb]{$-q^{\gamma-1/2}$}}\ARCSLEFT{1}{2}{3}}
\end{pspicture}
\begin{pspicture}(-4,1)(4,-0.5)
\newcommand{\ARCSLEFT}[3]{
\psline[linewidth=1pt]{->}(-1.5,0)(1.5,0)
\psarc[linecolor=blue, linewidth=2pt](-0.7,0){0.4}{0}{180}
\psarc[linecolor=blue, linewidth=2pt](0,0){0.3}{0}{180}
\psarc[linecolor=blue, linewidth=2pt](0.4,0){0.7}{0}{180}
\rput(-0.7,0.45){\makebox(0,0)[cb]{\tiny $#1$}}
\rput(0.0,0.25){\makebox(0,0)[ct]{\tiny $#2$}}
\rput(0.4,0.75){\makebox(0,0)[cb]{\tiny $#3$}}
\rput(-1.1,-0.05){\makebox(0,0)[ct]{\tiny $m$}}
\rput(-0.3,-0.05){\makebox(0,0)[ct]{\tiny $i$}}
\rput(0.3,-0.05){\makebox(0,0)[ct]{\tiny $i{+}1$}}
\rput(1.1,-0.05){\makebox(0,0)[ct]{\tiny $l$}}
}
\newcommand{\ARCSRIGHT}[3]{
\psline[linewidth=1pt]{->}(-1.5,0)(1.5,0)
\psarc[linecolor=blue, linewidth=2pt](-0.4,0){0.7}{0}{180}
\psarc[linecolor=blue, linewidth=2pt](0,0){0.3}{0}{180}
\psarc[linecolor=blue, linewidth=2pt](0.7,0){0.4}{0}{180}
\rput(-0.4,0.75){\makebox(0,0)[cb]{\tiny $#1$}}
\rput(0.0,0.25){\makebox(0,0)[ct]{\tiny $#2$}}
\rput(0.7,0.45){\makebox(0,0)[cb]{\tiny $#3$}}
\rput(-1.1,-0.05){\makebox(0,0)[ct]{\tiny $m$}}
\rput(-0.3,-0.05){\makebox(0,0)[ct]{\tiny $i$}}
\rput(0.3,-0.05){\makebox(0,0)[ct]{\tiny $i{+}1$}}
\rput(1.1,-0.05){\makebox(0,0)[ct]{\tiny $l$}}
}
\rput(-2.2,0){\rput(-1.7,0){\makebox(0,0)[cb]{$-q^{\alpha}$}}\ARCSLEFT{2}{1}{3}
}
\rput(2.2,0){\rput(-2,0){\makebox(0,0)[cb]{$+q^{\gamma+3/2}$}}\ARCSLEFT{1}{2}{3}}
\end{pspicture}
$$
Setting $\alpha=\beta=\gamma+1/2$ and again using the chord commutation relations, we transform the first line into
$$
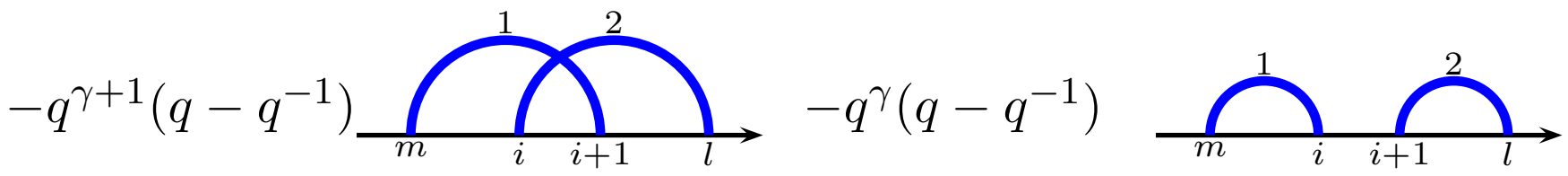
\begin{pspicture}(-4,1)(4,-0.5)
\newcommand{\ARCSCROSS}[2]{
\psline[linewidth=1pt]{->}(-1.5,0)(1.5,0)
\psarc[linecolor=blue, linewidth=2pt](-0.4,0){0.7}{0}{180}
\psarc[linecolor=blue, linewidth=2pt](0.4,0){0.7}{0}{180}
\rput(-0.4,0.75){\makebox(0,0)[cb]{\tiny $#1$}}
\rput(0.4,0.75){\makebox(0,0)[cb]{\tiny $#2$}}
\rput(-1.1,-0.05){\makebox(0,0)[ct]{\tiny $m$}}
\rput(-0.3,-0.05){\makebox(0,0)[ct]{\tiny $i$}}
\rput(0.3,-0.05){\makebox(0,0)[ct]{\tiny $i{+}1$}}
\rput(1.1,-0.05){\makebox(0,0)[ct]{\tiny $l$}}
}
\newcommand{\ARCSSEP}[2]{
\psline[linewidth=1pt]{->}(-1.5,0)(1.5,0)
\psarc[linecolor=blue, linewidth=2pt](-0.7,0){0.4}{0}{180}
\psarc[linecolor=blue, linewidth=2pt](0.7,0){0.4}{0}{180}
\rput(-0.7,0.45){\makebox(0,0)[cb]{\tiny $#1$}}
\rput(0.7,0.45){\makebox(0,0)[cb]{\tiny $#2$}}
\rput(-1.1,-0.05){\makebox(0,0)[ct]{\tiny $m$}}
\rput(-0.3,-0.05){\makebox(0,0)[ct]{\tiny $i$}}
\rput(0.3,-0.05){\makebox(0,0)[ct]{\tiny $i{+}1$}}
\rput(1.1,-0.05){\makebox(0,0)[ct]{\tiny $l$}}
}
\rput(-2.7,0){\rput(-2.8,0){\makebox(0,0)[cb]{$-q^{\gamma+1}(q-q^{-1})$}}\ARCSCROSS{1}{2}
}
\rput(3.2,0){\rput(-3,0){\makebox(0,0)[cb]{$-q^{\gamma}(q-q^{-1})$}}\ARCSSEP{1}{2}}
\end{pspicture}
$$
whereas the second line gives
$$
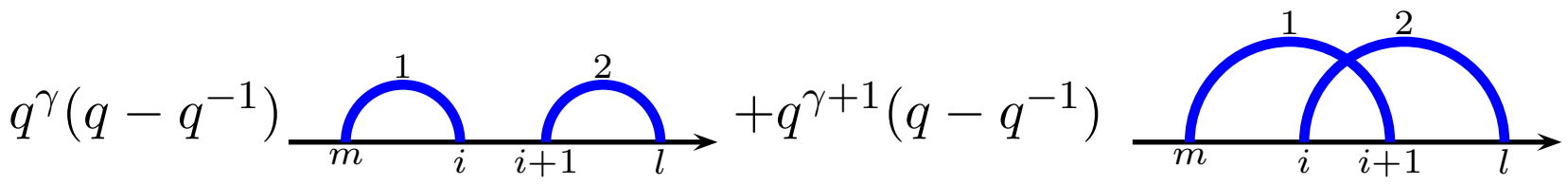
\begin{pspicture}(-4,1)(4,-0.5)
\newcommand{\ARCSCROSS}[2]{
\psline[linewidth=1pt]{->}(-1.5,0)(1.5,0)
\psarc[linecolor=blue, linewidth=2pt](-0.4,0){0.7}{0}{180}
\psarc[linecolor=blue, linewidth=2pt](0.4,0){0.7}{0}{180}
\rput(-0.4,0.75){\makebox(0,0)[cb]{\tiny $#1$}}
\rput(0.4,0.75){\makebox(0,0)[cb]{\tiny $#2$}}
\rput(-1.1,-0.05){\makebox(0,0)[ct]{\tiny $m$}}
\rput(-0.3,-0.05){\makebox(0,0)[ct]{\tiny $i$}}
\rput(0.3,-0.05){\makebox(0,0)[ct]{\tiny $i{+}1$}}
\rput(1.1,-0.05){\makebox(0,0)[ct]{\tiny $l$}}
}
\newcommand{\ARCSSEP}[2]{
\psline[linewidth=1pt]{->}(-1.5,0)(1.5,0)
\psarc[linecolor=blue, linewidth=2pt](-0.7,0){0.4}{0}{180}
\psarc[linecolor=blue, linewidth=2pt](0.7,0){0.4}{0}{180}
\rput(-0.7,0.45){\makebox(0,0)[cb]{\tiny $#1$}}
\rput(0.7,0.45){\makebox(0,0)[cb]{\tiny $#2$}}
\rput(-1.1,-0.05){\makebox(0,0)[ct]{\tiny $m$}}
\rput(-0.3,-0.05){\makebox(0,0)[ct]{\tiny $i$}}
\rput(0.3,-0.05){\makebox(0,0)[ct]{\tiny $i{+}1$}}
\rput(1.1,-0.05){\makebox(0,0)[ct]{\tiny $l$}}
}
\rput(2.2,0){\rput(-3,0){\makebox(0,0)[cb]{$+q^{\gamma+1}(q-q^{-1})$}}\ARCSCROSS{1}{2}
}
\rput(-3.7,0){\rput(-2.5,0){\makebox(0,0)[cb]{$q^{\gamma}(q-q^{-1})$}}\ARCSSEP{1}{2}}
\end{pspicture}
$$
that is, they are fully mutually canceled and the whole expression therefore commutes with $a_{i,i+1}$. 

We next consider the case where we are at the leftmost part of the cycle.
$$
    q^\alpha [[i+1,p, \dots,r]]+q^\beta [[i,p, \dots,r]]-q^\gamma [[i,i+1,p, \dots,r]]
    $$
    with all other indices coinciding for all three expressions. Graphically, we present this sum as
$$
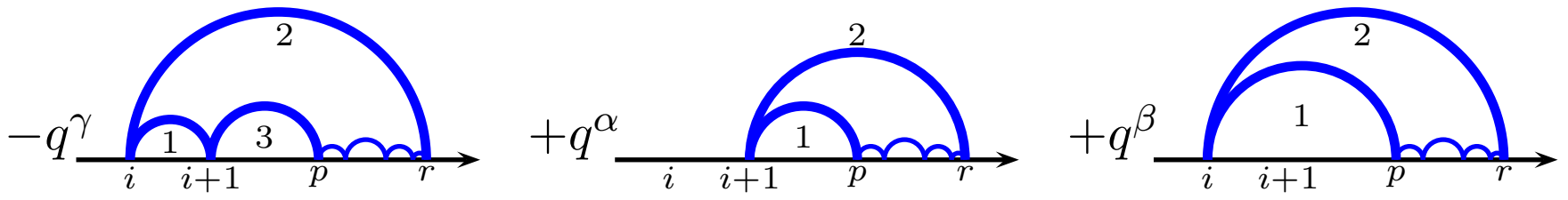
\begin{pspicture}(-5,1)(5,-0.5)
\newcommand{\ARCSRIGHT}[3]{
\psline[linewidth=1pt]{->}(-1.5,0)(1.5,0)
\psarc[linecolor=blue, linewidth=2pt](-0.8,0){0.3}{0}{180}
\psarc[linecolor=blue, linewidth=2pt](-0.1,0){0.4}{0}{180}
\psarc[linecolor=blue, linewidth=2pt](0.3,0){0.8}{0}{180}
\rput(-0.8,0.35){\makebox(0,0)[cb]{\tiny $#1$}}
\rput(-0.1,0.25){\makebox(0,0)[ct]{\tiny $#2$}}
\rput(0.3,0.85){\makebox(0,0)[cb]{\tiny $#3$}}
\rput(-1.1,-0.05){\makebox(0,0)[ct]{\tiny $i$}}
\rput(-0.5,-0.05){\makebox(0,0)[ct]{\tiny $i{+}1$}}
\rput(0.3,-0.05){\makebox(0,0)[ct]{\tiny $p$}}
\rput(1.1,-0.05){\makebox(0,0)[ct]{\tiny $r$}}
}
\newcommand{\ARCSLEFT}[3]{
\psline[linewidth=1pt]{->}(-1.5,0)(1.5,0)
\psarc[linecolor=blue, linewidth=2pt](-0.8,0){0.3}{0}{180}
\psarc[linecolor=blue, linewidth=2pt](-0.4,0){0.7}{0}{180}
\psarc[linecolor=blue, linewidth=2pt](0,0){1.1}{0}{180}
\rput(-0.8,0.1){\makebox(0,0)[cb]{\tiny $#1$}}
\rput(-0.4,0.25){\makebox(0,0)[ct]{\tiny $#2$}}
\rput(0.0,1.15){\makebox(0,0)[cb]{\tiny $#3$}}
\rput(-1.1,-0.05){\makebox(0,0)[ct]{\tiny $i$}}
\rput(-0.5,-0.05){\makebox(0,0)[ct]{\tiny $i{+}1$}}
\rput(0.3,-0.05){\makebox(0,0)[ct]{\tiny $p$}}
\rput(1.1,-0.05){\makebox(0,0)[ct]{\tiny $r$}}
}
\newcommand{\FOURARCS}[4]{
\psline[linewidth=1pt]{->}(-1.5,0)(1.5,0)
\psarc[linecolor=blue, linewidth=2pt](-0.8,0){0.3}{0}{180}
\psarc[linecolor=red, linewidth=2pt](-0.8,0){0.35}{0}{180}
\psarc[linecolor=blue, linewidth=2pt](-0.1,0){0.4}{0}{180}
\psarc[linecolor=blue, linewidth=2pt](0,0){1.1}{0}{180}
\rput(-0.8,0.05){\makebox(0,0)[cb]{\tiny $#1$}}
\rput(-0.8,0.35){\makebox(0,0)[cb]{\tiny $#2$}}
\rput(-0.1,0.25){\makebox(0,0)[ct]{\tiny $#3$}}
\rput(0.3,0.85){\makebox(0,0)[cb]{\tiny $#4$}}
\rput(-1.1,-0.05){\makebox(0,0)[ct]{\tiny $i$}}
\rput(-0.5,-0.05){\makebox(0,0)[ct]{\tiny $i{+}1$}}
\rput(0.3,-0.05){\makebox(0,0)[ct]{\tiny $p$}}
\rput(1.1,-0.05){\makebox(0,0)[ct]{\tiny $r$}}
}
\newcommand{\TWOARCSRIGHT}[2]{
\psline[linewidth=1pt]{->}(-1.5,0)(1.5,0)
\psarc[linecolor=blue, linewidth=2pt](-0.1,0){0.4}{0}{180}
\psarc[linecolor=blue, linewidth=2pt](0.3,0){0.8}{0}{180}
\psarc[linecolor=blue, linewidth=1pt](0.4,0){0.1}{0}{180}
\psarc[linecolor=blue, linewidth=1pt](0.65,0){0.15}{0}
{180}
\psarc[linecolor=blue, linewidth=1pt](0.9,0){0.1}{0}{180}
\psarc[linecolor=blue, linewidth=1pt](1.05,0){0.05}{0}{180}
\rput(-0.1,0.25){\makebox(0,0)[ct]{\tiny $#1$}}
\rput(0.3,0.85){\makebox(0,0)[cb]{\tiny $#2$}}
\rput(-1.1,-0.05){\makebox(0,0)[ct]{\tiny $i$}}
\rput(-0.5,-0.05){\makebox(0,0)[ct]{\tiny $i{+}1$}}
\rput(0.3,-0.05){\makebox(0,0)[ct]{\tiny $p$}}
\rput(1.1,-0.05){\makebox(0,0)[ct]{\tiny $r$}}
}
\newcommand{\TWOARCSLEFT}[2]{
\psline[linewidth=1pt]{->}(-1.5,0)(1.5,0)
\psarc[linecolor=blue, linewidth=2pt](-0.4,0){0.7}{0}{180}
\psarc[linecolor=blue, linewidth=2pt](0,0){1.1}{0}{180}
\psarc[linecolor=blue, linewidth=1pt](0.4,0){0.1}{0}{180}
\psarc[linecolor=blue, linewidth=1pt](0.65,0){0.15}{0}
{180}
\psarc[linecolor=blue, linewidth=1pt](0.9,0){0.1}{0}{180}
\psarc[linecolor=blue, linewidth=1pt](1.05,0){0.05}{0}{180}
\rput(-0.4,0.25){\makebox(0,0)[cb]{\tiny $#1$}}
\rput(0.05,0.85){\makebox(0,0)[cb]{\tiny $#2$}}
\rput(-1.1,-0.05){\makebox(0,0)[ct]{\tiny $i$}}
\rput(-0.5,-0.05){\makebox(0,0)[ct]{\tiny $i{+}1$}}
\rput(0.3,-0.05){\makebox(0,0)[ct]{\tiny $p$}}
\rput(1.1,-0.05){\makebox(0,0)[ct]{\tiny $r$}}
}
\newcommand{\THREEARCS}[3]{
\psline[linewidth=1pt]{->}(-1.5,0)(1.5,0)
\psarc[linecolor=blue, linewidth=2pt](-0.8,0){0.3}{0}{180}
\psarc[linecolor=blue, linewidth=2pt](-0.1,0){0.4}{0}{180}
\psarc[linecolor=blue, linewidth=2pt](0,0){1.1}{0}{180}
\psarc[linecolor=blue, linewidth=1pt](0.4,0){0.1}{0}{180}
\psarc[linecolor=blue, linewidth=1pt](0.65,0){0.15}{0}
{180}
\psarc[linecolor=blue, linewidth=1pt](0.9,0){0.1}{0}{180}
\psarc[linecolor=blue, linewidth=1pt](1.05,0){0.05}{0}{180}
\rput(-0.8,0.05){\makebox(0,0)[cb]{\tiny $#1$}}
\rput(-0.1,0.25){\makebox(0,0)[ct]{\tiny $#2$}}
\rput(0.05,0.85){\makebox(0,0)[cb]{\tiny $#3$}}
\rput(-1.1,-0.05){\makebox(0,0)[ct]{\tiny $i$}}
\rput(-0.5,-0.05){\makebox(0,0)[ct]{\tiny $i{+}1$}}
\rput(0.3,-0.05){\makebox(0,0)[ct]{\tiny $p$}}
\rput(1.1,-0.05){\makebox(0,0)[ct]{\tiny $r$}}
}

\rput(-4,0){\rput(-1.7,0){\makebox(0,0)[cb]{$-q^{\gamma}$}}\THREEARCS{1}{3}{2}}
\rput(0,0){\rput(-1.8,0){\makebox(0,0)[cb]{$+q^{\alpha}$}}\TWOARCSRIGHT{1}{2}}
\rput(4,0){\rput(-1.8,0){\makebox(0,0)[cb]{$+q^{\beta}$}}\TWOARCSLEFT{1}{2}}
%
\end{pspicture}
$$
We first consider the term proportional to $q^\gamma$:
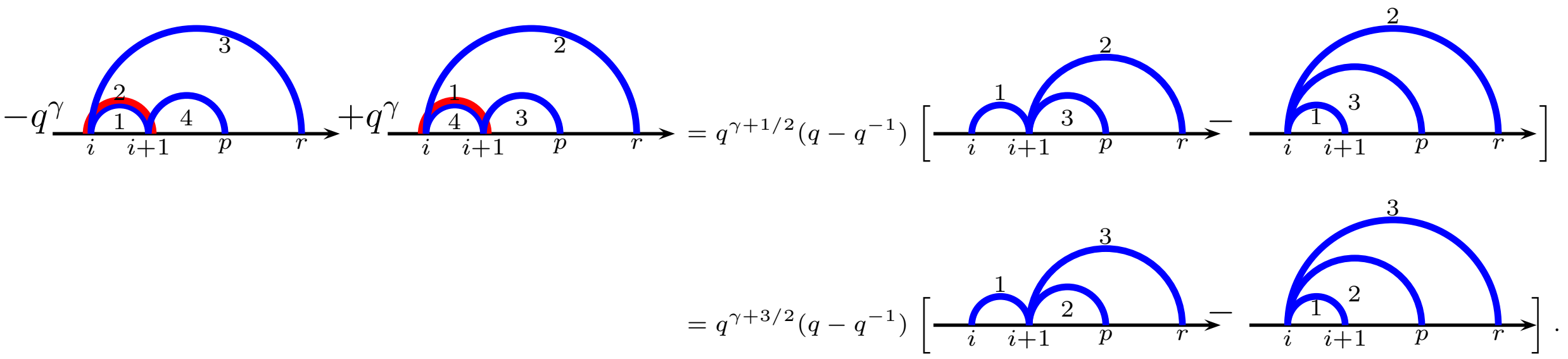
$$
\begin{pspicture}(-7,1.2)(7,-2.5)
\newcommand{\ARCSRIGHT}[3]{
\psline[linewidth=1pt]{->}(-1.5,0)(1.5,0)
\psarc[linecolor=blue, linewidth=2pt](-0.8,0){0.3}{0}{180}
\psarc[linecolor=blue, linewidth=2pt](-0.1,0){0.4}{0}{180}
\psarc[linecolor=blue, linewidth=2pt](0.3,0){0.8}{0}{180}
\rput(-0.8,0.35){\makebox(0,0)[cb]{\tiny $#1$}}
\rput(-0.1,0.25){\makebox(0,0)[ct]{\tiny $#2$}}
\rput(0.3,0.85){\makebox(0,0)[cb]{\tiny $#3$}}
\rput(-1.1,-0.05){\makebox(0,0)[ct]{\tiny $i$}}
\rput(-0.5,-0.05){\makebox(0,0)[ct]{\tiny $i{+}1$}}
\rput(0.3,-0.05){\makebox(0,0)[ct]{\tiny $p$}}
\rput(1.1,-0.05){\makebox(0,0)[ct]{\tiny $r$}}
}
\newcommand{\ARCSLEFT}[3]{
\psline[linewidth=1pt]{->}(-1.5,0)(1.5,0)
\psarc[linecolor=blue, linewidth=2pt](-0.8,0){0.3}{0}{180}
\psarc[linecolor=blue, linewidth=2pt](-0.4,0){0.7}{0}{180}
\psarc[linecolor=blue, linewidth=2pt](0,0){1.1}{0}{180}
\rput(-0.8,0.1){\makebox(0,0)[cb]{\tiny $#1$}}
\rput(-0.4,0.25){\makebox(0,0)[cb]{\tiny $#2$}}
\rput(0.0,1.15){\makebox(0,0)[cb]{\tiny $#3$}}
\rput(-1.1,-0.05){\makebox(0,0)[ct]{\tiny $i$}}
\rput(-0.5,-0.05){\makebox(0,0)[ct]{\tiny $i{+}1$}}
\rput(0.3,-0.05){\makebox(0,0)[ct]{\tiny $p$}}
\rput(1.1,-0.05){\makebox(0,0)[ct]{\tiny $r$}}
}
\newcommand{\FOURARCS}[4]{
\psline[linewidth=1pt]{->}(-1.5,0)(1.5,0)
\psarc[linecolor=blue, linewidth=2pt](-0.8,0){0.3}{0}{180}
\psarc[linecolor=red, linewidth=2pt](-0.8,0){0.35}{0}{180}
\psarc[linecolor=blue, linewidth=2pt](-0.1,0){0.4}{0}{180}
\psarc[linecolor=blue, linewidth=2pt](0,0){1.1}{0}{180}
\rput(-0.8,0.05){\makebox(0,0)[cb]{\tiny $#1$}}
\rput(-0.8,0.35){\makebox(0,0)[cb]{\tiny $#2$}}
\rput(-0.1,0.25){\makebox(0,0)[ct]{\tiny $#3$}}
\rput(0.3,0.85){\makebox(0,0)[cb]{\tiny $#4$}}
\rput(-1.1,-0.05){\makebox(0,0)[ct]{\tiny $i$}}
\rput(-0.5,-0.05){\makebox(0,0)[ct]{\tiny $i{+}1$}}
\rput(0.3,-0.05){\makebox(0,0)[ct]{\tiny $p$}}
\rput(1.1,-0.05){\makebox(0,0)[ct]{\tiny $r$}}
}
\newcommand{\TWOARCSRIGHT}[2]{
\psline[linewidth=1pt]{->}(-1.5,0)(1.5,0)
\psarc[linecolor=blue, linewidth=2pt](-0.1,0){0.4}{0}{180}
\psarc[linecolor=blue, linewidth=2pt](0.3,0){0.8}{0}{180}
\rput(-0.1,0.25){\makebox(0,0)[ct]{\tiny $#1$}}
\rput(0.3,0.85){\makebox(0,0)[cb]{\tiny $#2$}}
\rput(-1.1,-0.05){\makebox(0,0)[ct]{\tiny $i$}}
\rput(-0.5,-0.05){\makebox(0,0)[ct]{\tiny $i{+}1$}}
\rput(0.3,-0.05){\makebox(0,0)[ct]{\tiny $p$}}
\rput(1.1,-0.05){\makebox(0,0)[ct]{\tiny $r$}}
}
\newcommand{\TWOARCSLEFT}[2]{
\psline[linewidth=1pt]{->}(-1.5,0)(1.5,0)
\psarc[linecolor=blue, linewidth=2pt](-0.4,0){0.7}{0}{180}
\psarc[linecolor=blue, linewidth=2pt](0,0){1.1}{0}{180}
\rput(-0.4,0.25){\makebox(0,0)[cb]{\tiny $#1$}}
\rput(0.05,0.85){\makebox(0,0)[cb]{\tiny $#2$}}
\rput(-1.1,-0.05){\makebox(0,0)[ct]{\tiny $i$}}
\rput(-0.5,-0.05){\makebox(0,0)[ct]{\tiny $i{+}1$}}
\rput(0.3,-0.05){\makebox(0,0)[ct]{\tiny $p$}}
\rput(1.1,-0.05){\makebox(0,0)[ct]{\tiny $r$}}
}
\newcommand{\THREEARCS}[3]{
\psline[linewidth=1pt]{->}(-1.5,0)(1.5,0)
\psarc[linecolor=blue, linewidth=2pt](-0.8,0){0.3}{0}{180}
\psarc[linecolor=blue, linewidth=2pt](-0.1,0){0.4}{0}{180}
\psarc[linecolor=blue, linewidth=2pt](0,0){1.1}{0}{180}
\rput(-0.8,0.05){\makebox(0,0)[cb]{\tiny $#1$}}
\rput(-0.1,0.25){\makebox(0,0)[ct]{\tiny $#2$}}
\rput(0.05,0.85){\makebox(0,0)[cb]{\tiny $#3$}}
\rput(-1.1,-0.05){\makebox(0,0)[ct]{\tiny $i$}}
\rput(-0.5,-0.05){\makebox(0,0)[ct]{\tiny $i{+}1$}}
\rput(0.3,-0.05){\makebox(0,0)[ct]{\tiny $p$}}
\rput(1.1,-0.05){\makebox(0,0)[ct]{\tiny $r$}}
}

\rput(-6,0){\rput(-1.7,0){\makebox(0,0)[cb]{$-q^{\gamma}$}}\FOURARCS{1}{2}{4}{3}
}
\rput(-2.5,0){\rput(-1.7,0){\makebox(0,0)[cb]{$+q^{\gamma}$}}\FOURARCS{4}{1}{3}{2}}
\rput(3.2,0){\rput(-2.8,0){\makebox(0,0)[cc]{\tiny $=q^{\gamma+1/2}(q-q^{-1})\biggm[$}}\ARCSRIGHT{1}{3}{2}}
\rput(6.5,0){\rput(-1.8,0){\makebox(0,0)[cb]{$-$}}\ARCSLEFT{1}{3}{2}\rput(1.6,0){\makebox(0,0)[cc]{\tiny $\biggm]$}}}
\rput(3.2,-2){\rput(-2.8,0){\makebox(0,0)[cc]{\tiny $=q^{\gamma+3/2}(q-q^{-1})\biggm[$}}\ARCSRIGHT{1}{2}{3}}
\rput(6.5,-2){\rput(-1.8,0){\makebox(0,0)[cb]{$-$}}\ARCSLEFT{1}{2}{3}\rput(1.6,0){\makebox(0,0)[cc]{\tiny $\biggm].$}}}
\end{pspicture}
$$
Note that in the last line, when commuting the arcs numbered $2$ and $3$, two terms of the form $
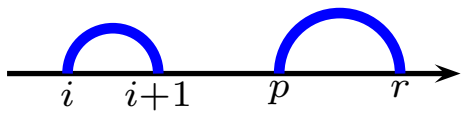
\begin{pspicture}(-1.5,0.5)(1.5,0)
\psline[linewidth=1pt]{->}(-1.5,0)(1.5,0)
\psarc[linecolor=blue, linewidth=2pt](-0.8,0){0.3}{0}{180}
\psarc[linecolor=blue, linewidth=2pt](0.7,0){0.4}{0}{180}
\rput(-1.1,-0.05){\makebox(0,0)[ct]{\tiny $i$}}
\rput(-0.5,-0.05){\makebox(0,0)[ct]{\tiny $i{+}1$}}
\rput(0.3,-0.05){\makebox(0,0)[ct]{\tiny $p$}}
\rput(1.1,-0.05){\makebox(0,0)[ct]{\tiny $r$}}
\end{pspicture}$ appear that are mutually canceled.
The terms with $q^\alpha$ transform into
$$
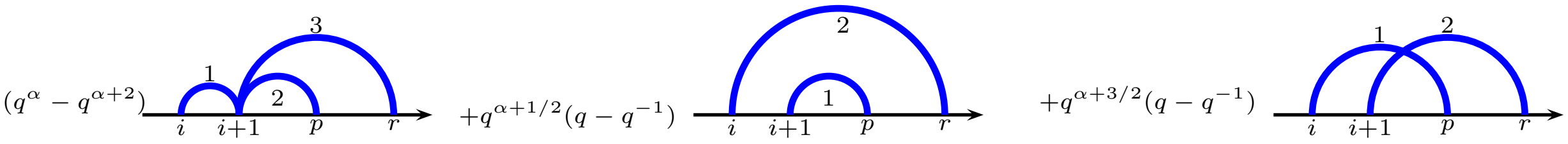
\begin{pspicture}(-7,1.2)(7,-0.5)
\newcommand{\ARCSRIGHT}[3]{
\psline[linewidth=1pt]{->}(-1.5,0)(1.5,0)
\psarc[linecolor=blue, linewidth=2pt](-0.8,0){0.3}{0}{180}
\psarc[linecolor=blue, linewidth=2pt](-0.1,0){0.4}{0}{180}
\psarc[linecolor=blue, linewidth=2pt](0.3,0){0.8}{0}{180}
\rput(-0.8,0.35){\makebox(0,0)[cb]{\tiny $#1$}}
\rput(-0.1,0.25){\makebox(0,0)[ct]{\tiny $#2$}}
\rput(0.3,0.85){\makebox(0,0)[cb]{\tiny $#3$}}
\rput(-1.1,-0.05){\makebox(0,0)[ct]{\tiny $i$}}
\rput(-0.5,-0.05){\makebox(0,0)[ct]{\tiny $i{+}1$}}
\rput(0.3,-0.05){\makebox(0,0)[ct]{\tiny $p$}}
\rput(1.1,-0.05){\makebox(0,0)[ct]{\tiny $r$}}
}
\newcommand{\ARCSLEFT}[3]{
\psline[linewidth=1pt]{->}(-1.5,0)(1.5,0)
\psarc[linecolor=blue, linewidth=2pt](-0.8,0){0.3}{0}{180}
\psarc[linecolor=blue, linewidth=2pt](-0.4,0){0.7}{0}{180}
\psarc[linecolor=blue, linewidth=2pt](0,0){1.1}{0}{180}
\rput(-0.8,0.1){\makebox(0,0)[cb]{\tiny $#1$}}
\rput(-0.4,0.25){\makebox(0,0)[cb]{\tiny $#2$}}
\rput(0.0,1.15){\makebox(0,0)[cb]{\tiny $#3$}}
\rput(-1.1,-0.05){\makebox(0,0)[ct]{\tiny $i$}}
\rput(-0.5,-0.05){\makebox(0,0)[ct]{\tiny $i{+}1$}}
\rput(0.3,-0.05){\makebox(0,0)[ct]{\tiny $p$}}
\rput(1.1,-0.05){\makebox(0,0)[ct]{\tiny $r$}}
}
\newcommand{\FOURARCS}[4]{
\psline[linewidth=1pt]{->}(-1.5,0)(1.5,0)
\psarc[linecolor=blue, linewidth=2pt](-0.8,0){0.3}{0}{180}
\psarc[linecolor=red, linewidth=2pt](-0.8,0){0.35}{0}{180}
\psarc[linecolor=blue, linewidth=2pt](-0.1,0){0.4}{0}{180}
\psarc[linecolor=blue, linewidth=2pt](0,0){1.1}{0}{180}
\rput(-0.8,0.05){\makebox(0,0)[cb]{\tiny $#1$}}
\rput(-0.8,0.35){\makebox(0,0)[cb]{\tiny $#2$}}
\rput(-0.1,0.25){\makebox(0,0)[ct]{\tiny $#3$}}
\rput(0.3,0.85){\makebox(0,0)[cb]{\tiny $#4$}}
\rput(-1.1,-0.05){\makebox(0,0)[ct]{\tiny $i$}}
\rput(-0.5,-0.05){\makebox(0,0)[ct]{\tiny $i{+}1$}}
\rput(0.3,-0.05){\makebox(0,0)[ct]{\tiny $p$}}
\rput(1.1,-0.05){\makebox(0,0)[ct]{\tiny $r$}}
}
\newcommand{\TWOARCSRIGHT}[2]{
\psline[linewidth=1pt]{->}(-1.5,0)(1.5,0)
\psarc[linecolor=blue, linewidth=2pt](-0.1,0){0.4}{0}{180}
\psarc[linecolor=blue, linewidth=2pt](0.3,0){0.8}{0}{180}
\rput(-0.1,0.25){\makebox(0,0)[ct]{\tiny $#1$}}
\rput(0.3,0.85){\makebox(0,0)[cb]{\tiny $#2$}}
\rput(-1.1,-0.05){\makebox(0,0)[ct]{\tiny $i$}}
\rput(-0.5,-0.05){\makebox(0,0)[ct]{\tiny $i{+}1$}}
\rput(0.3,-0.05){\makebox(0,0)[ct]{\tiny $p$}}
\rput(1.1,-0.05){\makebox(0,0)[ct]{\tiny $r$}}
}
\newcommand{\TWOARCSLEFT}[2]{
\psline[linewidth=1pt]{->}(-1.5,0)(1.5,0)
\psarc[linecolor=blue, linewidth=2pt](-0.4,0){0.7}{0}{180}
\psarc[linecolor=blue, linewidth=2pt](0,0){1.1}{0}{180}
\rput(-0.4,0.25){\makebox(0,0)[cb]{\tiny $#1$}}
\rput(0.05,0.85){\makebox(0,0)[cb]{\tiny $#2$}}
\rput(-1.1,-0.05){\makebox(0,0)[ct]{\tiny $i$}}
\rput(-0.5,-0.05){\makebox(0,0)[ct]{\tiny $i{+}1$}}
\rput(0.3,-0.05){\makebox(0,0)[ct]{\tiny $p$}}
\rput(1.1,-0.05){\makebox(0,0)[ct]{\tiny $r$}}
}
\newcommand{\TWOARCS}[2]{
\psline[linewidth=1pt]{->}(-1.5,0)(1.5,0)
\psarc[linecolor=blue, linewidth=2pt](-0.1,0){0.4}{0}{180}
\psarc[linecolor=blue, linewidth=2pt](0,0){1.1}{0}{180}
\rput(-0.1,0.25){\makebox(0,0)[ct]{\tiny $#1$}}
\rput(0.05,0.85){\makebox(0,0)[cb]{\tiny $#2$}}
\rput(-1.1,-0.05){\makebox(0,0)[ct]{\tiny $i$}}
\rput(-0.5,-0.05){\makebox(0,0)[ct]{\tiny $i{+}1$}}
\rput(0.3,-0.05){\makebox(0,0)[ct]{\tiny $p$}}
\rput(1.1,-0.05){\makebox(0,0)[ct]{\tiny $r$}}
}
\newcommand{\TWOARCSCROSS}[2]{
\psline[linewidth=1pt]{->}(-1.5,0)(1.5,0)
\psarc[linecolor=blue, linewidth=2pt](0.3,0){0.8}{0}{180}
\psarc[linecolor=blue, linewidth=2pt](-0.4,0){0.7}{0}{180}
\rput(-0.4,0.75){\makebox(0,0)[cb]{\tiny $#1$}}
\rput(0.3,0.85){\makebox(0,0)[cb]{\tiny $#2$}}
\rput(-1.1,-0.05){\makebox(0,0)[ct]{\tiny $i$}}
\rput(-0.5,-0.05){\makebox(0,0)[ct]{\tiny $i{+}1$}}
\rput(0.3,-0.05){\makebox(0,0)[ct]{\tiny $p$}}
\rput(1.1,-0.05){\makebox(0,0)[ct]{\tiny $r$}}
}

\rput(-4.7,0){\rput(-2.2,0){\makebox(0,0)[cb]{\tiny $(q^{\alpha}-q^{\alpha+2})$}}\ARCSRIGHT{1}{2}{3}
}
\rput(1,0){\rput(-2.8,0){\makebox(0,0)[cc]{\tiny $+q^{\alpha+1/2}(q-q^{-1})$}}\TWOARCS{1}{2}}
\rput(7,0){\rput(-2.8,0){\makebox(0,0)[cb]{\tiny $+q^{\alpha+3/2}(q-q^{-1})$}}\TWOARCSCROSS{1}{2}}
\end{pspicture}
$$
whereas the terms with $q^{\beta}$ transform into
$$
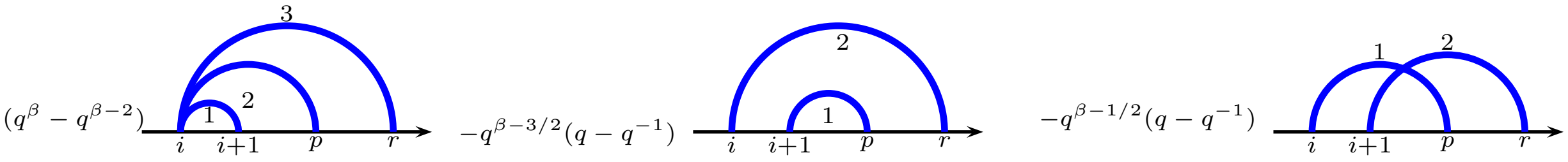
\begin{pspicture}(-7,1.2)(7,-0.5)
\newcommand{\ARCSRIGHT}[3]{
\psline[linewidth=1pt]{->}(-1.5,0)(1.5,0)
\psarc[linecolor=blue, linewidth=2pt](-0.8,0){0.3}{0}{180}
\psarc[linecolor=blue, linewidth=2pt](-0.1,0){0.4}{0}{180}
\psarc[linecolor=blue, linewidth=2pt](0.3,0){0.8}{0}{180}
\rput(-0.8,0.35){\makebox(0,0)[cb]{\tiny $#1$}}
\rput(-0.1,0.25){\makebox(0,0)[ct]{\tiny $#2$}}
\rput(0.3,0.85){\makebox(0,0)[cb]{\tiny $#3$}}
\rput(-1.1,-0.05){\makebox(0,0)[ct]{\tiny $i$}}
\rput(-0.5,-0.05){\makebox(0,0)[ct]{\tiny $i{+}1$}}
\rput(0.3,-0.05){\makebox(0,0)[ct]{\tiny $p$}}
\rput(1.1,-0.05){\makebox(0,0)[ct]{\tiny $r$}}
}
\newcommand{\ARCSLEFT}[3]{
\psline[linewidth=1pt]{->}(-1.5,0)(1.5,0)
\psarc[linecolor=blue, linewidth=2pt](-0.8,0){0.3}{0}{180}
\psarc[linecolor=blue, linewidth=2pt](-0.4,0){0.7}{0}{180}
\psarc[linecolor=blue, linewidth=2pt](0,0){1.1}{0}{180}
\rput(-0.8,0.1){\makebox(0,0)[cb]{\tiny $#1$}}
\rput(-0.4,0.25){\makebox(0,0)[cb]{\tiny $#2$}}
\rput(0.0,1.15){\makebox(0,0)[cb]{\tiny $#3$}}
\rput(-1.1,-0.05){\makebox(0,0)[ct]{\tiny $i$}}
\rput(-0.5,-0.05){\makebox(0,0)[ct]{\tiny $i{+}1$}}
\rput(0.3,-0.05){\makebox(0,0)[ct]{\tiny $p$}}
\rput(1.1,-0.05){\makebox(0,0)[ct]{\tiny $r$}}
}
\newcommand{\FOURARCS}[4]{
\psline[linewidth=1pt]{->}(-1.5,0)(1.5,0)
\psarc[linecolor=blue, linewidth=2pt](-0.8,0){0.3}{0}{180}
\psarc[linecolor=red, linewidth=2pt](-0.8,0){0.35}{0}{180}
\psarc[linecolor=blue, linewidth=2pt](-0.1,0){0.4}{0}{180}
\psarc[linecolor=blue, linewidth=2pt](0,0){1.1}{0}{180}
\rput(-0.8,0.05){\makebox(0,0)[cb]{\tiny $#1$}}
\rput(-0.8,0.35){\makebox(0,0)[cb]{\tiny $#2$}}
\rput(-0.1,0.25){\makebox(0,0)[ct]{\tiny $#3$}}
\rput(0.3,0.85){\makebox(0,0)[cb]{\tiny $#4$}}
\rput(-1.1,-0.05){\makebox(0,0)[ct]{\tiny $i$}}
\rput(-0.5,-0.05){\makebox(0,0)[ct]{\tiny $i{+}1$}}
\rput(0.3,-0.05){\makebox(0,0)[ct]{\tiny $p$}}
\rput(1.1,-0.05){\makebox(0,0)[ct]{\tiny $r$}}
}
\newcommand{\TWOARCSRIGHT}[2]{
\psline[linewidth=1pt]{->}(-1.5,0)(1.5,0)
\psarc[linecolor=blue, linewidth=2pt](-0.1,0){0.4}{0}{180}
\psarc[linecolor=blue, linewidth=2pt](0.3,0){0.8}{0}{180}
\rput(-0.1,0.25){\makebox(0,0)[ct]{\tiny $#1$}}
\rput(0.3,0.85){\makebox(0,0)[cb]{\tiny $#2$}}
\rput(-1.1,-0.05){\makebox(0,0)[ct]{\tiny $i$}}
\rput(-0.5,-0.05){\makebox(0,0)[ct]{\tiny $i{+}1$}}
\rput(0.3,-0.05){\makebox(0,0)[ct]{\tiny $p$}}
\rput(1.1,-0.05){\makebox(0,0)[ct]{\tiny $r$}}
}
\newcommand{\TWOARCSLEFT}[2]{
\psline[linewidth=1pt]{->}(-1.5,0)(1.5,0)
\psarc[linecolor=blue, linewidth=2pt](-0.4,0){0.7}{0}{180}
\psarc[linecolor=blue, linewidth=2pt](0,0){1.1}{0}{180}
\rput(-0.4,0.25){\makebox(0,0)[cb]{\tiny $#1$}}
\rput(0.05,0.85){\makebox(0,0)[cb]{\tiny $#2$}}
\rput(-1.1,-0.05){\makebox(0,0)[ct]{\tiny $i$}}
\rput(-0.5,-0.05){\makebox(0,0)[ct]{\tiny $i{+}1$}}
\rput(0.3,-0.05){\makebox(0,0)[ct]{\tiny $p$}}
\rput(1.1,-0.05){\makebox(0,0)[ct]{\tiny $r$}}
}
\newcommand{\TWOARCS}[2]{
\psline[linewidth=1pt]{->}(-1.5,0)(1.5,0)
\psarc[linecolor=blue, linewidth=2pt](-0.1,0){0.4}{0}{180}
\psarc[linecolor=blue, linewidth=2pt](0,0){1.1}{0}{180}
\rput(-0.1,0.25){\makebox(0,0)[ct]{\tiny $#1$}}
\rput(0.05,0.85){\makebox(0,0)[cb]{\tiny $#2$}}
\rput(-1.1,-0.05){\makebox(0,0)[ct]{\tiny $i$}}
\rput(-0.5,-0.05){\makebox(0,0)[ct]{\tiny $i{+}1$}}
\rput(0.3,-0.05){\makebox(0,0)[ct]{\tiny $p$}}
\rput(1.1,-0.05){\makebox(0,0)[ct]{\tiny $r$}}
}
\newcommand{\TWOARCSCROSS}[2]{
\psline[linewidth=1pt]{->}(-1.5,0)(1.5,0)
\psarc[linecolor=blue, linewidth=2pt](0.3,0){0.8}{0}{180}
\psarc[linecolor=blue, linewidth=2pt](-0.4,0){0.7}{0}{180}
\rput(-0.4,0.75){\makebox(0,0)[cb]{\tiny $#1$}}
\rput(0.3,0.85){\makebox(0,0)[cb]{\tiny $#2$}}
\rput(-1.1,-0.05){\makebox(0,0)[ct]{\tiny $i$}}
\rput(-0.5,-0.05){\makebox(0,0)[ct]{\tiny $i{+}1$}}
\rput(0.3,-0.05){\makebox(0,0)[ct]{\tiny $p$}}
\rput(1.1,-0.05){\makebox(0,0)[ct]{\tiny $r$}}
}

\rput(-4.7,0){\rput(-2.2,0){\makebox(0,0)[cb]{\tiny $(q^{\beta}-q^{\beta-2})$}}\ARCSLEFT{1}{2}{3}
}
\rput(1,0){\rput(-2.8,0){\makebox(0,0)[cc]{\tiny $-q^{\beta-3/2}(q-q^{-1})$}}\TWOARCS{1}{2}}
\rput(7,0){\rput(-2.8,0){\makebox(0,0)[cb]{\tiny $-q^{\beta-1/2}(q-q^{-1})$}}\TWOARCSCROSS{1}{2}}
\end{pspicture}
$$
The sum of the above three expressions becomes zero upon imposing the conditions
$$
\beta=\alpha+2=\gamma+5/2.
$$
The case when $a_{i,i+1}$ is commuting with 
$$
    q^\alpha [[p, \dots,r,i+1]]+q^\beta [[p, \dots,r,i]]-q^\gamma [[p, \dots,r,i,i+1]]
 $$
can be treated in the same way; in this case, we obtain
$$
\alpha=\beta+2=\gamma+1/2.
$$
The powers of $q$ in all three cases are consistent with those in the theorem assertion, so we have proved that it is a commuting element

When addressing conjugacy properties, note that $\widehat C_1$ is a ``simplest'' Casimir in the hierarchy of quantum Markov elements, so it must be self-adjoint up to an overall $q$-factor to be found. For this, consider the longest element of $\widehat C_1$:
\begin{equation}\label{eq:longest}
a_{1,2}a_{1,n}a_{2,3}a_{3,4}\cdots a_{n-1,n}.
\end{equation}
Its conjugate is $a_{n-1,n}a_{n-2,n-1}\cdots a_{2,3}a_{1,n} a_{1,2}$. When pushing the $a_{i,j}$'s to the left with the aim of reconstructing their original order in (\ref{eq:longest}), we obtain powers of $q$ and monomials in $a_{\alpha, \beta}$ of exactly shorter length (all constituting cycles), so we need only to calculate the corresponding total power of $q$, which is $q^{n-4}$. We therefore need to multiply the monomial (\ref{eq:longest}), and therefore the whole expression for $\widehat C_1$ by $q^{\frac n2-2}$, which agrees with the formula in the theorem, completing the proof.
\end{proof}

\section{Full-dimensional algebra}\label{sec:full_algebra}
We now proceed to the algebra of $a_{i,j}$ for the full-dimensional matrix $\mathbb A$.
The commutation relations for the full-dimensional algebra written in concise form read \cite[Eq.~(6.45)]{ChekhovMazzocco2013}
\begin{equation}\label{eq:FullQuantumCommutator}
\begin{split}
    q^{\delta_{s,j}+\delta_{i,j}}a_{i,s}a_{j,t} - q^{\delta_{s,t}+\delta_{i,t}}a_{j,t}a_{i,s} &= (q-q^{-1})q^{\delta_{s,i}}(\delta_{t>s}-\delta_{i>j})a_{j,s}a_{i,t}\\
    &+(q-q^{-1})(q^{\delta_{s,t}}\delta_{t>i}a_{j,i}a_{t,s}-q^{\delta_{i,j}}\delta_{s>j}a_{i,j}a_{s,t})\\
    &+(q-q^{-1})^2\delta_{s>i}(\delta_{t>s}-\delta_{i>j})a_{j,i}a_{s,t},
\end{split}
\end{equation}
but for the reader's convenience we present the full list of these relations in Appendix \ref{appendix:fullrelations}.

\begin{lem}
Assuming $a_{n,1}$ is invertible and $q\ne\pm1$, the full set of $a_{i,j}$ is generated by the $2n-1$ elements $a_{i,1}$ and $a_{n,i}$ with $i=2,\dots,n-1$, $a_{1,1}$, $a_{n,n}$, and $a_{n,1}$.
\end{lem}

The {\it proof} immediately follows from the commutation relations for $\{i,j\}\in [2,n-1]$:
\begin{equation}\label{eq:an1}
-(q-q^{-1})a_{n,1} a_{i,j} = \begin{cases}[a_{n,j},a_{i,1}]&\hbox{for}\ 1<j<i<n,\\ [a_{n,j},a_{i,1}]+(q-q^{-1})a_{n,i}a_{j,1}&\hbox{for}\ 1<i<j<n,\\ qa_{n,j}a_{i,1}-a_{i,1}a_{n,j}&\hbox{for}\ 1<i=j<n. \end{cases}
\end{equation}
Note that for the network in Fig.~\ref{fi:fullCasimirs}, the element $a_{n,1}$ is monomial in cluster variables and is therefore always invertible.

\begin{remark}
It is easy to see that the relations in (\ref{eq:an1}) are compatible with the conjugation relations (\ref{eq:conjugation_law_2}) (see also Lemma~\ref{lm:conj}).
\end{remark}

\subsection{Comparing expressions for the quantum Pfaffian \texorpdfstring{$\widehat{Pf}_\square$}{Pf_square}}\label{s:full_pfaffian}

Before proceeding further, we put forward the following conjecture motivated by the Kontsevich deformation quantization construction.

\begin{conjecture}\label{con:comm}
\begin{itemize}
\item[{\bf(a)}] Consider a semiclassical polynomial expression $X$ that is one of the ``lowest'' Casimirs $Pf$ or $C_1$. This expression then admits a uniquely defined \emph{quantum deformation} in which each monomial of $X$ is transformed into a product of the corresponding quantum operators $a_{i,j}$ taken in a prescribed (e.g., lexicographic) ordering whereas the integer coefficient $\pm r>0$ becomes the polynomial $p_r(q)=\sum_{i=1}^r q^{s_i}$.  
\item[{\bf(b)}] The thus obtained self-adjoint operator $\widehat X$ is a Casimir of the quantum algebra.
\end{itemize}
\end{conjecture}
Note that if part (b) holds true, then part (a) follows automatically from the fact that a lowest quantum Casimir operator must be self-adjoint.

We use the permutation subsets $\Pi,\widehat{\Pi},\Pi',\wp$ and the inversion number
from Section~\ref{subsubsec:notation}. We also introduce the following technical notation.

For $i<j$, let 
\begin{equation}\label{eq:bc}
  b_{i,j}:=a_{i,j}-qa_{j,i}\quad \hbox{and}\quad c_{i,j}:=a_{i,j}+q^{-1}a_{j,i}.
\end{equation}

We have the following formulas for adjoint operators compatible with the algebra:
\begin{equation}\label{eq:conjugation_law_2}
a_{i,j}^*=a_{i,j}-(q-q^{-1})a_{j,i}\quad \hbox{and}\quad a_{j,i}^*=a_{j,i}\quad \hbox{for}\ i<j;\qquad
a_{i,i}^* = q^{-1} a_{i,i}
\end{equation}


The conjugation law \eqref{eq:conjugation_law_2} differs from that in \cite[Eq.~(6.46)]{ChekhovMazzocco2013} because it was assumed in that paper that diagonal elements $a_{i,i}$ are self-adjoint. We depart from this assumption since declaring these elements non-self-adjoint is consistent with preservation of algebraic relations under conjugation.

The conjugation relations (\ref{eq:conjugation_law_2}) imply that
\begin{equation}
    b^*_{i,j}=b_{i,j}\quad \hbox{and}\quad c^*_{i,j}=c_{i,j},\quad i<j,\label{bstar}
\end{equation}
that is, both $b_{i,j}$ and $c_{i,j}$ are self-adjoint operators.

The aim of this subsection is to prove the following theorem.
\begin{thm}\label{thm-Pf}
The quantum Pfaffian $\widehat{Pf}_\square$  can be written in four equivalent ways:
\begin{align}
\widehat{Pf}_\square=q^{-n(n-1)}&\sum_{\pi\in \wp} (-q)^{l(\pi)} a_{\pi_1,\pi_2}\cdots a_{\pi_{2n-1},\pi_{2n}}\\
\widehat{Pf}_\square=q^{-n(n-1)}&\sum_{\pi\in \Pi'} (-q)^{l(\pi)} b_{\pi_1,\pi_2}\cdots b_{\pi_{2n-1},\pi_{2n}}\\
\widehat{Pf}_\square=q^{-n(n-1)}&\sum_{\pi\in \Pi} F_n(q) (-q)^{l(\pi)} \bigl( a_{\pi_1,\pi_2}-q^{2n-1}a_{\pi_2,\pi_1}\bigr)\bigl( a_{\pi_3,\pi_4}-q^{2n-3}a_{\pi_4,\pi_3}\bigr)\cdots\times\nonumber\\
&\times\cdots \bigl( a_{\pi_{2n-3},\pi_{2n-2}}-q^3 a_{\pi_{2n-2},\pi_{2n-3}}\bigr) \bigl( a_{\pi_{2n-1},\pi_{2n}}-q a_{\pi_{2n},\pi_{2n-1}}\bigr),\\
\widehat{Pf}_\square=q^{-n(n-1)}&\sum_{\pi\in \widehat{\Pi}} F_n(q) (-q)^{l(\pi)} \bigl( a_{\pi_1,\pi_2}-q a_{\pi_2,\pi_1}\bigr)\bigl( a_{\pi_3,\pi_4}-q^{3}a_{\pi_4,\pi_3}\bigr)\cdots\times\nonumber\\
&\times\cdots \bigl( a_{\pi_{2n-3},\pi_{2n-2}}-q^{2n-3} a_{\pi_{2n-2},\pi_{2n-3}}\bigr) \bigl( a_{\pi_{2n-1},\pi_{2n}}-q^{2n-1} a_{\pi_{2n},\pi_{2n-1}}\bigr),
\end{align}
where $F_n(q)$ is a Laurent polynomial in $q$ depending only on $n$. For its explicit form, see (\ref{Fn}).
\end{thm}

We begin with the example for $n=2$.
\begin{exm}
Direct calculations show that
$$
\sum_{\pi\in \wp=S_4}(-q)^{l(\pi)}a_{\pi_1,\pi_2}a_{\pi_3,\pi_4}=(1+q^2)\sum_{\pi\in \Pi\subset S_4}(-q)^{l(\pi)}\bigl(a_{\pi_1,\pi_2}-q^3a_{\pi_2,\pi_1}\bigr)\bigl(a_{\pi_3,\pi_4}-qa_{\pi_4,\pi_3}\bigr).
$$
\end{exm}

\begin{proof}[Proof of Theorem~\ref{thm-Pf}]
So far, we leave aside the common factor $q^{-n(n-1)}$. The first two expressions are obviously equivalent. We are now about to prove the equivalence of the second and third expressions by induction. The idea of the proof is inspired by the proof of Lemma~2.9 and Theorem~2.10 in~\cite{Jing2014}, but the technical relations involved are more tricky.

We need two technical formulas that can be proven by explicit calculations using the quantum algebra:
\begin{equation}\label{eq:b-c}
q c_{14}b_{23}-c_{13}b_{24}+q^{-1}c_{12}b_{34} = 
q b_{34}c_{12}-b_{24}c_{13}+q^{-1} b_{23} c_{14}
\end{equation}
and
\begin{equation}\label{eq:a-b}
q[qa_{41}b_{23}-a_{31}b_{24}+q^{-1}a_{21}b_{34}]
= q^{-1}[qb_{34}a_{21}-b_{24}a_{31}+q^{-1}b_{23} a_{41}]
\end{equation}
These two relations can be written in a more ``uniform'' way using inversion indices:
$$
\sum_{\pi\in \Pi'}(-q)^{l(1,\pi_2,\pi_3<\pi_4)-1} c_{1,\pi_2}b_{\pi_3,\pi_4}
=
\sum_{\pi\in \Pi'}(-q)^{l(\pi_3<\pi_4,1,\pi_2)-3} b_{\pi_3,\pi_4} c_{1,\pi_2}
$$
and
$$
q\sum_{\pi\in \Pi'}(-q)^{l(1,\pi_2,\pi_3<\pi_4)-1} a_{\pi_2,1}b_{\pi_3,\pi_4}
=
q^{-1}\sum_{\pi\in \Pi'}(-q)^{l(\pi_3<\pi_4,1,\pi_2)-3} b_{\pi_3,\pi_4} a_{\pi_2,1}
$$
Note that we intentionally keep inversion indices the same in both expressions although the indices of $a_{i,j}$ come in the opposite order in the second expression.

Of course, the same relations hold for any choice of four distinct ordered indices.

We now prove the statement of the theorem by induction. For $n=1$, $\Pi=\Pi'$, and both expressions  coincide. 

Using notation from Jing and Zhang~\cite{Jing2014}, let
$$
[1,2,\dots,\hat i,\dots \hat j\dots 2n]:=\sum_{\pi \in \Pi'\subset \textrm{Sym}([2n]\setminus\{i,j\})} (-q)^{l(\pi)}\prod_{i=1}^{n-1} b_{\pi_{2i-1},\pi_{2i}}
$$
where we exclude indices $i$ and $j$ (they can be equal to $1$ or $2n$ and we do not fix their order in this expression, we only assume $i\ne j$).

Note that the other way of rewriting the statement of the theorem is
\begin{equation}
[1,2,\dots,2n]=f_n(q) \sum_{k=2}^{2n} (-q)^{k-2} \bigl( a_{1,k}-q^{2n-1}a_{k,1}\bigr)[\hat 1,2,\dots,\hat k,\dots, 2n]
\end{equation}
with $F_n(q)=\prod_{m=1}^n f_m(q)$.

Note, first, that
\begin{align}
[1,2,\dots, 2n]&=\sum_{1\le i<j\le 2n} (-q)^{i+j-3}b_{i,j}[1,2,\dots,\hat i,\dots,\hat j \dots 2n]\nonumber\\
&=\sum_{1<j\le 2n}(-q)^{j-2} b_{1,j}[\hat 1,2,\dots,\hat j \dots 2n]\label{lineone}\\
&+\sum_{1<i<j\le 2n}(-q)^{i+j-3}b_{i,j}[1,2,\dots,\hat i,\dots,\hat j\dots,2n]\label{linetwo}
\end{align}

We now apply the induction assumption to the last term; note that the factors of $-q$ will depend on the mutual arrangement of indices $i<j$ and $k$:
\begin{align*}
b_{i,j}[1,2,\dots,\hat i,\dots,\hat j\dots,2n]=& f_{n-1}(q)\sum_{k>1,k\notin \{i,j\}}(-q)^{k-6+l(i,j,1,k)}
b_{i,j}\bigl(c_{1,k}-q^{-1}a_{k,1}-q^{2n-3}a_{k,1}\bigr)\times \\
&\times [\hat 1,2,\dots,\hat i,\dots,\hat j,\dots \hat k\dots,2n]
\end{align*}
We can therefore rewrite the whole expression in the last line as a sum over triples  $\{i,j,k\}$ of distinct numbers greater than one:
$$
f_{n-1}(q)\sum_{1<\{i<j,k\}}(-q)^{i+j+k-9}(-q)^{l(i,j,1,k)}b_{i,j}\bigl(c_{1,k}-q^{-1}a_{k,1}-q^{2n-3}a_{k,1}\bigr)
[\hat 1,2,\dots,\hat i,\dots,\hat j,\dots \hat k\dots,2n]
$$
and use relations (\ref{eq:b-c}) and (\ref{eq:a-b}) to rewrite the same expression as 
\begin{align}
&f_{n-1}(q)\sum_{1<\{i<j,k\}}(-q)^{i+j+k-9}(-q)^{l(1,k,i,j)+2}\bigl(c_{1,k}-qa_{k,1}-q^{2n-1}a_{k,1}\bigr)b_{i,j}
[\hat 1,2,\dots,\hat i,\dots,\hat j,\dots \hat k\dots,2n]=\\
&f_{n-1}(q)\sum_{k>1} (-q)^k \bigl(c_{1,k}-qa_{k,1}-q^{2n-1}a_{k,1}\bigr) [\hat 1,2,\dots \hat k\dots,2n],
\end{align}
where in the second line we incorporated $b_{i,j}$ back into the expression for the Pfaffian of size $2n-2$.

Using the expressions for $c_{1,k}$ and $b_{1,k}$ and unifying with the first line (\ref{lineone}), we obtain
\begin{equation}
[1,2,\dots,2n]=\sum_{1<k}(-q)^{k-2}\Bigl(a_{1,k}-qa_{k,1}+f_{n-1}(q)q^2\bigl(a_{1,k}+q^{-1}a_{k,1}-qa_{k,1}-q^{2n-1}a_{k,1}\bigr)\Bigr)[\hat 1,2,\dots,\hat k,\dots,2n].
\end{equation}
This expression implies that the theorem holds true if the expression in the brackets can be written as  $f_n(q)\bigl(a_{1,k}-q^{2n-1}a_{k,1}\bigr)$, which requires that two relations must hold simultaneously:
$$
f_{n}(q)=q^2f_{n-1}(q)+1
$$
and
$$
q^{2n-1}-q +f_{n-1}(q)(q-q^3)=0,
$$
which immediately gives
$$
f_{n-1}(q)=\frac{q^{2n-2}-1}{q^2-1},
$$
or
$$
f_{n-1}(q)=[n-1]_{q^2}=1+q^2+\cdots + q^{2(n-2)}.
$$
The recurrence relation holds then as well, which completes the proof.

Therefore,
\begin{equation}
    F_n(q)=\prod_{i=1}^{n}\bigl(1+q^2+\cdots +q^{2(i-1)}\bigr)=q^{n(n-1)/2}\prod_{i=1}^n\frac{q^i-q^{-i}}{q-q^{-1}}\label{Fn}
\end{equation}

\medskip

Now we proceed by induction in $n$ to show that the fourth expression is equivalent to the second expression. The base case, $n=1$, is trivial as $\Pi' = \widehat\Pi$.

Now assume that the theorem holds for $n-1$. Then, by induction, to show the theorem for $n$ we need to show that
\begin{equation}
    [1,\ldots,2n] = f_n(q) \sum_{k=1}^{2n-1} (-q)^{2n-k-1}[1,\ldots,\hat{k},\ldots,2n-1,\hat{2n}](a_{k,2n}-q^{2n-1}a_{2n,k})
\end{equation}
where $f_n(q)$ is as above. Observe that
\begin{align}
    [1,\ldots,2n] &= \sum_{1 \le i<j \le 2n} (-q)^{4n-i-j-1}[1,\ldots,\hat{i},\ldots,\hat{j},\ldots,2n]b_{i,j} \\
    &= \sum_{1 \le i < 2n} (-q)^{2n-i-1}[1,\ldots,\hat{i},\ldots,2n-1,\hat{2n}]b_{i,2n}\label{lineoneprime} \\
    &+ \sum_{1 \le i < j < 2n} (-q)^{4n-i-j-1}[1,\ldots,\hat{i},\ldots,\hat{j},\ldots,2n]b_{i,j}
\end{align}
We expand the second sum above using the inductive hypothesis:
\begin{align*}
    &\sum_{1 \le i < j < 2n} (-q)^{4n-i-j-1}[1,\ldots,\hat{i},\ldots,\hat{j},\ldots,2n]b_{i,j} \\
    &\qquad= f_{n-1}(q) \sum_{\{i<j,k\}<2n}(-q)^{4n-i-j-1+2n-k-3-I(i,j,k,2n)}[1,\ldots,\hat{i},\ldots,\hat{j},\ldots,\hat{k},\ldots,\hat{2n}](a_{k,2n}-q^{2n-3}a_{2n,k})b_{i,j}
\end{align*}
which we can write as
\begin{align*}
    f_{n-1}(q) \sum_{\{i<j,k\}<2n}(-q)^{6n-i-j-k-4-I(i,j,k,2n)}[1,\ldots,\hat{i},\ldots,\hat{j},\ldots,\hat{k},\ldots,\hat{2n}](c_{k,2n}-q^{-1}a_{2n,k}-q^{2n-3}a_{2n,k})b_{i,j}.
\end{align*}
Now we apply two technical relations, which are dual to (\ref{eq:b-c}) and (\ref{eq:a-b}):
\begin{align*}
q^{-2} c_{14}b_{23}-q^{-1}c_{24}b_{13}+c_{34}b_{12}
&= 
b_{23}c_{14}-q^{-1}b_{13}c_{24}+q^{-2} b_{12} c_{34} \\
q^{-1}[q^{-2} a_{41}b_{23}-q^{-1}a_{42}b_{13}+a_{43}b_{12}]
&= 
q[b_{23}a_{41}-q^{-1}b_{13}a_{42}+q^{-2} b_{12} a_{43}] \\
\end{align*}
which we can express as
\begin{align*}
\sum_{\pi\in \Pi'}(-q)^{-l(\pi_1<\pi_2,\pi_3,2n)+1} c_{\pi_3,2n}b_{\pi_1,\pi_2}
&=
\sum_{\pi\in \Pi'}(-q)^{-l(\pi_3,2n,\pi_1<\pi_2)+3} b_{\pi_1,\pi_2} c_{\pi_3,2n} \\
q^{-1}\sum_{\pi\in \Pi'}(-q)^{-l(\pi_1<\pi_2,\pi_3,2n)+1} a_{2n,\pi_3}b_{\pi_1,\pi_2}
&=
q\sum_{\pi\in \Pi'}(-q)^{-l(\pi_3,2n,\pi_1<\pi_2)+3} b_{\pi_1,\pi_2} a_{2n,\pi_3}
\end{align*}
Applying these relations to our expression, we can rewrite it as
\begin{align*}
    &f_{n-1}(q) \sum_{\{i<j,k\}<2n}(-q)^{6n-i-j-k-4-l(k,2n,i,j)+2}[1,\ldots,\hat{i},\ldots,\hat{j},\ldots,\hat{k},\ldots,\hat{2n}]b_{i,j}(c_{k,2n}-qa_{2n,k}-q^{2n-1}a_{2n,k}) \\
    &= f_{n-1}(q) \sum_{\{i<j,k\}<2n}(-q)^{2n-k+1}[1,\ldots,\hat{k},\ldots,\hat{2n}](c_{k,2n}-qa_{2n,k}-q^{2n-1}a_{2n,k}). \\
\end{align*}
Combining this sum with the first line \ref{lineoneprime} and expanding $c_{k,2n}$ and $b_{k,2n}$, we obtain
\begin{align*}
    &[1,2,\ldots,2n] \\
    &= \sum_{1 \le k < 2n} (-q)^{2n-k-1} [1,\ldots,\hat{k},\ldots,\hat{2n}] \times\\
   &\quad\times  \left( a_{k,2n}-qa_{2n,k} +q^2f_{n-1}(q)(a_{k,2n}+q^{-1}a_{2n,k}-qa_{2n,k}-q^{2n-1}a_{2n,k}) \right) .
\end{align*}
So the theorem holds true if the expression in parentheses can be written as $f_n(q)(a_{k,2n}{-}q^{2n-1}a_{2n,k})$, which implies that
\begin{align*}
    &f_n(q) = 1+q^2f_{n-1}(q); \\
    &q^{2n-1}-q+f_{n-1}(q)(q-q^3) = 0.
\end{align*}
These relations are exactly the same as for the previous formulation, and are satisfied once again by $f_n(q)=[n]_{q^2}$.
\end{proof}

\begin{lem}\label{self-adjoint-Pf}
The Pfaffian operator in Theorem~\ref{thm-Pf} is self-adjoint.
\end{lem}

\begin{proof}

We prove the self-adjointness for the second form of writing the Pfaffian. Using that $b_{i,j}$ are self-adjoint and $q^*=q^{-1}$, we obtain
$$
\widehat{Pf}_\square^*=q^{n(n-1)}\sum_{\pi\in \Pi'} (-q)^{-l(\pi)} b_{\pi_{2n-1},\pi_{2n}}b_{\pi_{2n-3},\pi_{2n-2}}\cdots b_{\pi_1,\pi_2}
$$
Let $\pi^{\text{inv}}$ denote the inverse permutation $(\pi_{2n-1},\pi_{2n},\pi_{2n-3},\pi_{2n-2},\dots,\pi_1,\pi_2)$. It is easy to see that 
$$
l(\pi^{\text{inv}})=\frac{2n(2n-1)}{2}-l(\pi)-n=2n(n-1)-l(\pi)
$$
In this expression, $\frac{2n(2n-1)}{2}$ is the total number of pairs $(\pi_i,\pi_j)$; we invert the orders in all pairs except exactly $n$ pairs $(\pi_{2i-1},\pi_{2i})$, which explains the additional subtraction of $n$. Note that $\pi\to \pi^{\text{inv}}$ is a bijection in $\Pi'$, so we can rewrite
\begin{align*}
\widehat{Pf}_\square^*=&q^{-n(n-1)} \sum_{\pi\in \Pi'}(-q)^{2n(n-1)-l(\pi)} b_{\pi_{2n-1},\pi_{2n}}b_{\pi_{2n-3},\pi_{2n-2}}\cdots b_{\pi_1,\pi_2}\\
=&q^{-n(n-1)} \sum_{\pi^{\text{inv}}\in \Pi'}(-q)^{l(\pi^{\text{inv}})} b_{\pi^{\text{inv}}_{1},\pi^{\text{inv}}_{2}}b_{\pi^{\text{inv}}_{3},\pi^{\text{inv}}_{4}}\cdots b_{\pi^{\text{inv}}_{2n-1},\pi^{\text{inv}}_{2n}}=\widehat{Pf}_\square
\end{align*}
\end{proof}
\begin{remark}
Note that for the multiplier $F_n$ (\ref{Fn}) we have that 
$$
F_n^* = q^{-n(n-1)} F_n,\quad\hbox{or}\quad \bigl(q^{-n(n-1)/2}F_n\bigr)^*=q^{-n(n-1)/2}F_n,
$$
so we can remove the factor $q^{-n(n-1)/2}F_n$ when discussing conjugation properties of the Pfaffian written in the third form. This explains why, in the triangular case, we obtain the overall factor $q^{-n(n-1)/2}$, not $q^{-n(n-1)}$, in the corresponding expression.
\end{remark}

\begin{remark}\label{rem:deformation}
If we discard the factor $F_n(q)$ (\ref{Fn}), the third and fourth forms of writing $\widehat{Pf}_\square$ satisfy the assumptions of Conjecture~\ref{con:comm}, being quantum deformations of the classical Pfaffian expression, so we expect them to be quantum Casimir functions. In Appendix~\ref{appendix:commutator} we partially prove it by verifying the commutativity of $\widehat{Pf}_\square$ with $a_{2n,2}$ for a general $n$.
\end{remark}

\subsection{Minors and Casimirs}\label{ss:minors}

We now address properties of quantum minors associated with the second set of Casimirs of the full-dimensional algebra $\mathbb A_n$. It is now convenient to switch to the $R$-matrix realization of the corresponding commutation relations and use the technique of direct products of multiple classical spaces sharing the same quantum space, which we do not indicate. In all expressions below we assume that the order of multiplication of quantum operators $a_{i,j}$ in the quantum space is ``natural'', i.e., it is prescribed by the order of multiplication in the corresponding expressions.

Using notation from \cite{JingZhangSklyanin}, let $\mathfrak A_m$ denote the $q$-antisymmetrizer \cite[equation 3.4]{JingZhangSklyanin}:
\begin{align}\label{eq:Am}
   \mathfrak  A_m := \frac{1}{[m]_{q^2}!} \sum_{\substack{c_1 < \cdots < c_m \\ \sigma,\tau \in S_m}} (-q)^{l(\sigma)+l(\tau)} \stackrel{1}{e}_{c_{\sigma(1)}c_{\tau(1)}} \otimes \cdots \otimes \stackrel{m}{e}_{c_{\sigma(m)}c_{\tau(m)}}
\end{align}
where $[n]_{q^2} = 1+q^2+\cdots+q^{2(n-1)}$ and $[m]_{q^2}!=[1]_{q^2} [2]_{q^2} \cdots [m]_{q^2}$. Define (\cite[equation 3.6]{JingZhangSklyanin})
\begin{align}\label{eq:Ang}
    \langle \stackrel{1}{\mathbb A},\ldots,\stackrel{m}{\mathbb A}\rangle &\coloneqq \stackrel{1}{\mathbb A}(R_{12}^{t_1} \cdots R_{1,m}^{t_1})\stackrel{2}{\mathbb A}(R_{23}^{t_2} \cdots R_{2,m}^{t_2}) \cdots \stackrel{m-1}{\mathbb A}R^{t_{m-1}}_{m-1,m}\stackrel{m}{\mathbb A}.
\end{align}
Then (\cite[equation 3.9]{JingZhangSklyanin}) 
\begin{align}\label{eq:Sklyanin_minor}
  [m]_{q^2}!\mathfrak A_m\langle \stackrel{1}{\mathbb A},\ldots,\stackrel{m}{\mathbb A}\rangle = \sum_{\substack{i_1,\ldots,i_m \in [n] \\j_1,\ldots,j_m \in [n]}} \mathbb A^{i_1,\ldots,i_m}_{j_1,\ldots,j_m} \stackrel{1}{e}_{i_1,j_1} \otimes \cdots \otimes \stackrel{m}{e}_{i_m,j_m}.
\end{align}
We call $\mathbb A^{i_1,\ldots,i_m}_{j_1,\ldots,j_m}$ the Sklyanin  minor associated to the rows $i_1,\ldots,i_m$ and the columns $j_1,\ldots,j_m$. Note that $\mathbb A^{i_1,\ldots,i_m}_{j_1,\ldots,j_m}=0$ if any of the $i$'s repeat or any of the $j$'s repeat. We will sometimes write $\mathbb A^{i_1,\ldots,i_m}_{j_1,\ldots,j_m}$ as $\mathbb A^I_J$ or $\widehat{M}^I_J$.

Note also the remarkable relations \cite{MolevBook}, \cite{JingZhangSklyanin}
\begin{equation}\label{eq:q-Am}
 \mathfrak  A_m^2=\mathfrak  A_m;\qquad   \mathfrak  A_m  \langle \stackrel{1}{\mathbb A},\ldots,\stackrel{m}{\mathbb A}\rangle =\langle \stackrel{1}{\mathbb A},\ldots,\stackrel{m}{\mathbb A}\rangle  \mathfrak  A_m= \mathfrak  A_m  \langle \stackrel{1}{\mathbb A},\ldots,\stackrel{m}{\mathbb A}\rangle  \mathfrak  A_m ,
\end{equation}
so we can apply the antisymmetrizer from either side of $\langle \stackrel{1}{\mathbb A},\ldots,\stackrel{m}{\mathbb A}\rangle$, or from both sides.

\begin{defn}
    For any $m \le n$, define
    $$
    \widehat{M}{}_{m}^{\text{BL}} \coloneqq ([m]_{q^2}!)\stackrel{1} e_{n-m+1} \otimes \cdots \otimes \stackrel{m} e_{n} \mathfrak A_m \langle \stackrel{1}{\mathbb A},\ldots,\stackrel{m}{\mathbb A} \rangle \stackrel{1}e_1\otimes \cdots\otimes \stackrel{m}e_m
    $$
    to be the $m\times m$ Sklyanin minor based in the bottom left corner. In the notation above, $\widehat{M}_{m}^{\text{BL}} = \mathbb A^I_J$ with $I=\{n{-}m{+}1,\dots, n\}$ and $J=\{1,\dots, m\}$.
\end{defn}

\begin{thm}\label{thm-mcas}
    For any $m \le n/2$,
    \begin{align*}
        \widehat{M}_{n-m}^{\text{BL}}[\widehat{M}_m^{\text{BL}}]^{-1}
    \end{align*}
    is a Casimir of the full algebra.
\end{thm}

\begin{proof}
Equation (3.27) of \cite{JingZhangSklyanin} reads
\begin{align}\label{eq:JZ327}
  R_{0m} \cdots R_{01} \stackrel{0}{\mathbb A} R_{01}^{t_0} \cdots R_{0m}^{t_0} \mathfrak A_m \langle \stackrel{1}{\mathbb A},\ldots,\stackrel{m}{\mathbb A} \rangle = \mathfrak A_m\langle \stackrel{1}{\mathbb A},\ldots,\stackrel{m}{\mathbb A} \rangle R_{01}^{t_0} \cdots R_{0m}^{t_0} \stackrel{0}{\mathbb A} R_{0m} \cdots R_{01}.
\end{align}

We expand each side of this equation and then multiply on the left by $\stackrel{0}{e}_a \otimes \stackrel{1}{e}_{i_1} \otimes \cdots \otimes \stackrel{m}{e}_{i_m}$ and on the right by $\stackrel{0}{e}_b \otimes \stackrel{1}{e}_{j_1} \otimes \cdots \otimes \stackrel{m}{e}_{j_m}$ for some chosen $a,b,i_1,\ldots,i_m,j_1,\ldots,j_m$.

Consider first the right action of $R_{0m} \cdots R_{01}$ on $\stackrel{0}{e}_a \otimes \stackrel{1}{e}_{i_1} \otimes \cdots \otimes \stackrel{m}{e}_{i_m}$: each $R_{0s}$ contains only $\stackrel{s}e_{j,i}$ with $i\ge j$ and $\stackrel{0}e_{i,j}$ with $i\ge j$, so the result of this action is a linear combination of vectors 
$$
\stackrel{0}{e}_{a-t_0} \otimes \stackrel{1}{e}_{i_1+t_1} \otimes \cdots \otimes \stackrel{m}{e}_{i_m+t_m}
$$
with all $t_i\ge 0$ for $i=0,1,\dots, m$.
Correspondingly, since
$$
R_{01}^{t_0} \cdots R_{0m}^{t_0} =\bigl[ R_{0m} \cdots R_{01} \bigr]^{t_0},
$$
the right action of $R_{01}^{t_0} \cdots R_{0m}^{t_0}$ on $\stackrel{0}{e}_{a-t_0} \otimes \stackrel{1}{e}_{i_1+t_1} \otimes \cdots \otimes \stackrel{m}{e}_{i_m+t_m}$ is then a linear combination of vectors
$$
\stackrel{0}{e}_{a-t_0+r_0} \otimes \stackrel{1}{e}_{i_1+t_1+r_1} \otimes \cdots \otimes \stackrel{m}{e}_{i_m+t_m+r_m}
$$
where all $r_i\ge 0$ and $t_i\ge 0$. While the first index can be shifted in both directions, all indices $i_s$ do not decrease. 

If we now choose $i_1=n-m+1,\ i_2=n-m+2,\dots,\ i_m=n$, then if any of $t_s$, $r_s$ is nonzero, and since the index cannot exceed $n$, in the set 
$$
i_s+t_s+r_s,\ s=1,\dots,m
$$
we unavoidably have two coinciding indices, which makes the corresponding Sklyanin minor 
$$
\stackrel{1}{e}_{i_1+t_1+r_1} \otimes \cdots \otimes \stackrel{m}{e}_{i_m+t_m+r_m}\mathfrak A_m\langle \stackrel{1}{\mathbb A},\ldots,\stackrel{m}{\mathbb A}\rangle \stackrel{1}{e}_{j_1}\otimes \cdots\otimes \stackrel{m}{e}_{j_m}
$$
equal zero irrespective of the choice of indices $j_1,\dots,j_m$. So the only nonvanishing contribution for this choice of $i_1,\dots,i_m$ comes from the diagonal part of each of the products $R_{0m} \cdots R_{01}$ and $\bigl[ R_{0m} \cdots R_{01}\bigr]^{t_0}$. This diagonal part $\bigl[ R_{0m} \cdots R_{01}\bigr]_{\text{diag}} $ is preserved by transpositions in any of the matrix spaces, so it is the same for both matrix products. Note that 
$$
\stackrel{0}{e}_a \otimes \stackrel{1}{e}_{i_1} \otimes \cdots \otimes \stackrel{m}{e}_{i_m} \bigl[ R_{0m} \cdots R_{01}\bigr]_{\text{diag}} \stackrel{0}{e}_a \otimes \stackrel{1}{e}_{i_1} \otimes \cdots \otimes \stackrel{m}{e}_{i_m}=\begin{cases}q,& a\in I\\ 1,&a\notin I \end{cases}\eqqcolon q^{\delta_{a,I}}
$$
Therefore,
\begin{align}
&\stackrel{0}{e}_a \otimes \stackrel{1}{e}_{n-m+1} \otimes \cdots \otimes \stackrel{m}{e}_{n} \bigl[ R_{0m} \cdots R_{01}\bigr]_{\text{diag}} \stackrel{0} {\mathbb A} \bigl[ R_{0m} \cdots R_{01}\bigr]^{t_0}_{\text{diag}}\nonumber\\
&= \sum_{b=1}^n a_{a,b}\, \stackrel{0}{e}_b \otimes \stackrel{1}{e}_{n-m+1} \otimes \cdots \otimes \stackrel{m}{e}_{n}\, q^{\delta_{a,I}+\delta_{b,I}},\ I=\{n{-}m{+}1,\dots, n\}.
\end{align}

The same logic works when we multiply from the right by the vector $\stackrel{0} e_b\otimes \stackrel{1}e_1\otimes \cdots\otimes \stackrel{m}e_m$: the indices of $\stackrel{s}e_{s}$ are nonincreasing by the left action of  $R_{0m} \cdots R_{01}$, so, again, only the diagonal part gives a nonzero contribution, and multiplying both parts of (\ref{eq:JZ327}) by $\stackrel{0} e_a \otimes \stackrel{1} e_{n-m+1} \otimes \cdots \otimes \stackrel{m} e_{n}$ from the left and by 
$\stackrel{0} e_b\otimes \stackrel{1}e_1\otimes \cdots\otimes \stackrel{m}e_m$ from the right, we obtain
\begin{equation}\label{eq:hom}
q^{\delta_{a,I}+\delta_{b,I}} a_{a,b} \widehat{M}_{m}^{\text{BL}} = q^{\delta_{a,J}+\delta_{b,J}}  \widehat{M}_{m}^{\text{BL}} a_{a,b}.
\end{equation}
The sign pattern in this relation (for $m\le \lfloor n/2\rfloor$) reads $\begin{array}{c|c|c|c|}\cline{2-4} m\updownarrow & 2&1&0\\ \cline{2-4} &1&0&-1\\ \cline{2-4} m\updownarrow & 0&-1&-2\\ \cline{2-4}\end{array}$\ , and it stays exactly the same upon replacing $m$ by $n-m$.

Note that for any choice of $a,b$, homogeneous commutation relations (\ref{eq:hom}) result in identical $q$-factors for $m$ and $n-m$, and all matrix elements of $\widehat{M}_{m}^{\text{BL}}$ with $m\le n/2$ commute with $\widehat{M}_{n-m}^{\text{BL}}$. Thus, these two minors commute, and the ratio
$$
\widehat{M}_{n-m}^{\text{BL}}[\widehat{M}_{m}^{\text{BL}}]^{-1}
$$
is well defined and commutes with all $a_{a,b}$, so it is therefore a quantum Casimir function.
\end{proof}


\begin{lem}\label{lm:mbl}
For the quantum Sklyanin minor, we have
    \begin{align*}
       \widehat{M}_m^{BL} = \sum_{\sigma \in S_m} (-q)^{l(\sigma)} q^{\#\{\sigma(i)>n-m+i, 1 \le i \le m\}} a_{n-m+1,\sigma(1)}a_{n-m+2,\sigma(2)} \cdots a_{n,\sigma(m)}
    \end{align*}
\end{lem}

\begin{proof}
First, we obtain that all terms in all $R$-matrices that strictly increase vector indices do not contribute because our set of vector indices $\{n-m+1,\dots,n\}$ is maximal and increasing any of the indices will result in a minor with two identical rows, which is zero. We present the details in Lemma~\ref{lm:increase} below.

Consider now the diagonal part of $R^{t_i}_{i,j}$. In the $i$th space, the only nonzero contribution is from the term $a_{n-m+i,\sigma(i)}\stackrel{i}{e}_{n-m+i,\sigma(i)}$, and for the corresponding part of the expression, we obtain
\begin{align*}
&\stackrel{i}{e}_{n-m+i}\otimes \stackrel{i+1}{e}_{n-m+i+1}\otimes\cdots\otimes \stackrel{m}{e}_n \bigl( a_{n-m+i,\sigma(i)}\stackrel{i}{e}_{n-m+i,\sigma(i)} \bigr)\bigl[R^{t_i}_{i,i+1}\cdots R^{t_i}_{i,m} \bigr]^{\text{diag}}\\ &=a_{n-m+i,\sigma(i)} \stackrel{i}{e}_{\sigma(i)}\otimes \stackrel{i+1}{e}_{n-m+i+1}\otimes\cdots\otimes \stackrel{m}{e}_n\bigl[R^{t_i}_{i,i+1}\cdots R^{t_i}_{i,m} \bigr]^{\text{diag}}\\
&=a_{n-m+i,\sigma(i)} \stackrel{i}{e}_{\sigma(i)}\otimes \stackrel{i+1}{e}_{n-m+i+1}\otimes\cdots\otimes \stackrel{m}{e}_n \cdot \begin{cases} q  & \hbox{if}\ \sigma(i)\in [i+1,n]\\ 1& \hbox{if}\ \sigma(i)\notin [i+1,n].
\end{cases}
\end{align*}
Together with the factor $(-q)^{l(\sigma)}$ coming from the skew-symmetrizer, taking the sum over $i$ from $1$ to $m$ we obtain the statement of the lemma.
\end{proof}

\begin{remark}
    One can repeat the above proof by considering the expression $\mathfrak A_m \langle \stackrel{1}{\mathbb A},\ldots, \stackrel{m}{\mathbb A} \rangle$ instead of $\langle \stackrel{1}{\mathbb A},\ldots,\stackrel{m}{\mathbb A} \rangle\mathfrak A_m$. This will yield an alternate formulation for the quantum Sklyanin minor:
    \begin{align*}
        {\widehat M}_m^{BL} = \sum_{\tau \in S_m} (-q)^{l(\tau)} q^{\#\{n-m+\tau(i)<i, 1 \le i \le m\}} a_{n-m+\tau(1),1}a_{n-m+\tau(2),2} \cdots a_{n-m+\tau(m),m}.
    \end{align*}
\end{remark}

\begin{cor}
    For $m \le \lfloor \frac{n}{2} \rfloor$,
    \begin{align*}
        {\widehat M}_m^{BL} &= \sum_{\sigma \in S_m} (-q)^{l(\sigma)}  a_{n-m+1,\sigma(1)}a_{n-m+2,\sigma(2)} \cdots a_{n,\sigma(m)} \\
        {\widehat M}_m^{BL} &= \sum_{\tau \in S_m} (-q)^{l(\tau)}a_{n-m+\tau(1),1}a_{n-m+\tau(2),2} \cdots a_{n-m+\tau(m),m}.
    \end{align*}
\end{cor}

\begin{lem}\label{lm:increase}
    For any set $J \subseteq [n]$ with $m$ elements $j_1 < j_2 < \cdots < j_m$,
    \begin{align*}
        \widehat{M}^{n-m+1,\ldots,n}_{J} \coloneqq \widehat{M}^{n-m+1,\ldots,n}_{j_1,\ldots,j_m} = \sum_{\sigma \in \textrm{Sym}(J)} (-q)^{l(\sigma)} q^{\#\{\sigma(i)>n-m+i, 1 \le i \le m\}} a_{n-m+1,\sigma(1)}a_{n-m+2,\sigma(2)} \cdots a_{n,\sigma(m)}.
    \end{align*}
    \end{lem}
\begin{proof}
    We proceed by induction on $m$. The base case $m=1$ is trivial, as the right-hand side simplifies to $a_{n,1}$ (which equals $\widehat{M}^n_1$).

    Now let $m \ge 2$, and assume that the claim holds for smaller $m$. Expanding
    \[\stackrel{1}{e}_{n-m+1} \otimes \cdots \otimes \stackrel{m}{e}_n\langle \stackrel{1}{\mathbb{A}}, \ldots, \stackrel{m}{\mathbb{A}} \rangle ([m]_{q^2}!)\mathfrak{A}_m \stackrel{1}{e}_{j_1} \otimes \cdots \otimes \stackrel{m}{e}_{j_m}\]
    yields
    \begin{align*}
        \widehat{M}^{n-m+1,\ldots,n}_J = \sum_{\sigma \in \textrm{Sym}(J)} (-q)^{\ell(\sigma)} \stackrel{1}{e}_{n-m+1} \otimes \cdots \otimes \stackrel{m}{e}_n\langle \stackrel{1}{\mathbb{A}}, \ldots, \stackrel{m}{\mathbb{A}} \rangle \stackrel{1}{e}_{\sigma(1)} \otimes \cdots \otimes \stackrel{m}{e}_{\sigma(m)}
    \end{align*}
    Consider a summand above in space 1:
    \begin{align*}
        e_{n-m+1} \mathbb{A} R_{12}^{t_1} R_{13}^{t_1} \cdots R_{1m}^{t_1} e_{\sigma(1)}.
    \end{align*}
    Expanding each $R$-matrix turns this expression into a sum of terms of the form $a_{n-m+1,i}$ (with some $q$ coefficients). In the full product of spaces, expanding $R_{1k}^{t_1}$ yields a sum of terms of the form $a_{n-m+1,i} \widehat{M}^I_{J \setminus \{i\}}$ (once again, ignoring $q$ coefficients). There are two cases to consider: either $i = \sigma(1)$, or $i < \sigma(1)$ (note that the case $i > \sigma(1)$ is impossible due to the entries of the $R^t$-matrices). If $i < \sigma(1)$, then we have an $R$-matrix $R_{1k}^{t_1}$ contributing a term of the form $(q-q^{-1}) \stackrel{1}{e}_{ji} \otimes \stackrel{k}{e}_{ji}$, where $j < i$. Note that $k$ is necessarily the largest $\ell$ such that $R_{1\ell}^{t_1}$ contributes a non-diagonal (i.e., of the form $e_{\alpha\beta}$, $\alpha < \beta$) term to this summand. In order for the contribution of this $R_{1k}^t$ term to be nonzero, the product in the $k$th space
    \begin{align*}
        e_{n-m+k} R_{1k}^{t_1} \cdots R_{k-1,k}^{t_{k-1}} \mathbb{A} R_{k,k+1}^{t_k} \cdots R_{k,m}^{t_k}
    \end{align*}
    must be nonzero. Thus, $j = n-m+k$, so $n-m+k < i$. This implies that the minor $\widehat{M}^I_{J \setminus \{i\}}$ associated with this summand will have $i$ appearing at least twice in $I$: one contribution from space $k$ (where the product $R_{2,k}^{t_2} \cdots R_{k-1,k}^{t_{k-1}} \mathbb{A} R_{k,k+1}^{t_k} \cdots R_{k,m}^{t_k}$ is multiplied on the left by $e_i$), and one contribution from space $i+m-n$ (since $i+m-n>k$, $R_{1,i+m-n}^{t_1}$ contributes a diagonal term). Since Sklyanin minors with repeated rows vanish, the contribution of all summands of the form $a_{n-m+1,i} \widehat{M}^I_{J \setminus \{i\}}$ with $i > \sigma(1)$ vanish as well.

    Thus, every $R$-matrix $R_{1,k}^{t_1}$ contributes only its diagonal entries. Considering the products in space 1 and space $k$, we can see that the only contributing terms of $R_{1,k}^{t_1}$ are $\stackrel{1}{I} \otimes \stackrel{k}{I}$ and $(q-1)\stackrel{1}{e}_{n-m+k,n-m+k} \otimes \stackrel{k}{e}_{n-m+k,n-m+k}$. This latter term only contributes if $\sigma(1) = n-m+k$; thus, the product $R_{12}^{t_1} \cdots R_{1,m}^{t_1}$ simplifies to
    \begin{align*}
        R_{12}^{t_1} \cdots R_{1,m}^{t_1} = I \otimes \cdots \otimes I + \sum_{k=2}^m \delta_{n-m+k=\sigma(1)} (q-1)\stackrel{1}{e}_{n-m+k,n-m+k} \otimes \stackrel{k}{e}_{n-m+k,n-m+k}.
    \end{align*}
    This contributes an overall factor of $q$ if such a $k$ exists, and an overall factor of $1$ otherwise. Note that all row and column indices to the left and right of $\langle \stackrel{1}{\mathbb{A}}, \ldots, \stackrel{m}{\mathbb{A}} \rangle$ remain unchanged. Thus, summing over all possible values of $\sigma(1)$,
    \begin{align*}
        \widehat{M}^{n-m+1,\ldots,n}_J = \sum_{i=1}^m (-q)^{i-1} q^{\delta(j_i > n-m+1)} a_{n-m+1,j_i} M^{n-m+2,\ldots,n}_{J \setminus \{j_i\}}.
    \end{align*}
    Applying the inductive hypothesis completes the proof.
\end{proof}

\begin{cor} Because we can interchange rows and columns, a similar statement holds for minors $\widehat{M}^{i_1,i_2,\dots,i_m}_{1,2,\dots,m} $ for any set of indices $1\le i_1<i_2<\dots<i_m\le n$. In particular,
\begin{align*}
    \widehat{M}^{i_1,i_2,\ldots,i_m}_{1,2,\ldots,m} = \sum_{\tau \in \textrm{Sym}(I)} (-q)^{l(\tau)}q^{\#\{\tau(i)<i, 1 \le i \le m\}} a_{\tau(1),1} a_{\tau(2),2} \cdots a_{\tau(m),m}.
\end{align*}
\end{cor}

\begin{defn}\label{def:qdet}
    We define the $n \times n$ quantum determinant to be
    \begin{align*}
        q^{-n/2}\sum_{\pi \in S_n}(-q)^{l(\pi)}q^{\exc(\pi)}a_{1,\pi_1}a_{2,\pi_2} \cdots a_{n,\pi_n}.
    \end{align*}
\end{defn}

\begin{remark}
    The same expression for a quantum determinant appears in other quantum reflection equation algebras, e.g., \cite{JordanWhite2020}.
\end{remark}

\begin{exm}
    The $2\times 2$ and $3 \times 3$ quantum determinants are
    \begin{align*}
    \begin{vmatrix}
         a_{11} & a_{12}  \\ a_{21} & a_{22} 
     \end{vmatrix}
     &=q^{-1}a_{11}a_{22}-qa_{12}a_{21},\\
    \begin{vmatrix}
         a_{11} & a_{12} & a_{13} \\ a_{21} & a_{22} & a_{23} \\ a_{31} & a_{32} & a_{33}
     \end{vmatrix}
     &=
     q^{-3/2}a_{11}a_{22}a_{33} -q^{1/2}a_{11}a_{23}a_{32} -q^{1/2}a_{12}a_{21}a_{33} \\
     &\qquad+q^{5/2}a_{12}a_{23}a_{31} +q^{3/2}a_{13}a_{21}a_{32} -q^{5/2}a_{13}a_{22}a_{31}.
    \end{align*}
\end{exm}

\begin{lem}\label{lem-msa}
    The $n \times n$ quantum determinant and all quantum minors ${\widehat M}_m^{BL}$ become self-adjoint quantum operators when being weighted by global factors $q^{\min\{n/2-m,0\}}$. Therefore, the quantum Casimirs $\widehat{\mathcal K}_m:=q^{m-n/2}{\widehat M}_{n-m}^{BL}\bigl[{\widehat M}_m^{BL}\bigr]^{-1}$ with $m<n/2$ are self adjoint.
\end{lem}

\begin{proof} 
We interchange the order of proving statements and address first the problem of the overall factor. For this,  consider the term $a_{n-m+1,1}a_{n-m+2,2}\cdots a_{n,m}$ that contains the ``most intersections'' in the corresponding minor. This term does not appear in the right-hand sides of commutation relations, so it must be self-adjoint modulo ``lower-order'' terms. The power of the term $q^{\min\{n/2-m,0\}}$ just counts how many $a_{i_k,j_k}$ in this expression have coinciding indices $i_k=j_l$, which results in $q$-factors when implementing conjugation. Note that for $m<n$, all $a_{i_k,j_k}$ in this monomial are self-adjoint, whereas $a_{i,i}^*=q^{-1}a_{i,i}$, which results in the proper factor $q^{-n/2}$ for the quantum determinant itself. 

We next use the cluster-algebra representation from Sec.~\ref{s:clusters} to analyze structures of quantum minors.  
Considering a non-normalized (to the determinants of transport matrices) quantum $a_{i,j}$, we use that each $a_{i,j}\in \mathbb Z_+[x_\alpha,q^{\pm 1/2}]$, i.e., each $a_{i,j}$ can be represented as a polynomial in the cluster variables $x_\alpha$ of the quiver in Fig.~\ref{fi:fullCasimirs} with positive coefficients that are powers of $q^{\pm 1/2}$. Therefore, any polynomial combination $P(a_{i,j})$ of $a_{i,j}$, including all quantum minors, is a polynomial in $x_\alpha$ and $q^{\pm 1/2}$. We also have a local isomorphism between $\{a_{i,j}\}$ with $1\le i,j\le n$ and cluster coordinates of the amalgamated quiver, the number of these coordinates being exactly $n^2$. We assume that all $x_\alpha$ are self-adjoint and independent. This means that, in the quantum case, if $P(a_{i,j}(x_\alpha))$ is self-adjoint as a function of $x_\alpha$, then $P(a_{i,j})$ will be self-adjoint as a function of $a_{i,j}$ for which we use conjugation relations (\ref{eq:conjugation_law_2}). 

 We now turn to the quantum minors  ${\widehat M}_m^{BL}$. Recall that these minors have homogeneous commutation relations with all $a_{i,j}$, and the powers of $q$ in all these relations depends only on $i$, $j$, and $m$. This indicates that the only cluster variables in the quiver in Fig.~\ref{fi:fullCasimirs} with which the corresponding minor ${\widehat M}_m^{BL}$ does \emph{not} commute are frozen cluster variables on the left side of the left FG quiver. If we require all powers of $x_\alpha$ to be nonnegative, then there are exactly two monomials in cluster variables of the right FG quiver that satisfy the corresponding homogeneous commutation relations: those that are quantum deformations of $\widetilde{\mathfrak M}_m \prod\limits_{i=n-m+1}^n [\widetilde {M}_1]_{i,i}$ and of $\widetilde{\mathfrak M}_{n-m} \prod\limits_{i=m+1}^n [\widetilde {M}_1]_{i,i}$. Both these monomial expressions do not commute only with variables at corners of the quiver and with exactly two other (frozen) variables situated at distance $m$ from the corners. In Figure~\ref{fi:frozen}, we present the example of these monomials for $n=6$ and $m=2$ using the same color code as in Example~\ref{ex:full-quiver}: only cluster variables at colored vertices do not commute with the corresponding monomial expression,
 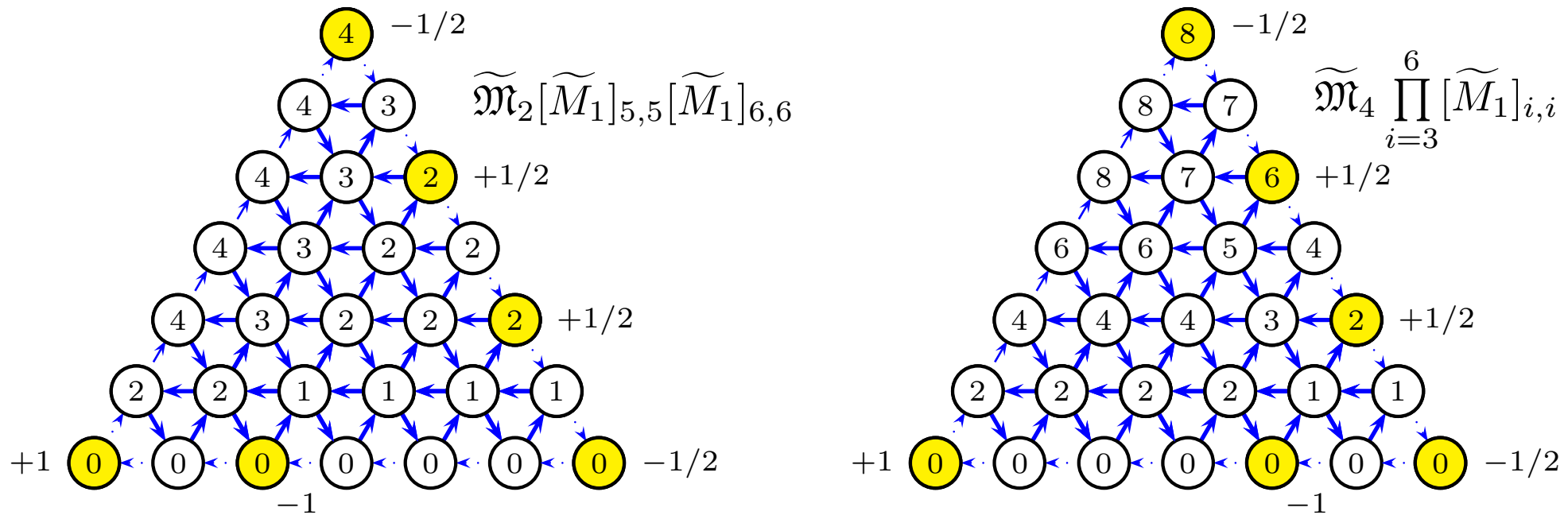
\begin{figure}[h]
\begin{pspicture}(-6,-1.8)(6,2.5){\psset{unit=0.7}
		\newcommand{\PATGEN}{%
			{\psset{unit=1}
				\rput(0,0){\psline[linecolor=blue,linewidth=1pt]{->}(0,0)(.35,.6)}
				\rput(0,0){\psline[linecolor=blue,linewidth=1pt]{->}(0.7,0)(0.3,0)}
				\rput(0,0){\psline[linecolor=blue,linewidth=1pt]{->}(0,0)(.35,-.6)}
				\put(0,0){\pscircle[fillstyle=solid,fillcolor=white]{.3}}
                \put(0,0){\makebox(0,0){\hbox{{\tiny $0$}}}}
		}}
		\newcommand{\PATLEFT}{%
			{\psset{unit=1}
				\rput(0,0){\psline[linecolor=blue,linewidth=0.5pt,linestyle=dashed]{->}(0.05,0.085)(.35,.6)}
				\rput(0,0){\psline[linecolor=blue,linewidth=1pt]{->}(0.7,0)(0.3,0)}
				\rput(0,0){\psline[linecolor=blue,linewidth=1pt]{->}(0,0)(.35,-.6)}
				\put(0,0){\pscircle[fillstyle=solid,fillcolor=white]{.3}}
                \put(0,0){\makebox(0,0){\hbox{{\tiny $0$}}}}
		}}
		\newcommand{\PATRIGHT}{%
			{\psset{unit=1}
				\rput(0,0){\psline[linecolor=blue,linewidth=0.5pt,linestyle=dotted]{->}(0.05,-0.085)(.35,-.6)}
				\put(0,0){\pscircle[fillstyle=solid,fillcolor=white]{.3}}
                   \put(0,0){\makebox(0,0){\hbox{{\tiny $0$}}}}
		}}
		\newcommand{\PATBOTTOM}{%
			{\psset{unit=1}
				\rput(0,0){\psline[linecolor=blue,linewidth=1pt]{->}(0,0)(.35,.6)}
				\rput(0,0){\psline[linecolor=blue,linewidth=0.5pt,linestyle=dotted]{->}(0.7,0)(0.3,0)}
				\put(0,0){\pscircle[fillstyle=solid,fillcolor=white]{.3}}
                       \put(0,0){\makebox(0,0){\hbox{{\tiny $0$}}}}
		}}
		\newcommand{\PATTOP}{%
			{\psset{unit=1}
				\rput(0,0){\psline[linecolor=blue,linewidth=1pt]{->}(0.7,0)(0.3,0)}
				\rput(0,0){\psline[linecolor=blue,linewidth=1pt]{->}(0,0)(.35,-.6)}
				\put(0,0){\pscircle[fillstyle=solid,fillcolor=white]{.3}}
                   \put(0,0){\makebox(0,0){\hbox{{\tiny $0$}}}}
		}}
		\newcommand{\PATBOTRIGHT}{%
			{\psset{unit=1}
				\rput(0,0){\psline[linecolor=blue,linewidth=1pt]{->}(0,0)(.35,.6)}
				\put(0,0){\pscircle[fillstyle=solid,fillcolor=white]{.3}}
                \put(0,0){\makebox(0,0){\hbox{{\tiny $0$}}}}
                \put(.5,0.85)
                {\pscircle[fillstyle=solid,fillcolor=white]{.3}}
                \put(0.5,0.85){\makebox(0,0){\hbox{{\tiny $0$}}}}
		}}
		\newcommand{\PATNORTH}{%
			{\psset{unit=1}
				\rput(0,0){\psline[linecolor=blue,linewidth=0.5pt,linestyle=dotted]{->}(-.35,-.6)(-0.15,-0.255)}
				\rput(0,0){\psline[linecolor=blue,linewidth=0.5pt,linestyle=dotted]{<-}(.35,-.6)(0.15,-0.255)}
				\put(0,0){\pscircle[fillstyle=solid,fillcolor=white]{.3}}
            \put(0,0){\makebox(0,0){\hbox{{\tiny $0$}}}}
            }}
		\newcommand{\PATSW}{%
			{\psset{unit=1}
				\rput(0,0){\psline[linecolor=blue,linewidth=0.5pt,linestyle=dotted]{->}(0.15,0.255)(.35,.6)}
				\rput(0,0){\psline[linecolor=blue,linewidth=0.5pt,linestyle=dotted]{<-}(0.3,0)(0.7,0)}
				\put(0,0){\pscircle[fillstyle=solid,fillcolor=white]{.3}}
            \put(0,0){\makebox(0,0){\hbox{{\tiny $0$}}}}
		}}
		\newcommand{\PATSE}{%
			{\psset{unit=1}
				\rput(0,0){\psline[linecolor=blue,linewidth=0.5pt,linestyle=dotted]{<-}(-0.15,0.255)(-.35,.6)}
				\rput(0,0){\psline[linecolor=blue,linewidth=0.5pt,linestyle=dotted]{<-}(-0.7,0)(-0.3,0)}
				\put(0,0){\pscircle[fillstyle=solid,fillcolor=white]{.3}}
            \put(0,0){\makebox(0,0){\hbox{{\tiny $0$}}}}
		}}
        \rput(-5,0){
		\multiput(-2.5,-0.85)(0.5,0.85){4}{\PATLEFT}
		\multiput(-2,-1.7)(1,0){4}{\PATBOTTOM}
		\put(-0.5,2.55){\PATTOP}
		\put(0,3.4){\PATNORTH}
		\multiput(-1.5,-0.85)(1,0){4}{\PATGEN}
		\multiput(-1,0)(1,0){3}{\PATGEN}
		\multiput(-.5,0.85)(1,0){2}{\PATGEN}
		\put(0,1.7){\PATGEN}
		\multiput(-1.5,-0.85)(1,0){4}{\PATGEN}
		\multiput(0.5,2.55)(0.5,-0.85){4}{\PATRIGHT}
		\put(2,-1.7){\PATBOTRIGHT}
		\put(-3,-1.7){\PATSW}
		\put(3,-1.7){\PATSE}
            \multiput(-2,0)(0.5,0.85){5}{\pscircle[fillstyle=solid,fillcolor=white]{.3}
            \put(0,0){\makebox(0,0){\hbox{{\tiny $4$}}}}} 
            \multiput(-1,0)(0.5,0.85){4}{\pscircle[fillstyle=solid,fillcolor=white]{.3}
            \put(0,0){\makebox(0,0){\hbox{{\tiny $3$}}}}}
            \multiput(-2.5,-0.85)(1,0){2}{\pscircle[fillstyle=solid,fillcolor=white]{.3}
            \put(0,0){\makebox(0,0){\hbox{{\tiny $2$}}}}}
            \multiput(0,0)(1,0){3}{\pscircle[fillstyle=solid,fillcolor=white]{.3}
            \put(0,0){\makebox(0,0){\hbox{{\tiny $2$}}}}}
            \multiput(0.5,0.85)(1,0){2}{\pscircle[fillstyle=solid,fillcolor=white]{.3}
            \put(0,0){\makebox(0,0){\hbox{{\tiny $2$}}}}}
            \multiput(1,1.7)(1,0){1}{\pscircle[fillstyle=solid,fillcolor=white]{.3}
            \put(0,0){\makebox(0,0){\hbox{{\tiny $2$}}}}}
            \multiput(-0.5,-0.85)(1,0){4}{\pscircle[fillstyle=solid,fillcolor=white]{.3}
            \put(0,0){\makebox(0,0){\hbox{{\tiny $1$}}}}}
            			\put(0,3.4){\pscircle[fillstyle=solid,fillcolor=yellow]{.3}}
            \put(0,3.4){\makebox(0,0){\hbox{{\tiny $4$}}}}
            			\put(1,1.7){\pscircle[fillstyle=solid,fillcolor=yellow]{.3}}
            \put(1,1.7){\makebox(0,0){\hbox{{\tiny $2$}}}}
                        			\put(2,0){\pscircle[fillstyle=solid,fillcolor=yellow]{.3}}
            \put(2,0){\makebox(0,0){\hbox{{\tiny $2$}}}}
                        			\put(-3,-1.7){\pscircle[fillstyle=solid,fillcolor=yellow]{.3}}
            \put(-3,-1.7){\makebox(0,0){\hbox{{\tiny $0$}}}}
                        			\put(-1,-1.7){\pscircle[fillstyle=solid,fillcolor=yellow]{.3}}
            \put(-1,-1.7){\makebox(0,0){\hbox{{\tiny $0$}}}}
                        			\put(3,-1.7){\pscircle[fillstyle=solid,fillcolor=yellow]{.3}}
            \put(3,-1.7){\makebox(0,0){\hbox{{\tiny $0$}}}}
        \put(1.5,2.3){\makebox(0,0)[bl]{\hbox{{\small $\widetilde{\mathfrak M}_2[\widetilde {M}_1]_{5,5}[\widetilde{M}_1]_{6,6}$}}}}
        \put(0.5,3.5){\makebox(0,0)[lc]{\hbox{{\tiny $-1/2$}}}}
	\put(1.5,1.7){\makebox(0,0)[lc]{\hbox{{\tiny $+1/2$}}}}
	 \put(2.5,0){\makebox(0,0)[lc]{\hbox{{\tiny $+1/2$}}}}
	 \put(3.5,-1.7){\makebox(0,0)[lc]{\hbox{{\tiny $-1/2$}}}}
	 \put(-3.5,-1.7){\makebox(0,0)[rc]{\hbox{{\tiny $+1$}}}}
	 \put(-0.85,-2.2){\makebox(0,0)[lc]{\hbox{{\tiny $-1$}}}}
			}
                    \rput(5,0){
		\multiput(-2.5,-0.85)(0.5,0.85){4}{\PATLEFT}
		\multiput(-2,-1.7)(1,0){4}{\PATBOTTOM}
		\put(-0.5,2.55){\PATTOP}
		\put(0,3.4){\PATNORTH}
		\multiput(-1.5,-0.85)(1,0){4}{\PATGEN}
		\multiput(-1,0)(1,0){3}{\PATGEN}
		\multiput(-.5,0.85)(1,0){2}{\PATGEN}
		\put(0,1.7){\PATGEN}
		\multiput(-1.5,-0.85)(1,0){4}{\PATGEN}
		\multiput(0.5,2.55)(0.5,-0.85){4}{\PATRIGHT}
		\put(2,-1.7){\PATBOTRIGHT}
		\put(-3,-1.7){\PATSW}
		\put(3,-1.7){\PATSE}
            \multiput(-1,1.7)(0.5,0.85){3}{\pscircle[fillstyle=solid,fillcolor=white]{.3}
            \put(0,0){\makebox(0,0){\hbox{{\tiny $8$}}}}} 
            \multiput(0,1.7)(0.5,0.85){2}{\pscircle[fillstyle=solid,fillcolor=white]{.3}
            \put(0,0){\makebox(0,0){\hbox{{\tiny $7$}}}}}
            \multiput(-1.5,0.85)(1,0){2}{\pscircle[fillstyle=solid,fillcolor=white]{.3}
            \put(0,0){\makebox(0,0){\hbox{{\tiny $6$}}}}}
            \multiput(1,1.7)(0,1){1}{\pscircle[fillstyle=solid,fillcolor=white]{.3}
            \put(0,0){\makebox(0,0){\hbox{{\tiny $6$}}}}}
            \multiput(0.5,0.85)(0,1){1}{\pscircle[fillstyle=solid,fillcolor=white]{.3}
            \put(0,0){\makebox(0,0){\hbox{{\tiny $5$}}}}}
            \multiput(-2,0)(1,0){3}{\pscircle[fillstyle=solid,fillcolor=white]{.3}
            \put(0,0){\makebox(0,0){\hbox{{\tiny $4$}}}}}
            \multiput(1.5,0.85)(0,1){1}{\pscircle[fillstyle=solid,fillcolor=white]{.3}
            \put(0,0){\makebox(0,0){\hbox{{\tiny $4$}}}}}
            \multiput(1,0)(0,1){1}{\pscircle[fillstyle=solid,fillcolor=white]{.3}
            \put(0,0){\makebox(0,0){\hbox{{\tiny $3$}}}}}
            \multiput(-2.5,-0.85)(1,0){4}{\pscircle[fillstyle=solid,fillcolor=white]{.3}
            \put(0,0){\makebox(0,0){\hbox{{\tiny $2$}}}}}
            \multiput(2,0)(1,0){1}{\pscircle[fillstyle=solid,fillcolor=white]{.3}
            \put(0,0){\makebox(0,0){\hbox{{\tiny $2$}}}}}
            \multiput(1.5,-0.85)(1,0){2}{\pscircle[fillstyle=solid,fillcolor=white]{.3}
            \put(0,0){\makebox(0,0){\hbox{{\tiny $1$}}}}}
            			\put(0,3.4){\pscircle[fillstyle=solid,fillcolor=yellow]{.3}}
            \put(0,3.4){\makebox(0,0){\hbox{{\tiny $8$}}}}
            			\put(1,1.7){\pscircle[fillstyle=solid,fillcolor=yellow]{.3}}
            \put(1,1.7){\makebox(0,0){\hbox{{\tiny $6$}}}}
                        			\put(2,0){\pscircle[fillstyle=solid,fillcolor=yellow]{.3}}
            \put(2,0){\makebox(0,0){\hbox{{\tiny $2$}}}}
                        			\put(-3,-1.7){\pscircle[fillstyle=solid,fillcolor=yellow]{.3}}
            \put(-3,-1.7){\makebox(0,0){\hbox{{\tiny $0$}}}}
                        			\put(1,-1.7){\pscircle[fillstyle=solid,fillcolor=yellow]{.3}}
            \put(1,-1.7){\makebox(0,0){\hbox{{\tiny $0$}}}}
                        			\put(3,-1.7){\pscircle[fillstyle=solid,fillcolor=yellow]{.3}}
            \put(3,-1.7){\makebox(0,0){\hbox{{\tiny $0$}}}}
        \put(1.5,2){\makebox(0,0)[bl]{\hbox{{\small $\widetilde{\mathfrak M}_4 \prod\limits_{i=3}^6 [\widetilde {M}_1]_{i,i}$}}}}
        \put(0.5,3.5){\makebox(0,0)[lc]{\hbox{{\tiny $-1/2$}}}}
	\put(1.5,1.7){\makebox(0,0)[lc]{\hbox{{\tiny $+1/2$}}}}
	 \put(2.5,0){\makebox(0,0)[lc]{\hbox{{\tiny $+1/2$}}}}
	 \put(3.5,-1.7){\makebox(0,0)[lc]{\hbox{{\tiny $-1/2$}}}}
	 \put(-3.5,-1.7){\makebox(0,0)[rc]{\hbox{{\tiny $+1$}}}}
	 \put(1.15,-2.2){\makebox(0,0)[lc]{\hbox{{\tiny $-1$}}}}
			}
            }
	\end{pspicture}
  \caption{\small
  Setting zeros for powers of cluster variables on the lowest row and imposing deficiencies $\{+1,-1,-1/2\}$ at the zeroth, $m$th, and the last ($n$th) positions of variables at this row together with the requirement that \emph{all} other cluster variables except frozen variables at the right boundary have zero deficiencies (and therefore commute with the corresponding monomial) uniquely fixes all weights. The only variables that do not commute then happen to be exactly the $0$th, $m$th, $(n-m)$th, and $n$th frozen variables at the right boundary with the respective deficiencies $\{-1/2,1/2,1/2,-1/2\}$ indicated in the figure.
  }
  \label{fi:frozen}
\end{figure}
Since we have only two monomial expressions (shown in Fig.~\ref{fi:frozen}) exhibiting the required homogeneous commutation relations with all quantum $a_{i,j}$, only these monomials can appear in the quantum polynomial expressions for ${\widehat M}_m^{BL}$. But, say, the highest power of $x_{n00}$ (the variable at the upper corner) is $2m$, so the second case never appears in the expression for ${\widehat M}_m^{BL}$ with $m< n/2$, and the resulting expression must be some $q$-multiple of the first monomial alone.

For ${\widehat M}_{n-m}^{BL}$ with $m< n/2$, we cannot exclude the first monomial, so, in principle, both monomials may appear in the expression for the quantum minor. But even if this is the case, with the proper $q$ factor, $\bigl[{\widehat M}_{n-m}^{BL}\bigr]^*={\widehat M}_{n-m}^{BL}+f(q) {\widehat M}_{m}^{BL}$, where ${\widehat M}_{m}^{BL}$ is self-adjoint. For the corresponding quantum Casimir, we then have
$$
\Bigl[{\widehat M}_{n-m}^{BL}\bigl[{\widehat M}_m^{BL}\bigr]^{-1}\Bigr]^*=\bigl[{\widehat M}_m^{BL}\bigr]^{-1}\bigl[{\widehat M}_{n-m}^{BL}+f(q) {\widehat M}_{m}^{BL}\bigr]={\widehat M}_{n-m}^{BL}\bigl[{\widehat M}_m^{BL}\bigr]^{-1}+f(q),
$$
where $f(q)$ is some polynomial in $q^{\pm 1/2}$ with $f^*(q)=-f(q)$. Then the combination ${\widehat M}_{n-m}^{BL}\bigl[{\widehat M}_m^{BL}\bigr]^{-1}+f(q)/2$ is a self-adjoint quantum Casimir operator. Furthermore, considering the expression $\bigl[{\widehat M}_{n-m}^{BL}\bigr]^*={\widehat M}_{n-m}^{BL}+f(q) {\widehat M}_{m}^{BL}$ in terms of the matrix entries $a_{i,j}$, observe that ${\widehat M}_{n-m}^{BL}$ and $\bigl[{\widehat M}_{n-m}^{BL}\bigr]^*$ contain monomials with homogeneous degree $n-m$, whereas $f(q) {\widehat M}_{m}^{BL}$ contains monomials with homogeneous degree $m$. Since conjugation and commutation relations in the full algebra preserve homogeneous degree, $f(q)$ must equal zero.
\end{proof}

\subsection{Higher order Casimirs}
\begin{exm}\label{exm-Markov}
    The following expression is a self-adjoint Casimir in $\mathbb{A}_3$.
    \begingroup
    \allowdisplaybreaks
    \begin{align*}
        &(q^{-1/2}+q^{-5/2})a_{1,1}a_{2,2}a_{3,3}
        -q^{-3/2}a_{1,1}a_{2,3}^2\\
        &-q^{-1/2}(q-q^{-1})^2a_{1,1}a_{2,3}a_{3,2}
        -(2q^{1/2}-q^{-3/2})a_{1,1}a_{3,2}^2\\
        &-q^{-3/2}a_{1,2}^2a_{3,3}
        +q^{-1/2}a_{1,2}a_{1,3}a_{2,3}
        +q^{-3/2}a_{1,2}a_{1,3}a_{3,2}
        +(0)a_{1,2}a_{2,1}a_{3,3}\\
        &-q^{3/2}a_{1,2}a_{2,3}a_{3,1}
        +q^{3/2}a_{1,2}a_{3,1}a_{3,2}
        -q^{1/2}a_{1,3}^2a_{2,2}\\
        &+q^{3/2}a_{1,3}a_{2,1}a_{2,3}
        -q^{5/2}a_{1,3}a_{2,1}a_{3,2}
        +(0)a_{1,3}a_{2,2}a_{3,1}\\
        &-q^{5/2}a_{2,1}^2a_{3,3}
        +q^{9/2}a_{2,1}a_{2,3}a_{3,1}
        +q^{5/2}a_{2,1}a_{3,1}a_{3,2}
        -q^{9/2}a_{2,2}a_{3,1}^2
    \end{align*}
    \endgroup
    If we denote the above expression by $D$, observe that $D$ is the first Markov element (plus a scalar) in the upper-triangular reduction $A_3$, and $\lim_{q \to 1} D+\det_q$ is the semiclassical Markov element. However, $D$ is not a quantum deformation of the semiclassical Markov element in the sense of Definition \ref{def:deformatiom}. Indeed, the semiclassical Markov element shares monomials with the quantum determinant, and is therefore not a ``lowest'' Casimir in the full algebra, so Conjecture \ref{con:comm} does not apply. Note also that the expression for $D$ has an alternating $q$-coefficient in front of $a_{1,1}a_{3,2}^2$, which will not be removed by adding/subtracting the other two Casimirs of $\mathbb{A}_3$. Additionally, at least one of the monomials $a_{1,2}a_{2,3}a_{3,1}$ or $a_{1,3}a_{2,1}a_{3,2}$ must appear with nonzero coefficient in any (not even self-adjoint) Casimir deformation of the semiclassical Markov element, despite not appearing in its semiclassical expression.

    There is an expression for $D$ which is a quantum deformation in the sense of Definition \ref{def:deformatiom}, with the important caveat that not all terms are lexicographically ordered. For emphasis, we write the two non-lexicographically ordered terms first.
    \begingroup
    \allowdisplaybreaks
    \begin{align*}
        &q^{-1/2}a_{1,1}a_{3,3}a_{2,2}
        +q^{-1/2}a_{1,3}a_{1,2}a_{3,2}
        +q^{-5/2}a_{1,1}a_{2,2}a_{3,3}\\
        &-q^{-3/2}a_{1,1}a_{2,3}^2
        -q^{-3/2}a_{1,1}a_{3,2}^2
        -q^{-3/2}a_{1,2}^2a_{3,3}
        +q^{-1/2}a_{1,2}a_{1,3}a_{2,3}\\
        &+(0)a_{1,2}a_{2,1}a_{3,3}
        -q^{3/2}a_{1,2}a_{2,3}a_{3,1}
        +q^{3/2}a_{1,2}a_{3,1}a_{3,2}
        -q^{1/2}a_{1,3}^2a_{2,2}\\
        &+q^{3/2}a_{1,3}a_{2,1}a_{2,3}
        -q^{5/2}a_{1,3}a_{2,1}a_{3,2}
        +(0)a_{1,3}a_{2,2}a_{3,1}\\
        &-q^{5/2}a_{2,1}^2a_{3,3}
        +q^{9/2}a_{2,1}a_{2,3}a_{3,1}
        +q^{5/2}a_{2,1}a_{3,1}a_{3,2}
        -q^{9/2}a_{2,2}a_{3,1}^2
    \end{align*}
    \endgroup
    It is tempting to conjecture from the first expression for the Pfaffian (Theorem \ref{thm-Pf}) that there exists a quantum deformation of the first Markov element (or of the first Markov element plus the determinant) which sums over all possible monomial reorderings: however, for $n \ge 3$ any such expression will necessarily have coefficients which are either $0$ or combinations of multiple powers of $q$. (For instance, for $n=3$ there are $6$ terms contributing to the lexicographically ordered monomial $a_{1,2}a_{1,3}a_{2,3}$, which appears in the first Markov element with coefficient 1 and does not appear in the determinant.)
\end{exm}

This behavior is reminiscent of the situation for the second Markov element in the upper-triangular reduction (Proposition \ref{thmn=52}).

\subsection{Coisotropic reductions of the full algebra}\label{ss:reductions}

Note, first, that the commutation relations for $a_{i,j}$ imply the following general statement (shown, e.g., in \cite[Remark 2.2]{ChekhovMazzocco2013})
\begin{lem}
    Let us represent the matrix $\mathbb A$ as an $n\times n$ collection of unit squares with the element $a_{i,j}$ written in the $(i,j)$-square. Draw \emph{any} path $p$ that is a broken line going only down and right between the unit squares (likewise splitting the $n\times n$ square table into two Young diagrams: one from the bottom--left, another from the top--right in Fig.~\ref{fi:Young}). Then imposing the condition $a_{i,j}=0$ in all squares of the ``bottom--left'' Young diagram is a coisotropic reduction.
\end{lem}

\begin{co}
    For any coisotropic reduction, the quantum Pfaffian and other quantum Casimirs with this reduction imposed remain Casimirs for the restricted algebra, but new Casimirs may appear.
\end{co}

Let us consider several important examples of coisotropic reduction.

\subsubsection{The $O_{2N}$ algebra}
The $O_{2N}$ algebra corresponds to the case in which all $a_{j,i}=0$ for $i>j$ (the case of the triangular algebra $A$). In this case, the Pfaffian takes the form in Theorem~\ref{t1}; this form is the one derived by Jing and Zhang~\cite{Jing2014}. Note that in this case, $a_{i,j}^*=a_{i,j}$ for $i<j$, but $a_{i,i}=q^{1/2}$, i.e., the operators on the diagonal are not self-adjoint, which is the feature of the quantum $A$-matrix observed in \cite{Ch-Maz-Advances}.

\subsubsection{The $Sp_{2N}$ algebra}
In the $Sp_{2N}$ case studied first by Molev, Ragoucy, and Sorba~\cite{MRS2003}, the $A$-matrix has block form: it is composed out of $2\times 2$ blocks $A_{I,J}$, $1\le I,J\le n$, and all blocks with $I>J$ are null matrices. In the Pfaffian expression, all $a_{j,i}$ with $i>j$ except exactly $a_{2i,2i-1}$, are zeros. Casimirs in this case are $\lfloor n/2\rfloor=N$ Markov elements (including the Pfaffian $\widetilde{Pf}$) and $N$ quantum $2\times 2$ determinants 
$$
q^{-1} a_{2i-1,2i-1}a_{2i,2i} - qa_{2i-1,2i}a_{2i,2i-1},\quad i=1,\dots, N.
$$
The total Poisson dimension is $2N(N+1)-2N=2N^2$.

\subsubsection{The skew-symmetric matrix}
One coisotropic reduction that is not of ``Young diagram'' type is determined by the condition that 
$$
c_{i,j}=a_{i,j}+q^{-1}a_{j,i}\equiv 0\quad \forall i,j: i<j,\quad a_{i,i}=0
$$
which in the semiclassical limit means that the matrix $A$ is skew-symmetric. The Pfaffian then exists only for even $n$ and coincides with the one in the upper-triangular case. This is the $Sp_{2N}$ case in \cite{JingZhangSklyanin} In this case, $A^T\sim A$, so no nontrivial Markov element appears besides the quantum determinant itself for even $n=2N$, but all $\widehat{\mathcal K}_m$ from Lemma~\ref{lem-msa} may still exist, so, in general, we have $\bigl\lfloor \frac{n-1}{2}\bigr\rfloor+1=N$ Casimirs for even $n=2N$ and  $\bigl\lfloor \frac{n-1}{2}\bigr\rfloor=N$ Casimirs for odd $n=2N+1$. The Poisson dimension is $N(2N-1)-N=2N(N-1)$ in the first case and $N(2N+1)-N=2N^2$ in the second case.

\subsubsection{The symmetric matrix}
Another coisotropic reduction that is not of ``Young diagram'' type is governed by the condition
$$
b_{i,j}=a_{i,j}-qa_{j,i}\equiv 0\quad \forall i,j: i<j,
$$
which in the semiclassical limit just means that the matrix $A$ is symmetric.  This is the $O_n$ case in \cite{JingZhangSklyanin}. The Pfaffian is then identically equal to zero, and no nontrivial Markov elements appear as in the skew-symmetric case, but the quantum determinant can be nontrivial, as well as all $\widehat{\mathcal K}_m$ from Lemma~\ref{lem-msa}, so we have $\bigl\lfloor \frac{n-1}{2}\bigr\rfloor+1$ Casimirs. The total Poisson dimension is 
$$
\frac{n(n+1)}{2}-\Bigl\lfloor \frac{n-1}{2}\Bigr\rfloor-1=\begin{cases}2N^2&\hbox{for}\ n=2N,\\ 2N(N+1)&\hbox{for}\ n=2N+1. \end{cases}
$$

\section{Conclusion}
In conclusion, we briefly indicate perspectives and unsolved problems.

First, it is interesting to construct quantum Markov elements $\widehat C_i$ with $i=1,\dots,\lfloor n/2\rfloor$ for the full-size matrix $\mathbb A$ and higher 
$\widehat C_i$ for the triangular matrix $A$. 

Second, despite the fact that an independent proof of centrality of the full quantum Pfaffian in Appendix~\ref{app:centrality} makes our conjugation conjecture obsolete regarding the quantum Pfaffian case, we nevertheless want to attract attention to this conjecture, which may result in a deeper understanding of the relation between conjugation and commutativity in quantum algebras.

Third, it is worth studying the relation to quantum cluster algebras initiated in this paper. In our accounting, so far, mutations, which are Poisson and quantum morphisms of these algebras, did not play any role. We however expect that the matrix elements $a_{i,j}$ can be universal polynomials (elements of an upper cluster algebra), as happens both in the case of the triangular FG network studied in \cite{Chekhov2020} and in the case of the symplectic groupoid network explored in \cite{ChekhovShapiro2023}.

Fourth, it might be interesting to understand quantum Casimirs $\widehat{\mathcal K}_m$ from the standpoint of Gelfand--Retakh quasideterminants.

\section*{Acknowledgments}

This project started in the spring of 2025 under the umbrella of the ``Discovering America'' program at Michigan State University, hosting one-semester joint research teams of visiting students from China and groups of USA students. The authors are grateful to the Math Department of Michigan State University for its hospitality, support, and inspiring research atmosphere, which allowed us to continue our research. Special thanks are to Michael Shapiro, who took part in this project at its early stages. We are grateful to Naihuan Jing for the useful discussion.

\section*{Statement on AI usage}

During an exploratory stage of the project, GPT-5.5 was used to simplify a conjectured expression for the $q$-coefficient present in the quantum determinant (Definition \ref{def:qdet}) from a recurrence relation to a closed form. The final manuscript was written and edited entirely by the authors.

Upon completion of the text, GPT-6 provided a proof of centrality of the Pfaffian of the full algebra. The authors added Appendix~\ref{app:centrality} in which the main idea of this proof, namely, using polarization operators $K_k,E_k,F_k$ to prove centrality, was developed adding a special set of $R$-matrix relations that was not a part of the A.I.-assisted text, which can be found in its entirety in \cite{AI-text}. The authors also added a conjecture that quantum Markov elements should also enjoy invariance w.r.t. the action of quantum polarization operators and another proof of centrality of the quantum determinant based on technique of polarization operators.

\appendix

\section{Algorithms for Constructing Quantum Casimir Functions}\label{appendix:algorithms}

\subsection{An algorithm for polynomial simplification}

To compute the quantum Casimir functions, we design a symbolic algorithm that brings noncommutative polynomials into lexicographically ordered forms. 

\begin{algorithm}[H]\allowdisplaybreaks
\caption{Simplify Noncommutative Polynomial}\label{algorithm1}
\begin{algorithmic}[1]
\State Fix a positive integer \( n \), representing the size of the unipotent upper triangular matrix
\State Define the following noncommutative transformation rules:
\begin{itemize}
    \item Scalar extraction rules: pull out scalar coefficients (e.g., powers of \( q \), constants) from monomials
    \item Distributive rule: distribute noncommutative multiplication over addition
\end{itemize}
\State Apply quantum commutation relations:
\begin{itemize}
    \item For index triples \( (i,j,k) \): apply quantum commutation relations to reorder terms involving \( a_{i,j}a_{j,k} \)
    \item For index quadruples \( (i,j,k,l) \) satisfying \((1 \leq i < j < k < l \leq n)\): apply quantum commutation relations to reorder terms involving \( a_{i,k}a_{j,l} \)
    \item For all other index combinations not covered above: apply a general commutation rule to enforce lexicographic ordering
\end{itemize}
\State Combine all rules into a unified set \( \mathcal{R} \)
\State Given a noncommutative polynomial \( M \), repeatedly apply all rules in \( \mathcal{R} \) until no further rewriting is possible (i.e., until a fixed point is reached)
\State Output the simplified polynomial
\end{algorithmic}
\end{algorithm}

The rule \( \mathcal{R} \) is designed to transform any noncommutative monomial into a lexicographically ordered form. By recursively applying these rules, every polynomial is reduced to a linear combination of lexicographically ordered monomials.

\subsection{An algorithm for computing self-adjoint quantum Casimir functions}

Given a candidate polynomial \( \widehat{M} \) with undetermined scalar coefficients \( \beta_1, \dots, \beta_m \), we aim to determine whether \( \widehat{M} \) is a self-adjoint quantum Casimir function. This is done by testing whether \( \widehat{M} \) commutes with all generators \( a_{i,j} \) and is invariant under conjugation. The procedure is outlined in Algorithm~\ref{algorithm2}.

\begin{algorithm}[H]
\caption{Determine Whether a Polynomial \( \widehat{M} \) is a Self-Adjoint Quantum Casimir}
\label{algorithm2}
\begin{algorithmic}[1]
\State \textbf{Input:} A candidate polynomial \( \widehat{M}(\beta_1, \dots, \beta_m) \) with parameters to determine
\State Define a set of noncommutative rules:
\begin{itemize}
    \item Scalar extraction rules (to factor out \( q \)-powers)
    \item Distributive rules over addition
    \item Quantum commutation relations for triples and quadruples
    \item General commutation rules to enforce lexicographic order
\end{itemize}
\State Define the conjugation map \( \texttt{conj} \) by:
\[
\texttt{conj}(a_{i,j}) = a_{i,j}, \quad \texttt{conj}(q^\alpha) = q^{-\alpha}, \quad \texttt{conj}(XY) = \texttt{conj}(Y) \texttt{conj}(X)
\]
\State Define commutators for all generators:
\[
P_{i,j} := \widehat{M}  a_{i,j} - a_{i,j}  \widehat{M}
\]
\State Define self-conjugacy error:
\[
P_c := \widehat{M} - \texttt{conj}(\widehat{M})
\]
\State Apply Algorithm~\ref{algorithm1} to simplify \( P_{i,j} \) and \( P_c \) into standard lexicographically ordered monomials
\State Identify a basis of standard monomials (e.g., all lexicographically ordered degree \(\leq 3\) words in \( a_{i,j} \))
\State Extract the coefficient of each monomial in \( P_{i,j} \) and \( P_c \)
\State Set all coefficients equal to zero to form a system of equations in \( \beta_k \)
\State \textbf{Solve} the system to obtain all possible \( \beta_k \)
\end{algorithmic}
\end{algorithm}

\begin{remark}
In certain cases, the candidate polynomial \( \widehat{M}(\beta_1, \dots, \beta_m) \) may lead to an overdetermined system in Algorithm~\ref{algorithm2}, yielding no solution. However, if the self-adjointness condition (i.e., the requirement \( P_c = 0 \)) is removed from the algorithm, a solution may still exist. Such a situation will occur in Section~\ref{subsecn=5}, where we compute the quantum Casimir function for the case \( n = 5 \).
\end{remark}

\section{Relations for the full algebra \texorpdfstring{$\mathbb A_n$}{An}}\label{appendix:fullrelations}

Now we present the expansion of \ref{eq:FullQuantumCommutator} for varying $i$, $s$, $j$, and $t$.

\subsection{Two distinct indices}
In the case of two distinct indices (say 1 and 2), the relations are
%

\begin{align*}
    a_{12}a_{11} &= a_{11}a_{12}-(q-q^{-1})a_{11}a_{21} \\
    a_{22}a_{12} &= a_{12}a_{22}-(q-q^{-1})a_{21}a_{22} \\
    a_{21}a_{11} &= q^{-2}a_{11}a_{21} \\
    a_{22}a_{21} &= q^{-2}a_{21}a_{22} \\
    a_{21}a_{12} &= a_{12}a_{21} \\
    a_{22}a_{11} &= a_{11}a_{22}-q(q-q^{-1})a_{12}a_{21}-(q-q^{-1})a_{21}^2
\end{align*}

\subsection{Three distinct indices}
In the case of three distinct indices (say 1, 2, and 3), the relations are

\begin{align*}
    a_{23}a_{11} 
    &= a_{11}a_{23}-(q-q^{-1})a_{13}a_{21}-(q-q^{-1})^2a_{12}a_{31}-(q-q^{-1})a_{21}a_{31} \\
    a_{32}a_{11} 
    &= a_{11}a_{32}-(q-q^{-1})a_{12}a_{31}-q^{-1}(q-q^{-1})a_{21}a_{31} \\
    a_{22}a_{13} &= a_{13}a_{22}+(q-q^{-1})a_{12}a_{32}-(q-q^{-1})a_{21}a_{23} \\
    a_{31}a_{22} &= a_{22}a_{31} \\
    a_{33}a_{12} &= a_{12}a_{33}-q(q-q^{-1})a_{13}a_{32}-(q-q^{-1})a_{31}a_{32} \\
    a_{33}a_{21} 
    &= a_{21}a_{33}-q(q-q^{-1})a_{23}a_{31}-q^{-1}(q-q^{-1})a_{31}a_{32} \\
    a_{13}a_{12} &= q^{-1}a_{12}a_{13}+q^{-1}(q-q^{-1})a_{11}a_{23}-(q-q^{-1})a_{11}a_{32} \\
    a_{31}a_{12} &= q^{-1}a_{12}a_{31} \\
    a_{21}a_{13} &= q^{-1}a_{13}a_{21}+q^{-1}(q-q^{-1})a_{12}a_{31} \\
    a_{31}a_{21} &= q^{-1}a_{21}a_{31} \\
    a_{23}a_{12} 
    &=qa_{12}a_{23}-(q-q^{-1})^2a_{12}a_{32}-(q-q^{-1})a_{13}a_{22}-(q-q^{-1})a_{21}a_{32} \\
    a_{32}a_{12} 
    &= q^{-1}a_{12}a_{32}-(q-q^{-1})a_{22}a_{31} \\
    a_{23}a_{21} &= q^{-1}a_{21}a_{23}-(q-q^{-1})a_{22}a_{31} \\
    a_{32}a_{21} 
    &= q^{-1}a_{21}a_{32}-q^{-1}(q-q^{-1})a_{22}a_{31} \\
    a_{23}a_{13} &= q^{-1}a_{13}a_{23}+q^{-1}(q-q^{-1})a_{12}a_{33}-(q-q^{-1})a_{21}a_{33} \\
    a_{32}a_{13} 
    &= qa_{13}a_{32}-q(q-q^{-1})a_{23}a_{31} \\
    a_{31}a_{23} &= qa_{23}a_{31} \\
    a_{32}a_{31} &= q^{-1}a_{31}a_{32}
\end{align*}

\subsection{Four distinct indices}
In the case of four distinct indices (say 1, 2, 3, and 4), the relations are

\begin{align*}
    a_{34}a_{12} &= a_{12}a_{34}-(q-q^{-1})a_{14}a_{32}-(q-q^{-1})a_{31}a_{42}-(q-q^{-1})^2a_{13}a_{42} \\
    a_{43}a_{12} 
    &= a_{12}a_{43}-(q-q^{-1})a_{13}a_{42}-(q-q^{-1})a_{32}a_{41}  \\
    a_{24}a_{13} &= a_{13}a_{24}-(q-q^{-1})a_{14}a_{23}+(q-q^{-1})a_{12}a_{34}-(q-q^{-1})a_{21}a_{43}-(q-q^{-1})^2a_{12}a_{43} \\
    a_{42}a_{13} 
    &= a_{13}a_{42}-(q-q^{-1})a_{23}a_{41} \\
    a_{23}a_{14} &= a_{14}a_{23}-(q-q^{-1})a_{21}a_{34}+(q-q^{-1})a_{12}a_{43} \\
    a_{32}a_{14} 
    &= a_{14}a_{32}-(q-q^{-1})a_{24}a_{31}-(q-q^{-1})^2a_{23}a_{41}+(q-q^{-1})a_{13}a_{42} \\
    a_{34}a_{21} 
    &= a_{21}a_{34}-(q-q^{-1})a_{24}a_{31}-(q-q^{-1})^2a_{23}a_{41}-(q-q^{-1})a_{32}a_{41} \\
    a_{43}a_{21} 
    &= a_{21}a_{43}-(q-q^{-1})a_{23}a_{41}-(q-q^{-1})a_{31}a_{42}+(q-q^{-1})^2a_{32}a_{41} \\
    a_{41}a_{23} &= a_{23}a_{41} \\
    a_{31}a_{24} &= a_{24}a_{31}+(q-q^{-1})a_{23}a_{41} \\
    a_{42}a_{31} 
    &= a_{31}a_{42}-(q-q^{-1})a_{32}a_{41} \\
    a_{41}a_{32} &= a_{32}a_{41}
\end{align*}

\section{The commutator of \texorpdfstring{$a_{2n,2}$}{a2n,2} with \texorpdfstring{$[1,2,\dots,2n]$}{[1,2,...,2n]}}\label{appendix:commutator}
It is easy to see that the commutator $\bigl[a_{2n,1},[1,2,\dots,2n]\bigr]$ vanishes because $a_{2n,1}$ has homogeneous commutation relations with all $a_{i,j}$ with $1\le i,j\le 2n$. But already the next case turns out to be rather nontrivial.
\begin{align*}
    &a_{2n,2}[1,2,\dots,2n]=a_{2n,2}\sum_{k>1}(-q)^{k-2}\bigl(a_{1,k}-q^{2n-1}a_{k,1}\bigr)[\widehat 1,2,\dots \widehat k\dots,2n]\\
    &=a_{2n,2}(a_{1,2}-q^{2n-1}a_{2,1} [\widehat 1,\widehat 2,3,\dots, 2n]\\
    &\quad + \sum_{2<k<2n}(-q)^{k-2} a_{2n,2}\bigl(a_{1,k}-q^{2n-1}a_{k,1}\bigr)[\widehat 1,2,\dots \widehat k\dots,2n]\\
    &\quad + (-q)^{2n-2}a_{2n,2}(a_{1,2n}-q^{2n-1}a_{2n,1} [\widehat 1,2,3,\dots, \widehat{2n}]\\
    &=\bigl(q^{-1}a_{1,2}a_{2n,2}-(q-q^{-1})a_{2n,1}a_{2,2}-q^{2n-1}(q^{-1}a_{2,1}a_{2n,2}-(q-q^{-1})a_{2n,1}a_{2,2})[\widehat 1,\widehat 2,3,\dots, 2n]\\
    &\quad + \sum_{2<k<2n}(-q)^{k-2} 
    \bigl(a_{1,k}a_{2n,2}-(q-q^{-1})a_{2n,1}a_{2,k}-q^{2n-1}(a_{k,1}a_{2n,2}-(q-q^{-1})a_{2n,1}a_{k,2})
    \bigr)[\widehat 1,2,\dots \widehat k\dots,2n]\\
    &\quad + (-q)^{2n-2}\bigl(qa_{1,2n}a_{2n,2}-(q-q^{-1})a_{2n,1}a_{2,2n}-q^{2n-1}\cdot q^{-1}a_{2n,1}a_{2n,2}\bigr)[\widehat 1,2,3,\dots, \widehat{2n}]\\
    &=[1,2,\dots 2n]a_{2n,2}+(q-q^{-1})a_{2n,1}\Bigl\{ (-a_{2,2}+q^{2n-2}a_{2,2})[\widehat 1,\widehat 2,3,\dots, 2n]\Bigr.\\
    &+\sum_{2<k<2n}(-q)^{k-2} 
    \bigl(-a_{2,k}+q^{2n-1}a_{k,2}\bigr)[\widehat 1,2,\dots \widehat k\dots,2n]\\
    &\Bigl.+(-q)^{2n-2}\bigl(-a_{2,2n}+q^{2n-1}a_{2n,2}\bigr)[\widehat 1,2,3,\dots, \widehat{2n}]\Bigr\}
\end{align*}
We now consider the expression in the curly brackets. In its first term, we apply the second (unordered) form of the Pfaffian expansion, whereas in the second and third terms we apply the third, ordered, form (with the coefficient $f_{n-1}(q)$ appearing due to Theorem~\ref{thm-Pf}. We have
\begin{align*}
    &a_{2,2}(q^{2n-2}-1)\sum_{3\le i<j\le 2n}b_{i,j}(-q)^{i+j-7}[\widehat 1,\widehat 2,3,\widehat i, \widehat j\dots, 2n]\\
    &+\sum_{3\le j<k<2n} (-q)^{k-2} 
    \bigl(-a_{2,k}+q^{2n-1}a_{k,2}\bigr)f_{n-1}(q)(-q)^{j-3} \bigl(a_{2,j}-q^{2n-3}a_{j,2}\bigr)[\widehat 1,\widehat 2,3,\widehat i, \widehat j\dots, 2n]\\
    &+\sum_{3\le k<j\le 2n} (-q)^{k-2} 
    \bigl(-a_{2,k}+q^{2n-1}a_{k,2}\bigr)f_{n-1}(q)(-q)^{j-4} \bigl(a_{2,j}-q^{2n-3}a_{j,2}\bigr)[\widehat 1,\widehat 2,3,\widehat i, \widehat j\dots, 2n]\\
    &+(-q)^{2n-2}\bigl(-a_{2,2n}+q^{2n-1}a_{2n,2}\bigr)f_{n-1}(q)\sum_{3\le j< 2n} (-q)^{j-3} \bigl(a_{2,j}-q^{2n-3}a_{j,2}\bigr)[\widehat 1,\widehat 2, \widehat j\dots, \widehat{2n}]\\
    &=a_{2,2}(q^{2n-2}-1)\sum_{3\le j<k\le 2n}(a_{j,k}-qa_{k,j})(-q)^{k+j-7}[\widehat 1,\widehat 2,3,\widehat j, \widehat k\dots, 2n]\\
    &+f_{n-1}(q)\sum_{3\le j<k\le 2n} (-q)^{k+j-5} \bigl(-a_{2,k}+q^{2n-1}a_{k,2}\bigr) \bigl(a_{2,j}-q^{2n-3}a_{j,2}\bigr)[\widehat 1,\widehat 2,3,\widehat j, \widehat k\dots, 2n]\\
    &+f_{n-1}(q)\sum_{3\le j<k\le 2n} (-q)^{k+j-5}(-q)^{-1} \bigl(-a_{2,j}+q^{2n-1}a_{j,2}\bigr) \bigl(a_{2,k}-q^{2n-3}a_{k,2}\bigr)[\widehat 1,\widehat 2,3,\widehat j, \widehat k\dots, 2n]
\end{align*}
Fix now $j$ and $k$. The coefficient of $[\widehat 1,\widehat 2,3,\widehat j, \widehat k\dots, 2n]$ reads
\begin{align*}
    &a_{2,2}(q^{2n-2}-1)(-q)^{k+j-7}(a_{j,k}-qa_{k,j}) + f_{n-1}(q)(-q)^{k+j-5}\Bigl[ -(a_{2,k}a_{2,j}+q^{-1}a_{2,j}a_{2,k}) \Bigr.\\
    &\Bigl.\quad +q^{2n-1}a_{k,2}a_{2,j}-q^{2n-3}\cdot q^{-1} a_{2,j}a_{k,2}+(q^{2n-3}a_{2,k}a_{j,2}-q^{2n-2}a_{j,2}a_{2,k})+(q^{4n-4}a_{k,2}a_{j,2}-q^{4n-5}a_{j,2}a_{k,2}) \Bigr]\\
    =&a_{2,2}(a_{j,k}-qa_{k,j})(-q)^{k+j-7}\bigl[ q^{2n-2}-1 -(q^2-1)f_{n-1}(q) \bigr]\\
    &+f_{n-1}(q) (-q)^{k+j-5}\bigl[q^{2n-1}a_{k,2}a_{2,j}-q^{2n-3}\cdot q^{-1} a_{2,j}a_{k,2}-q^{2n-3}(q-q^{-1})a_{2,j}a_{k,2} \bigr]
\end{align*}
The first line in the last expression is zero because
$$
f_{n-1}(q)=\frac{q^{2n-2}-1}{q^2-1}
$$
and the second line is zero because $a_{k,2}a_{2,j}=q^{-1}a_{2,j}a_{k,2}$ for $2<j<k$. So, the whole expression in curly brackets vanishes, and we obtain the desired commutation relation
$$
a_{2n,2}[1,2,\dots,2n]=[1,2,\dots,2n]a_{2n,2}.
$$

Analogously, but a little easier, is to prove that
$$
a_{2n,2}[1,\widehat 2,\dots,\widehat r,\dots,2n]=q [1,\widehat 2,\dots,\widehat r,\dots,2n]a_{2n,2}-(q-q^{-1})a_{2n,1}[\widehat 1,2,\dots,\widehat r,\dots,2n].
$$

\rem{

\begin{lem}
    Commutativity relations
    \begin{align*}
        &\Bigl[ a_{r,1},[1,2,\dots,2n]\Bigr]=\Bigl[ a_{2n,r},[1,2,\dots,2n]\Bigr]=0\quad\forall r: 1<r<2n
    \end{align*}
    imply that $[1,2,\dots,2n]$ is a Casimir.
\end{lem}
\begin{proof}
    Assuming $a_{2n,1}$ is invertible, we can obtain all other $a_{r,p}$ (with $1<r,p<2n$ from the relations
    \begin{align*}
        &[a_{r,1},a_{2n,p}]=(q-q^{-1})a_{2n,1}a_{r,p}\quad \hbox{for}\quad r\ne p.
    \end{align*}
Then, the Jacobi relations imply
\begin{align*}
    0=-\bigl[[a_{2n,p},Pf],a_{r,1}\bigr]-\bigl[[Pf,a_{r,1}],a_{2n,p}\bigr]=\bigl[[a_{r,1},a_{2n,p}],Pf\bigr]=(q-q^{-1})a_{2n,1}[a_{r,p}, Pf],
\end{align*}
so if $a_{2n,1}^{-1}$ exists, the commutator must vanish.
\end{proof}

}

\section{Centrality of the full quantum Pfaffian \texorpdfstring{$\widehat{Pf}_\square$}{Pf_square}}\label{app:centrality}

For \(N=2n\), let $\mathcal V_{N}$ be a linear span of all quantum monomials $a_{\pi_1,\pi_2}\cdots a_{\pi_{N-1},\pi_{N}}$ with $\pi\in S_{N}$. Then the set of \emph{oriented perfect matchings} $\pi\in\Pi$, i.e.,
\[
 m=a_{\pi_1,\pi_2}\cdots a_{\pi_{2n-1},\pi_{2n}},\qquad \pi_1<\pi_3<\cdots<\pi_{2n-3}<\pi_{2n-1},
\]
constitutes a basis in $\mathcal V_{N}$.

Define
\begin{equation}
 \widehat{Pf}_n=q^{-n(n-1)} \sum_{\pi\in S_{2n}}(-q)^{l(\pi)}
 a_{\pi_1\pi_2}\cdots a_{\pi_{2n-1}\pi_{2n}},
\label{eq:P}
\end{equation}
where \(l(\pi)\) is the ordinary inversion number. This is the first expression in Theorem~\ref{thm-Pf}.

\begin{thm}\label{thm:main}
For every \(n\ge1\),  $\widehat{Pf}_n$ is a nonzero central element of
\(\mathbb A_{2n}\).
\end{thm}

\subsection{Quantum polarizations}

For \(1\le k<2n\), let
\(\chi_k(i)=\delta_{i,k}-\delta_{i,k+1}\). On the free algebra set
\begin{align}
 K_k(a_{ij})&=q^{\chi_k(i)+\chi_k(j)}a_{ij},& K_k(\mathbb A)&=\mathbb K_k \mathbb A\mathbb K_k  \label{eq:K}\\
 E_k(a_{ij})&=\delta_{i,k}a_{k+1,j}+q^{\chi_k(i)}\delta_{j,k}a_{i,k+1}, & E_k(\mathbb A)&=e_{k,k+1}\mathbb A + \mathbb K_k \mathbb A e_{k+1,k}
       \label{eq:E}\\
 F_k(a_{ij})&=q^{-\chi_k(j)}\delta_{i,k+1}a_{k,j} +\delta_{j,k+1}a_{i,k}, & F_k(\mathbb A)&=e_{k+1,k}\mathbb A\mathbb K_k^{-1} + \mathbb A e_{k,k+1} .\label{eq:F}
\end{align}
where $\mathbb K_k$ is the diagonal matrix with entries $e_{i,i}q^{\chi_k(i)}$.

The maps are \(L\)-linear. Extend \(K_k\) multiplicatively and set
\begin{align}
 E_k(xy)&=E_k(x)y+K_k(x)E_k(y),\label{eq:Eleibniz}\\
 F_k(xy)&=F_k(x)K_k^{-1}(y)+xF_k(y).
\label{eq:Fleibniz}
\end{align}
Thus \(E_k\) raises a label \(k\) to \(k+1\), in the right-action convention. 

The above relations are known in the literature (see \cite{LSS24} and references therein). They are commonly referred to as the Drinfeld--Jimbo quantized enveloping algebra for $U_q(gl_N)$ (see \cite{LSS24}, Sec.3.1 where the above polarization operators action on coproducts has the same form as in the above formulas). Proposition 5.10 in \cite{LSS24} claims the ideal preservation, but it is not immediately clear whether the authors used in their proof additional relations formulated in the same proposition, so we present a detailed proof of the ideal preservation below.

\begin{lem}\label{lm:ideal}
The mappings $K_k,E_k,F_k$ preserve the ideal generated by the quantum algebra $\mathbb A_n$ for any $k=1,\dots,n-1$.
\end{lem}

\begin{remark}
The original A.I.-assisted proof of this statement was by checking all relations for a sufficiently large $N=6$: since all relations include only four (distinct or coinciding) indices, this case incorporates all possible \(1296=6^4\) choices of indices and by Bergman's Diamond Lemma \cite{Bergman} yields the assertion. Below we present a different, human-assisted proof based on $R$-matrix relations.
\end{remark}

\begin{proof}
we begin with proving four relations for the quantum $R$-matrix (\ref{eq:R12}):
\begin{align} 
&R_{12}\Bigl[\stackrel{1}{e}_{k,k+1}+\stackrel{1}{\mathbb K}_k \stackrel{2}{e}_{k,k+1}  \Bigr] = \Bigl[\stackrel{2}{e}_{k,k+1}+\stackrel{2}{\mathbb K}_k \stackrel{1}{e}_{k,k+1}  \Bigr]R_{12}\label{rel-I}  \\
&R_{12}\Bigl[\stackrel{1}{e}_{k,k+1}\stackrel{2}{\mathbb K}{}_k^{-1}+ \stackrel{2}{e}_{k,k+1}  \Bigr] = \Bigl[\stackrel{1}{e}_{k,k+1}+\stackrel{1}{\mathbb K}{}_k^{-1} \stackrel{2}{e}_{k,k+1}  \Bigr]R_{12}\label{rel-II}  \\
&R_{12}\Bigl[\stackrel{1}{e}_{k+1,k}+\stackrel{1}{\mathbb K}_k \stackrel{2}{e}_{k+1,k}  \Bigr] = \Bigl[\stackrel{2}{e}_{k+1,k}+\stackrel{2}{\mathbb K}_k \stackrel{1}{e}_{k+1,k}  \Bigr]R_{12}\label{rel-III}  \\
&R_{12}\Bigl[\stackrel{1}{e}_{k+1,k}\stackrel{2}{\mathbb K}{}_k^{-1}+ \stackrel{2}{e}_{k+1,k}  \Bigr] = \Bigl[\stackrel{2}{e}_{k+1,k}\stackrel{1}{\mathbb K}{}_k^{-1}+ \stackrel{1}{e}_{k+1,k}  \Bigr]R_{12}\label{rel-IV} 
\end{align}
We also use that, for any diagonal matrix $D$,
\begin{equation}\label{rel-D}
\stackrel{1}D\stackrel{2}D R_{12}=R_{12} \stackrel{1}D\stackrel{2}D.
\end{equation}
We obtain the corresponding relations for $R_{12}^{t_1}$ by transposition in the first space. For example, relation (\ref{rel-I}) implies
$$
\stackrel{1}{e}_{k+1,k} R^{t_1}_{12}+\stackrel{1}{\mathbb K}_k R^{t_1}_{12} \stackrel{2}{e}_{k,k+1}  = \stackrel{2}{e}_{k,k+1}R^{t_1}_{12}+\stackrel{2}{\mathbb K}_k R^{t_1}_{12} \stackrel{1}{e}_{k+1,k}.
$$
We now prove relation (\ref{rel-I}):
\begin{align*}
R_{12}\Bigl[\stackrel{1}{e}_{k,k+1}+\stackrel{1}{\mathbb K}_k \stackrel{2}{e}_{k,k+1}  \Bigr]&=\stackrel{1}{e}_{k,k+1}\stackrel{2}{\mathbb I}+\stackrel{1}{\mathbb K}_k \stackrel{2}{e}_{k,k+1}+(q-1)\stackrel{1}{e}_{k,k+1}\stackrel{2}{e}_{k,k}+q(q-1)\stackrel{1}{e}_{k,k}\stackrel{2}{e}_{k,k+1}\\
&+(q-q^{-1})\sum_{i>k} \stackrel{1}{e}_{i,k+1}\stackrel{2}{e}_{k,i}+(q-q^{-1})\sum_{j<k}\stackrel{1}{e}_{k,j}\stackrel{2}{e}_{j,k+1}
\end{align*}
\begin{align*}
\Bigl[\stackrel{2}{e}_{k,k+1}+\stackrel{2}{\mathbb K}_k \stackrel{1}{e}_{k,k+1}  \Bigr]R_{12}&=\stackrel{1}{\mathbb I}\stackrel{2}{e}_{k,k+1}+\stackrel{1}{e}_{k,k+1}\stackrel{2}{\mathbb K}_k +(q-1)\stackrel{1}{e}_{k+1,k+1}\stackrel{2}{e}_{k,k+1}+q^{-1}(q-1)\stackrel{1}{e}_{k,k+1}\stackrel{2}{e}_{k+1,k+1}\\
&+(q-q^{-1})\sum_{i>k+1} \stackrel{1}{e}_{i,k+1}\stackrel{2}{e}_{k,i}+(q-q^{-1})\sum_{j<k+1}\stackrel{1}{e}_{k,j}\stackrel{2}{e}_{j,k+1}q^{\delta_{j,k}}.
\end{align*}
Comparing the above two expressions, we see that the whole possible difference can be only in the direct product of two $2\times 2$ blocks with $\{i,j\}\in \{k,k+1\}$. Subtracting the second expression from the first one, we obtain
\begin{align*}
&(1-q)\stackrel{1}{e}_{k,k+1}\stackrel{2}{e}_{k,k}+(1-q^{-1}) \stackrel{1}{e}_{k,k+1}\stackrel{2}{e}_{k+1,k+1}+(q-1)\stackrel{1}{e}_{k,k}\stackrel{2}{e}_{k,k+1}+(q^{-1}-1) \stackrel{1}{e}_{k+1,k+1}\stackrel{2}{e}_{k,k+1}\\
&+(q-1)\stackrel{1}{e}_{k,k+1}\stackrel{2}{e}_{k,k} +q(q-1) \stackrel{1}{e}_{k,k}\stackrel{2}{e}_{k,k+1}+(1-q) \stackrel{1}{e}_{k+1,k+1}\stackrel{2}{e}_{k,k+1} +(q^{-1}-1) \stackrel{1}{e}_{k,k+1}\stackrel{2}{e}_{k +1,k+1}\\
&+(q-q^{-1})\stackrel{1}{e}_{k+1,k+1}\stackrel{2}{e}_{k,k+1} -(q-q^{-1})q \stackrel{1}{e}_{k,k}\stackrel{2}{e}_{k,k+1}\\
&= \stackrel{1}{e}_{k,k+1}\stackrel{2}{e}_{k,k+1}\bigl[q-1+q(q-1)-q(q-q^{-1}) \bigr] + \stackrel{1}{e}_{k+1,k+1}\stackrel{2}{e}_{k,k+1} \bigl[q^{-1}-1+1-q+(q-q^{-1}) \bigr] =0.
\end{align*}
We obtain relation (\ref{rel-II}) by multiplying (\ref{rel-I}) by $\stackrel{1}{\mathbb K}{}_k^{-1}\stackrel{2}{\mathbb K}{}_k^{-1}$ from the right with accounting for condition (\ref{rel-D}) and that $\stackrel{\alpha}{e}_{k,k+1}\stackrel{\alpha}{\mathbb K}{}_k^{-1}=q \stackrel{\alpha}{e}_{k,k+1}$ for $\alpha=1,2$. Relation (\ref{rel-III}) can be obtained analogously to relation (\ref{rel-I}) and relation (\ref{rel-IV}) can be obtained from (\ref{rel-III}) upon multiplication  by $\stackrel{1}{\mathbb K}{}_k^{-1}\stackrel{2}{\mathbb K}{}_k^{-1}$ from the right. Therefore, in order to verify any of the above four relations  we can restrict our consideration merely to the case of $k=1$, $N=2$ with 
$$
R_{12}=\left[\begin{array}{cccc} q&0&0&0\\ 0&1&q-q^{-1}&0\\0&0&1&0\\0&0&0&q \end{array} \right]
$$
and, say,
$$
\Bigl[\stackrel{1}{e}_{1,2}+\stackrel{1}{\mathbb K}_1 \stackrel{2}{e}_{1,2}  \Bigr] = \left[\begin{array}{cccc} 0&1&q&0\\ 0&0&0&q^{-1}\\0&0&0&1\\0&0&0&0 \end{array} \right],
\qquad
\Bigl[\stackrel{2}{e}_{1,2}+\stackrel{2}{\mathbb K}_1 \stackrel{1}{e}_{1,2}  \Bigr] = \left[\begin{array}{cccc} 0&q&1&0\\ 0&0&0&1\\0&0&0&q^{-1}\\0&0&0&0 \end{array} \right].
$$

We next use the above $R$-matrix relations when acting by $K_k,E_k,F_k$ on two sides of the reflection equation (\ref{RARA}). We show it for $F_k$: when acting on the right-hand side, we obtain
\begin{align}
&F_k(R_{12}\stackrel{1}{\mathbb A}R_{12}^{t_1}\stackrel{2}{\mathbb A})=R_{12}\Bigl[ \stackrel{1}{e}_{k+1,k} \stackrel{1}{\mathbb A} \stackrel{1}{\mathbb K}{}_k^{-1}+\stackrel{1}{\mathbb A} \stackrel{1}{e}_{k,k+1} \Bigr] R_{12}^{t_1} \stackrel{2}{\mathbb K}{}_k^{-1}\stackrel{2}{\mathbb A} \stackrel{2}{\mathbb K}{}_k^{-1}+R_{12}\stackrel{1}{\mathbb A}R_{12}^{t_1} \Bigl[ \stackrel{2}{e}_{k+1,k} \stackrel{2}{\mathbb A} \stackrel{2}{\mathbb K}{}_k^{-1}+\stackrel{2}{\mathbb A} \stackrel{2}{e}_{k,k+1} \Bigr]\nonumber\\
&=R_{12}\stackrel{1}{e}_{k+1,k} \stackrel{2}{\mathbb K}{}_k^{-1} \stackrel{1}{\mathbb A} R_{12}^{t_1} \stackrel{2}{\mathbb A} \stackrel{1}{\mathbb K}{}_k^{-1} \stackrel{2}{\mathbb K}{}_k^{-1}  +R_{12}\stackrel{1}{\mathbb A}\Bigl[ \stackrel{1}{e}_{k,k+1}  R_{12}^{t_1} \stackrel{2}{\mathbb K}{}_k^{-1} + R_{12}^{t_1} \stackrel{2}{e}_{k+1,k} \Bigr] \stackrel{2}{\mathbb A} \stackrel{2}{\mathbb K}{}_k^{-1} + R_{12}\stackrel{1}{\mathbb A}R_{12}^{t_1}\stackrel{2}{\mathbb A}\stackrel{2}{e}_{k,k+1}\nonumber\\
&\stackrel{(\ref{rel-IV})}{=}R_{12}\stackrel{1}{e}_{k+1,k} \stackrel{2}{\mathbb K}{}_k^{-1} \stackrel{1}{\mathbb A} R_{12}^{t_1} \stackrel{2}{\mathbb A} \stackrel{1}{\mathbb K}{}_k^{-1} \stackrel{2}{\mathbb K}{}_k^{-1}  +R_{12}\stackrel{1}{\mathbb A}\Bigl[ \stackrel{2}{e}_{k+1,k} R_{12}^{t_1} \stackrel{1}{\mathbb K}{}_k^{-1} + R_{12}^{t_1} \stackrel{1}{e}_{k,k+1} \Bigr] \stackrel{2}{\mathbb A} \stackrel{2}{\mathbb K}{}_k^{-1} + R_{12}\stackrel{1}{\mathbb A}R_{12}^{t_1}\stackrel{2}{\mathbb A}\stackrel{2}{e}_{k,k+1}\nonumber\\
&=R_{12}\Bigl[ \stackrel{1}{e}_{k+1,k} \stackrel{2}{\mathbb K}{}_k^{-1} +  \stackrel{2}{e}_{k+1,k}\Bigr] \stackrel{1}{\mathbb A} R_{12}^{t_1} \stackrel{2}{\mathbb A} \stackrel{1}{\mathbb K}{}_k^{-1} \stackrel{2}{\mathbb K}{}_k^{-1}  +R_{12}\stackrel{1}{\mathbb A}R_{12}^{t_1} \stackrel{2}{\mathbb A} \Bigl[ \stackrel{2}{e}_{k,k+1} + \stackrel{1}{e}_{k,k+1}\stackrel{2}{\mathbb K}{}_k^{-1}\Bigr]\nonumber\\
&\stackrel{(\ref{rel-IV})}{=} \Bigl[ \stackrel{2}{e}_{k+1,k} \stackrel{1}{\mathbb K}{}_k^{-1} +  \stackrel{1}{e}_{k+1,k}\Bigr] \bigl( R_{12} \stackrel{1}{\mathbb A} R_{12}^{t_1} \stackrel{2}{\mathbb A} \bigr)\stackrel{1}{\mathbb K}{}_k^{-1} \stackrel{2}{\mathbb K}{}_k^{-1}  +\bigl(R_{12}\stackrel{1}{\mathbb A}R_{12}^{t_1} \stackrel{2}{\mathbb A}\bigr) \Bigl[ \stackrel{2}{e}_{k,k+1} + \stackrel{1}{e}_{k,k+1}\stackrel{2}{\mathbb K}{}_k^{-1}\Bigr]\label{Fk-left}.
\end{align}
When acting on the left-hand side of (\ref{RARA}), we obtain
\begin{align}
&F_k(\stackrel{2}{\mathbb A}R_{12}^{t_1}\stackrel{1}{\mathbb A}R_{12})=\Bigl[ \stackrel{2}{e}_{k+1,k} \stackrel{2}{\mathbb A} \stackrel{2}{\mathbb K}{}_k^{-1}+\stackrel{2}{\mathbb A} \stackrel{2}{e}_{k,k+1} \Bigr] R_{12}^{t_1} \stackrel{1}{\mathbb K}{}_k^{-1}\stackrel{1}{\mathbb A} \stackrel{1}{\mathbb K}{}_k^{-1}R_{12}+\stackrel{2}{\mathbb A}R_{12}^{t_1} \Bigl[ \stackrel{1}{e}_{k+1,k} \stackrel{1}{\mathbb A} \stackrel{1}{\mathbb K}{}_k^{-1}+\stackrel{1}{\mathbb A} \stackrel{1}{e}_{k,k+1} \Bigr]R_{12}\nonumber\\
&=\stackrel{2}{e}_{k+1,k} \stackrel{1}{\mathbb K}{}_k^{-1} \stackrel{2}{\mathbb A} R_{12}^{t_1} \stackrel{1}{\mathbb A}R_{12} \stackrel{1}{\mathbb K}{}_k^{-1} \stackrel{2}{\mathbb K}{}_k^{-1}  +\stackrel{2}{\mathbb A}\Bigl[ \stackrel{2}{e}_{k,k+1}  R_{12}^{t_1} \stackrel{1}{\mathbb K}{}_k^{-1} + R_{12}^{t_1} \stackrel{1}{e}_{k+1,k} \Bigr] \stackrel{1}{\mathbb A} \stackrel{1}{\mathbb K}{}_k^{-1}R_{12} + \stackrel{2}{\mathbb A}R_{12}^{t_1}\stackrel{1}{\mathbb A}\stackrel{1}{e}_{k,k+1}R_{12}\nonumber\\
&\stackrel{(\ref{rel-II})}{=}\stackrel{2}{e}_{k+1,k} \stackrel{1}{\mathbb K}{}_k^{-1} \stackrel{2}{\mathbb A} R_{12}^{t_1} \stackrel{1}{\mathbb A}R_{12} \stackrel{1}{\mathbb K}{}_k^{-1} \stackrel{2}{\mathbb K}{}_k^{-1}  +\stackrel{2}{\mathbb A}\Bigl[ \stackrel{1}{e}_{k+1,k}  R_{12}^{t_1} \stackrel{2}{\mathbb K}{}_k^{-1} + R_{12}^{t_1} \stackrel{2}{e}_{k,k+1} \Bigr] \stackrel{1}{\mathbb A} \stackrel{1}{\mathbb K}{}_k^{-1}R_{12} + \stackrel{2}{\mathbb A}R_{12}^{t_1}\stackrel{1}{\mathbb A}\stackrel{1}{e}_{k,k+1}R_{12}\nonumber\\
&=\Bigl[  \stackrel{2}{e}_{k+1,k} \stackrel{1}{\mathbb K}{}_k^{-1} + \stackrel{1}{e}_{k+1,k} \Bigr] \stackrel{2}{\mathbb A} R_{12}^{t_1} \stackrel{1}{\mathbb A}R_{12} \stackrel{1}{\mathbb K}{}_k^{-1} \stackrel{2}{\mathbb K}{}_k^{-1}  +  \stackrel{2}{\mathbb A} R_{12}^{t_1} \stackrel{1}{\mathbb A} \Bigl[  \stackrel{2}{e}_{k,k+1} \stackrel{1}{\mathbb K}{}_k^{-1} + \stackrel{1}{e}_{k,k+1} \Bigr] R_{12}\nonumber\\
&\stackrel{(\ref{rel-II})}{=} \Bigl[  \stackrel{2}{e}_{k+1,k} \stackrel{1}{\mathbb K}{}_k^{-1} + \stackrel{1}{e}_{k+1,k} \Bigr]\bigl( \stackrel{2}{\mathbb A} R_{12}^{t_1} \stackrel{1}{\mathbb A}R_{12}\bigr) \stackrel{1}{\mathbb K}{}_k^{-1} \stackrel{2}{\mathbb K}{}_k^{-1}  + \bigl( \stackrel{2}{\mathbb A} R_{12}^{t_1} \stackrel{1}{\mathbb A} R_{12}\bigr)\Bigl[  \stackrel{1}{e}_{k,k+1} \stackrel{2}{\mathbb K}{}_k^{-1} + \stackrel{2}{e}_{k,k+1} \Bigr], \label{Fk-right}
\end{align}
and the above two expressions coincide provided (\ref{RARA}) holds. The proof for $E_k$ is similar and uses relations (\ref{rel-I}) and (\ref{rel-III}). 
\end{proof}

\subsection{Proof of centrality}

\begin{lem}\label{lem:cyclic}
Suppose \(B\in\mathbb A_{2n}\) satisfies
\[
 K_k(B)=B,\qquad E_k(B)=F_k(B)=0\quad(1\le k<2n),
 \qquad [B,a_{2n,1}]=0.
\]
Then \(B\) is central.
\end{lem}
\begin{proof}
Let \(W=\operatorname{span}_K\{a_{ij}\}\) and
\(C_B=\{x\in W:[B,x]=0\}\). The skew Leibniz rules give
\[
 E_k([B,x])=[B,E_k(x)],\qquad
 F_k([B,x])=[B,F_k(x)].
\]
Thus \(C_B\) is stable under these operators and contains \(a_{2n,1}\).
For \(2\le r<2n\),
\[
 F_r(a_{r+1,1})=a_{r,1},\qquad F_1(a_{21})=q^{-1}a_{11}.
\]
All first-column generators consequently lie in \(C_B\).
For \(i>j\), acting by
\[
 E_j(a_{i,j})=q^{\chi_j(i)}a_{i,j+1},
\]
we add to the set all lower-triangular and diagonal entries.
Next, \(E_i(a_{i,i})=a_{i+1,i}+qa_{i,i+1}\) adds all entries adjacent to the diagonal, and \(E_j(a_{i,j})=a_{i,j+1}\) for \(j>i\) supplies the rest.
Hence \(C_B=W\).
\end{proof}

\begin{prop}\label{prop:central}
The element \(\widehat{Pf}_n\) is central.
\end{prop}
\begin{proof}
Every summand of \(\widehat{Pf}_n\) 
contains exactly one index $k$ and exactly one index $k+1$, so \(K_k(\widehat{Pf}_n)=\widehat{Pf}_n\), and since each term in the Pfaffian contains one index $1$ and one index $2n$, it commutes with $a_{2n,1}$, so \([\widehat{Pf}_n,a_{2n,1}]=0\). We now prove \(E_k(\widehat{Pf}_n)=F_k(\widehat{Pf}_n)=0\) already in the free algebra.

Regard a word as \(N=2n\) index slots, grouped into successive pairs. Applying \(E_k\) replaces its unique label \(k\) by \(k+1\), with coefficient
\[
 q^{\#\{k\text{ before the slot}\}-
       \#\{k+1\text{ before the slot}\}} = q^{-\#\{k+1\text{ before the slot}\}}
\]
Pair each permutation \(\pi\) in which \(k\) precedes \(k+1\) with the permutation obtained by interchanging these labels. Upon action by $E_k$, both occurrences of $k$ are replaced by $k+1$ producing identical expressions with the respective coefficients $\{1,q^{-1}\}$ while the second permutation has inversion number $l(\pi)+1$, so the total coefficient is
\[
 (-q)^{l(\pi)}+q^{-1}(-q)^{l(\pi)+1}=0.
\]
For \(F_k\), the unique label \(k+1\) is replaced by \(k\) with the coefficient
$
 q^{-\#\{k\text{ after the slot}\}}.
$
The same paired permutations again come with coefficients $\{1,q^{-1}\}$ and cancel. Lemma~\ref{lem:cyclic} proves centrality.
\end{proof}



\begin{lem}
  Let $B \in \mathbb{A}_n$ be a Casimir which does not contain inverted elements (i.e., it is a polynomial in $\{a_{ij}\}_{i,j \in [n]}$). Then $B$ almost satisfies the conditions of Lemma~\ref{lem:cyclic}. Precisely,
  \begin{align*}
    K_k(B)=B, \qquad E_k(B)=F_k(B)=0
  \end{align*}
  for all $1 \le k < n$, unless $n$ is even and $k=\frac{n}{2}$, in which case only one of $E_k(B)$ and $F_k(B)$ is necessarily zero.
\end{lem}

\begin{proof}
  For any monomial $a_{i_1,j_1}^{e_1} \cdots a_{i_m,j_m}^{e_m} \in \mathbb{A}_n$, define the index multiset of that monomial be the multiset which, for every $1 \le k \le n$, contains
  \begin{align*}
    \sum_{l: i_l=k} e_l + \sum_{l: j_l=k} e_l
  \end{align*}
  copies of $k$. (Note this number of copies may be negative.) This definition extends to expressions of the form $m_1/m_2$, where $m_1$ and $m_2$ are monomials, or more generally, where $m_1$ and $m_2$ are sums of monomials with the same index multisets. Note that for each of our independent semiclassical Casimirs $D$, every monomial in $D$ has the same index multiset. (In this case, we write ``the index multiset of $D$'' to refer to the index multiset of any monomial in $D$.) Specifically, when $D$ is the Pfaffian, the quantum determinant, or a coefficient in $\det(\mathbb{A}+\lambda \mathbb{A}^T)$, the index multiset of $D$ is some number of copies of $\{1,\ldots,n\}$. When $D$ is a ratio $\mathfrak{M}_{n-m}/\mathfrak{M}_m$ of minors, the index multiset of $D$ is two copies of $\{m+1,\ldots,n-m\}$.
  
  Observe that the index multiset of a monomial in the quantum algebra $\mathbb{A}_n$ is invariant under commutation relations (and hence well-defined) and invariant under conjugation relations. If $B$ contains monomials with distinct index multisets, let $B_I$ be the projection of $B$ onto one of those multisets. Then $B = B_I + (B-B_I)$. For any $a_{ij}$, the index multiset of monomials in $[a_{ij},B_I]$ is disjoint from the index multisets of monomials in $[a_{ij},B-B_I]$. Since $[a_{ij},B]=0$, it follows that $[a_{ij},B_I]$ and $[a_{ij},B-B_I]$ must each be $0$, so both $B_I$ and $B-B_I$ are Casimirs. Thus, it suffices to consider the case where every monomial in $B$ shares the same index multiset.

  Since $B$ is a Casimir, $\lim_{q \to 1} B$ is a semiclassical Casimir. If we divide out by any common factor of $(q-1)$ in $B$, then $\lim_{q \to 1} B \ne 0$, so it is a Casimir whose index multiset is the same as $B$'s. Since $B$ contains no inverted elements, the index multiset of $B$ is some number of copies of $\{1,\ldots,n\}$. Thus, $K_k$ acts identically on $B$ for every $k$ (since $K_k$ multiplies $B$ by $q$ to the power of the number of times $k$ appears in the multiset minus the number of times $k+1$ appears in the multiset).

  Now observe that, since $B$ is a Casimir, for any $x \in \mathbb{A}_n$ we have
  \begin{align*}
    E_k(B)x + BE_k(x) = E_k(B)x + K_k(B)E_k(x) = E_k(Bx) = E_k(xB) = E_k(x)B + K_k(x)E_k(B) \\
    F_k(B)K_k^{-1}(x) + BF_k(x) = F_k(Bx) = F_k(xB) = F_k(x)K_k^{-1}(B) + xF_k(B) = F_k(x)B + xF_k(B)
  \end{align*}
  Thus,
  \begin{align*}
    E_k(B) x &= K_k(x) E_k(B) \\
    F_k(B) x &= K_k(x) F_k(B)
  \end{align*}
  for every $x \in \mathbb{A}_n$.
  
  Now consider the situation when $x=\widehat{M}_{n-m}^{BL}[\widehat{M}_m^{BL}]^{-1}$. Since the index multiset of $\widehat{M}_{n-m}^{BL}[\widehat{M}_m^{BL}]^{-1}$ is two copies of $\{m+1,\ldots,n-m\}$,
  \begin{align*}
    K_m(\widehat{M}_{n-m}^{BL}[\widehat{M}_m^{BL}]^{-1}) &= q^{-2}\widehat{M}_{n-m}^{BL}[\widehat{M}_m^{BL}]^{-1} \\
    K_{n-m}(\widehat{M}_{n-m}^{BL}[\widehat{M}_m^{BL}]^{-1}) &= q^2\widehat{M}_{n-m}^{BL}[\widehat{M}_m^{BL}]^{-1}
  \end{align*}
  so $E_m(B)$ and $F_m(B)$ both $q^{-2}$ commute with $\widehat{M}_{n-m}^{BL}[\widehat{M}_m^{BL}]^{-1}$, and $E_{n-m}(B)$ and $F_{n-m}(B)$ both $q^2$ commute with $\widehat{M}_{n-m}^{BL}[\widehat{M}_m^{BL}]^{-1}$. But we know $\widehat{M}_{n-m}^{BL}[\widehat{M}_m^{BL}]^{-1}$ is a Casimir, and hence commutes with $E_m(B)$, $F_m(B)$, $E_{n-m}(B)$, and $F_{n-m}(B)$. Thus, we must have $E_m(B) = F_m(B) = E_{n-m}(B) = F_{n-m}(B) = 0$. This argument holds for any $1 \le m < \lfloor\frac{n}{2}\rfloor$, so $E_k(B)=F_k(B)=0$ for every $k$, unless $n$ is even and $k=\frac{n}{2}$.

  Let $n$ be even, and assume towards a contradiction that $E_{n/2}(B)$ and $F_{n/2}(B)$ are both nonzero. Then $E_{n/2}(B)[F_{n/2}(B)]^{-1}$ is a Casimir, and its index multiset has $1$ copy of $\frac{n}{2}+1$, and $-1$ copies of $\frac{n}{2}$. But no combination of semiclassical Casimirs can achieve this index multiset, a contradiction. Thus, at least one of $E_{n/2}(B)$ and $F_{n/2}(B)$ must be zero.
\end{proof}

\begin{conjecture}
Note that in the semiclassical limit, $K_k$ acts identically whereas $E_k$ and $F_k$ act as
$$
E_k(\mathbb A)=e_{k,k+1}\mathbb A + \mathbb A e_{k+1,k},\qquad F_k(\mathbb A)=e_{k+1,k}\mathbb A + \mathbb A e_{k,k+1},
$$
and it is easy to see that
$$
E_k\bigl(\mathbb A+\lambda \mathbb A^{\text{T}}\bigr)=e_{k,k+1}\bigl(\mathbb A+\lambda \mathbb A^{\text{T}}\bigr) + \bigl(\mathbb A+\lambda \mathbb A^{\text{T}}\bigr) e_{k+1,k}.
$$
and
$$
F_k\bigl(\mathbb A+\lambda \mathbb A^{\text{T}}\bigr)=e_{k+1,k}\bigl(\mathbb A+\lambda \mathbb A^{\text{T}}\bigr) + \bigl(\mathbb A+\lambda \mathbb A^{\text{T}}\bigr) e_{k,k+1}.
$$
It then follows immediately that
$$
E_k\Bigl[ \det\bigl(\mathbb A+\lambda \mathbb A^{\text{T}}\bigr)\Bigr]=F_k\Bigl[ \det\bigl(\mathbb A+\lambda \mathbb A^{\text{T}}\bigr)\Bigr]=0,
$$
so all classical Markov elements are invariant under the semiclassical action of $E_k$ and $F_k$ for all $k=1,\dots,n-1$. We therefore expect that the corresponding quantum Markov elements will enjoy the conditions of Lemma~\ref{lem:cyclic}. We next prove this conjecture for the quantum determinant itself.
\end{conjecture}

\subsection{Centrality of the quantum determinant via polarization operators}

\begin{thm}\label{thm-EFK}
We have
\begin{equation}\label{eq:EF}
F_k\bigl(\mathfrak A_m    \langle \stackrel{1}{\mathbb A},\ldots,\stackrel{m}{\mathbb A}\rangle \mathfrak A_m \bigr)=
E_k\bigl(\mathfrak A_m    \langle \stackrel{1}{\mathbb A},\ldots,\stackrel{m}{\mathbb A}\rangle \mathfrak A_m \bigr)=0\quad k=1,\dots,m-1.
\end{equation}
\end{thm}
The proof for the action of  $F_k$ is based on three technical lemmas.
\begin{lem}\label{lm:Kk-Ang}
$
K_k^{-1} \bigl(  \langle \stackrel{1}{\mathbb A},\ldots,\stackrel{m}{\mathbb A}\rangle \bigr) =   \stackrel{1} {\mathbb K}{}^{-1}_k \stackrel{2} {\mathbb K}{}^{-1}_k\cdots \stackrel{m} {\mathbb K}{}^{-1}_k \langle \stackrel{1}{\mathbb A},\ldots,\stackrel{m}{\mathbb A}\rangle   \stackrel{1} {\mathbb K}{}^{-1}_k \stackrel{2} {\mathbb K}{}^{-1}_k\cdots \stackrel{m} {\mathbb K}{}^{-1}_k.
$
\end{lem}
The proof is by induction: for $m=1$, 
$$K_k^{-1}\stackrel{1}{\mathbb A}= \stackrel{1} {\mathbb K}{}^{-1}_k \stackrel{1}{\mathbb A}\stackrel{1} {\mathbb K}{}^{-1}_k.$$
Assume the statement holds for $m-1$, then
\begin{align*}
&K_k^{-1}\Bigl[\stackrel{1}{\mathbb A} R_{12}^{t_1} \cdots R_{1,m}^{t_1} \langle \stackrel{2}{\mathbb A},\ldots,\stackrel{m}{\mathbb A}\rangle\Bigr]\\
&= \stackrel{1} {\mathbb K}{}^{-1}_k \stackrel{1}{\mathbb A}\stackrel{1} {\mathbb K}{}^{-1}_k R_{12}^{t_1} \cdots R_{1,m}^{t_1} \stackrel{2} {\mathbb K}{}^{-1}_k\cdots \stackrel{m} {\mathbb K}{}^{-1}_k  \langle \stackrel{2}{\mathbb A},\ldots,\stackrel{m}{\mathbb A}\rangle  \stackrel{2} {\mathbb K}{}^{-1}_k\cdots \stackrel{m} {\mathbb K}{}^{-1}_k \\
&= \stackrel{1} {\mathbb K}{}^{-1}_k \stackrel{1}{\mathbb A} \stackrel{2} {\mathbb K}{}^{-1}_k\cdots \stackrel{m} {\mathbb K}{}^{-1}_k R_{12}^{t_1} \cdots R_{1,m}^{t_1} \stackrel{1} {\mathbb K}{}^{-1}_k  \langle \stackrel{2}{\mathbb A},\ldots,\stackrel{m}{\mathbb A}\rangle  \stackrel{2} {\mathbb K}{}^{-1}_k\cdots \stackrel{m} {\mathbb K}{}^{-1}_k \\
&= \stackrel{1} {\mathbb K}{}^{-1}_k\stackrel{2} {\mathbb K}{}^{-1}_k\cdots \stackrel{m} {\mathbb K}{}^{-1}_k  \stackrel{1}{\mathbb A} R_{12}^{t_1} \cdots R_{1,m}^{t_1}   \langle \stackrel{2}{\mathbb A},\ldots,\stackrel{m}{\mathbb A}\rangle \stackrel{1} {\mathbb K}{}^{-1}_k  \stackrel{2} {\mathbb K}{}^{-1}_k\cdots \stackrel{m} {\mathbb K}{}^{-1}_k,
\end{align*}
which implies the lemma statement.

\begin{lem}\label{lm:Fk-Ang}
Denote
\begin{equation}\label{Fk-mat}
\mathbb F_k[1,m]:= \stackrel{1}{e}_{k+1,k}\stackrel{2} {\mathbb K}{}^{-1}_k \stackrel{3} {\mathbb K}{}^{-1}_k\cdots \stackrel{m} {\mathbb K}{}^{-1}_k + \stackrel{2}{e}_{k+1,k}\stackrel{3} {\mathbb K}{}^{-1}_k \cdots \stackrel{m} {\mathbb K}{}^{-1}_k +\cdots +\stackrel{m-1}{e}_{k+1,k}\stackrel{m} {\mathbb K}{}^{-1}_k + \stackrel{m}{e}_{k+1,k} 
\end{equation}
Then
\begin{align}
F_k\bigl(  \langle \stackrel{1}{\mathbb A},\ldots,\stackrel{m}{\mathbb A}\rangle \bigr)&= \mathbb F_k[1,m]
\langle \stackrel{1}{\mathbb A},\ldots,\stackrel{m}{\mathbb A}\rangle  \stackrel{1} {\mathbb K}{}^{-1}_k \stackrel{2} {\mathbb K}{}^{-1}_k\cdots \stackrel{m} {\mathbb K}{}^{-1}_k + \langle \stackrel{1}{\mathbb A},\ldots,\stackrel{m}{\mathbb A}\rangle \mathbb F_k^{\text{T}}[1,m],
\end{align}
where the transposition is taken in all spaces. 
\end{lem}

\begin{proof} The proof is by induction. The relation for $m=2$ is contained in (\ref{Fk-left}). Assume the statement holds true for the set of indices running from $2$ to $m$. The main technical computation is based on the relation 
$$
R_{1,i}^{t_1}  \stackrel{i}{e}_{k+1,k}=\stackrel{i}{e}_{k+1,k} R_{1,i}^{t_1} \stackrel{1} {\mathbb K}{}^{-1}_k + R_{1,i}^{t_1} \stackrel{1}{e}_{k,k+1} - \stackrel{1}{e}_{k,k+1} R_{1,i}^{t_1} \stackrel{i} {\mathbb K}{}^{-1}_k 
$$ 
and reads
\begin{align*}
&R_{12}^{t_1} \cdots R_{1,m}^{t_1} \mathbb F_k[2,m] =\Bigl( \stackrel{2}{e}_{k+1,k} R_{12}^{t_1} \stackrel{1} {\mathbb K}{}^{-1}_k R_{13}^{t_1} \cdots R_{1,m}^{t_1}   \stackrel{3} {\mathbb K}{}^{-1}_k \cdots \stackrel{m} {\mathbb K}{}^{-1}_k \Bigr. \\
&+\Bigl. R_{12}^{t_1} \stackrel{1}{e}_{k,k+1} R_{13}^{t_1} \cdots R_{1,m}^{t_1} \stackrel{3} {\mathbb K}{}^{-1}_k \cdots \stackrel{m} {\mathbb K}{}^{-1}_k - \stackrel{1}{e}_{k,k+1} R_{12}^{t_1} \stackrel{2} {\mathbb K}{}^{-1}_k R_{13}^{t_1} \cdots R_{1,m}^{t_1}   \stackrel{3} {\mathbb K}{}^{-1}_k \cdots \stackrel{m} {\mathbb K}{}^{-1}_k\Bigr)\\
&+ R_{12}^{t_1} \Bigl( \stackrel{3}{e}_{k+1,k} R_{13}^{t_1} \stackrel{1} {\mathbb K}{}^{-1}_k R_{14}^{t_1} \cdots R_{1,m}^{t_1}   \stackrel{4} {\mathbb K}{}^{-1}_k \cdots \stackrel{m} {\mathbb K}{}^{-1}_k \Bigr. \\
&+\Bigl. R_{13}^{t_1} \stackrel{1}{e}_{k,k+1} R_{14}^{t_1} \cdots R_{1,m}^{t_1} \stackrel{4} {\mathbb K}{}^{-1}_k \cdots \stackrel{m} {\mathbb K}{}^{-1}_k - \stackrel{1}{e}_{k,k+1} R_{13}^{t_1} \stackrel{3} {\mathbb K}{}^{-1}_k R_{14}^{t_1} \cdots R_{1,m}^{t_1}   \stackrel{4} {\mathbb K}{}^{-1}_k \cdots \stackrel{m} {\mathbb K}{}^{-1}_k\Bigr)\\
&+\cdots +  R_{12}^{t_1} \cdots R_{1,m-2}^{t_1} \Bigl( \stackrel{m-1}{e}_{k+1,k} R_{1,m-1}^{t_1} \stackrel{1} {\mathbb K}{}^{-1}_k R_{1,m}^{t_1}  \stackrel{m} {\mathbb K}{}^{-1}_k \Bigr. \\
&+\Bigl. R_{1,m-1}^{t_1} \stackrel{1}{e}_{k,k+1} R_{1,m}^{t_1}  \stackrel{m} {\mathbb K}{}^{-1}_k - \stackrel{1}{e}_{k,k+1} R_{1,m-1}^{t_1} \stackrel{m-1} {\mathbb K}{}^{-1}_k R_{1,m}^{t_1}  \stackrel{m} {\mathbb K}{}^{-1}_k\Bigr)\\
&+R_{12}^{t_1} \cdots R_{1,m-1}^{t_1} \Bigl( \stackrel{m}{e}_{k+1,k} R_{1,m}^{t_1} \stackrel{1} {\mathbb K}{}^{-1}_k 
+R_{1,m}^{t_1} \stackrel{1}{e}_{k,k+1}- \stackrel{1}{e}_{k,k+1} R_{1,m}^{t_1} \stackrel{m} {\mathbb K}{}^{-1}_k\Bigr).
\end{align*}
In this sum, the second term in each bracket is canceled by the third term in the next bracket, the first terms are combined into $\mathbb F_k[2,m]$, and we obtain a compact expression
$$
\mathbb F_k[2,m] R_{12}^{t_1} \cdots R_{1,m}^{t_1} \stackrel{1} {\mathbb K}{}^{-1}_k + R_{12}^{t_1} \cdots R_{1,m}^{t_1} \stackrel{1}{e}_{k,k+1} - \stackrel{1}{e}_{k,k+1} R_{12}^{t_1} \cdots R_{1,m}^{t_1}  \stackrel{2} {\mathbb K}{}^{-1}_k \cdots \stackrel{m} {\mathbb K}{}^{-1}_k.
$$
Therefore
\begin{align*}
&F_k\bigl( \stackrel{1}{\mathbb A} R_{12}^{t_1} \cdots R_{1,m}^{t_1}  \langle \stackrel{2}{\mathbb A},\ldots,\stackrel{m}{\mathbb A}\rangle \bigr)= F_k(\stackrel{1}{\mathbb A})  R_{12}^{t_1} \cdots R_{1,m}^{t_1}  K_k^{-1}\bigl(  \langle \stackrel{2}{\mathbb A},\ldots,\stackrel{m}{\mathbb A}\rangle \bigr) + \stackrel{1}{\mathbb A} R_{12}^{t_1} \cdots R_{1,m}^{t_1}  F_k \bigl(  \langle \stackrel{2}{\mathbb A},\ldots,\stackrel{m}{\mathbb A}\rangle \bigr)\\
&=\bigl(\stackrel{1}{e}_{k+1,k} \stackrel{1}{\mathbb A} \stackrel{1} {\mathbb K}{}^{-1}_k +   \stackrel{1}{\mathbb A} \stackrel{1}{e}_{k,k+1}   \bigr) R_{12}^{t_1} \cdots R_{1,m}^{t_1}   \stackrel{2} {\mathbb K}{}^{-1}_k \cdots \stackrel{m} {\mathbb K}{}^{-1}_k \langle \stackrel{2}{\mathbb A},\ldots,\stackrel{m}{\mathbb A}\rangle   \stackrel{2} {\mathbb K}{}^{-1}_k \cdots \stackrel{m} {\mathbb K}{}^{-1}_k\\
&+ \stackrel{1}{\mathbb A} R_{12}^{t_1} \cdots R_{1,m}^{t_1}  \mathbb F_k[2,m]  \langle \stackrel{2}{\mathbb A},\ldots,\stackrel{m}{\mathbb A}\rangle   \stackrel{2} {\mathbb K}{}^{-1}_k \cdots \stackrel{m} {\mathbb K}{}^{-1}_k +  \langle \stackrel{1}{\mathbb A},\ldots,\stackrel{m}{\mathbb A}\rangle  \mathbb F^{\text{T}}_k[2,m] \\
&= \stackrel{1}{e}_{k+1,k} \stackrel{2} {\mathbb K}{}^{-1}_k \cdots \stackrel{m} {\mathbb K}{}^{-1}_k \langle \stackrel{1}{\mathbb A},\ldots,\stackrel{m}{\mathbb A}\rangle   \stackrel{1} {\mathbb K}{}^{-1}_k \cdots \stackrel{m} {\mathbb K}{}^{-1}_k \\ 
&+ \stackrel{1}{\mathbb A}  \stackrel{1}{e}_{k,k+1} R_{12}^{t_1} \cdots R_{1,m}^{t_1}  \stackrel{2} {\mathbb K}{}^{-1}_k \cdots \stackrel{m} {\mathbb K}{}^{-1}_k  \langle \stackrel{2}{\mathbb A},\ldots,\stackrel{m}{\mathbb A}\rangle   \stackrel{2} {\mathbb K}{}^{-1}_k \cdots \stackrel{m} {\mathbb K}{}^{-1}_k \\
&+\mathbb F_k[2,m]  \langle \stackrel{1}{\mathbb A},\ldots,\stackrel{m}{\mathbb A}\rangle   \stackrel{1} {\mathbb K}{}^{-1}_k \cdots \stackrel{m} {\mathbb K}{}^{-1}_k +  \langle \stackrel{1}{\mathbb A},\ldots,\stackrel{m}{\mathbb A}\rangle \stackrel{1}{e}_{k,k+1} \stackrel{2} {\mathbb K}{}^{-1}_k \cdots \stackrel{m} {\mathbb K}{}^{-1}_k \\
&- \stackrel{1}{\mathbb A}  \stackrel{1}{e}_{k,k+1} R_{12}^{t_1} \cdots R_{1,m}^{t_1}  \stackrel{2} {\mathbb K}{}^{-1}_k \cdots \stackrel{m} {\mathbb K}{}^{-1}_k  \langle \stackrel{2}{\mathbb A},\ldots,\stackrel{m}{\mathbb A}\rangle   \stackrel{2} {\mathbb K}{}^{-1}_k \cdots \stackrel{m} {\mathbb K}{}^{-1}_k +  \langle \stackrel{1}{\mathbb A},\ldots,\stackrel{m}{\mathbb A}\rangle \mathbb F^{\text{T}}_k[2,m] \\
&=\Bigl[   \stackrel{1}{e}_{k+1,k} \stackrel{2} {\mathbb K}{}^{-1}_k \cdots \stackrel{m} {\mathbb K}{}^{-1}_k  +  \mathbb F_k[2,m] \Bigr] \langle \stackrel{1}{\mathbb A},\ldots,\stackrel{m}{\mathbb A}\rangle   \stackrel{1} {\mathbb K}{}^{-1}_k \cdots \stackrel{m} {\mathbb K}{}^{-1}_k \\
&  + \langle \stackrel{1}{\mathbb A},\ldots,\stackrel{m}{\mathbb A}\rangle \Bigl[ \stackrel{1}{e}_{k,k+1} \stackrel{2} {\mathbb K}{}^{-1}_k \cdots \stackrel{m} {\mathbb K}{}^{-1}_k +  \mathbb F^{\text{T}}_k[2,m] \Bigr] \\
&= \mathbb F_k[1,m] \langle \stackrel{1}{\mathbb A},\ldots,\stackrel{m}{\mathbb A}\rangle   \stackrel{1} {\mathbb K}{}^{-1}_k \cdots \stackrel{m} {\mathbb K}{}^{-1}_k +  \langle \stackrel{1}{\mathbb A},\ldots,\stackrel{m}{\mathbb A}\rangle  \mathbb F^{\text{T}}_k[1,m],
\end{align*}
which completes the proof.
\end{proof}

\begin{lem}\label{lm:Fk-Zero}
We have $\mathfrak A_m \mathbb F_k[1,m]= \mathbb F^{\text{T}}_k[1,m] \mathfrak A_m =0$ for all $m$ and $k=1,\dots,m-1$.
\end{lem}
The \emph{proof} just uses the explicit form (\ref{eq:Am}) of $\mathfrak A_m$: there are exactly two terms with the right indices $k$ and $k+1$, their sum (in which all other indices coincide) reads
\begin{align*}
&(-q)^{l(\sigma)+l(\tau)} \cdots \stackrel{i} e_{\sigma(i),k=\tau(i)} \otimes\cdots\otimes \stackrel{j} e_{\sigma(j),k+1=\tau(j)}\cdots \\ 
&+(-q)^{l(\sigma)+l(\tau')} \cdots \stackrel{i} e_{\sigma(i),k+1=\tau'(i)} \otimes\cdots\otimes \stackrel{j} e_{\sigma(j),k=\tau'(j)}\cdots,
\end{align*}
where the permutation $\tau'$ differs from $\tau$ only by interchanging $k$ and $k+1$, so $l(\tau')=l(\tau)+1$. Acting by $\mathbb F_k[1,m]$ on the left raises the unique right index $k$ to $k+1$ producing identical terms; for the second term no additional factors appear whereas for the first term we have additional factor $q$ due to the action $ \stackrel{j} e_{\sigma(j),k+1} \stackrel {j} {\mathbb K}{}^{-1}_k=q \stackrel{j} e_{\sigma(j),k+1}$, so the total coefficient is $q-q=0$. Because we can always split the sum in the expression for $\mathfrak A_m$ into the sum over such pairs, we obtain that the left action of $\mathbb F_k[1,m]$ annihilates the whole expression, which completes the proof of the first statement.

Since $\mathfrak A_m^\text{T}=\mathfrak A_m$, transposing the obtained equality we obtain the second part, which completes the proof.

We have analogous statements for the action of  $E_k$. The statement of Theorem~\ref{thm-EFK} now follows immediately from Lemmas~\ref{lm:Fk-Ang} and~\ref{lm:Fk-Zero}.

\begin{cor}\label{cor-q-det}
The quantum determinant in Definition~\ref{def:qdet} is central. 
\end{cor}


\subsection{Action of polarization operators on Sklyanin minors}
Fix $1 \le k \le n-1$. Let $I=\{i_1,i_2,\ldots,i_m\} \subset [n]$ with $i_1<i_2<\cdots<i_m$. If $k+1 \in I$ and $k\notin I$, let $I'$ be the set $I$ with $k+1$ replaced by $k$. Let $J=\{j_1,j_2,\ldots,j_m\} \subset [n]$ with $j_1<j_2<\cdots<j_m$, and when $k+1 \in J$ and $k\notin J$, define $J'$ with $k+1$ replaced by $k$. Recall the definition of the Sklyanin minor $\mathbb{A}^I_J$ \ref{eq:Sklyanin_minor}:
\begin{align*}
    \mathbb{A}^I_J = \mathbb{A}^{i_1,\ldots,i_m}_{j_1,\ldots,j_m} \coloneqq [m]_{q^2}! \stackrel{1}{e}_{i_1} \otimes \cdots \otimes \stackrel{m}{e}_{i_m} \mathfrak{A}_m \langle \stackrel{1}{\mathbb{A}},\ldots,\stackrel{m}{\mathbb{A}}\rangle\mathfrak{A}_m \stackrel{1}{e}_{j_1} \otimes \cdots \otimes \stackrel{m}{e}_{j_m}
\end{align*}
For shorthand, we will denote $\stackrel{1}{e}_{i_1} \otimes \cdots \otimes \stackrel{m}{e}_{i_m}$ by $e_I$ and $\stackrel{1}{e}_{j_1} \otimes \cdots \otimes \stackrel{m}{e}_{j_m}$ by $e_J$. Then
\begin{align}
    F_k(\mathbb{A}^I_J) &= F_k(e_I) K_k^{-1}( \mathfrak{A}_m \langle \stackrel{1}{\mathbb{A}},\ldots,\stackrel{m}{\mathbb{A}}\rangle\mathfrak{A}_m) K_k^{-1}(e_J)
     + e_IF_k(\mathfrak{A}_m \langle \stackrel{1}{\mathbb{A}},\ldots,\stackrel{m}{\mathbb{A}}\rangle\mathfrak{A}_m)K_k^{-1}(e_J)\nonumber \\
    &\quad + e_I\mathfrak{A}_m \langle \stackrel{1}{\mathbb{A}},\ldots,\stackrel{m}{\mathbb{A}}\rangle\mathfrak{A}_mF_k(e_J) \nonumber \\
    &= \delta_{k+1 \in I} e_{I'} \mathfrak{A}_m \langle \stackrel{1}{\mathbb{A}},\ldots,\stackrel{m}{\mathbb{A}}\rangle\mathfrak{A}_m q^{-\delta_{k \in J}+\delta_{k+1 \in J}}e_J
     + 0
    + e_I \mathfrak{A}_m \langle \stackrel{1}{\mathbb{A}},\ldots,\stackrel{m}{\mathbb{A}}\rangle\mathfrak{A}_m \delta_{k+1 \in J}e_{J'}\nonumber \\
    &= \delta_{k+1 \in I}(1-\delta_{k\in I}) q^{-\delta_{k \in J}+\delta_{k+1} \in J} \mathbb{A}^{I'}_J + \delta_{k+1 \in J} (1-\delta_{k\in J}) \mathbb{A}^I_{J'}.\label{eq:Sylvester}
\end{align}
Notably, for any $1 \le m \le n-1$,
\begin{align*}
    F_{n-m}(\widehat{M}_m^{BL}) = q^{-\delta_{2m=n}} \mathbb{A}^{n-m,n-m+2,\ldots,n}_{1,\ldots,m}
\end{align*}
and $F_k(\widehat{M}_m^{BL})=0$ for any $k \ne n-m$. Note that for any $x \in \mathbb{A}_{n}$,
\begin{align*}
    0 = F_k(1) = F_k(x^{-1}x) = F_k(x^{-1})K_k^{-1}(x) + x^{-1}F_k(x).
\end{align*}
Thus,
\begin{align*}
    F_k(x^{-1}) = -x^{-1}F_k(x)[K_k^{-1}(x)]^{-1}.
\end{align*}
Therefore, for any $m < \frac{n}{2}$,
\begin{align*}
    F_m([\widehat{M}_m^{BL}]^{-1}\widehat{M}_{n-m}^{BL}) &= [\widehat{M}_m^{BL}]^{-1} \mathbb{A}^{m,m+2,\ldots,n}_{1,\ldots,n-m} \\
    F_{n-m}(\widehat{M}_{n-m}^{BL}[\widehat{M}_m^{BL}]^{-1}) &= -q\widehat{M}_{n-m}^{BL}[\widehat{M}_m^{BL}]^{-1}\mathbb{A}^{n-m,n-m+2,\ldots,n}_{1,\ldots,m}[\widehat{M}_m^{BL}]^{-1}
\end{align*}
and $F_k(\widehat{M}_{n-m}^{BL}[\widehat{M}_m^{BL}]^{-1})=0$ for all other $k$. Thus, ratios of minors are not annihilated by all $E$ and $F$ operators, despite being Casimirs. 

Similar computations hold for the action of the $E$'s. Formula (\ref{eq:Sylvester}) shows that any $m \times m$ Sklyanin minor can be obtained from $\widehat{M}_m^{BL}$ by repeated action of the $E$ and $F$ operators. We are about to explore this dynamics of minors in subsequent studies with the aim of developing Sylvester-type relations.

\end{document}